\documentclass[plain,biber,numericm]{nowfnt} % creates the journal version, needs biber version
\usepackage{tikz}

\usepackage[T1]{fontenc}
\usepackage[utf8]{inputenc}
\usepackage{kpfonts}
\usepackage{booktabs}
\usepackage{siunitx}
\usepackage{pgfplots}
\usepackage{floatrow}
\usepackage{caption}
\usepackage{microtype}
\usepackage{varioref}
\usepackage{framed}
\usepackage{tikz}
\usepackage{pgf,interval}
\usepackage{algorithm} % For algorithms
\usetikzlibrary{positioning}
\usepackage{booktabs} % For formal tables
\usepackage{amsmath}
\usepackage{amsthm}
\usepackage{amssymb,boldline}
\usepackage{amsfonts}       % blackboard math symbols
\usepackage{nicefrac}       % compact symbols for 1/2, etc.
\usepackage{floatrow}
\usepackage{graphicx}
\usepackage{multirow}
\usepackage{dsfont}
\usepackage{xspace}

\usepackage{amsbsy}

\definecolor{mygray}{gray}{0.9}

   \newtheorem{theorem2}{Theorem}[chapter]
 \newtheorem{lemma2}[theorem2]{Lemma}
 \newtheorem{corollary2}[theorem2]{Corollary}
 \newtheorem{proposition2}[theorem2]{Proposition}
 \newtheorem*{remark2}{Remark}
 \newtheorem{definition2}[theorem2]{Definition}
  \newtheorem*{example2}{Example}

\newcommand{\vValue}{\psi}
\newcommand{\indicator}[1]{\mathds{1}\{#1\}}

\newcommand{\nek}[1]{{\color{blue} NEK: #1}}

\newcommand{\shadingFunc}{\beta}

\newcommand{\Exp}[1]{\mathds{E}\left(#1\right)}

\newcommand{\tendsto}{\rightarrow}

\newcommand{\lE}{\mathds{E}}

\newcommand{\lP}{\mathds{P}}

\newcommand{\lR}{\mathds{R}}

\newcommand{\bx}{\mathbf{x}}
\newcommand{\bb}{\mathbf{b}}
\newcommand{\bz}{\mathbf{z}}
\newcommand{\bcB}{\boldsymbol{\mathcal{B}}}
\newcommand{\bc}{\mathbf{c}}
\newcommand{\bq}{\mathbf{q}}

\newcommand{\bF}{\mathbf{F}}
\newcommand{\bbeta}{\boldsymbol{\beta}}

\usepackage{empheq}

\DeclareMathOperator*{\smax}{\textrm{smax}}

\DeclareMathOperator*{\argmax}{\textrm{argmax}}

\DeclareMathOperator{\R}{\mathds{R}}

\newcommand{\cA}{\mathcal{A}}
\newcommand{\cB}{\mathcal{B}}

\newcommand{\cD}{\mathcal{D}}
\newcommand{\cE}{\mathcal{E}}

\newcommand{\cH}{\mathcal{H}}
\newcommand{\cI}{\mathcal{I}}

\newcommand{\cN}{\mathcal{N}}
\newcommand{\cO}{\mathcal{O}}

\newcommand{\cR}{\mathcal{R}}

\newcommand{\cU}{\mathcal{U}}

\newcommand{\cX}{\mathcal{X}}

\newcommand{\cav}[1]{\mathrm{cav}\left(#1\right)}

\newcommand{\Expxi}[1]{\mathds{E}_{x_{i} \sim F_{i}}\left[#1\right]}

\usepackage{algpseudocode}

\title{Learning in repeated auctions}

\maintitleauthorlist{
Thomas Nedelec\\
ENS Paris Saclay\\
Criteo AI Lab\\
thomas.nedelec@polytechnique.edu\and
Cl\'ement Calauz\`enes\\
Criteo AI Lab\\
c.calauzenes@criteo.com\and
Noureddine El Karoui \\
Work done while at \\
UC, Berkeley\\
Criteo AI Lab\\
nkarouiprof@gmail.com
\and
Vianney Perchet\\
ENSAE\\
Criteo AI Lab\\
vianney.perchet@normalesup.org
}

\issuesetup
{%
 copyrightowner={A.~Heezemans and M.~Casey},
 volume        = xx,
 issue         = xx,
 pubyear       = 2018,
 isbn          = xxx-x-xxxxx-xxx-x,
 eisbn         = xxx-x-xxxxx-xxx-x,
 doi           = 10.1561/XXXXXXXXX,
 firstpage     = 1, %Explain
 lastpage      = 18
 }

\usepackage{mwe}

\author[1]{Nedelec,Thomas}
\author[2]{Calauz\`enes, Cl\'ement}
\author[3]{El Karoui, Noureddine}

\author[4]{Perchet,Vianney}

\affil[1]{ENS Paris Saclay, Criteo AI Lab; nedelec@cmla.ens-cachan.fr}
\affil[2]{Criteo AI Lab; c.calauzenes@criteo.com}
\affil[3]{Worked done while at UC, Berkeley, and Criteo AI Lab;nkarouiprof@gmail.com}
\affil[4]{ENSAE, Criteo AI Lab;vianney.perchet@normalesup.org}

\articledatabox{\nowfntstandardcitation}

\begin{document}

\makeabstracttitle
\begin{abstract}

Online auctions are one of the most fundamental facets of the modern economy and power an industry generating hundreds of billions of dollars a year in revenue. Auction theory has historically focused on the question of designing the best way to sell a single item to potential buyers, with the concurrent objectives of  maximizing revenue generated or  welfare created. Theoretical results in this area have typically relied on some prior Bayesian knowledge agents were assumed to have on each-other. This assumption is no longer satisfied in new markets such as  online advertising: similar items are sold repeatedly, and agents are unaware of each other or might try to manipulate each-other. On the other hand, statistical learning theory now provides tools to supplement those missing pieces of information given enough data, as agents can learn from their environment to improve their strategies.

This survey covers recent advances in learning in repeated auctions, starting from the traditional economic study of optimal one-shot auctions with a Bayesian prior. We then focus on the question of learning optimal mechanisms from a dataset of bidders' past values. The sample complexity as well as the computational efficiency of different methods will be studied. We will also investigate online variants where gathering data has a cost to be accounted for, either by seller or buyers ("earning while learning"). Later in the survey, we will further assume that bidders are also adaptive to the mechanism as they interact repeatedly with the same seller. We will show how strategic agents can actually manipulate repeated auctions, to their own advantage. A particularly interesting example is that of reserve price improvements for strategic buyers in second price auctions.

All the questions discussed in this survey are grounded in real-world applications and many of the ideas and algorithms we describe are used every day to power the Internet economy.

\end{abstract}
\clearpage
\tableofcontents
\clearpage
\begin{refsegment}
%!TEX root = ../main.tex
\chapter{Introduction: scope and motivation}
\label{Chapter:Intro}
The main purpose of auction theory is to construct a set of rules that will be used by a seller to sell one or several items to a group of potential buyers, that will send messages  (or \emph{bids}) to the seller  -- usually indicating how much they value the item or how much they are willing to pay to acquire it.  In almost all cases, it is sufficient  to define only two rules. First, the \emph{allocation rule} describes which buyer wins the auction (if a unique non-divisible item is sold), depending on the different messages received; if the item is divisible, the allocation rule describes how the item is shared between winners.  Second, the \emph{payment rule} indicates to buyers how much they are going to pay to the seller, again based on the different messages. Those rules are known publicly before the auction starts, and they influence the behavior, or strategy, of the different buyers.

When choosing an allocation and a payment rule, the seller might have several constraints to respect: {\bf 1)} maximizing the revenue she is getting from the auction (\emph{revenue maximization}); {\bf 2)} ensuring the participation of buyers to the auction and making sure they have an incentive to participate (\emph{individual rationality}); {\bf 3)} ensuring that given the rules of the auction, it is in the best interest of buyers to reveal how much they truly value an item (\emph{incentive compatibility}) as it may make revenue maximization easier. On the other side of the game, the buyers adapt strategically the bids sent to the seller depending on auction rules in order to maximize their own utility.

Historically, auctions have often been designed so that buyers have an incentive to  bid in a way that  reflects how much they truly value the items that are for sale. This constraint still leaves plenty of choices for auction design, and a large part of the literature has focused on designing auctions that  maximize the seller's revenue, assuming  buyers are rational. However, with the advent of the Internet and the automation of auctions, the landscape of possible applications has changed drastically, necessitating more complex settings to accurately study the incentives and behaviors at play. More recently, the auction literature has aimed at understanding how the design of an auction platform impacts seller's revenue, the global welfare and the behavior of buyers and sellers in contexts where sellers (and sometimes buyers) participate in a very large number of auctions each day. These setups reflect situations appearing in modern online marketplaces.

\section{Bayesian mechanism design}
Auction theory has focused for a long time on the simplest case: there is a single, non-divisible item to be sold to a set of predefined buyers in a one-shot auction.  The chosen mechanism indicates which buyer (if any) gets the item and at which price.  The seminal works of \citet{vickrey1961counterspeculation},  \citet{Myerson81} and \citet{RilSam81} emphasize the importance of the information structure of an auction system. It consists in the information owned privately by the buyers and the information that the seller has on each buyer. This information owned privately by the buyers is the value they give to the item, i.e, the highest price they are willing to pay to get the item. The uncertainties upon these different values lie at the gist of the seller's optimization problem: otherwise, she would just have to sell the item to the buyer with the highest value, at this price or infinitesimally less. 

To handle this deficit of information about buyers, it is standard to take a ``Bayesian" viewpoint and assume that the seller has some probabilistic prior on the values  given to the item by each bidder. This prior distribution is usually called the value distribution and it encompasses the seller's uncertainty on a specific bidder's values. There are of course several possibilities for how this value distribution is constructed. For instance, in wine or art auctions, it often comes from expert knowledge about an admissible price for a good wine bottle or for an important piece of art.
 
\section{Learning theory and auction design}
%Recently, the information structure of the game as auction mechanisms are used on some of the main Internet platforms. 
It is now possible for Internet platforms to run billions of auctions a day and store most of the historical data coming from them. This digitization of auction mechanisms was the first step into gathering data to optimize selling mechanisms. Auctions are now used in most Internet platforms to organize interactions between the different stakeholders. Ebay was one of the first big online platforms to use ascending auction to sell objects on the platform. Google and most search engines companies started to use auctions to sell ad opportunities on their front page. For instance, they let advertisers bid on some keywords to get sponsored links above the first results for a certain user query. Nowadays, Facebook and LinkedIn are also using them to determine which ad to display, Amazon and most e-commerce marketplaces decides which products are going to be sponsored (and/or advertised) through an auction mechanism and auctions are also used to sell carbon permits by the European union or to run large electricity markets. 

To exploit this new source of available information (i.e., enormous datasets of past bids), practitioners used advanced statistical learning algorithms in connection with the classical Bayesian theory. Indeed, beyond the AI hype, machine learning algorithms are now widely applied in the industry for numerous applications: the value distribution is no longer coming from some given and fixed prior, but learned (hopefully accurately and efficiently) on historical -bidding - data.  The first large-scale field experiment in production showed how engineers at Yahoo could handle their huge datasets to learn an optimal reserve price per key word \citep{OstSch11}. This results in data-driven mechanisms whose design use techniques coming from a large variety of fields, including statistics, machine learning, game theory and Economics. Similarly,  bidders on these online platforms also gather data and use new statistical learning techniques to improve their bidding strategies  against automated mechanisms. This flood of data and the associated paradigm shift it constitutes opens many new interesting practical problems, new theoretical questions and new interesting games to study.

\subsection{Repeated auctions only from a seller's standpoint}
The first natural repeated game setting consists in understanding how the seller can learn a revenue-maximizing auction mechanism from a dataset of bids or values.
In the example of Ebay marketplace, the seller (Ebay)  observes numerous auctions a day for similar items. Hence, from its point of view, the mechanism is repeated and she can aim at optimizing some long-term revenue. On the contrary, buyers are individuals that participate in a few, if not a single, auctions at best. Then, from their point of the view, the mechanism still looks like a one-shot auction and they are bound to implement myopic short-term strategies, optimizing point-wise their utility  (by opposition to long-term and effectively in expectation). Let us consider the simplifying assumption where  bidder values on the  platform are sampled from a certain unknown distribution, that  encompasses the variability in their readiness to pay a certain price. Assuming the bidders actually bid their true value (for instance, if the mechanism chosen is fixed and ``incentive-compatible'', i.e., bidding one's value is optimal for buyers), the seller has then access at the end of the day to a dataset of buyer values.

Inspired by the computational learning formalism, \citet{elkind2007designing, balcan2008reducing, ColRou14} initiated a line of research aiming at finding approximations of the revenue-maximizing auction, if possible, efficiently, with approximation guarantees depending on the size of the dataset gathered (a.k.a., the sample complexity). This setting is called the \emph{batch learning setting}.  A variant considers the case where the flow of buyers is continuously coming on the platform and the seller can update continuously her mechanism. This is \emph{the online learning setting} introduced in \citep{cesa2015regret}. In all these problems, it is crucial that the samples gathered in the dataset do have the same distribution as the samples that will be gathered and treated in the future.

\subsection{Repeated auctions from seller and bidder standpoint}
The crucial assumption of myopic/short-sighted/impatient bidders facing a patient seller is unfortunately not necessarily satisfied, depending on the setting. In modern-day practice, typically large online ad platforms, such as Google DoubleClick or AppNexus, are selling ad opportunities for large publishers such as some of the biggest online newspapers. The main difference with the aforementioned Ebay example is that only a few companies are actually bidding in these auctions. They are furthermore doing so repeatedly and participating in massive number of auctions. 

Indeed, most companies willing to display ads actually rely on third-parties,  \emph{demand-side platforms} (DSP), that are buying and displaying ads for them (because of technical constraints, even sending bids in real-time might actually be quite complex). These aggregated bidders are repeatedly interacting with the (same) seller, billions of times a day. Consequently, this type of buyers can also optimize for long-term utility and need not be myopic. Thus, even if the seller is using one-shot incentive compatible auctions - for instance to gather data in order to later design and switch to a revenue maximizing mechanism -, the bidder might have an interest in not bidding ``truthfully", as classical theory would suggest is optimal for them. Indeed, if buyers do not bid their values, this will modify the distribution of ``values'' observed by the seller. Subsequently, the mechanism chosen to optimize her revenue will be different from what it would have been had bidders been na\"{i}ve, to the advantage of the buyers \citep{tang2016manipulate,nedelec2019learning}.

%{\color{red} VP: je comprends pas le paragraphe suivant} In this setting, the value distribution encompasses the variability of valuations for one specific bidder that could exist from one auction to the other.
%Indeed, he is participating in billions of auctions a day corresponding to slightly different items, i.e., various ad opportunities for the online advertising example. In this setup, the goal of the seller is to learn this value distribution to maximize her revenue.
%\redtext{Clem: It's not clear here where the valuation distribution comes from. I think it is the right place to introduce information asymetry and private vs public information.}

Intuitively, this is possible because the information asymmetry that arose in the Ebay example between the seller and the bidders -- one optimizing over the long-term, the other over the short-term  --  is almost reversed. If the seller must commit to a specific mechanism or a family of mechanisms, for instance for contractual reasons, and  buyers have this information, they can strategically leverage it by e.g. changing their bidding behavior.  In the end, the respective  utilities of the seller and buyers will somehow depend on the underlying amount of asymmetry between them. Several works have started studying various intermediate settings, for example when bidders are (almost) identical  \citep{KanNaz14}, or are patient, but not as patient as the seller \citep{amin2013learning}, etc.

\section{Organization of the survey}
In this survey, our overarching objective is to provide a widely accessible introduction to the fascinating topics of classical and modern auction theory while bringing to the fore the statistical and machine learning lenses to the topic.  We will very clearly state the differences between the different information-asymmetry settings we will review, and point to cutting edge theoretical and practical solutions adapted to them. We will also show how new statistical tools can be used to tackle some important and well-known problems from Economics. Furthermore, those questions open many new interesting  problems in Economics since algorithms are replacing classical sellers and buyers. We believe that modern auction theory offers a nice framework to understand what data and Computer Science can bring to modern Economics.

In Chapter \ref{Chapter:Bayes}, we survey the main results of the Bayesian auction literature, initiated with the seminal works of Vickrey and Myerson. Those results form some of the backbone of classical auction theory and are widely used in Internet practice. We will recall what is the revenue-maximizing auction once the seller has a prior on bidder's valuations and introduce some approximations of the revenue-maximizing auction when the seller must use simpler auctions. In Chapter \ref{Chapter:Seller}, we focus on the setting derived from the Ebay use case and tackle both the batch learning setting and the online learning setting. We recall some key concepts of statistical learning theory, derive the sample complexity of some of the learning algorithms used to compute a revenue-maximizing auction and show their computational complexity. In Chapter \ref{Chapter:Learning}, we focus on the less studied but crucially important setting where bidders can be strategic regarding the mechanism itself since they have multiple interactions with the seller. We review some of the main methods that have been devised to keep bidders from being strategic in that context, show their limitations and introduce some very new results and approaches developed for bidders to take advantage of the seller's learning process. 

This survey only assumes basic familiarity with standard notions of Machine Learning, Statistics and Data Science and is written with a reader having this background in mind. We hope our survey will be useful to engineers and researchers looking for an introduction to the beautiful and fast developing topics of modern auction theory and applications.

\printbibliography[segment=1, heading=subbibintoc]
\end{refsegment}
\clearpage
\begin{refsegment}
%!TEX root = ../main.tex
\chapter{Bayesian mechanism design}
\label{Chapter:Bayes}

\centerline{\colorbox{mygray}{ \begin{minipage}{11cm}
\begin{center} \textbf{First read of this chapter, key concepts and ideas}\end{center}
This chapter introduces the Bayesian setting of auction theory. For a reader who is new to the topic, we think the key results and ideas of this chapter are: second price auctions and associated optimal bidding strategy (Theorem \ref{thm:OptimalStratSecondPrice}); the notion of ``truthful'' bidding, p. \pageref{def:truthfulBidding}; first price auctions and associated bidding strategy at a symmetric equilibrium (Theorem \ref{thm:NashFirstPrice}); the revenue equivalence theorem (Theorem \ref{Th:RevEquivalence}); the notion of virtual value (Definition \ref{def:VirtualValue}) and its implications for optimal reserve prices in second price auctions (Theorem \ref{thm:optimalReservesLazySecondPrice}), optimal ``truthful"/Myerson auctions (Definition \ref{def:MyersonAuction} and Theorem \ref{theorem_regular_distributions}) and the conceptually fundamental Myerson Lemma (Theorem \ref{Th:MyersonLemma}). We also recommend focusing on the symmetric setting on first reading. 
\end{minipage}
}
}

\vspace{0.5cm}

Auctions mechanisms involve many different agents, sellers and/or buyers, with possibly different and conflicting objectives as they all seek to optimize of their own utility functions. These interactions can be modeled using game-theoretic concepts. More specifically, we are going to focus on a specific type of games with \emph{incomplete information} that are called \textit{mechanisms}. In those games, each player has some private information (i.e., unknown from everyone else), and send a \emph{message} to a central authority.  Based on those gathered messages, the latter decides on the final outcome.  The utility of each player then solely depends on this outcome. 

Mechanisms model appropriately  many practical situations such as the celebrated  problems of assigning students to schools, or matchings in organ-transplant applications.  In these problems, a central authority forms pairs between school and students or donors and recipients. In selling mechanisms, the central authority is the seller of a specific item and the players are the buyers.  \emph{Auctions} are mechanisms used to sell a particular item. In the case of a sale of a single non-divisible item, they have the following specific features.

\medskip

%and decides each player's outcome according to these messages. In these games, players' actions correspond to the message they send to the central authority. 

%In a game of incomplete information, every player's payoff function depends on a variable $X$ called private information and which is only known to each specific player. If we denote by, the payoff function of player $i$ is defined as 

%In this survey, we are focusing on a certain type of games with incomplete information where there exists a central authority which gathers messages from the different players and decides each player's outcome according to these messages. In these games, players' actions correspond to the message they send to the central authority. This type of centralized gamed is called mechanism. 
%\begin{definition2}[Mechanism]
%A mechanism is a game of incomplete information where players send a message to a central authority who does not know the private information of each player. The central authority decides based on messages it receives and some predefined rules every player's outcome.  
%\end{definition2}

Auctions are games of incomplete information as each buyer has some private valuation for the item to be sold, i.e., the highest price they are willing to  pay to acquire  this item. This valuation might be different from one buyer to another. We denote by $\mathcal{N}$ the set of buyers, of cardinality $n \in \mathds{N}$, by $\mathcal{X}_i \subset \R $ the set of possible private values of bidder $i \in \mathcal{N}$, by  $x_i \in \mathcal{X}_i$ the actual private value of bidder $i$ and by  $\bx = (x_1,\dots,x_n)\in \boldsymbol{\mathcal{X}}:= \prod_i \mathcal{X}_i \subset \mathds{R}^n$, the so-called profile of private values; the bold notation will refer to vectors for the sake of clarity. 
The possible messages (or actions) of buyer $i\in \mathcal{N}$ are called ``bids'' and the set of bids of buyer $i$ is denoted by $\mathcal{B}_i$. The outcome of an auction mechanism is defined by two different rules:
\begin{enumerate}
\item an \textit{allocation rule} $q: \bcB = \mathcal{B}_1 \times \dots \times  \mathcal{B}_n \rightarrow \Delta_n$ where $\Delta_n$ is the set of probability distributions over the set of $n$ buyers. It specifies the probability that each player gets the item.
\item a \textit{payment rule} $p: \bcB =\mathcal{B}_1 \times \dots \times  \mathcal{B}_n \rightarrow \mathds{R}^{n}$. It specifies the expected payment of each player, whether or not they get the item.
\end{enumerate}
We will denote by  $\mathfrak{A}$ the set of all auction mechanisms, i.e., the set of pairs of allocation/payment rules.
 
 \medskip
 
The utility of bidder $i$ is simply the difference between the item value (if he won the auction, and 0 otherwise) and his payment (that can be positive even if the auction is lost). If we denote by $\bb = (b_1,\dots,b_n) \in \cB$ the vector of bids of the bidders, the expected utility of bidder $i \in \mathcal{N}$, given bids and values , is then defined as
%$$u_i(b,x_i) = q_i(b)\big(x_i -p_i(b)\nig)\;;$$
$$u_i(\bb,x_i) = q_i(\bb)x_i -p_i(\bb)\;;$$
on the other hand, the seller aims at maximizing the expectation of her revenue defined, given the bids, by: %$$\sum_{i=1}^n q_i(b)p_i(b)\;.$$
$$\sum_{i=1}^n p_i(\bb)\;.$$

\section{The Bayesian setting}

The auction literature has often considered  a Bayesian setting where the value $x_i$ of bidder $i \in \mathcal{N}$ is random, drawn according to his \emph{value distribution} that ``represents the seller assessment of the probability of bidder $i$ having a value estimate of $x_i$ or less''  \citep{Myerson81}. The value distribution quantifies the uncertainty of the seller on the maximum price that one buyer is willing to pay for the item and is represented as a Bayesian prior on the private information of each buyer.

A widely made assumption is that value distributions are common knowledge among bidders and seller.  As a consequence, we denote by $F_i$ the value distribution - i.e., the cumulative distribution function of the value -  of buyer $i \in \mathcal{N}$ and by $f_i$ its corresponding density function, assuming it exists, an assumption we are going to make repeatedly (removing this assumption is almost always a matter of technicalities and has little to no impact on conceptual questions). We implicitly identify the distribution $F_i$ with its cumulative distribution function (cdf) and use both terms exchangeably. We are also going to assume that values $x_i \in {\mathcal X}_i$ are non-negative, i.e., the support of $F_i$ is included in (but not necessarily equal to)  $\mathds{R}_+ = [0,\infty)$. Unless otherwise noted, we assume that $\mathds{E}_{x_i \sim F_i} [x_i] < +\infty$ to get optimality results (and not just $\varepsilon$-optimality). A crucial assumption throughout this survey is that, unless otherwise noted, the values $x_i$ are drawn independently for different $i$'s and hence they are statistically independent as random variables. We shall  denote by $\bF =  F_1\otimes \ldots\otimes F_n$ the joint - and hence - product distribution of $\bx=(x_1,\ldots,x_n)$, the vector of values.

\medskip

\paragraph{Examples} Typical examples of value distributions are the \emph{uniform} distribution, which is widely used in textbook examples due to its simplicity,  the \emph{exponential} and the \emph{log-normal} distributions as they are similar to some empirical distributions encountered on modern internet platforms. Power law distributions (also known as \emph{Pareto} distributions) are also widely used, as they capture the idea that the value in real time bidding and online advertising comes from few matches of very high quality, such as consumers who recently viewed a product \citep{arnosti2016adverse} (this situation where 20\% of individuals own/generate 80\% of wealth is also referred as the Pareto principle in economics). \emph{Generalized Pareto} distributions are also often used as examples because their virtual value - an important concept we will define later - is linear. 
\begin{definition}[Symmetric setting]
The auction setting is  called \textit{symmetric} if all bidders have the same value distribution.
\end{definition} 

For the sake of clarity, we use pronouns her/she for the seller and he/his for one specific bidder.

\subsubsection*{Assumptions on Value Distributions and Notations}
If $\mathbf{v}$ is a vector (of scalars or functions) in $\mathds{R}^n$, we call $\mathbf{v}_{-i}$ the vector in $\mathds{R}^{n-1}$ that contains all entries of $\mathbf{v}$ except the $i$-th one, $v_i$. 
In other words, $\mathbf{v}_{-i}=(v_1,\ldots,v_{i-1},v_{i+1},\ldots,v_n)$.  With a slight abuse of notations, the vector $(v_i,\mathbf{v}_{-i})$ is identified to the vector $\mathbf{v}$. 
%Finally, we shall also use the notation $\mathbf{v}_{-i}(j)$ to denote the $j$-th entry of $\mathbf{v}_{-i}$. 
Similarly, the notation  $\bF_{-i}$ will denote the product distribution of values of all bidders except $i$ and $\mathcal{X}_{-i}$ is the cartesian product of all $\mathcal{X}_j$ apart the $i$-th one. 

We are always going to make the following key default assumption on distributions, unless explicitly noted. Each value distribution $F_{i}$ is assumed to be continuous, supported on some interval $[0,H]$, with $H$ being possibly infinite. Moreover, we assume that $f_i>0$ on $(0,H)$, except possibly on a set of Lebesgue measure 0. Finally, unless otherwise noted $F_i$ is assumed to have at least one moment (i.e., the associated random variable $x_i$ has finite expectation). This will also imply that $\psi_i$, the function defined  later on in Section \ref{def:VirtualValue} and called the virtual value function associated with $F_i$, also has one moment.

\medskip

We recall that if ${G}_i$ is the cdf of the largest value of a vector $\bx_{-i}$ drawn according to $\bF_{-i}$, then  ${ G}_i(t)=\prod_{j\neq i}F_j(t)$. In particular, with our key default assumption on distributions, ${ G}_i$ is differentiable and its density is strictly positive except possibly on a set of measure zero. We also use the standard convention that a  function $\beta$ is increasing,  if  for all $l < u$, $\beta(l) < \beta(u)$; on the other hand,  a non-decreasing function is such that if $l < u$, $\beta(l)\leq  \beta(u)$. We use $\partial f$ to denote the subdifferential of a convex function $f$ (see \cite{HULLbook01}, Chapter D). Finally $\langle \cdot ,\cdot \rangle$ is the standard dot product between two vectors and for any integer $N \in \mathds{N}$, we define $[N]=\{1,\ldots,N\}$.

%{\color{red} VP: Je pense qu'il faut elaborer un peu sur le "typical". Ca veut dire quoi exactement ? Soit on met des refs, soit on vire}

%The notion of information owned by the seller and value distribution have progressively changed with several new applications of auction systems. One of the goal of this thesis is to clarify this change and show how sellers and bidders can adapt to this new information structure of the game. For instance, in a repeated setting, where multiple different items are sold successively, the value distribution could also represent the variety of values that a specific buyer can give to one item. 

\subsection{Properties of auction systems}
\label{subsec_prop_auction}
A bidder's \textit{strategy} is the mapping indicating which bids he sends to the seller to buy one specific item, conditional on his private information, a.k.a., his value. Stated otherwise, a strategy for bidder $i$ is a mapping    that maps private values in $\mathcal{X}_i$ to bids in $\mathcal{B}_i$. %\nek{technical detail: is this mapping possibly random? or is it a function?}
%Bidders have the choice on the bids they send to the seller to get one specific item. A strategy defines how these bids depend on buyer's original private value.
%\begin{definition2}[Strategy]
%A strategy is a function $\beta: \mathcal{X} \rightarrow \mathcal{B}$ that maps private values to bids.
%\end{definition2}
We will denote by $\mathds{B}_i$ the set of all strategies of player $i$, $\beta_i$ the strategy chosen by $i$ and $\boldsymbol{\beta}_{-i}$ the profile of strategies corresponding to all bidders except bidder $i$. %COMMENTED (deja introduit plus haut) Similarly, the notation  $F_{-i}$ will denote the product distribution of values of all bidders except $i$ and $b_{-i}$ is the vector of all bids except bidder's $i$. 

In Bayesian games, it is important to differentiate between \textit{ex ante}, \textit{interim} and \textit{ex post} properties \citep{hartline2013bayesian}. These notions depend on the information available to buyers when they decide to participate in the game and choose their strategy.  
\begin{itemize}
\item For  \emph{ex-ante} properties, bidders do not know yet their own value for the item; i.e., they only know $F_i$ and $\bF_{-i}$.
\item For \emph{interim} properties, bidders know their valuation but do not know the values of other players, i.e., they know $x_i$ and $\bF_{-i}$.
\item For \emph{ex-post} properties, bidders know both their and the other players' valuations, i.e., they know  $x_i$ and $\bx_{-i}$.
\end{itemize}
In the next sections, we will mostly focus on  interim properties. It is the starting point of most of the auction literature: we assume that value distributions are common knowledge and that exact valuations are private information to each bidder.  We will mention  explicitly when  we refer to ex-ante or ex-post properties.

We are also assuming that bidders are risk-neutral. In other words, they seek to maximize their expected utility and use utility-maximizing strategies: %over all other bidders' possible valuations and with respect to their strategies\nek{awkward; need to rephrase; need to def for interim and ex-ante. review that.}.  
Given the strategy $\bbeta_{-i}$ of other players, the  expected utility  of the strategy $\beta_i$  given bidder $i$'s value $x_i$ is denoted by
$$
U_{i}(\beta_i,\bbeta_{-i},x_i) = \mathds{E}_{\bx_{-i} \sim \bF_{-i}}\Big[u_{i}\big((\beta_1(x_1), \dots,\beta_{i-1}(x_{i-1}),\beta_i(x_i),\beta_{i+1}(x_{i+1}),\beta_n(x_n)), x_i\big)\Big]\;.
$$
%\nek{this is awkward as interim doesn't mean we know other people's strategies... and a strategy is a mapping but technically here we just need to know $\beta_i(x_i)$; also if $\beta_i(x_i)$ included in the stuff?}

\paragraph{Optimality and characterization of strategies} Maybe one of the most central concepts in game theory is \emph{(Bayesian) Nash equilibrium}. %It is a characteristic of the vector of strategies $(\beta_1,\ldots,\beta_n)$ and the valuation distributions $F_1,\ldots,F_n$.
 At a Bayesian Nash Equilibrium, for any bidder $i$, his strategy $\beta_i$ maximizes his expected utility, given his valuation distribution $F_i$, and given the strategies of his opponents and their valuation distributions, i.e., $\boldsymbol{\beta}_{-i}$ and $\bF_{-i}$. A stronger concept is that of \emph{weak dominance}: a strategy  $\beta_i$ is weakly dominant when it is optimal in terms of expected utility of bidder $i$ against any strategies used in $\boldsymbol{\beta}_{-i}$ and not only those at a Bayesian Nash Equilibrium. The strongest concept is ex-post dominance, where optimality is achieved at any possible profile of valuations.
\begin{definition2}
\begin{itemize}
\item \emph{Nash equilibrium:} 
A profile of strategies $(\beta_{1},\dots,\beta_{n})$ is \textit{a Bayesian Nash equilibrium} if for all players $i \in \mathcal{N}$, 
\begin{equation*}
\forall \beta \in \mathds{B}_i, \forall x_i \in \mathcal{X}_i, \quad U_{i}(\beta_i,\bbeta_{-i},x_i)  \geq U_{i}(\beta,\bbeta_{-i},x_i) \;.
\end{equation*} 
\item \emph{Weak dominance:} A strategy $\beta_i$ is  weakly dominant for player $i \in \mathcal{N}$ if
\begin{equation*}
\forall \beta \in \mathds{B}_i, \forall x_i \in \mathcal{X}_i, \forall \bbeta_{-i} \in  \boldsymbol{\mathds{B}}_{-i}, \quad U_{i}(\beta_i,\bbeta_{-i},x_i)  \geq U_{i}(\beta,\bbeta_{-i},x_i) \;.
\end{equation*}
\item \emph{Ex-Post Weak dominance:} A strategy $\beta_i$ is  ex-post weakly dominant for player $i \in \mathcal{N}$ if
\begin{equation*}
%\forall \beta \in \mathds{B}_i, \forall x_i \in \mathcal{X}, \forall \beta_{-i} \in  \mathds{B}_{-i}, \forall x_{-i} \in \mathcal{X}^{N-1},\quad u_{i}((\beta_1(x_1),\dots,\beta_i(x_i),\dots,\beta_N(x_N)),x_i)  \geq u_{i}((\beta_1(x_1),\dots,\beta(x_i),\dots,\beta_N(x_N)),x_i) \;.
\forall \beta \in \mathds{B}_i, \forall x_i \in \mathcal{X}_i, \forall \bbeta_{-i} \in  \mathds{B}_{-i}, \forall x_{-i} \in \boldsymbol{\mathcal{X}}_{-i},\quad u_{i}((\beta_i(x_i),\bbeta_{-i}(\bx_{-i})),x_i)  \geq u_{i}((\beta(x_i),\bbeta_{-i}(\bx_{-i})),x_i) \;.
\end{equation*}
\end{itemize}
\end{definition2}

%
%
%A strategy which is optimising the expected utility of one bidder for any valuation and for any strategies of the other bidders is said dominant. 
%\begin{definition2}[(Weakly) dominant strategy]
%A strategy $\beta^{*}$ is  ex-post weakly dominant for player $i$ if for all $\beta_{-i} \in  \mathds{B}^{N-1}$, for all $\beta \in \mathds{B}$ and for all $x \in X^{N}$:
%
%\end{definition2}

Those properties of strategies are classical concepts in game theory. On the other hand, it is also possible to introduce and study different properties of mechanisms. Some of them require the concept of \emph{``truthful bidding''}\label{def:truthfulBidding} which correspond to the specific strategy $\beta_i(x)= x$. We will denote by $\beta_{i,\text{tr}}$ this truthful strategy.

\paragraph{Characterization of mechanisms}
\begin{definition2} A mechanism is
\begin{description}
\item[(BIC)] Bayesian Incentive-Compatible:  if bidding truthfully for all bidders is a Bayesian Nash equilibrium.
\item[(DSIC)] Dominant Strategy Incentive-Compatible:  if bidding truthfully is a weakly dominant strategy for all bidders.
\item[(Standard)] if it allocates the item to the buyer with the highest bid.
\item[(Efficient)] if it allocates the item to the buyer with the highest valuation (at least at some equilibrium).
\item[(IR)]  interim Individually-Rational: if 
\begin{equation*}
\forall \, i \in \mathcal{N}, \; \forall x_i \in \mathcal{X}_i, \quad U_{i}(\beta_{i,\text{tr}}, \bbeta_{-i,\text{tr}},x_i) \geq 0\;.
 \end{equation*}
 and ex-post Individually-Rational: if 
 \begin{equation*}
\forall \, i \in \mathcal{N},\; \forall x_i \in \mathcal{X}_i,  \forall \bx_{-i} \in \boldsymbol{\mathcal{X}}_{-i}, \quad u_{i}\big((x_i, \bx_{-i}), x_i\big) \geq 0\;.
\end{equation*}
\end{description}
\end{definition2}

A DSIC mechanism is obviously a BIC mechanism. More generally,  Incentive Compatible (IC) auctions have the nice property of being ``simple'' for the buyers from a strategic standpoint:  bidding their (known in the interim setting) valuation is optimal for them. Notice  that this unfortunately does not ensure the uniqueness of the equilibrium where each bidder bids truthfully (we will call this equilibrium the  \textsl{truthful equilibrium}). See Section \ref{sec:NashIsDifficult} for more details. Like most authors we restrict attention to the truthful equilibrium from now and leave more pathological equilibria aside. As we will see later, being DSIC is one of the main reasons explaining the tremendous success of second-price auctions in practice. Another reason is that if bidders are bidding truthfully, then the seller can, in a first step, elicit their value distributions through a DSIC mechanism and then  move to another  mechanism that maximizes her revenue (this is detailed in Section \ref{Section:RevenueMax}).

% - even if it is at the cost of efficiency\nek{efficiency as in what?}.

\medskip
Finally, before presenting and analyzing two classical types of auctions, we indicate that individual rationality simply ensures that bidders have an interest in taking part in these auctions.

\section{Sealed-bid auctions}
In a sealed-bid auction bidders privately send their bid to the seller. We present below two of the most well-known sealed-bid auctions, the \emph{second-price} and the \emph{first-price} auctions. This class of auctions does not include the well-known - in popular culture -  ascending auction, a.k.a., English auction, where bidders can observe bids from other bidders and progressively choose to increase their bids until only one bidder remains. On the other hand,  if bidders' valuations are independent, there exists a strategic-equivalence between the ascending and the second-price auction. 
\subsection{The sealed-bid second-price auction}
The second-price auction allocates the item to the highest bidder who pays the highest bid among other bidders, i.e., the second highest bid. A key property of this auction is the following result \citep{vickrey1961counterspeculation}, which does not require any assumption on bidders' value distributions. 

\begin{theorem2}\label{thm:OptimalStratSecondPrice}
The second-price auction is DSIC. In other words, bidding truthfully is weakly dominant. 
\end{theorem2}
\begin{proof}
Let us denote by $x_i$ the private value of bidder $i$ and by $b^*_{-i}$ the highest bid of the competition. We are going to compare the utility of bidding  $b$ instead of $x_i$.
\begin{itemize}
\item Case $b > x_i$. The only case where his (ex-post) utility is changed is when $x_i < b^*_{-i} < b$. With a bid $b$, he now wins the auction but his utility is negative since $x_i - b^*_{-i} < 0 $.
\item  Case $b < x_i$. The only case where his (ex-post) utility is changed is when $x_i > b^*_{-i} > b$. With a bid $b$, he now loses a profitable  (i.e., with a positive utility) auction.
\end{itemize}
Hence, bidder $i$ has no incentive to deviate  from truthful bidding. 
\end{proof}
The classical second-price auction was used until recently \citep{feng2020reserve} by most of the biggest online platforms to sell ad placements on publishers' websites. Another widely used and studied sealed-bid auction is the first-price auction.
\subsection{The sealed-bid first-price auction} 
\label{first-price-auction}
The first-price auction allocates the item to the highest bidder who pays his own bid.

\medskip

Before studying Nash equilibria of symmetric first-price auction, we derive the general best reply of player $i$ to the bid distribution of the competition, specifically the distribution of the maximum bid of the competition. This result is of increasing interest to practitioners as many online auctions are now first price auctions.

\begin{proposition2}\label{Prop:BestReply1st}
	Let $G_i$ be the cdf of the highest bid of the competition of bidder $i$, i.e., $\max_{j\neq i} b_j$. 
In a sealed-bid first price auction, a best response of bidder $i$  to $G_i$ is any mapping $\beta_i$ satisfing
\begin{equation*}\label{eq:bestresponsefirstpriceSemiAbstract}
%\beta(x_i):  \beta(x_i)=b(x_i) \text{ where } b(x_i)\in \argmax_{b \in \mathds{R}} G_i(b)(x_i-b)\;.    COMMENT: trop complique
\beta_i(x_i) \in \argmax_{b \in \mathds{R}} G_i(b)(x_i-b)\;.   
\end{equation*}
When $G_i$ is log-concave and $G_i(x_i)>0$, the best response is unique.  If we further assume that $G_i$ has a pdf $g_i$, first order conditions also give, if $G_i(x_i)>0$, 
\begin{equation*} \label{eq:bestresponsefirstprice}
\beta_i(x_i)  \text{ is a solution (in }b \text{) of } b + \frac{G_{i}(b)}{g_{i}(b)}= x_i\,.
\end{equation*}
If $G_i(x_i)=0$ a best response is $\beta_i(x_i)=x_i$.

 \end{proposition2}
Calling $Y_{-i}$ a random variable with cdf $G_i$, it can also be shown that under mild technical conditions that $\beta_i$ is increasing and satisfies the equation $\beta_i(x)=\Exp{\beta_i^{-1}(Y_{-i})|Y_{-i}\leq \beta_i(x)}$. We also have the following interesting corollary.

\begin{corollary2}\label{coro:1stPriceNotBIC}
Proposition \ref{Prop:BestReply1st} implies that the first price auction is in general not BIC.	
\end{corollary2}
The corollary simply follows by  showing that the best response of bidder $i$ when all other bidders bid truthfully (and hence the top bid of the competition is the largest value of the other bidders) consists in bidding something else than $x_i$. If $G_i(x_i)>0$ it follows immediately than they are better strategies than bidding $b_i=x_i$: for instance, take any $\tilde{b}_i$ such that $G_i(\tilde{b}_i)=G_i(x_i)/2$ ($\tilde{b}_i$ exists by continuity of $G_i$ and $G_i(a_i)=0$). The utility of bidder $i$ is strictly positive at $\tilde{b}_i$ and is 0 at $x_i$. 
\begin{proof}
Let us denote by $Y_{-i}$ the random variable corresponding to the maximum bid of the competition of bidder $i$, so that its cdf and pdf are $G_i$ and $g_i$.

When bidder $i$ has private value is $x_i$ and bids $b_i$, the utility he derives from the auction is $u_i(b_i,x_i)=(x_i-b_i)\mathds{1}\{b_i>Y_{-i}\}$; in other words, it is his value minus his cost when he wins the auction and zero otherwise. 

We denote by $\mathcal{U}_i(b_i,x_{i}):\mathds{R}^{+}\times\mathds{R}^{+}\rightarrow \mathds{R}$ the associated expected utility of bidder $i$ when his private value is $x_i$ and he bids $b_i$. We have 
$$
\mathcal{U}_i(b_i,x_i) = \mathds{E}_{Y_{-i}}\Big[(x_i-b_i)\mathds{1}\{b_i>Y_{-i}\}\Big] = G_i(b_i)(x_i-b_i)\;.
$$
A best response is therefore any $b(x_i)\in \argmax_{b\in \mathds{R}^+}G_i(b)(x_i-b)$. Note that when $G_i$ is log-concave and $G_i(x_i)>0$, we can verify by inspection that $\mathcal{U}_i(b_i,x_i)$ is strictly log-concave in $b_i$ on the support of $F_i$  and therefore it has a unique maximum smaller than $x_i$ \citep{boydvandenberghe04}. This property follows also immediately from the definition of a strictly concave function.

Since $G_i$ has a pdf, it is continuous and differentiable and therefore so is $\mathcal{U}_i(b_i,x_i)$ as a function of $b_i$. 
The derivative with respect to $b_i$ is then equal to 
$$
\frac{\partial \mathcal{U}_i}{\partial b_i}(b_i,x_i) = g_i(b_i)(x_i-b_i) - G_i(b_i).
$$
Let us assume that $G_i(x_i) >0$, then $\mathcal{U}_i(x_i,x_i)=0$ but $\left.\frac{\partial \mathcal{U}_i(b_i,x_i)}{\partial b_i}\right|_{b=x_i} <0$. This first implies by continuity that there exists  bids where the utility is positive. Rolle's theorem applied to $t\mapsto G_i(t)(x_i-t)$ also gives the existence of a stationary point $b_i^* \leq x_i$ where  $\frac{\partial \mathcal{U}_i}{\partial b_i}(b^*_i,x_i) =0$, as $\mathcal{U}_i(0,x_i)=0$, too (since we assumed non-negative bids). 
Since we showed above that $\mathcal{U}_i(b_i,x_i)$ is  positive somewhere in a neighborhood of $x_i$, then necessarily  $\mathcal{U}_i(b_i^*,x_i)>0$. Finally, if $g_i(b^*_i)=0$ then this  would imply that  $G_i(b^*_i)=0$ to satisfy the first order condition and therefore $\mathcal{U}_i(b_i^*,x_i)$ would be equal to 0 which is impossible.

The case where $G_i(x_i)=0$ is trivial as bidder $i$ cannot have a positive utility (recall that $G_i$ is a non-decreasing and non-negative function, so $0\leq G_i(b_i)\leq G_i(x_i)$ if $b_i\leq x_i$). Bidding $x_i$ is then optimal.
\end{proof}

There is a very rich line of work focusing on deriving Nash equilibria in first-price auctions when bidders have different value distributions. This involves solving complex systems of coupled first-order differential equations (at least with continuous value distributions, see \cite{krishna2009auction}, Section 4.3; see also p. \pageref{thm:SystemEquiAsymetricFirstPrice}). On the other hand, with symmetric bidders, i.e., with identical value distributions, it is possible to solve explicitly this system of equations and to derive the unique symmetric  Nash equilibrium with increasing strategy. From now on, we will call a Nash equilibrium  \emph{increasing} if the  strategies  are all increasing mappings. 
\begin{theorem2}\label{thm:NashFirstPrice}%[Nash equilibrium in a first-price auction]
In the symmetric case,  if the common pdf $f$ is such that $f(x)>0$ (except  on a set of Lebesgue measure 0 within the support of $F$), there exists a symmetric increasing Nash-equilibrium whose strategy is described by:  
\begin{equation*}
\beta(x) = \mathds{E}[x^{(1)}_{-i} | x^{(1)}_{-i} < x]\;.
\end{equation*}
where $x^{(1)}_{-i}$ is the highest value among bidders except bidder $i$.
\end{theorem2}
% \nek{tech question: if we assume $f>0$ then $\beta$ is increasing and we have zero problems below. If $f\geq 0$, $\beta$ is only non-decreasing; then we have to deal with possible ties, how to break them and make sure that the PWin is $G_i(x)$. when they all bid $\beta(x_j)$ I haven't done this. this minor tech stuff would preferably done in an appendix; probably need to replace $x_j$ by $x_j^-$, i.e., the $inf_x F(x)=t$. I leave the proof with non-decreasing but worth a check}.
This bidding strategy can be interpreted as bidding the expectation of the largest value of the competition, conditionally on the fact that this value is smaller than bidder $i$'s value. We note that this bidding strategy can be derived from the proof of the revenue-equivalence Theorem \ref{Th:RevEquivalence}, and specifically the expected payment formula. This is another common method for finding equilibrium bidding strategies. The proof presented below might lead more directly to the solution.  

\begin{example2}
Suppose there are $n\geq 2$ bidders, and they all have uniform [0,1] value distribution, i.e., $F(x)=x$ on $[0,1]$. Then  a symmetric increasing Nash equilibrium exists in 1st price auctions where all bidders bid using the strategy
$$
\beta(x)=\frac{n-1}{n}x\;.
$$	
\end{example2}

\begin{proof}
We assume that $n\geq 2$ and that all bidders are using the function $\beta$ described above. As we will show below, this function is increasing on the support of $F$ under our assumptions. Furthermore,  when all bidders are using the same \emph{increasing} strategy on the union of the support of their value distributions, the probability that bidder $i$ wins the auction is the same as the probability that he has the highest value; this would not always be true if the strategy were only non-decreasing. 

Furthermore, elementary properties of conditional expectations give, if $G_i$ and $g_i$ are the cdf and pdf of $x_{-i}^{(1)}$,
$$
\beta(x)=\beta_i(x)=\frac{\int_{0}^{x} y g_i(y) dy}{G_i(x)}=x-\frac{\int_0^x G_i(u)du}{G_i(x)}\;, \text{ whenever } G_i(x)>0\;.
$$
Under our assumptions, $G_i=F^{n-1}$, thus $g_i(x)=(n-1)f(x)F^{n-2}(x)$ and $g_i(x)>0$ on the support of $F$, except possibly on a set of Lebesgue measure 0. As a consequence,   $$\beta'(x)=\frac{g_i(x)\int_0^x G_i(u)du}{[G_i(x)]^2}$$ and therefore, restricted to the support of $F$, $\beta$ is a non-decreasing function whose derivative is 0 on a set of Lebesgue measure 0 . We conclude that $\beta$ is actually increasing on the support of $F$. Since the latter is supposed to be $[0,H]$, with $H$ possibility infinite, bidder $i$ has no incentive to bid higher than $\beta(H)$. %VP: je pense que c'est suffisament evident pour etre supprime : suppose he did bid $b>\beta(H)$. Clearly since the other bidders bid according to $\beta$, their maximum bid is lower than $\beta(H)$. Then, bidding $b$, bidder $i$'s probability of winning the auction would be 1 and his cost $x-b$. He could achieve the same probability of winning by bidding $\beta(H)$ but his cost would be lower~: $x-\beta(H)$. 
Then, any other bid $b$ will satisfy $b \in [0,\beta(H)]$ and since $\beta(0) =0$ and $\beta$ continuous , because $F$ is continuous, there must exist $z \in [0,H]$ such that $b = \beta(z)$. Finally, note that the probability that bidder $i$ wins the auction when bidding $\beta(z)$ is just $G_i(z)$, since $\beta$ is increasing on the support of $F$. Therefore,
\begin{align*}
\mathcal{U}_i(\beta(x),x) - \mathcal{U}_i(b,x) &= \mathcal{U}_i(\beta(x),x) - \mathcal{U}_i(\beta(z),x)
\\&= G_i(x)\left(x - \frac{\int_{0}^{x} y g_i(y) dy}{G_i(x)}\right) - G_i(z)\left(x - \frac{\int_{0}^{z} y g_i(y) dy}{G_i(z)}\right)
\\&= x(G_i(x) - G_i(z)) + \int_{x}^{z} y g_i(y) dy = 
\int_x^z(y-x)g_i(y)dy 
\geq 0\;.
\end{align*}
This shows that $\beta$ is the  best response, and thus   $(\beta,\dots,\beta)$ is a symmetric (increasing) Nash equilibrium.   We will prove unicity of this increasing differentiable symmetric Nash equilibrium in symmetric first-price auctions in Section \ref{sec_rev_equi}. 
\end{proof}

Unlike second-price auctions, first price auctions are not incentive compatible (see e.g. Corollary \ref{coro:1stPriceNotBIC}). As a consequence, the strategy of a bidder at an equilibrium  depends on the bidding strategy of the other bidders, and ultimately on other bidders' valuation distributions (see Proposition \ref{Prop:BestReply1st}). So, in practice, computing a good or optimal bidding strategy would require estimating the distribution of the highest bid of the competition, which can be very challenging. Nevertheless, because of their relative transparency for bidders (who know ahead of time what they might pay if they win), first-price auctions are increasingly used in online advertising auctions \citep{feng2020reserve}. However, optimal bidding becomes much more complex for bidders than it is in second price or other BIC/DSIC/``truthful'' auctions.
\paragraph{Nash equilibrium in the asymmetric case}

Asymmetric first price auctions are much more intricate than symmetric ones as  the equilibrium strategy of each  bidder depends in a very subtle manner of the other bidders' strategies  \citep{LebrunFirstPriceAsymmetric1999}. Indeed, let us assume that the distributions $F_1,\ldots,F_n$ are supported on $[h,H]$, have a density  bounded away from 0 on $(h,H)$ and possibly have a point mass  at $h$. 
%This assumption comes from practical  -- as many auctions have 0 value to online advertisers -- and technical -- if a common reserve price is set at $r>h$, then distributions can be assumed to be  supported on $[r,H]$ with a point mass at $r$  \citep{LebrunFirstPriceAsymmetric1999}. 
\begin{theorem2}[\cite{LebrunFirstPriceAsymmetric1999}]\label{thm:SystemEquiAsymetricFirstPrice}
Under these  assumptions, there exist deterministic Nash equilibrium strategies that are increasing. Let us denote them by $\beta_i$ and by $x_i(b)=\beta_i^{-1}(b)\geq b$ their inverse, i.e., the value inducing the bid $b$. Then the increasing functions $b\rightarrow x_i(b)$ solve  the system of differential equations: 
\begin{equation}\label{eq:equilibrium1stPriceAsymmetric}
\forall \, i \in \cN,\, \frac{\partial}{\partial b}\log(F_i(x_i(b)))=\frac{-1}{x_i(b)-b}+\frac{1}{n-1}\sum_{j=1}^n \frac{1}{x_j(b)-b}\;. 
\end{equation}	
When the distributions are without atoms, the boundary conditions are 
$x_i(h)=h$ for all $i \in \mathcal{N}$, and there exists $\eta>0$ such that for all $i$, $\beta_i(H)=\eta$. 
\end{theorem2}
The boundary conditions mean that bidders bid their value at $h$ and have a common maximal bid \citep{FibichGaviousAsymmetricFirstPricePerturbApproach2003}.
\paragraph{Numerical issues in computing Nash equilibrium}
Finding the solution to the differential system \eqref{eq:equilibrium1stPriceAsymmetric} is considered hard essentially because the solutions are unstable near the boundary \citep{MarshallNumericalAnalysis1stPrice94}. The case of $n=2$ bidders has received a fair amount of attention both from both theoretical and numerical perspectives \citep{FibichGaviousDynamicalSystem2012}. Another approach finds yet another form of the differential system of equations and expands the functions appearing in it in a fixed polynomial basis.  \citep{NumericalSolutionsFirstPriceGayle2008}. 
\section{A revenue equivalence theorem}
\label{sec_rev_equi}
Revenue equivalence theorems are general results showing that different auction systems  -- sharing nonetheless some properties -- are equivalent in terms of expected revenue \emph{at a specific equilibrium}.  The first revenue equivalence theorem showed that the sealed-bid second-price auction and the ascending/English auction lead to the same revenue for the seller \citep{vickrey1961counterspeculation}. This result was later extended to all standard auctions - under rather minimal assumptions.

\begin{theorem2}%[Revenue equivalence]
\label{Th:RevEquivalence}
Consider the family of auctions that are both

 i) \textsl{standard} (i.e., the winner is the bidder with the highest bid)

 ii) \textsl{0-rational} (i.e., winning the auction with a bid of 0 induces a payment of 0).

\noindent When all bidders have the same value distribution, %and their bids are statistically independent,
 i.e., in the symmetric case, the seller's expected revenue and bidders' expected utilities at a symmetric increasing Nash equilibrium are independent of the  specific payment rule. 
\end{theorem2}
\begin{remark2} The revenue equivalence Theorem \ref{Th:RevEquivalence} applies in particular to first and second price auctions, in the symmetric case where all bidders have the same value distribution \citep{Myerson81} and use an increasing strategy at equilibrium. The revenue equivalence theorem assumes that bidders have all the same value distributions. There exists asymmetric cases where the first-price auction brings more revenue than the second-price auction and vice-versa (\cite{krishna2009auction}, Section 4.3.2). We give in Section \ref{sec:NashIsDifficult} an example showing that the assumption that $\beta$ is increasing is crucial and cannot be dispensed with. 
\end{remark2}
\begin{proof} 
Given a specific standard auction, let $\beta$  be a  strategy at an increasing symmetric Nash equilibrium. Let us denote by $P(t)$ the corresponding expected payment of bidder~$i$ when he bids $\beta(t)$ and the other bidders are using the same strategy $\beta$. As before, we  denote by $G_i$ the distribution of $x^{(1)}_{-i}$, the highest value among all the bidders except bidder~$i$. Using the fact that $\beta$ is increasing, that all players use this strategy - at the Nash equilibrium -  and the fact that the auction is standard, the probability that bidder $i$ wins when he bids $\beta(t)$ is $G_i(t)$.

We can still assume that a deviation of bidder $i$ consists in  bidding $\beta(z)$ instead of $\beta(x_i)$ because the auction is standard and $\beta$ is increasing.  In particular, the expected utility for bidder $i$ of this deviation can be written as
\begin{align*}
\mathcal{U}_i(z,x_i) = x_iG_{i}(z) - P_i(z)\;.
\end{align*}
% If $p_i(t,x_{-i})$ is the payment of bidder $i$ when bidding $\beta(t)$ and the other players have value $x_{-i}$,
% $$P_{i}(t) =  \mathds{E}_{F_{-i}}\Big[p_i(t_i,x_{-i})\Big] = \int_{\mathcal{X}_{-i}} p_i(t,x_{-i})f_{-i}(x_{-i})dx_{-i} \;.$$

Since $\beta$ is a strategy corresponding to a Nash equilibrium, bidder $i$'s expected utility is maximized when he bids $\beta(x_i)$ and hence 
$$
\mathcal{U}_i(x_i,x_i) =\max_{z} \mathcal{U}_i(z,x_i) =\max_{z}  \, x_iG_i(z) - P_i(z)  \;.
$$
We now introduce the mapping ${\mathcal V}_i(x_i)=\mathcal{U}_i(x_i,x_i)$ that is convex,  as the maximum of linear and functions, and hence almost everywhere differentiable \citep{HULLbook01}). Recall also that (Lemma 4.4.1 in \citep{HULLbook01}) the subdifferential at some $x$ of a supremum of convex functions contains the convex hull of the subdifferentials of the functions achieving this supremum (and is empty if the supremum is not achieved). Before proving formally the result, let us give some intuitions. The function $\mathcal{V}_i$ we consider is the maximum (over $z$) of linear mapping in $x$, whose differential are simply $G_i(z)$. As a consequence,  when $\mathcal{V}_i$ is differentiable at $x$ it holds that ${\mathcal V}'_i(x)=G_i(z^*(x))$, where $z^*(x)=\argmax_z \mathcal{U}_i(z,x)$, which suggests that  ``${\mathcal V}'_i(x)=G_i(x)$". 

This intuition is formalized thanks to the envelop theorem \citep{MilgromSegal2002} that holds because $\mathcal{U}_i(z,\cdot)$ is linear for all $z$ and hence differentiable and therefore absolutely continuous. Furthermore, $\frac{\partial \mathcal{U}_i(z,x)}{\partial x} =G_i(z)\in[0,1]$ for all $z$. Since the maximum of $z\mapsto \mathcal{U}_i(z,x_i)$ is attained for $z=x_i$ at the Nash equilibrium, the envelope theorem finally states 
$$
{\mathcal V}_i(x)-{\mathcal V}_i(0)=\int_0^{x}G_i(u)du\;.
$$
However, it also holds that ${\mathcal V}_i(x)-{\mathcal V}_i(0)=xG_i(x)-P_i(x)+P_i(0)$. 
Thus, since $P_i(0)=0$ because of 0-rationality, we finally get that,  integrating by parts,
$$
P_i(x)=P_i(0)+xG_i(x)-\int_0^x G_i(u)du = \int_0^x tdG_i(t) dt=G_i(x)\mathds{E}[x^{(1)}_{-i}|x^{(1)}_{-i} < x]\;.
$$
%\iffalse
%\nek{I changed the argument to above}
%Assuming differentiability, it  holds that, since $\mathcal{U}_i(z,x_i)$ attains its maximum at $x_i$, 
%\begin{equation*}
%\left.\frac{\partial \mathcal{U}_i}{\partial z} (\cdot,x_i)\right|_{z=x_i} = x_ig_i(x_i) - P_i'(x_i) = 0\;.  
%\end{equation*}
%Thus, since $P_i(0)=0$ because of 0-rationality, we finally get that
%\begin{equation*}
%P_i(x) = P_i(0) + \int_{0}^{x} sg_i(s)ds = G_i(x)\mathds{E}[X^{(1)}_{-i}|X^{(1)}_{-i} < x]\;.
%\end{equation*}
%In the case where $\mathcal{U}_i$ is not everywhere differentiable, the same argument holds using standard mean-value theorems for the subdifferential \citep{HULLbook01}, pp.178-180. \nek{check with Thomas, Vianney, Clem} \nek{note to thomas: there's a slight confusion around the function of the first variable and the function $\gamma(x_i)=\mathcal{U}_i(x_i,x_i)$}
%\nek{keep below}
%\fi
Hence the expected payment of bidder $i$ is independent of the specific auction format. As a consequence, so are  the expected seller's revenue and the expected utility of bidder $i$, because the auction is standard.
%\iffalse
%Finally, we assumed in the proof that $P$ is  differentiable (for the sake of simplifying the exposition). This assumption is not necessary and can be removed using the convexity of $x_i \mapsto \mathcal{U}_i(x_i,x_i)$ in a way that is similar to the proof of the more general Theorem \ref{Th:MyersonLemma}. \nek{je pense qu'il faut faire très gaffe; l'exemple d'Hartline a un payoff discontinu (a l'equilibre je sais pas parce que son contre example est faux)}
%\fi

\end{proof}

\begin{remark2} 
This proof shows a principled way to get necessary conditions on bidding strategies forming an increasing Nash equilibrium, through the payment formula derived above.
% COMMENT : c'est fait juste apres
%Finally, a maybe more informal but perhaps more intuitive derivation of this result is the following. Assuming differentiability of $P_i(z)$ and $G_i(z)$, it holds that, since $\mathcal{U}_i(z,x_i)$ attains its maximum at $x_i$, 
%\begin{equation*}
%\left.\frac{\partial \mathcal{U}_i}{\partial z} (\cdot,x_i)\right|_{z=x_i} = x_ig_i(x_i) - P_i'(x_i) = 0\;.  
%\end{equation*}
%Thus, since $P_i(0)=0$ because of 0-rationality, we get 
%\begin{equation*}
%P_i(x) = P_i(0) + \int_{0}^{x} sg_i(s)ds = G_i(x)\mathds{E}[X^{(1)}_{-i}|X^{(1)}_{-i} < x]\;.
%\end{equation*}
\end{remark2}

\subsubsection{Informal derivation of symmetric first-price auction equilibrium strategy} The proof of Theorem \ref{Th:RevEquivalence} is a bit formal and technical as it relies on convex analysis arguments; however, it provides insight on how to easily and informally derive symmetric equilibrium strategies. Denote by $\beta$ the common strategy of an increasing Nash equilibrium of the first-price auction, postulated at this point for this informal derivation to exist. Note that by symmetry, $G_j=G_i=G$ for all $i,j$ and similarly for the pdfs. So we use the notations $G$ and $g$ for cdf and pdf below. Conducting the same computations as in the proof of Theorem \ref{Th:RevEquivalence}, assume that all bidders  but $i$  follow $\beta$ - i.e., bidder $j\neq i$ bids $\beta(x_j)$ if $x_j$ is his value - and that bidder $i$ is bidding $\beta(z)$ instead of $\beta(x_i)$.  Note that because all players are using the strategy $\beta$ and $\beta$ is \emph{increasing}, the probability that bidder $i$ wins the auction is the exactly the probability that $z$ is higher than the largest value of the competition. In other words, the probability that he wins the auction is $G(z)$. His utility in the specific case of a first price auction can then be written as 
\begin{equation*}
\mathcal{U}_i(\beta(z),x_i) = (x_i - \beta(z))G(z)\;. 
\end{equation*}
Since $\bbeta$  is a  symmetric increasing Nash equilibrium,  the maximum utility is attained by bidding $b=\beta(x_i)$. Differentiating with respect to $z$, temporarily assuming that $\beta$ is differentiable, then yields 
\begin{equation*}
0=\left.\frac{\partial \mathcal{U}_i}{\partial z} (\beta(z),x_i)\right|_{z=x_i} = (x_i-\beta(x_i))g(x_i) - \beta'(x_i)G(x_i)\;.
\end{equation*}
Notice that the above equation can be rewritten in the more compact form
\begin{equation*}
(\beta(x)G(x))' = xg(x)\;.
\end{equation*}
Integrating the above equation and using the fact that $G(0)=0$, we can compute explicitly the symmetric equilibrium strategy
\begin{equation*}
\beta(x) = \frac{\int_{s=0}^{x}sg(s)ds}{G(x)}= \mathds{E}[X^{(1)}_{-1}|X^{(1)}_{-1}<x]\;.
\end{equation*}
We can verify as posteriori that $\beta$ is differentiable and increasing when $g>0$. 

This also proves the uniqueness of  increasing differentiable Nash equilibria  in a symmetric first-price auction.

% \begin{remark2} The system of equations corresponding to the Nash Equilibrium in the asymmetric setting presented in Theorem \ref{thm:SystemEquiAsymetricFirstPrice} can actually be easily deduced by applying the same perturbative approach.\nek{this is very unclear; we should drop it I think or do it} \end{remark2}

\section{Deriving revenue-maximizing auctions}
\label{Section:RevenueMax}
We now focus on how the seller can design her auction system to maximize her revenue. A large part of the recent literature on auctions have focused on this objective since most of the auctioneers have a choice in designing the rules of their respective auction platforms.

To compute the optimal revenue-maximizing auction, we assume that the seller has prior knowledge on the distribution $F_i$ on each bidders' valuations.  These value distributions quantify the information that the seller has on each bidder. 

\subsection{The role of reserve prices}
Setting reserve prices is a crucial tool used to improve or maximize seller's revenue in classical auctions. To illustrate its role, we focus on  second-price auctions. 
\begin{definition2}
The reserve price is the minimum price a bidder must pay to acquire an  item. The reserve price is  \textit{anonymous} (respectively, \emph{personalized}) if it is the same for all bidders (resp., if it is different from one bidder to the other).
\end{definition2}
In a second-price auction with reserve prices the buyer that wins the auction must have bid above his reserve price. He then pays the maximum between his reserve price and the second highest bid. A crucial and perhaps surprising point  is that, in expectation,  it can be beneficial for the seller to sometimes not allocate the item. However, whenever the item is sold, the payment is higher with this reserve price than without. 

To illustrate this point, we quickly recall a historical example, the New-Zealand radio spectrum rights auction \citep{milgrom2004putting}. In 1990, the NZ government decided to sell some radio spectrum rights through multiple simultaneous sealed-bid second-price auctions without reserve prices for the corresponding licenses. These auctions were expected to raise around NZ\$250 million. Instead, the government revenue was around NZ\$36 million. On many licenses, there was a huge discrepancy between the first and the second bid. For instance, a firm bid NZ\$100 000 and the second price was only NZ\$36\dots If reserve prices had been set beforehand, it would have ensured to the government that the firm who made a bid of NZ\$100 000 would have paid a price sufficiently high. This example illustrates the importance of setting reserve prices. A first way to do it would be to assume that the seller can compute an intrinsic value for keeping the item. This could be a reasonable assumption for housing or wine auctions. However, in many other situations (TV rights, radio spectrum rights auctions...), the seller does not have any intrinsic value for the item. 

For illustration purpose,  we will compute in the following the optimal reserve price in the simple case of a single buyer. As before, we denote by $F$ the cdf of the buyer valuation, and for the sake of simplicity, we assume it has a density $f$ (again, all the following results generalize to arbitrary distributions, but at the cost of technicalities) and has finite expectation.

%\iffalse and the need for a careful modelisation of the different incentives of the system. {\color{red}VP: la phrase suivante est moche, et il faut une reference et/ou plus d'explications} For instance, in the recent sell of French football TV rights, in 2017, the auctioneer was very careful on its definition of the reserve prices by setting very specific reserve prices for every TV rights packages. 
%\fi
\subsection{The posted price setting: monopoly pricing}
\label{monopoly-pricing}
In this  particularly simple through practically very common setting, designing an auction simply reduces to  a \textit{take-it-or-leave-it} offer, also called \emph{posted price}. In other words, a fixed selling price is offered and a rational bidder will accept to pay it to acquire the item if and only if the price is smaller than his valuation. 
\begin{lemma2}\label{TH:Monopoly}
In a posted price setting, the seller's expected revenue is 
\begin{equation} \label{eq:defPiOfr}
\Pi(r) = r(1-F(r))\;.
\end{equation}
In the same setting, when $F$ is differentiable and has finite expectation, the optimal reserve price, called \emph{the monopoly price}, is a solution of:
\begin{equation*}
0=-[r(1-F(r))]'=rf(r) - (1-F(r)) = f(r)\Big(r- \frac{1-F(r)}{f(r)}\Big)\;. 
\end{equation*}
\end{lemma2}
\begin{proof}
The seller's revenue can be written as a function of $r$ as: in expectation, the seller's revenue is just the fixed price $r$ multiplied by the probability that the buyer buys the item. This latter probability is just the probability that the value of the buyer is above $r$. Formally, if $x$ is the value of the item for the buyer, the revenue of the seller $\Pi(r)$ when selling at the fixed price $r$ is 
\begin{equation*} 
\Pi(r) = r \mathds{P}\{ x\geq r \}=r \int_{r}^{+\infty} f(x)dx = r(1-F(r))\;.
\end{equation*}
The result on the monopoly price follows from differentiating the previous relation, since
\begin{equation*} 
\frac{d\Pi(r)}{dr} = (1-F(r)) - rf(r) \;.
\end{equation*}
Furthermore, choosing $r=0$ or $r$ arbitrarily high gives 0 revenue, the latter because $F$ is assumed to have finite expectation, which then implies that $\lim_{t\tendsto \infty} t(1-F(t))=0$ by the dominated convergence theorem. Indeed, if $x$ has distribution $F$, $t(1-F(t))\leq \mathds{E}_{x\sim F}[x\indicator{x\geq t}]\leq \mathds{E}_{x\sim F}[x]<\infty$.  Hence a maximum of the function $\Pi$ exists among its stationary points, finishing the proof. 
\end{proof}
%\iffalse
%\begin{figure}[H]
%\includegraphics[width=.5\linewidth]{images/monopolistic_revenue}
%\caption{Seller's revenue as a function of the reserve price. The revenue is computed based on $r(1-F(r))$ with $F = \mathcal{U}([0,1])$.}
%\end{figure}
%{\color{red} VP: Quel est l'interet du paragraphe suivant ?} 
%The setting of posted price is very related to the setting where a monopolistic firm sells a service to different customers. She must define a single price for all her clients. She will choose the monopoly price corresponding to the distribution of values representing her customers. Sometimes, the firm prefers not to sell the item and sacrifies a bit of short-term revenue to insure that in most other cases, buyers will pay a price sufficiently high.  
%\fi
\bigskip
Notice that without the assumption that $F$ has a finite expectation, the optimal reserve price could be arbitrarily high: take for instance $F(r)=1-r^{-\alpha}$ with $0<\alpha<1$.
Such an example might actually be relevant in luxury items markets.
 
One of the purpose of discussing the single bidder case was to introduce organically the crucial concept of \textit{virtual value} \citep{Myerson81}. 
\begin{definition2}\label{def:VirtualValue}%[Virtual value]
The virtual value function $\psi : \mathcal{X} \to \mathds{R}$ of a distribution $F$ (with pdf $f$) is: 
\begin{equation} \label{eq:defVirtualValue}
%\boxed.  VP: Pourquoi une box ?
{
\psi(x) = x - \frac{1-F(x)}{f(x)}\;.
}
\end{equation}
\end{definition2}
The virtual value function can be either positive or negative, irrespective of the support of the value distribution. The expectation  under $F$ of the virtual value,  i.e., $\mathds{E}_{x \sim F} [ \psi(x)]$, is actually equal to the infimum of the support of $F$, when $F$ has finite expectation. In particular, it is equal to 0 if the support of $F$ ``starts" at 0. 
%\nek{maybe stop here; afterwards it kind of implies that $\psi_F$ is increasing...} and in that case, the virtual value is necessarily sometimes negative (at least around $x=0$) and sometimes positive (at least around the supremum of the support of $F$) \nek{latter is false unless virtual value is assumed to be increasing}.

The virtual value $\psi(x)$ is a crucial concept that can be interpreted as  \emph{virtual payment}, as we explain now. If the bidder has value $x$ and decides to buy the item, so $x\geq r$, his (virtual) payment can be thought of as $\psi(x)$,  \textsl{independently} of the price $r$ set by the seller. Indeed, the revenue generated by such a price $r$ is
\begin{gather*}
\mathds{E}_{x \sim F} [ \psi(x) \mathds{1}\{x \geq r\}]=
\int_r^\infty \psi(x) f(x)dx= \int_r^\infty xf(x) - (1-F(x)) dx \\
= -\int_r^\infty (x(1-F(x)))' dx=
r(1-F(r))=\Pi(r)\;.
\end{gather*}
The last equality comes from the definition of $\Pi(r)$, see Equation \eqref{eq:defPiOfr}. As a consequence, even though it is traditionally called virtual value, $\psi(x)$ could rather be understood as a virtual \textsl{payment}: the buyer pays on average $\psi(x)$ when his value is $x$ and he buys/wins the item, i.e., $x\geq r$. See also Proposition \ref{prop:virtualValueIsVirtualPaymentGeneralCase} for an explanation of why this interpretation holds for general auction systems and buyers optimizing their expected utility.

\subsubsection{Examples of virtual value functions:} 
\begin{itemize}
\item if $x\sim\mathfrak{U}([0,1])$, the uniform distribution over $[0,1]$, then $\psi(x)=2x-1$ for $x \in [0,1]$.
\item if $x\sim \mathfrak{E}\mathrm{xp}(\lambda)$, the exponential distribution, then  $\psi(x)=x-\lambda$ for $x \in \mathds{R}_+$ ($F(x)=1-\exp(-x/\lambda)$).
\item Generalized Pareto (GP) distributions, parametrized by $(\mu,\sigma,\xi)$ where $\sigma >0$ and $\xi\leq 0$, have cdf 
\begin{equation*}
F_{\mu,\xi,\sigma}(x)   = \begin{cases} 1-(1 +\frac{\xi (x-\mu)}{\sigma})^{-1/\xi} & \text{for } \xi < 0 \\
1-e^{-(x-\mu)/\sigma} & \text{for } \xi= 0\end{cases}\;.
\end{equation*} 
Their virtual value is affine \citep{balseiro2020multistage}
$$\psi_{\mu,\xi,\sigma}(x) = (1-\xi)x+\xi\mu-\sigma$$
\end{itemize}
%\begin{remark2}
%With $\xi = -1 , \mu = 0, \sigma = 1 $, we recover the uniform distribution. With $\xi = 0 , \mu = 0 , \sigma = 1$, we recover the exponential distribution. With  $ \mu = 1, \sigma = \xi$ , this corresponds the power law (a.k.a the Pareto distribution) with parameter $\alpha = 1/\xi$. Hence, the generalized Pareto distribution are widely used in auction theory due to their flexibility (they model a large spectrum of distributions) and the fact that their virtual value is easy to compute (the virtual value is affine).
%\end{remark2}
\subsubsection{Relationship between  monopoly price and virtual value/payment}
Lemma \ref{TH:Monopoly} states that the optimal reserve price against  a single bidder (which was called the monopoly price)  is, in the case where the virtual value/payment function  $\psi$ is increasing and changes sign, necessarily the root of $\psi$ (or the point where the sign changes if $\psi$ is not continuous): if $\Pi(t)$ is the expected revenue of the seller at reserve price $t$ (see Equation \eqref{eq:defPiOfr}), 
$$\Pi'(t)=-f(t)\psi(t)\;.
$$
If $\psi$ is strictly positive everywhere, which can happen if the  infimum of the support of $F_i$ is positive, then the optimal reserve price is that specific point (or equivalently 0).   Quite interestingly, even with multiple other bidders, the optimal reserve price for bidder $i$ is still $i$'s monopoly price, i.e., the same as if he were the only bidder. This is illustrated in the following Section \ref{optimal_reserve_for_lazy} under the same assumptions of $\psi$ being increasing and/or changing sign once. 

As a consequence, in the following,  our main focus will be on these distributions which are called \textsl{regular}. The results we will prove can be generalized to non-regular distribution with a technique called \textsl{ironing}, see Section \ref{Se:Ironing}.  
 \begin{definition2} 
The distribution $F$ is  regular if its corresponding virtual value $\psi$ is increasing.
\end{definition2}

The uniform, exponential and generalized Pareto distributions with $\xi <1$, are all regular distributions. % 
%\begin{proposition2}
%If $\psi$ increasing and crossing zero exactly once and F is defined over its full support, then r is the monopoly price if and only if
%$$\psi(r) = 0$$ 
%\end{proposition2}
%\begin{proof}
%The monopoly price r is defined as $$\argmax_r R(r) = r(1-F(r)).$$
%Then, 
%$$
%\frac{dR(r)}{dr} = -f(r)\psi(r)\;.
%$$
%Since $\psi$ increasing, R is concave and $\psi(r)=0$.
%\end{proof}
%If $\psi$ is crossing zero multiple times, the monopoly price also verifies $\psi(r)=0$. 
%One of the beauty of auction theory is that this notion of monopoly price can be extended to auctions with multiple buyers.
\subsection{Optimal reserve prices in a second-price auction}
\label{optimal_reserve_for_lazy}
We  focus in this section on second-price auctions with reserve prices \citep{riley1981optimal} where  $n$ buyers are asymmetric, i.e., their value distribution can be different. As a consequence, the seller might also set different, personalized, reserve prices so as to increase her revenue. When introducing the concept of reserve price, we mentioned that a bidder can only win the auction if his bid was higher than his reserve price and that the latter is the minimal payment that bidder might pay. There however remains some ambiguity on how the auction unfolds (depending on which condition ``highest bidders" or ``bid above reserve price" is checked first). As a consequence, there exist at least two different types of second-price auction with reserve prices.

\begin{description}
\item[``Lazy'' 2nd-price auction:] The winner can only be the highest bidder. He gets the item only if he  clears his reserve price (i.e., he bids above it), and pays the maximum between his reserve price and the second highest bid overall (regardless of whether the second highest bid cleared its reserve).
\item[``Eager'' 2nd-price auction:] Bidders that have not cleared their respective reserve price are  disregarded. Thus the winner is the highest bidder amongst those that have cleared their reserve price and he pays the maximum between his reserve price and the second highest cleared bid.
\end{description}

First of all, notice that if the reserve prices are anonymous, i.e., the same for all bidders, as they should be in the symmetric case for instance, both types of auctions coincide.  Optimal reserve prices are easy to compute in lazy auctions, as they have an explicit form. They are on the other hand hard to compute for eager auctions. Moreover, the eager 2nd-price auction is also not the revenue-maximizing auction for the seller. So this concept is neither simple (as is the lazy auction) nor optimal (as is the Myerson auction, see Section \ref{emp_Myerson}). As a consequence, we will not put too much emphasis on eager auctions. In practice, if one wishes to implement eager 2nd-price auctions, a good idea would be to use the reserve prices of the corresponding lazy 2nd-price auction.

\medskip

It is quite immediate to see that lazy and eager second price auctions are still DSIC mechanism (the proof follows the exact same lines as that without reserve prices), hence we shall again only consider the truthful equilibrium. We now derive the expected payment of a bidder at this equilibrium.
%\begin{theorem2}[Expected payment in a lazy second-price auction]\label{myerson_for_second_price}
%Consider bidder $i$ with virtual value $\psi_i$ and reserve price $r_i$. We denote by $G_i$ the cdf of $Y_{-i} = max_{j\neq i} X_j$. The expected payment $P_i(\beta_{tr})$ of bidder $i$ in a lazy second-price auction when all bidders are bidding truthfully is equal to:
%\begin{equation*}
%P_i(\beta_{tr}) = \mathds{E}_{x_i\sim F_{i}}\bigg(\psi_i(x_i)G_i(x_i)\textbf{1}_{x_i\geq r_i}\bigg)\;.
%\end{equation*}
%\end{theorem2}
\begin{theorem2}\label{myerson_for_second_price}
Let   $P_i$ be the expected payment of bidder $i$ facing reserve price $r_i$ at the truthful equilibrium of a lazy second-price auction. %Assume the bidders are statistically independent. 
Then 
\begin{equation*}
P_i = \mathds{E}_{x_i\sim F_{i}}\Big[\psi_i(x_i)G_i(x_i)\mathds{1}\{x_i\geq r_i\}\Big]\;,
\end{equation*} 
where $G_i$ still denotes the cdf of $\max_{j\neq i} x_j$.

\end{theorem2}
\begin{proof} 
Let us introduce the notation $y_{-i}= \max_{j\neq i} x_j$ so that the pointwise payment of bidder $i$, when he has value $x_i$, given all the values $x_j$ is equal to: 
\begin{equation*}
\max\{r_i,y_{-i}\}\mathds{1}\{x_i\geq y_{-i}\}\mathds{1}\{x_i\geq r_i\}\;,
\end{equation*}
We note that 
$$
\max\{r_i,y_{-i}\}\mathds{1}\{x_i\geq y_{-i}\}\mathds{1}\{x_i\geq r_i\}=
\mathds{1}\{x_i\geq r_i\}\big(r_i\mathds{1}\{y_{-i}\leq r_i\}+y_{-i}\mathds{1}\{x_i\geq y_{-i}\}\mathds{1}\{ y_{-i}\geq r_i\}\big)\;.
$$
As a consequence,
\allowdisplaybreaks
\begin{align*}
P_i&= \mathds{E}_{\bF}\Big[\max\{ r_i,y_{-i} \}\mathds{1}\{x_i\geq y_{-i}\}\mathds{1}\{x\geq r_i\}\Big]
\\& = \int_{x_i} \int_{y_{-i}} \max\{r_i,y_{-i}\}\mathds{1}\{x_i\geq y_{-i}\}\mathds{1}\{x_i\geq r_i\}f_i(x_i)g_i(y_{-i})dy_{-i}dx_i
\\& =\Big(1-F_i(r_i)\Big)G_i(r_i)r_i + \int_{x_i=r_i}^{+\infty}\int_{y_{-i}=r_i}^{+\infty}y_{-i}g_i(y_{-i})f_i(x_i)\textbf{1}(x_i \geq y_{-i})dy_{-i}dx_i
\\& = \Big(1-F_i(r_i)\Big)G_i(r_i)r_i + \int_{y_{-i} = r_i}^{+\infty}y_{-i}g_i(y_{-i})\int_{x_i=y_{-i}}^{+\infty}f_i(x_i)dy_{-i}dx_i \hspace{2cm} \text{ (by Fubini)}
\\& = \Big(1-F_i(r_i)\Big)G_i(r_i)r_i + \int_{y_{-i} = r_i}^{+\infty}y_{-i}g_i(y_{-i})(1-F_i(y_{-i}))dy_{-i} 
\\& = \int_{y_{-i} = r_i}^{+\infty} \Big(f(y_{-i})y_{-i} - (1-F_i(y_{-i}))\Big)G_i(y_{-i})dy_{-i} \hspace{2.5cm} \text{(by integration by parts)}
\\& = \mathds{E}_{F_i}\Big[\psi_i(x)G_i(x)\mathds{1}\{x\geq r_i\}\Big]\;.
\end{align*}
\end{proof}
We can therefore easily derive the optimal reserve prices, as a function of $\psi_i$ at least for regular distributions.%\begin{definition2} 
%The distribution F  is said regular if its corresponding virtual value $\psi$ is increasing.
%\end{definition2}
%\redtext{changement}
%{\color{red} VP: est-ce qu'on a besoin de ce graphique ? En plus celui en haut a droite deconne, le vert n'est pas aligne avec le trait pointille rouge}
%\begin{figure}[h!]
%\begin{tabular}{cc}
%\includegraphics[width=.50\linewidth]{images/r025}&
%\includegraphics[width=.50\linewidth]{images/r075}\\
%\multicolumn{2}{c}{\includegraphics[width=.50\linewidth]{images/r05}}
%\end{tabular}
%\caption{\textbf{Change in bidder payment as a function of the reserve price.} The value distribution of the bidder is uniform [0,1] with $\psi(x) = 2x -1$, and is represented by the blue line. The dashed red vertical line corresponds to the current reserve price. We picked $G=1$, i.e., no competition. The bidder payment is equal to the area under the curve. The seller has no incentive to set the reserve price lower than $\psi_i^{-1}(0)$ since it decreases her total expected payment (because of the negative contribution of the red area). It is also clear that $r \ge \psi_i^{-1}(0)$ is suboptimal since it results in lost revenue for the seller.}
%\label{fig:visualOptimalR} % I can do without the label too
%\end{figure}
\begin{theorem2}\label{thm:optimalReservesLazySecondPrice}%[Optimal reserve price in lazy second-price auction]
If  $F_i$ are regular, the optimal reserve prices in a lazy second-price auction are:
\begin{equation*}
\bigg(r_1,\dots,r_n\bigg) = \bigg(\psi_1^{-1}(0),\dots,\psi_n^{-1}(0)\bigg)\;,
\end{equation*}
with the convention that $\psi_1^{-1}(0)$ is the minimum of the support of $F_i$ is $\psi_i$ is  positive everywhere and the point where $\psi_i$ changes its sign if it is discontinuous.
\end{theorem2}
\begin{proof} 
The seller maximizes the sum of expected payment: 
\begin{align*}
\mathds{E}_{\bF}\Big[\sum_{i=1}^{n} P_i\Big] &= \sum_{i=1}^{n}  \mathds{E}_{F_i}\Big[\psi_i(x_i)G_i(x_i)\mathds{1}\{x_i\geq r_i\}\Big]\;.
\end{align*}
Since $F_i$ is regular, $\psi_i$ has one zero and is negative before  and positive after. Thus, the optimal choice for $r_i$ is  $\psi_i^{-1}(0)$ as the function $t\rightarrow \mathds{E}_{F_i}\Big[\psi_i(x_i)G_i(x_i)\mathds{1}\{x_i\geq t\}\Big]$ is increasing before $\psi_i^{-1}(0)$ and decreasing afterwards, owing to the sign of the integrand on both sides of $\psi_i^{-1}(0)$.
\end{proof}
The proof indicates that in a lazy second-price auction, the seller can safely maximize the payment of each bidder one by one independently. Indeed, in a lazy second price auction, changing the reserve price of one specific bidder does not change the probability of winning and the payment of the other bidders. This is not the case for eager second-price auctions on the other hand, which explains the complexity of computing the optimal reserve prices in them. Finally, the optimal reserve prices in a lazy second-price auction correspond to the monopoly prices of each bidder.

\begin{corollary2}The optimal reserve price for a bidder in a lazy second-price auction is independent of the presence, or not, of other bidders. In particular, it is the same as in the situation where he is the only bidder.
\end{corollary2}

\begin{figure}[H]
\includegraphics[width=.6\linewidth]{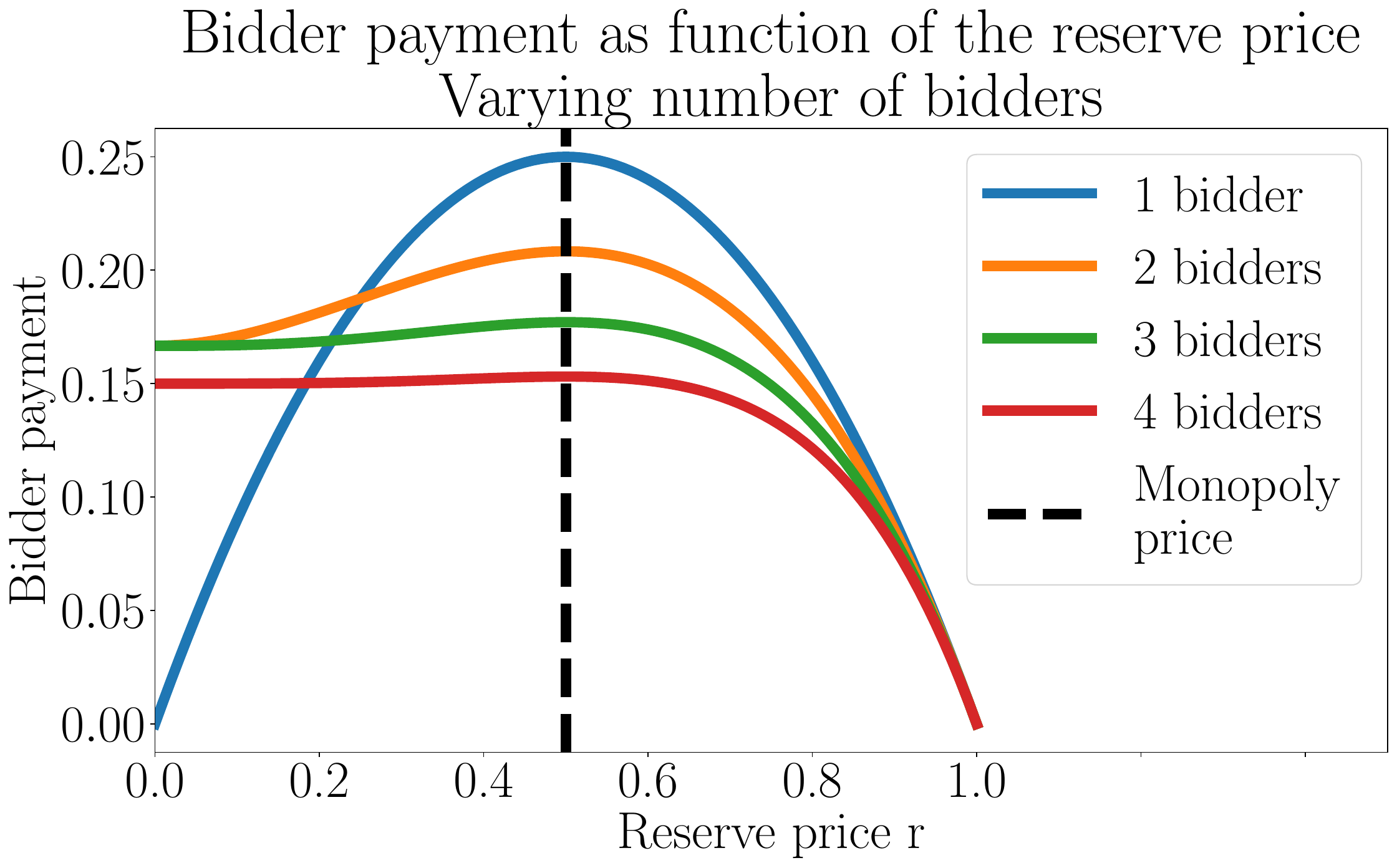}
\caption{Bidder's payment as a function of the reserve price depending on the number of players in second-price auction with bidders all having a uniform value distribution $\mathfrak{U}([0,1])$.}
\end{figure}
%The proof also shows that when the distribution is not regular, the optimal reserve prices for each bidder verifies: $r_i \in \argmax(\mathds{E}_{F_i}(\psi_i(x)G(x)\textbf{1}_{x\geq r_i}))$.

The seller's revenue increases using personalized reserve prices when bidders have very different value distributions. Intuitively, it is in  her best interest to set a high reserve price to bidders with high values most of the time  (or very high values sometimes) and low reserve prices to bidders with  low values most the time.

\bigskip
So far, we have only focused on the seller's revenue when designing auctions. An alternative objective can be the maximization of the  \textsl{global welfare} of the system, which is the sum of the seller's revenue and all bidders' utility.

Even though reserve prices largely increase the seller's revenue, they actually significantly decrease the expected total welfare, as the item will sometimes not be allocated. This happens when all bidders (or at least the highest one in lazy auctions) have values below their reserve prices.

\begin{example2}
To illustrate this decrease in welfare, we are going to consider a simple example.  There are $n=2$ symmetric bidders with a value drawn uniformly over $[0,1]$. Because of the  symmetry, the optimal reserve price is the same for both bidders hence lazy and eager auctions coincide (and we do not need to specify the rule).  In this simple case, $r^* = \psi^{-1}(0) = 1/2$. As a consequence, the item is allocated as soon as one  bidder bids above 1/2, which happens with probability $3/4$. Otherwise, the item is not sold (which obviously happens with probability $1/4$). Simple computations show that changing the design from a second-price auction without reserve price to a second-price auction with optimal reserve price yields 
\begin{itemize}
\item a {12,5 \% decrease} of the global welfare (from $2/3$ to $7/12$).
\item a {50 \% decrease} of every single  bidder's utility  (from $1/6$ to $1/12$).
\item a {25 \%  increase} of the seller's revenue (from $1/3$ to $5/12$).
\end{itemize}
\end{example2}
\subsection{Myerson's lemma and characterization of BIC and DSIC auctions}
We mentioned before that the virtual value could (and maybe should) be understood as a virtual payment in a single bidder auction and/or in lazy second price auctions. The following lemma (that will be referred to as the ``Myerson Lemma'') is a crucial result \citep{Myerson81}. It states that the virtual payment of a bidder is the correct quantity to study in any  incentive-compatible auctions, and not just lazy second-price ones. The proof, while conceptually profound, is not very hard technically and requires integration by part and Fubini's theorem as we used in the case of the lazy second-price auction. This result holds for any Bayesian Incentive Compatible auction (if bidding truthfully is weakly dominant) that is 0-rational (if bidding 0 ensures a payment of 0). The latter assumption can be weakened, at the cost of an additive constant in the payment formula. 
\begin{theorem2}%[Myerson lemma]
\label{Th:MyersonLemma}
For any BIC and 0-rational auction, the expected payment of bidder $i$ at the truthful equilibrium is 
\begin{equation*} 
\mathds{E}_{x_i\sim F_i}[P_i(x_i)] = \mathds{E}_{x_i \sim F_{i}}[\psi_i(x_i)Q_i(x_i)] = \mathds{E}_{x \sim \bF}[\psi_i(x_i)q_i(\bx)]\;.
\end{equation*}
Here $Q_i(x_i)$ is the winning probability  of bidder $i$  at the truthful equilibrium given his value  $x_i$ and $q_i(\bx)$ is the probability that the item is attributed to bidder $i$ when the values are $\bx=(x_1,\ldots,x_n)$. 
\label{myerson_lemma}
\end{theorem2}
\begin{proof}
Let us consider only  the truthful equilibrium of the BIC auction. In particular, we assume that all bidders except possibly $i$ bid their values, i.e., they bid truthful. Let us call $Q_i(z)$ the probability that bidder $i$ bidding $z$ wins the auction (when all other bidders bid truthful) and $P_i(z)$ his expected payment (when all other bidders bid truthful). The expected utility of bidder $i$ when he has value $x_i$ and bids $z$ is simply $Q_i(z)x_i - P_i(z)$. 
By definition of Bayesian incentive-compatibility, at the truthful equilibrium the auction must verify: 
\begin{equation*}
\forall z \in \mathds{R}^{+}, \forall x_i \in \cX_i,\ Q_i(x_i)x_i - P_i(x_i) \geq Q_i(z)x_i - P_i(z)\;,
\end{equation*}
since at the truthful equilibrium bidder $i$ gets maximum utility by bidding his value. 
Thus, if we still denote by $\mathcal{V}_i(x_i)$ the expected utility of bidder $i$ when he has value $x_i$, 
\begin{equation*}
\mathcal{V}_i(x_i) = \max_{z} Q_i(z)x_i - P_i(z) = x_i Q_i(x_i)-P_i(x_i)\;.
\end{equation*}
As a consequence, $\mathcal{V}_i$ is a  convex mapping (the maximum of affine mappings) and therefore is differentiable almost everywhere and absolutely continuous (i.e., it is equal to the integral of its derivative).

Bayesian-incentive compatibility also implies that
\begin{equation*}
\forall z \in \cX_i,\ \mathcal{V}_i(z) =\max_t zQ_i(t)-P_i(t)\geq z Q_i(x_i)-P_i(x_i)= \mathcal{V}_i(x_i) + Q_i(x_i)(z-x_i)\;.
\end{equation*}
Since $\mathcal{V}_i$ is convex, this means that $Q_i(x_i)$ belongs to the subdifferential of $\mathcal{V}_i$ at $x_i$, i.e.,
\begin{equation*}
Q_i(x_i) \in \partial \mathcal{V}_i(x_i)\;,
\end{equation*}
and  $\nabla \mathcal{V}_{i}(x_i) = Q_i(x_i)$  if  $\mathcal{V}_i$ differentiable at  $x_i$. Therefore, using Theorem  D.2.3.4 in \citep{HULLbook01}, 
$$
\mathcal{V}_i(x_i)-\mathcal{V}_i(0)=\int_0^{x_i}Q_i(z)dz\;,
$$
and since $\mathcal{V}_i(x_i)=x_i Q_i(x_i)-P_i(x_i)$, 
\begin{equation*}
P_i(x_i) = P_i(0) + Q_i(x_i)x_i - \int_{0}^{x_i}Q_i(z)dz\;.
\end{equation*}
Taking expectation over $x_i$ now gives: 
\begin{align*}
 \mathds{E}_{x_i\sim F_{i}}[P_i(x_i)] &= \int_{x_i}P_i(x_i)f_i(x_i)dx_i 
\\&=  P_i(0) + \int_{x_i} \Big(Q_i(x_i)x_i - \int_{0}^{x_i}Q_i(z)dz\Big)f_i(x_i)dx_i
\\& = P_i(0) + \int_{x_i} Q_i(x_i)x_if_i(x_i)dx_i - \int_{z}Q_i(z)\Big(\int_{z}^\infty f_i(x_i)dx_i\Big)dz \qquad \text{ (Fubini) }
\\& = P_i(0) + \int_{x_i} \Big(x_i - \frac{1-F_i(x_i)}{f_i(x_i)}\Big)Q_i(x_i)f_i(x_i)dx_i 
%\\&= P_i(0) + \int_{x} \Big(x_i - \frac{1-F_i(x_i)}{f_i(x_i)}\Big)q_i(x)f(x)dx
\\& =  
 P_i(0) + \mathds{E}_{x_i \sim F_i}[\psi_i(x_i)Q_i(x_i)]\;.
% P_i(0) + \mathds{E}_{x \sim \mathcal{F}}[\psi_i(x_i)q_i(x)] =
\end{align*}
The result follows from the fact that $P_i(0) =0$ since the auction is 0-rational.

The last equality comes from the fact that $Q_i(x_i)= \mathds{E}_{\bx \sim \bF}[q_i(\bx)|x_i]$ and the tower property of conditional expectations.
\end{proof}
\begin{remark2}
Theorem \ref{myerson_for_second_price} is a direct consequence  of this result with the specific choice of $Q_i(x_i) = G_i(x_i)\mathds{1}\{x_i\geq r_i\}$.
\end{remark2}
Myerson's lemma indicates that the expected payment of a 0-rational BIC auction only depends on the allocation rule and the virtual value; the proof actually gives a characterization of any incentive-compatible auction.  However, this characterization is slightly different for  BIC and DSIC auctions.

\begin{corollary2}%[characterization of BIC auctions]
[\cite{Myerson81}]
\label{characterization_inc_comp}
Using the notations of Theorem \ref{Th:MyersonLemma}, an auction is 0-rational and BIC if and only if
\begin{itemize}
\item[i)] the allocation rule is monotone, i.e.,  the probability of winning, as a function of the bid, is non-decreasing (for any fixed bids of others bidders)  and
\item[ii)] the expected payment verifies 
$$P_i(x_i) = Q_{i}(x_i)x_i - \int_{0}^{x_i} Q_{i}(z)dz.$$
\end{itemize}
\end{corollary2}
\begin{remark2}
Using the fact that $Q_i(x_i)= \mathds{E}_{\bx \sim \bF}[q_i(\bx)|x_i]$, we see that given an allocation rule $q_i(\bx)$, the expected payment requirement can be fulfilled by the requiring, auction by auction, an expected payment, given the vector of bids/values $\bx$, of $p_i(x)=xq_i(\bx)-\int_0^x q_i(\bx) dx_i$. (In the last integral all the bids $\bx_{-i}$ are fixed and the integral is performed over $x_i$ which varies from 0 to $x$.) 
\end{remark2}

\begin{proof} 
The proof of Theorem \ref{myerson_lemma} gives the first implication. For the reverse, 
let us assume that all bidders except $i$ bid truthfully; and let us show that $i$ has an incentive to also bid truthfully. This will show that truthful bidding constitutes a Nash equilibrium and hence the auction is BIC. 

Note that because we have assumed that all other bidders bid truthfully, if bidder $i$ bids $z$, the probability that he wins is $Q_i(z)$. Hence, the expected utility derived by bidder $i$ when bidding $z$ and his value is $x_i$ is 
$$
\mathcal{U}_i(z,x_i)=x_i Q_i(z)-P_i(z)=(x_i-z)Q_i(z)+\int_0^z Q_i(t) dt\;.
$$
The second equality comes from assumption ii). 
Let us call $b_i^*$, the optimal bid of bidder $i$. 
% \nek{version 1}
% $b_i^*$ satisfies
% $$
% b_i^*=\argmax_z (x_i-z)Q_i(z)+\int_0^z Q_i(t) dt\;.
% $$
% Let us now show that $b_i^*=x_i$, which would conclude the proof.
% Consider the function $H: y\mapsto \int_0^y Q_i(t) dt$\;. This function is differentiable, with derivative $H'(y)=Q_i(y)$; as a consequence of assumption i) it is convex, since $Q_i(y)$ is non-decreasing in $y$. Therefore it is above all of its tangents, i.e~:
% $$
% \forall z\;, H(x_i)\geq H(z)+(x_i-z)H'(z)=H(z)+(x_i-z)Q_i(z)\;.
% $$
% In other words,
% $$
% \forall z\;, \mathcal{U}_i(x_i,x_i)=\int_0^{x_i}Q_i(t)dt\geq (x_i-z)Q_i(z)+\int_0^z Q_i(t) dt=\mathcal{U}_i(z,x_i)\;.
% $$
% This shows that $b_i^*=x_i$ and the auction is BIC.
%
% \nek{alternative}
Let us now show that $b_i^*=x_i$. To do so, we simply need to establish that 
$$
\forall z\in \mathds{R}_+,\; \mathcal{U}_i(x_i,x_i)=\int_0^{x_i} Q_i(t) dt \geq (x_i-z)Q_i(z)+\int_0^z Q_i(t) dt\;.
$$
This is equivalent to showing that 
$$
\forall z \in \mathds{R}_+\;, \int_z^{x_i} Q_i(t) dt \geq (x_i-z)Q_i(z)\;.
$$
If $z\leq x_i$, since $Q_i$ is non-decreasing, $Q_i(t)\geq Q_i(z)$ on $[z,x_i]$ and hence 
$$
\int_z^{x_i} Q_i(t) dt \geq (x_i-z)Q_i(z)\;.
$$
If $z\geq x_i$, since $Q_i$ is non-decreasing, $Q_i(t)\leq Q_i(z)$ on $[z,x_i]$ and hence
$$
\int_{x_i}^z Q_i(t) dt \leq (z-x_i) Q_i(z)\;.
$$
Multiplying the previous inequality by $(-1)$ on both sides shows that if $z\leq x_i$, we also have 
$$
\int_z^{x_i} Q_i(t) dt \geq (x_i-z)Q_i(z)\;.
$$
So we have shown that 
$$
\forall z\in \mathds{R}_+,\ \mathcal{U}_i(x_i,x_i)\geq \mathcal{U}_i(z,x_i)\;.
$$
Therefore, bidding truthfully is an optimal strategy for bidder $i$ and the auction is BIC. %(Note that the inequalities needed also follow from realizing that the function $H: y\mapsto \int_0^y Q_i(t) dt$ is convex.) COMMENT: I think adding alternative arguments just complicate things and scramble the message

% if the auction verifies
% $$P_i(x_i) = Q_{i}(x_i)x_i - \int_{0}^{x_i} Q_{i}(z)dz\;,$$
% then it implies that the auction verifies  $\nabla \mathcal{U}_i(x_i) = Q_{i}(x_i)$.\nek{do you mean this with subdifferentials?}
% \nek{I don't understand below}
% Since the probability of winning is increasing as a function of the bid, $$\mathcal{U}_i(x_i) - \mathcal{U}_i(z) = \int_{x_i}^{z}Q_{i}(s)ds \geq Q_i(x_i)(z - x_i)$$ which entails that $$\mathcal{U}_{i}(x)\geq Q_i(z)x - P(z).$$ This is the definition of bayesian incentive-compatibility (BIC).
%
%
% \nek{alternative}
% not clear yet that we know the gradient of $U_i$.
%
% Increasing allocation+others are truthful means $PWin(z)=Q_i(z)$ if we use $z$ instead of $x_i$ in $\beta(z)$.
%
% Now optimal strategy means $U_i(x_i)=\max_z x_iQ_i(z)-P(z)$. use convexity.
%
% see krishna pp.64-65 and possibly 69.
%
% Since the probability of winning is increasing as a function of the bid, the function
% $\mathcal{U}_i(x_i)$ is convex since its derivative is increasing. It is in particular above all of its tangents, hence
% $$
% \forall z\;,\mathcal{U}_i(x_i)\geq \mathcal{U}_i(z)+\mathcal{Q}_i(z)(x_i-z)\;.
% $$
% Since $\mathcal{U}_i(z)=z\mathcal{Q}_i(z)-P(z)$, we have
% $$
% \forall z\;,\mathcal{U}_i(x_i)\geq x_i \mathcal{Q}_i(z)-P(z)\;.
% $$
% This is the definition of bayesian incentive-compatibility (BIC).

\end{proof}
\begin{corollary2}%[characterization of DSIC auctions]
[\cite{Myerson81}]
\label{corollary_charect_DSIC}
An auction is DSIC if and only if
\begin{itemize}
\item[i)] the allocation rule is monotone and
\item[ii)]  the payment  of the winning bidder is the minimum bid guaranteeing that he would still have won the auction.\end{itemize}
 Given a monotone allocation rule and assuming 0-rationality, the payment rule is unique. 
\end{corollary2}
\begin{proof} 
The proof is almost identical, one just needs to make the various computations pointwise (for any vector $x_{-i}$) instead of in expectation.
\end{proof}
This characterization can be extended to very general mechanisms \citep{archer2001truthful}.
\subsection{The Myerson auction: revenue maximization for BIC auctions}
After having established the Myerson lemma, it is now possible to derive the  revenue-maximizing auction among all BIC auctions. 
\begin{definition2}\label{def:MyersonAuction}%[Myerson auction]
The Myerson auction, for regular value distribution $F_i$ with associated virtual value $\psi_i$,  is defined by the two following rules: 
\begin{description}
\item[Allocation rule:] Given the bids $\bb=(b_1,\ldots,b_n)$, the winner is the bidder with the highest non-negative virtual value $\psi_i(b_i)$, i.e.,
$$
q_i(\bb) = \mathds{1}\Big\{ i = \arg\max \big\{ \psi_j(b_j) \, ; \, j \text{ s.t. } \psi_j(b_j) \geq 0 \big\} \Big\}
$$
with the convention that if  all virtual values are negative, then the item is not allocated and $q_i(\bb) = 0$. Ties are broken arbitrarily.
\item[Payment rule:] If bidder $i$ wins the auction, he pays
$$p_i(\bb)=\max \Big\{ \psi_i^{-1}(0), \psi_i^{-1}\Big(\max_{j\neq i} \big\{\psi_j(b_j)\big\}\Big) \Big\}$$
\end{description}
\end{definition2}
This auction amounts to running a second price auction with reserve prices 0 among the \emph{virtualized bids} $\psi_k(b_k)$ and converting back this ``virtual cost" in the original bid space of the winner $i$ through the function $\psi_{i}^{-1}$\;.
\begin{theorem2} \label{theorem_regular_distributions}
If $F_1,\dots,F_n$ are regular, the Myerson auction maximizes seller's revenue among all BIC and  interim-IR  auctions.
\end{theorem2}
\begin{proof}
The Myerson auction is BIC as it verifies the condition of Corollary \ref{characterization_inc_comp}. Since $\psi_i$ are non-decreasing (as $F_i$ are regular), the probability of winning is non-decreasing. %\nek{do we need to explain how to check the payment condition?}

To show individual-rationality, we remark that since the auction is BIC, $$\mathcal{V}_{i}(x_i) = \mathcal{V}_{i}(0) + \int_{0}^{x_i}Q_i(s_i)ds_i = -P_{i}(0) + \int_{0}^{x_i}Q_i(s_i)ds_i \geq 0,$$ because $P_i(0) = 0$.  Thanks to  Myerson's lemma, Theorem \ref{Th:MyersonLemma}, the payment of each BIC auction is equal to 
$$
 P_i(0) + \mathds{E}_{x_i \sim F_{i}}[\psi_i(x_i)Q_i(x_i)]\;. 
$$
The Myerson auction maximizes the two terms of this expression since for any rational auction, $P_i(0)\leq0$.  Since the winner in the Myerson auction is the bidder who verifies
\begin{equation*} 
\psi_i(x_i) = \max_{j \in S}\psi_j(x_j)\;,
\end{equation*}
and the item is not allocated when all $\psi_i$ are negative, the second term, is also maximized pointwise. 
Indeed, note that we can rewrite this second term
$$
\mathds{E}_{x_i \sim F_{i}}[\psi_i(x_i)Q_i(x_i)]=\mathds{E}_{\bx \sim \bF}[\langle\psi(\bx),q(\bx)\rangle]\;,
$$
where $\langle \cdot,\cdot\rangle $ is the standard inner product. 
\end{proof}
\begin{corollary2} In the symmetric case, the second-price auction with reserve prices set to  monopoly prices is the revenue-maximizing auction.
\end{corollary2}
\begin{remark2} 
The seller can increase her revenue if the mechanism is  only required to satisfy ex-ante rationality instead of interim rationality, as shown in 	
\citep{cremer1988full}. Indeed, there exists a BIC auction that is ex ante individually rational that accomplishes full-surplus extraction for the seller. In other words, the utility of bidders in this auction is equal to zero. This auction is not interim individually-rational since the expected utility when the bidder's value is zero is strictly negative. This setting of ex-ante individual rationality only makes sense when bidders have to decide to take part in the auction before understanding their value for the item. We shall come back in more details to this setting in Section \ref{sub_full_surplus_extraction}.
\end{remark2}
\subsection{Generalization of optimality result}
\label{SE:General_OptimalityResults}
\subsubsection{Non Incentive-Compatible mechanism: the revelation principle}
Myerson's optimality result can be extended to any incentive-compatible auctions, as long as there is a Nash equilibrium between bidders.
\begin{theorem2}%[Revelation principle]
Given a mechanism and a specific Nash equilibrium for this mechanism, there exists another BIC mechanism where the bidders' expected utility and seller's revenue at the truthful equilibrium are equal to the ones at the original Nash equilibrium. 
\end{theorem2}
\begin{proof}
Consider a mechanism and  $\bbeta = (\beta_1,\dots,\beta_n)$ a profile of strategies that is a Nash equilibrium. This mechanism is defined by an allocation rule $q:\bcB\rightarrow \mathds{R}^{n}$ and a payment rule  $p:\bcB\rightarrow \mathds{R}^{n}$. The mechanism $(q \circ \bbeta, p \circ \bbeta)$ is then clearly BIC and bidding truthful generate the same bids distributions, allocation and payment as in the original mechanism; hence utilities and revenue are unchanged.
\end{proof}
\begin{corollary2} 
If value distributions are regular, the Myerson auction is the revenue-maximizing mechanism among all individually-rational mechanisms which have a Nash equilibrium.
\end{corollary2} 
\begin{proof}
This is a direct application of the revelation principle.
\end{proof}
We now extend the Myerson auction to cases where value distributions are not regular. 
\subsubsection{Non-regular distribution: the ironing technique}\label{Se:Ironing}

If $F_i$'s are not regular, the Myerson auction is not always defined as $\psi_i$'s may not be invertible. From an allocation standpoint, since $\psi_i$ is not necessarily increasing,  the allocation rule may not be monotone. Hence a bidder might have incentive to ``shade" (or lower) his bid to increase his virtual bid and his probability of winning. Non-regular distributions are not uncommon, as a mixture of two distributions typically is not regular (for instance the mixture of two uniforms, Gaussians, etc...). It is therefore crucial to adapt the Myerson mechanism to non-regular distributions. The canonical way is to define a slightly different allocation rule based on a modified virtual value \citep{Myerson81} called the \textit{ironed virtual value}. 

One key consequence of having a non-decreasing virtual value is that the monopoly revenue $\Pi(r) = r(1-F(r)) = \int_r^\infty \psi(v)f(v)dv$ is ``almost''  concave, in the sense that $\Pi\circ F^{-1}$ is concave on $[0,1]$ with derivative  $-\psi\circ F^{-1}(\cdot)$. The ironing technique consists in replacing $\Pi\circ F^{-1}$, which is not necessarily concave, by its concavification, a.k.a., its least concave majorant.  %As a consequence, optimizing the monopoly revenue boils down to finding the root of $\psi$.  

\begin{definition2}\label{def:Concavification}
For a function $h$ defined on some set $\mathcal{E} \subset \mathds{R}^d$, we call  $\mathrm{cav}(h)$ is the concavification of the  function $h$, which is its smallest concave majorant, i.e., the smallest concave function above $h$: its hypograph is the convex hull of the hypograph of $h$. Moreover, this function is defined pointwise as 
$$
\cav{h}(x) = \sup \Big\{ \mathds{E}_\mu[h(z)] \, ;  \mu \text{ is a probability distribution on } \mathcal{E} \text{ such that } \ \mathds{E}_\mu [z] =x \Big\} 
$$
\end{definition2}
We refer to \citep{Rockafellar70convexanalysis}, p. 36, \citep{HULLbook01} pp.98-102 and \citep{GroeneboomNonParamEstimShapeConstraints2014} pp.55-57 for properties of least concave majorant, greatest convex minorant and convex hull of functions. In particular, if $h$ is bounded and attains its maximum, $\cav{h}$ has the same maximum attained (at least) on the convex hull of the set of maximizers of $h$. Moreover, $h$ and $\cav{h}$ are equal on the extreme points of the definition set of $h$; this implies that if $h$ is defined on $[a,b]$, then necessarily $\cav{h}(a)=h(a)$ and $\cav{h}(b)=h(b)$.

We can now define the ironed virtual value. 
\begin{definition2} \label{def:IronedVirtualValue}
For any non-regular distribution $F$, the \emph{ironed virtual value} of $\psi$, denoted by $\tilde{\psi}$ is defined by
$$
 \tilde{\psi}(x)= \partial \Big(-\mathrm{cav}(\Pi\circ F^{-1})\Big)(F(x))\;, \text{ where } \partial \text{ denotes the subdifferential}. 
$$
\end{definition2}

In general, the concavification of a function $g$ is either equal to $g$ at some point or linear on some interval otherwise. Luckily enough, the ironed virtual value has a closed form on intervals where it is not equal to the virtual value.

\begin{figure}[h!]
\centering
\begin{tabular}{cc}
\includegraphics[width=.5\linewidth]{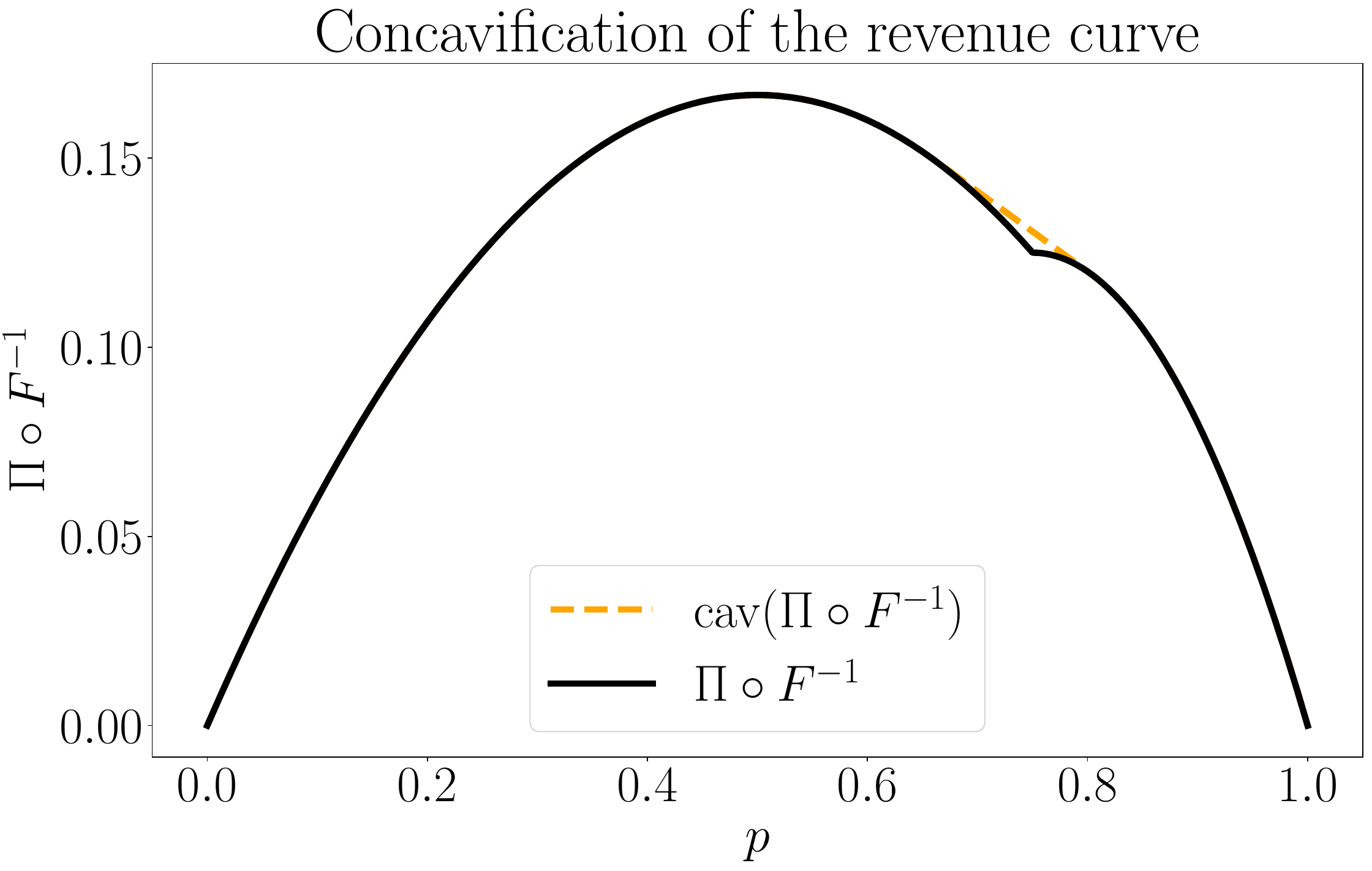}&
\includegraphics[width=.5\linewidth]{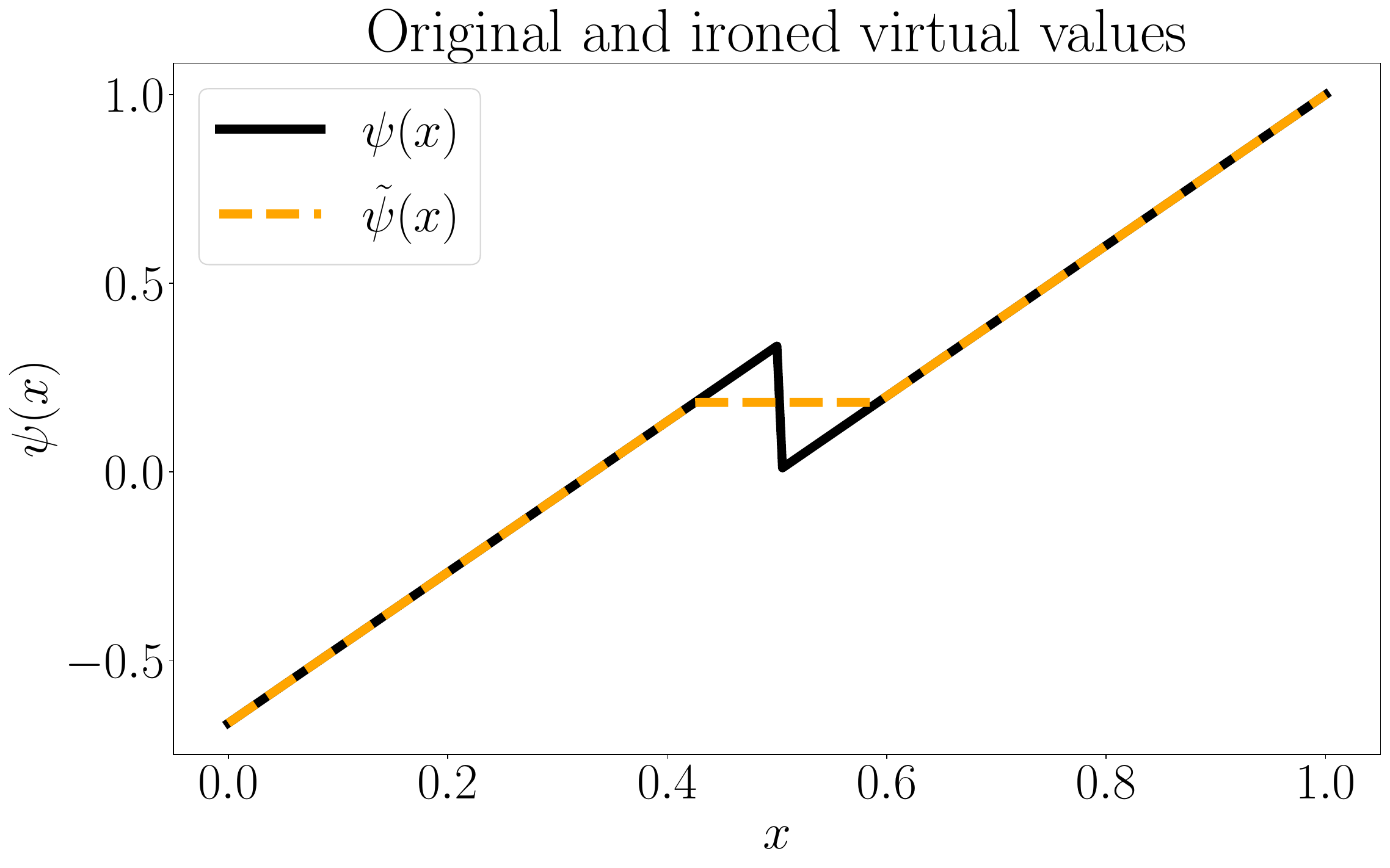}
\end{tabular}
\includegraphics[width=.55\linewidth]{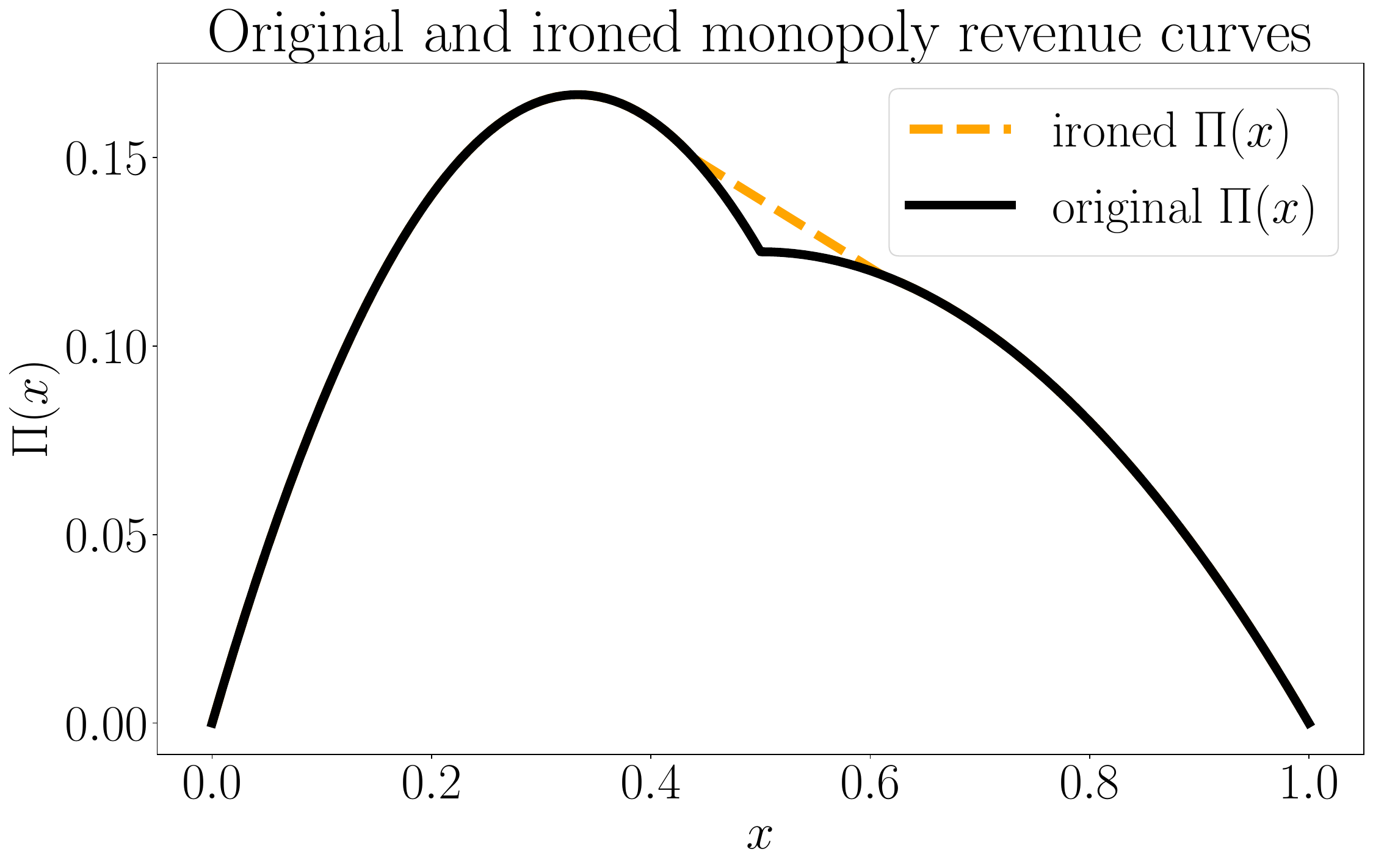}

\caption{\textbf{Left:} Concavification of $\Pi\circ F^{-1}$ for a  mixture of $\mathfrak{U}(0,0.5)$ and $\mathfrak{U}(0,1)$. \textbf{Right:} Associated virtual value (plain black) and ironed one (dashed orange).
\textbf{Center:} Original (plain black) and ironed (dashed orange) monopoly revenue.}
\label{Ironed}
\end{figure}

\begin{lemma2}%[Expression of ironed virtual value on [a,b]] 
Assume that for some $\alpha<\beta$, it holds that \begin{itemize}
\item[--] $\mathrm{cav}(\Pi\circ F^{-1})(F(\alpha))=\Pi (\alpha)$,  
\item[--]$\mathrm{cav}({\Pi}\circ F^{-1})(F(\beta))=\Pi(\beta)$
\item[--] $\mathrm{cav}({\Pi}\circ F^{-1})(F(x)) > \Pi (x)$ for $x \in(\alpha,\beta)$ 
\end{itemize}then
the ironed virtual value  of $\psi$ on $(\alpha,\beta)$ is constant and equal to
$$
\forall x \in (\alpha,\beta)\;, \tilde{\psi}(x)= \frac{\int_{\alpha}^\beta \psi(t) f(t) \mathrm{d} t}{F(\beta)-F(\alpha)}=\frac{\Pi(\alpha)-\Pi(\beta)}{F(\beta)-F(\alpha)}=\frac{\alpha(1-F(\alpha))-\beta(1-F(\beta))}{F(\beta)-F(\alpha)}\;.$$
\end{lemma2}
\begin{proof}
By definition of the concavification, $\mathrm{cav}(\Pi\circ F^{-1})$ is linear on $[F(\alpha),F(\beta)]$ and the results come from linear interpolation.
\end{proof}
We refer to \citep{note_ironings} for more technical details. In particular, the ironed virtual value, since  it is defined as a sub-differential, is not a function but a multi-valued mapping. On the other hand, selecting the aforementioned $\tilde{\psi}(x)$ as $\psi(x)$ when the sub-differential is not reduced to a singleton is also perfectly valid and implicitly used as convention from now on.  With this latter expression, either $\tilde{\psi}$ is equal to $\psi$ or it is constant on some interval around $x$. It is  non-decreasing everywhere and intervals where $\psi$ is  decreasing are ``flattened'', as illustrated in Figure \ref{Ironed}.

Recall that the purpose of ironing is to replace the - possibly somewhere decreasing - virtual value function $\psi$ in the Myerson auction (that might then not be BIC) by $\tilde{\psi}$. We now show that ironing the virtual value does not decrease the revenue of the Myerson auction.

\begin{lemma2} \label{lemma:ironingIncreasesExpectedRevenue}
The payment of bidder $i$ at the truthful equilibrium of any BIC auction satisfies 
$$ \mathds{E}_{x_i\sim F_{i}}[P_i(x_i)] \leq P_i(0) + \mathbf{E}_{x_{i} \sim F_{i}}\left[Q_{i}\left(x_{i}\right) \tilde{\psi}_{i}\left(x_{i}\right)\right]\;,$$
where $\tilde{\psi}_i$ is his ironed virtual value. %VP: that's always the case, provided $\psi_i(x_i)$ and $\tilde{\psi}_i(x_i)$ have finite mean. 
\end{lemma2} 
Our assumptions that $f_i>0$ almost everywhere and hence $F_i$ is increasing is important; so is the assumption that $F_i$ has 1 moment, which implies that  $\psi_i(x_i)$ and $\tilde{\psi}_i(x_i)$ have finite mean. Note that this assumption is really minimal as it just means that the expected payment under $Q_i(x_i)=1$ is finite. 

The papers \citep{Myerson81} and \citep{note_ironings} implicitly assume differentiability of $Q_i$ in the previous lemma without stating it explicitly. In the case of differentiable $Q_i$ the proof boils down to  integration by parts applied twice and the fact that $\cav{h}\geq h$ for any function $h$. At the level of generality of our statement, which is needed for the most important applications, it is more technical and we give the proof in Subsection \ref{subsec:ProofOfIroningIncreasesRevenue}.

We now generalize Theorem \ref{theorem_regular_distributions} to the case of non-regular value distributions.
 \begin{theorem2}[\cite{Myerson81}]%[Optimal auctions for general value distribution]
With general value distribution, in a revenue-maximizing auction,  the seller allocates the item to the bidder with the highest non-negative ironed virtual value $\tilde{\psi}_i(x_i)$, ties broken at random, with the payment rule of Corollary \ref{characterization_inc_comp}.
 \end{theorem2}
\begin{example2}
We consider the case of the Myerson auction with symmetric players but non-regular value distributions in Subsection \ref{subsec:MyersonNonRegular}, where we derive the payment and allocation rules. With non-regular value distributions the Myerson auction in the symmetric case is not a second price auction with reserves anymore. 
\end{example2} 
 \subsection{Reserve prices in first price auctions}
%\nek{need to read his assumptions carefully and state them} 
We start with an abstract result that we then apply to first price auctions. 

\begin{proposition2}\label{prop:virtualValueIsVirtualPaymentGeneralCase}
Suppose bidder $i$ participates in an auction such that the probability of winning the auction when bidding $b$ is $G_i(b)$ and the corresponding expected payment is $P_i(b)$. Let $\beta_i$ an optimal strategy for bidder $i$ in this setup that maximizes his utility. 	
Then the seller revenue coming from bidder $i$ is 
\begin{equation}\label{eq:RevSellerGeneralPriceWithVValue}
 \mathds{E}_{x_i\sim F_{i}}[P_i(\beta_i(x_i))]-P_i(\beta_i(0)) =\mathds{E}_{x_i\sim F_{i}} \big[G_i(\beta_i(x_i))\psi_i(x_i)\big]\;,
\end{equation}
where $\psi_i$ is the virtual value associated to $F_i$. 
\end{proposition2}
\begin{remark2}
This proposition helps explaining how our interpretation of the $\psi_i(x_i)$ as a virtual payment makes sense for utility-maximizing bidders in general and not only in the case of BIC or DSIC auctions we encountered previously. It can also be seen as a more quantitative version of the revelation principle. 
\end{remark2}
\begin{proof}
Note that since bidder $i$ tries to maximize his utility, 
$$
\beta_i(x_i)=\argmax_b x_iG_i(b)-P_i(b)\;. 
$$
Let us call 
$$
W(x)=x G_i(\beta_i(x))-P_i(\beta_i(x))
$$
The envelope theorem in the form of Theorem 2 of \cite{MilgromSegal2002} applies since $0\leq G_i(t)\leq 1$ for all $t$ and hence 
$$
W(x)-W(0)=\int_0^x G_i(\beta_i(t)) dt\;.
$$
Therefore, the expected payment satisfies
$$
P_i(\beta_i(x))-P_i(\beta_i(0))=xG_i(\beta_i(x))-\int_0^x G_i(\beta_i(t)) dt\;.
$$
Taking expectation with respect to $x_i$ with cdf $F_i$ in the previous equation gives
\begin{align*}
\mathds{E}_{x_i\sim F_{i}}[P_i(\beta_i(x_i))]-P_i(\beta_i(0))&=\int_0^{\infty} xG_i(\beta_i(x)) f_i(x)dx-\int_0^{\infty}\int_0^{\infty} f_i(x)\indicator{t\leq x}G_i(\beta_i(t))dt\, dx\;,\\
&=\int_0^{\infty} xG_i(\beta_i(x)) f_i(x)dx-\int_0^{\infty} (1-F_i(t))G_i(\beta_i(t))dt\;,\\
&=\mathds{E}_{x_i\sim F_{i}} \big[G_i(\beta_i(x_i))\psi_i(x_i)\big]\;.
\end{align*}
% $$
% \mathds{E}_{x_i\sim F_{i}}[P_i(x_i)] =\mathds{E}_{x_i\sim F_{i}} \big[G_i(\beta_i(x_i))\psi_i(x_i)\big]\;.
% $$

% This is not limited to first price... as long as $P(\beta_i(0))=0$\\
% En gros,  1st order conditions give $xG_i'(\beta_i(x))=P'(\beta_i(x))$. Multiply both sides by $\beta'_i(x)$ and integrate to get $P(\beta_i(x))-P(\beta_i(0))=xG_i(\beta_i(x))-\int_0^x G_i(\beta_i(u))du\;.$ Then doing the same as below, we get that
% $$
% \mathds{E}_{x_i\sim F_{i}}[P_i(x_i)]-P(\beta_i(0))=\mathds{E}_{x_i\sim F_{i}} \big[G_i(\beta_i(x_i))\psi_i(x_i)\big]\;.
% $$
% Ou: envelope theorem says that if $W_i(x)=\max_b xG_i(b)-P(b)$, and $\beta_i$ is an optimal strategy, $W_i'(x)=G(\beta_i(x))$. Hence, $W_i(x)-W_i(0)=\int_0^x G_i(\beta_i(u))du=xG_i(\beta_i(x))-P(\beta_i(x))+P(\beta_i(0))$. We conclude that
% $P(\beta_i(x))-P(\beta_i(0))=xG_i(\beta_i(x))-\int_0^x G_i(\beta_i(u))du\;.$ And the result follows as above.

\end{proof}

In the case of first price auctions, we have $P_i(\beta_i(x_i))=\beta_i(x_i)G_i(\beta_i(x_i))$, since the probability of winning is $G_i(\beta_i(x_i))$. 
So the arguments given in the proof above also implies the following result \citep{KirkegaardAsymmetricFirstPrice2009}: the optimal strategy for bidder $i$ when the top bid of the competition has cdf $G_i$ and $P_i(\beta_i(0))=0$ satisfies 
\begin{equation}\label{eq:OptStrat1stPriceEnvelope}
\beta_{i}(x)=x-\frac{\int_0^x G_i(\beta_i(t))dt}{G_i(\beta_i(x))}\;.
\end{equation}

%Because of the the envelope theorem and arguments similar to those used to derived the revenue maximizing auctions, the equilibrium strategy must satisfy 
\iffalse
Using the envelope theorem, one can show \citep{KirkegaardAsymmetricFirstPrice2009} that the optimal strategy for bidder $i$ when the top bid of the competition has cdf $G_i$ satisfies 
\begin{equation}\label{eq:OptStrat1stPriceEnvelope}
\beta_{i}(x)=x-\frac{\int_0^x G_i(\beta_i(t))dt}{G_i(\beta_i(x))}\;.
\end{equation}
\fi
The expected utility of player $i$ at $x$ is then equal to $\cU_i(\beta_{i}(x),x)=\int_0^x G_i(\beta_i(t))dt$ and  it also holds  that $\mathds{E}_{x_i\sim F_{i}}\big[\cU_i (\beta_i(x_i),x_i)\big]=\ \mathds{E}_{x_i\sim F_{i}}\big[G_i(\beta_i(x_i))(x_i-\psi(x_i))\big]$. 
\iffalse
\nek{I think with the envelope theorem we can show that if I'm a utility maximizer, i.e., I pick 
$$
\beta(x)=\argmax_b xG(b)-P(b)\;,
$$
regardless of the payment rule, then 
$$
\mathds{E}_{x_i\sim F_{i}}[P_i(x_i)] =\mathds{E}_{x_i\sim F_{i}} \big[G_i(\beta_i(x_i))\psi_i(x_i)\big]\;.
$$
This is not limited to first price... as long as $P(\beta_i(0))=0$\\
En gros,  1st order conditions give $xG_i'(\beta_i(x))=P'(\beta_i(x))$. Multiply both sides by $\beta'_i(x)$ and integrate to get $P(\beta_i(x))-P(\beta_i(0))=xG_i(\beta_i(x))-\int_0^x G_i(\beta_i(u))du\;.$ Then doing the same as below, we get that 
$$ 
\mathds{E}_{x_i\sim F_{i}}[P_i(x_i)]-P(\beta_i(0))=\mathds{E}_{x_i\sim F_{i}} \big[G_i(\beta_i(x_i))\psi_i(x_i)\big]\;.
$$
Ou: envelope theorem says that if $W_i(x)=\max_b xG_i(b)-P(b)$, and $\beta_i$ is an optimal strategy, $W_i'(x)=G(\beta_i(x))$. Hence, $W_i(x)-W_i(0)=\int_0^x G_i(\beta_i(u))du=xG_i(\beta_i(x))-P(\beta_i(x))+P(\beta_i(0))$. We conclude that 
$P(\beta_i(x))-P(\beta_i(0))=xG_i(\beta_i(x))-\int_0^x G_i(\beta_i(u))du\;.$ And the result follows as above. 
}
\fi
\begin{corollary2}\label{coro:RevSeller1stPriceWithVValue}
In a first price auction, when bidder $i$ uses the strategy implicitly defined in Equation \eqref{eq:OptStrat1stPriceEnvelope}, the seller revenue coming from bidder $i$ is 
\begin{equation}\label{eq:RevSeller1stPriceWithVValue}
 \mathds{E}_{x_i\sim F_{i}} \big[G_i(\beta_i(x_i))\psi_i(x_i)\big]\;,
\end{equation}
where $\psi_i$ is the virtual value associated with the value distribution of $x_i$. 
\end{corollary2}
The corollary follows from Proposition \ref{prop:virtualValueIsVirtualPaymentGeneralCase} after noticing that $P_i(\beta_i(0))=\beta_i(0)G_i(\beta_i(0))=0$ when bidder $i$ uses $\beta_i$ defined in Equation \eqref{eq:OptStrat1stPriceEnvelope}.

\iffalse
\begin{proof}
We note that 
$$
 \mathds{E}_{x_i\sim F_{i}}[P_i(x_i)] =\int_0^{\infty}\beta_{i}(x) G_i(\beta_i(x)) f_i(x) dx =\int_0^{\infty} xG_i(\beta_i(x))f_i(x)dx - \int_0^{\infty}f_i(x)\int_0^x G_i(\beta_i(t))dt dx\;.
$$	
Using Fubini's theorem, we have 
$$
\int_0^{\infty}f_i(x)\int_0^x G_i(\beta_i(t)))dt dx=\int_0^{\infty}f_i(x)G_i(\beta_i(t))dt \indicator{t\leq x} dx dt=\int_0^{\infty}(1-F_i(t))G_i(\beta_i(t))) dt\;.
$$
We conclude that 
$$
 \mathds{E}_{x_i\sim F_{i}}[P_i(x_i)] =\int_0^\infty G_i(\beta_i(x))\left[x-\frac{1-F_i(x)}{f_i(x)}\right]f_i(x) dx = \mathds{E}_{x_i\sim F_{i}} \big[G_i(\beta_i(x_i))\psi(x_i)\big]\;.
$$
\end{proof}
\fi

\paragraph{Good or optimal reserve prices} Proposition \ref{prop:virtualValueIsVirtualPaymentGeneralCase} and Corollary \ref{coro:RevSeller1stPriceWithVValue} suggest that from a seller revenue standpoint it would be good to avoid bids corresponding to values that have negative virtual values. In other words, setting individual reserve values at $\psi_i^{-1}(0)$ for bidders may have positive impact for seller revenue. However, this interpretation ignores the impact of setting such reserve values on the strategic response of the bidder that is optimizing his utility.

Finding optimal reserve prices for first price auctions is much more complicated than finding them for the second price auctions, even with $n=2$ symmetric bidders \citep{KotowskiAsymmetricReserves2018} for the following two reasons. First, if the distribution $F$ is not regular and its density is discontinuous at the monopoly price, giving two different reserve prices to the two symmetric buyers actually increases the seller revenue at equilibrium, at least when $r_1$ and $r_2$ are close.
Second, in the specific case where those two reserve prices $r_1$ and $r_2$ are sufficiently close,  the equilibrium bid distribution of the player with the lower reserve price becomes discontinuous.

Many further difficulties arise when the seller tries to learn good reserve prices for first price auctions from data and does not have access to the bidders' value distributions. We further detail them in Section \ref{learning_reserve_prices_first_price}.

\section{Prior-independent optimal auctions} 
\label{sec-Bulow}
In the previous section, we  derived the revenue-maximizing auction given the prior of the seller on the possible valuations of the different bidders. In some cases, the seller does not have access to a reliable prior and she has to choose an auction without this information. An associated challenge is to understand what auctions are robust to a lack of knowledge about (or some mis-specification in) the bidders' value distributions.

The following theorem shows that a basic second-price auction without reserve price -  referred to henceforth as the Vickrey auction - with just one extra bidder, leads to a higher revenue than the optimal Myerson auction \citep{bulow1996auction}. Only the symmetric case is considered, so that the comparison of two auctions with different set of bidders is possible.
\begin{theorem2}[\cite{bulow1996auction}]%[Bulow-Klemperer Theorem]
In the symmetric setting and with regular distributions, the revenue of the Myerson auction with $n$ bidders is lower than the revenue of the Vickrey auction with $n+1$ bidders. 
\end{theorem2}
\begin{figure}[H]
\includegraphics[width=.5\linewidth]{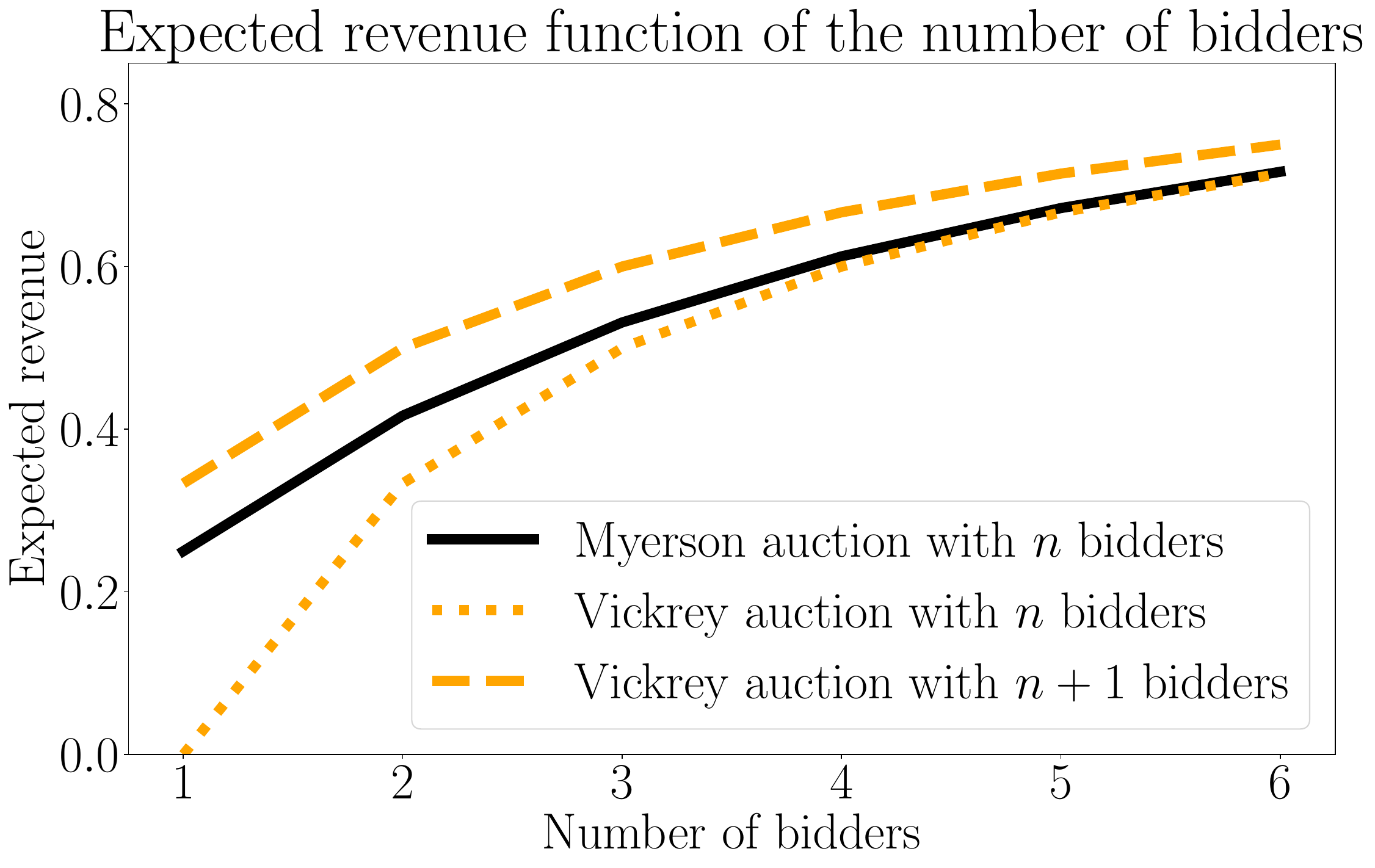}
\caption{Illustration of the Bulow-Klemperer theorem for the case with value distributions  $\mathfrak{U}([0,1])$.}
%{\color{red} Ca ne devrait pas etre $K$ bidders mais $n$ bidders ; choix des couleurs discutables ; on ne dit pas Nb bidders en Anglais }}
\label{fig:BK}
\end{figure}
Before proving the Bulow-Klemperer theorem, we introduce a useful lemma.
\begin{lemma2}\label{second_price_without_reserve}
In the symmetric case and with regular distributions, the Vickrey auction is revenue-maximizing in the class of individually-rational auctions where the item is always attributed. 
\end{lemma2}
\begin{proof}
Based on the revelation principle, we shall only consider incentive-compatible auctions. Theorem \ref{Th:MyersonLemma} then applies. Since the item must always be attributed, in order to maximize the seller's revenue, it should be allocated to the bidder with the highest virtual value, regardless of whether the highest virtual value is non-negative or not. This is a consequence of the proof of Theorem \ref{Th:MyersonLemma}. Since the distributions are regular and identical, the bidder with the highest virtual value is also the bidder with the highest value. As a consequence, the mechanism we just described is exactly the Vickrey auction.
\end{proof}
The proof of the Bulow-Klemperer theorem can now be derived from the previous lemma.
\begin{proof} 
Let us assume that there are $n+1$ bidders, and consider the following mechanism. First, the seller runs a Myerson auction on  $n$ bidders (chosen arbitrarily). If the item is not allocated by the Myerson auction, it is allocated for free (i.e., without any payment) the $(n+1)$-th bidder. The revenue of this auction is equal to the revenue of the Myerson auction with $n$ bidder, yet it is  an auction that always allocate an item amongst $n+1$ bidders. Lemma \ref{second_price_without_reserve} implies that the revenue of this auction is smaller than the one of the Vickrey auction. 
This gives the result.
\end{proof}

As a consequence, it is more interesting,  from a seller revenue maximization standpoint to have one more bidder in the auction than to implement a complex mechanism. This is the reason why one of the first recommendations of most economists dealing with institutions organizing auctions is to maximize the competition  before implementing complex mechanisms \citep{bulow1996auction,milgrom2004putting}. This is sometimes referred to as the Wilson doctrine.

\bigskip

Nonetheless, let us assume that a specific mechanism has been chosen, independently of the value distributions (which are unknown in this setup). A crucial  question that remains is the evaluation of this specific choice of mechanism in the worst-case analysis. We will restrict ourselves to DSIC auctions. We will use the notion of \textsl{competitive ratio} defined as the infimum, over all possible value distributions, of the revenue of this auction divided by the optimal revenue of the Myerson auction for these distributions. 

The competitive ratio is obviously smaller than 1, and the bigger the better. Unfortunately, if the class of value distributions is not restricted when computing this infimum, there does not exist any auction with a positive competitive ratio \citep{allouah2018prior}. As a consequence, we will use different types of restrictions to achieve non-trivial approximation results. Based on the Bulow-Klemperer theorem, we can derive some revenue guarantees on the second price auction without reserve price. 
\begin{corollary2} \label{compet-ratio}
The Vickrey auction  with $n$ symmetric bidders and regular value distributions is at least a $\left(\frac{n-1}{n}\right)$-approximation of the Myerson auction. 
\end{corollary2}
\begin{proof}
We denote  respectively by $R(\mathrm{Vickrey}_n)$ and $R(\mathrm{Myerson}_n)$ the expected revenue of the seller  in the Vickrey or  Myerson auction with $n$ bidders and by $P_i(\mathrm{Vickrey}_n)$ and $P_i(\mathrm{Myerson}_n)$ the expected payment of the bidder $i$ in those auctions (so that $R(\cdot) = \sum_{i=1}^n P_i(\cdot)$).

Let us first prove that $P_i(\mathrm{Myerson}_n)\geq P_i(\mathrm{Myerson}_{n+1})$.

By symmetry and with regular distributions, there exists a unique monopoly price $r^*$ which is independent of the number of players. Note that $\psi(x)\geq 0$ when $x\geq r^*$. Furthermore, in the symmetric case, the probability that bidder $i$ wins the auction when facing $k$ competitors is just $F^k(x)$. Hence, 
\begin{align*}
	P_i(\mathrm{Myerson}_{n})&= \int  \psi(x_i) F^{n-1}(x_i)f(x_i)\mathds{1}\{x_i\geq r^*\} dx_i\\&\geq \int \psi(x_i) F^{n}(x_i)f(x_i)\mathds{1}\{x_i\geq r^*\}dx_i\\&=P_i(\mathrm{Myerson}_{n+1}),\end{align*}
which proves the inequality.

Then, based on the Bulow-Klemperer theorem, 
$$R(\mathrm{Vickrey}_n)\geq R(\mathrm{Myerson}_{n-1}) = (n-1)  P_i(\mathrm{Myerson}_{n-1}) \geq (1-\frac{1}{n})  R(\mathrm{Myerson}_{n})$$
which gives the result. \end{proof}
We proved in Corollary \ref{compet-ratio}, through the Bulow-Klemperer theorem that the competitive ratio of the Vickrey auction is at least 0.5, when the distributions are restricted to regular ones. Interestingly, it is possible to do better with a slightly different auction. 
\begin{theorem2}[\cite{fu2015randomization}]
There exists an incentive-compatible auction with a competitive ratio of 0.512 against regular value distributions. 
\end{theorem2}
The mechanism considered was the first  with a higher competitive ratio than 0.5  against regular distributions. It is a slight modification  of the Vickrey auction where the seller inflates the second highest bid. Formally, the mechanism is the following: with probability $1-\varepsilon$,  a second price auction without reserve price is run. With the  probability $\varepsilon$, the mechanism allocates the object to the bidder with the highest valuation, but only if his valuation is greater than $1+\delta$  times the valuation of the second highest bidder and pays  $(1+\delta)$ the second highest bid. Otherwise, the mechanism does not allocate the item. 

The idea behind this theorem is the following. The Bulow-Klemperer bound of $0.5$ on the competitive ratio is rather tight for regular distributions that would induce a high optimal reserve price. On the other hand, it is rather loose for regular distribution with a low associated reserve price. Inflating the second highest bid has a positive effect for the former type of distribution (as it somehow emulates a high reserve price) and a negative effect for the later type (because it induces some reserve price, bigger than what it should be). As a consequence, the ratio of revenues increases in the first case, and decreases in the second one; but thanks to the looseness in the Bulow-Klemperer bound of $0.5$, the infimum globally increases. 

On the other hand, when restricted to MHR distributions, the ratio is equal to 0.7153 and this result is tight. 

For regular distribution, this result was then improved up to 0.519 \citep{allouah2018prior}  and further improved in \citep{hartline2020benchmark} which identified the optimal prior-independent mechanism. The optimal mechanism is a mixture between a second price auction and the same auction where the prices are scaled up by a factor of about 2.5. The authors find the worst-case family of distributions and use these distributions to derive the optimal mechanism and solve the problem. 

\section{Advanced material: non-unicity of Nash equilibria and related complications}
\label{sec:NashIsDifficult}
\subsection{The case of 2nd price auctions}
As we have explained while presenting them, second price auctions are DSIC, and hence there exists a truthful equilibrium. 

There however exist many other equilibria such as the following one. Suppose for concreteness that all the value distributions of the bidders are supported on $[0,1]$. Suppose now that every bidder always bids 0 except one of them - say,  bidder 1 - bids arbitrarily high, say 1. Clearly for bidder 1 this is a best response to having a competition of 0 since he wins all auctions and pays 0. For bidders other than 1, winning entails bidding more than 1 and paying 1; but their values are less than 1. So, the utility of winning any auction is non-positive and negative as soon as their value is strictly less than 1 and the maximum utility they can expect is 0. And this is achieved by many strategies but in particular by bidding 0 all the time, which is then clearly a best response. 

\subsection{No revenue equivalence when $\beta$ is only non-increasing}
We are grateful to an anonymous referee for bringing up this example while discussing Theorem \ref{Th:RevEquivalence}. Suppose bidders have the same value distribution on say $[0,1]$ and that the value distribution admits a density.

Consider the following 0-rational auction with standard allocation rule, i.e., the winner is the highest bidder; ties are broken at random among top bidders. Bidders pay 0 when they bid 0 and pay 1 otherwise.  Note that for any bidder bidding $x>0$ results in non-positive expected utility: either they win, and their utility is negative or they lose and their utility is 0. Hence an equilibrium is for all bidders to bid 0. This equilibrium is symmetric. The seller's expected revenue is therefore 0. However, Theorem \ref{Th:RevEquivalence} states that the expected payment for this type of auctions is independent of the payment rule at an \emph{increasing} symmetric equilibrium. The issue in this very interesting example is that the symmetric equilibrium strategy described here is not increasing: it is in fact constant, since it maps all values to 0. Looking at the proof of Theorem \ref{Th:RevEquivalence} it is clear that the expected utility ${\mathcal U}_i(z,x_i)$ is then not $x_iG_i(z)-P_i(z)$ - as was key to the proof. The utility when bidding $b$ when the other players use this strategy is $(x_i-1)\indicator{b>0}$ and $\frac{1}{n}x_i$ if $b=0$. 

\subsection{Example: the Myerson auction in the symmetric case with non-regular value distribution}
\label{subsec:MyersonNonRegular} 
We consider the symmetric case where the $n$  independent bidders still have a value distribution denoted by $F$ with a density $f$ with $f>0$. In particular, with probability 1 the values they draw are all different. For simplicity we suppose that there is a single interval $[\alpha,\beta]$ on which $\psi$ requires ironing. In this case, using the remark following Corollary \ref{characterization_inc_comp}, an optimal auction is the following: suppose bidder $i$ value is such that $\tilde{\psi}(x_i)\geq \max_{j}\{ \tilde{\psi}(x_j)\}$ and $\tilde{\psi}(x_i) \geq 0$, so that he might win the auction. As before, we call $\tilde{\psi}^{-1}(0)=\inf_t\{t:\tilde{\psi}(t)\geq 0\}$.
\begin{enumerate}
\item if $\max_{j\neq i} \tilde{\psi}(x_j)<0$, bidder $i$ wins the auction and pays $\tilde{\psi}^{-1}(0)$. For the other cases below, we assume that $\max_{j\neq i} \tilde{\psi}(x_j)\geq 0$.
\item if $\max_{j\neq i} x_j>\beta$, then bidder $i$ wins the auction and pays second price i.e., $\max_{j\neq i} x_j$. 
\item if $\max_{j\neq i} x_j<\alpha $, then bidder $i$ wins the auction and pays $\max_{j\neq i} x_j$, i.e., second price.
\item if $\max_{j\neq i} x_j \in (\alpha,\beta)$, let us call $K$ the number of bidders in $i$'s competition who have $x_j \in (\alpha,\beta)$. Then two situations arise: \begin{itemize}
\item[{4a)}] either $x_i>\beta$, in which case bidder $i$ wins the auction and pays $\beta-\frac{\beta-\alpha}{K+1}$. An  equivalently payment scheme would be to draw $K$ i.i.d.\ uniform random variables $u_k$  on $[\alpha,\beta]$ and to charge their maximal value $Y=\max_{1\leq k \leq K} u_k$, since  $\mathds{E}[Y]=\beta-\frac{\beta-\alpha}{K+1}$; 
\item[{4b)}] or $x_i \in (\alpha,\beta)$ in which case the winner is chosen uniformly at random among the $K+1$ bidders having value in $(\alpha,\beta)$ - and hence having the same virtualized bid $\tilde{\psi}(x_k)$. When bidder $i$ wins, he pays $\alpha$. 
\end{itemize}
\end{enumerate}

Recall that when $\psi$ is regular, the optimal auction is a second price auction with monopoly reserve. 	

\iffalse
Finally, we note that when $h$ defined on [0,1] is continuous, has a unique (strictly positive) maximum and has $\lim_{x\tendsto 0} h(x)=0=\lim_{x\tendsto 1} h(x)$, it is not hard to show that $\cav{h}$ also has a unique maximum, and we have $h(x^*)=\max_x h(x)=\cav{h}(x^*)=\max_x \cav{h}(x)$. Applying this remark to $h=\Pi\circ F^{-1}$, we see that when $F$ is increasing, has 1 moment and $\Pi(x)=x(1-F(x))$ has a unique maximum, then $\psi^{-1}(0)=\tilde{\psi}^{-1}(0)$. In particular, in the posted price setting, under the assumptions stated above, the monopoly price is always $\psi^{-1}(0)$ regardless of whether or not $\psi$ is regular. 
\fi

\subsection{Proof of Lemma \ref{lemma:ironingIncreasesExpectedRevenue}}\label{subsec:ProofOfIroningIncreasesRevenue}
\begin{proof} 
% The proof is based on  the Harris inequality \citep{harris1960lower}, a direct application of Chebyschev' sum inequality.
% \begin{lemma2}[Harris inequality]
% For any  non-decreasing mapping $\rho$ and  non-increasing mapping $\gamma$, the following holds for any random variable $X$,
% $$\mathds{E}[\rho(X)\gamma(X)] \leq \mathds{E}[\rho(X)]\mathds{E}[\gamma(X)]\;.$$
% \end{lemma2}
Consider a BIC auction. According to Theorem \ref{Th:MyersonLemma}, at the truthful equilibrium,
$$
\mathds{E}_{x_i\sim F_{i}}[P_i(x_i)] = P_i(0) + \mathbf{E}_{x_{i} \sim F_{i}}\left[Q_{i}\left(x_{i}\right) \psi_{i}\left(x_{i}\right)\right]\;.
$$
Furthermore, since the auction is BIC, Corollary \ref{characterization_inc_comp} implies that $Q_i$ must be non-decreasing. Almost by definition we also have $0\leq Q_i(x_i)\leq 1$ since $Q_i(t)$ is the probability that bidder $i$ wins the item in the auction when bidding $t$. 

So the Lemma will be shown is we can show that 
$$
I(Q_i;\psi_i)=\mathds{E}_{x_{i} \sim F_{i}}\left[Q_{i}\left(x_{i}\right) \psi_{i}\left(x_{i}\right)\right]
\leq 
\mathds{E}_{x_{i} \sim F_{i}}\left[Q_{i}\left(x_{i}\right) \tilde{\psi}_{i}\left(x_{i}\right)\right]
=I(Q_i;\tilde{\psi}_i)
$$
The papers \cite{Myerson81} and \cite{note_ironings} implicitly assume differentiability of $Q_i$ without stating it explicitly. We give a rigorous proof without this assumption, but the proof is a bit technical.

Let us call $x_{k,N}=\inf_{x} \{x:Q_i(x)\geq k/N\}$ for $0\leq k\leq N$ and by definition $x_{N+1,N}=\infty$; recall that $0\leq Q_i(x)\leq 1$.  Note that because $Q_i$ is non-decreasing, $\{x:Q_i(x)\geq k/N\}$ is a semi-infinite interval.

Let us call
$$
Q_{i,N}(t)=\frac{1}{N}\sum_{k=1}^N \indicator{t\geq x_{k,N}}\;.
$$
Suppose $(x_{k,N},x_{k+1,N})$ is not empty, so that if $x\in (x_{k,N},x_{k+1,N})$, then $k/N \leq Q_i(x)\leq (k+1)/N$.  %Because $x<x_{k+1,n}$ we have on the other hand $Q_i(x)\leq (k+1)/n$, for otherwise $x_{k+1,n}$ would not be $\inf_{x} \{x:Q_i(x)\geq (k+1)/n\}$.
This yields that 
$$
\forall k \in \{0,\ldots,N-1\} \;, \forall x \in (x_{k,N},x_{k+1,N})\;, |Q_i(x)-\frac{k}{N}|\leq \frac{1}{N}\;.
$$
Note that if $(x_{k,N},x_{k+1,N})$ is empty the statements are logically valid as the empty set has all universal properties. 
Now since $Q_i$ is non-decreasing and bounded by $1$, it has at most $N$ jump discontinuities of size greater than $1/N$. We conclude that under $F_i$, which has a density, the measure of the set where $|Q_i-Q_{i,N}|>1/N$ is 0. In particular, it is a (possibly empty) subset of $\cup_{k=1}^N\{x_{k,N}\}$ as our previous results show.

This implies that 
$$
\left|\int_{x_{k,N}}^{x_{k+1,N}} \psi_i(x)f_i(x) (Q_i(x)-\frac{k}{N})dx\right|
\leq \frac{1}{N} \int_{x_{k,N}}^{x_{k+1,N}}|\psi_i(x)|f_i(x) dx\;.
$$

Before starting the main argument of the proof, we now show that 
$\Expxi{|\psi_i(x_i)|}$ and $\Expxi{|\tilde{\psi}_i(x_i)|}$ are finite when $F_i$ has one moment.
We simply note that 
$$
\Expxi{|\psi_i(x_i)|}=\int_0^{\infty } |x_if_i(x_i)-(1-F_i(x_i))|dx_i \leq\int_0^{\infty } x_fi_i(x_i)+(1-F_i(x_i))dx_i=
2 \Expxi{x_i}\;.
$$
Now $\tilde{\psi}_i=\psi_i$ except on intervals where $\tilde{\psi}_i$ is constant and equal to the mean of $\psi_i$ on those intervals. It follows that on those intervals, the mean of $|\tilde{\psi}_i|$ is less that the mean on $|\psi_i|$. Hence we also have 
$$
\Expxi{|\tilde{\psi}_i(x_i)|}\leq 4 \Expxi{x_i}\;.
$$

Recall that we want to prove
$$\mathds{E}_{x_{i} \sim F_{i}}\left[Q_{i}\left(x_{i}\right) \psi_{i}\left(x_{i}\right)\right]
\leq 
\mathds{E}_{x_{i} \sim F_{i}}\left[Q_{i}\left(x_{i}\right) \tilde{\psi}_{i}\left(x_{i}\right)\right]$$
i.e.,
$$
\int_0^\infty Q_i(x)\psi_i(x)f_i(x) dx \leq \int_0^\infty Q_i(x)\tilde{\psi}_i(x)f_i(x) dx 
$$
We have
\begin{align*}
\int_0^\infty Q_i(x)\psi_i(x)f_i(x) dx &= \sum_{k=0}^{N} \int_{x_{k,N}}^{x_{k+1,N}} Q_i(x)\psi_i(x)f_i(x) dx  \\
&= \sum_{k=0}^{N} \int_{x_{k,n}}^{x_{k+1,N}} \psi_i(x)f_i(x) dx \frac{k}{N} + \sum_{k=0}^{N} \int_{x_{k,N}}^{x_{k+1,N}} \psi_i(x)f_i(x) (Q_i(x)-\frac{k}{N})dx \\
&=\sum_{k=0}^{N}  \frac{k}{N} \big(\Pi(x_{k,N})-\Pi(x_{k+1,N})\big) + \sum_{k=0}^{N} \int_{x_{k,N}}^{x_{k+1,N}} \psi_i(x)f_i(x) (Q_i(x)-\frac{k}{N})dx\\
&= \frac{1}{N} \sum_{k=0}^{N} \Pi(x_{k,N}) +  \sum_{k=0}^{N} \int_{x_{k,N}}^{x_{k+1,N}} \psi_i(x)f_i(x) (Q_i(x)-\frac{k}{N})dx\\
& \leq  \frac{1}{N} \sum_{k=0}^{N} \Pi(x_{k,N})  + \frac{1}{N} \int_0^\infty |\psi_i(x)|f_i(x)dx\\
&\leq  \frac{1}{N} \sum_{k=0}^{N} \Pi(x_{k,N})  + \frac{1}{N} \int_0^\infty xf_i(x)+(1-F(x)))dx\\
& \leq \frac{1}{N} \sum_{k=0}^{N} \Pi(x_{k,N})  + \frac{2\mathds{E}_{x_i \sim F_i} [x_i]}{N} 
\end{align*}
Using the exact same argument for $\tilde{\Pi}$ defined above, which is a primitive of $-\tilde{\psi}f$ that upperbounds $\Pi$, we finally get
\begin{align*}
\int_0^\infty Q_i(x)\psi_i(x)f_i(x) dx &\leq \frac{1}{N} \sum_{k=0}^{N} \Pi(x_{k,N})  + \frac{2\mathds{E}_{x_i \sim F_i} [x_i]}{N} \\
&\leq \frac{1}{N} \sum_{k=0}^{N} \tilde{\Pi}(x_{k,N})  + \frac{2\mathds{E}_{x_i \sim F_i} [x_i]}{N}\\
& \leq \int_0^\infty Q_i(x)\tilde{\psi}_i(x)f_i(x) dx  + \frac{6\mathds{E}_{x_i \sim F_i} [x_i]}{N}
\end{align*}

Hence, 
$$
\int_0^\infty Q_i(x)\psi_i(x)f_i(x) dx-\int_0^\infty Q_i(x)\tilde{\psi}_i(x)f_i(x) dx\leq 
\frac{6\mathds{E}_{x_i \sim F_i} [x_i]}{N}\;.
$$
As the left-hand side does not depend on $N$ we can take the limit as $N\tendsto \infty$ to conclude that 
$$
\int_0^\infty Q_i(x)\psi_i(x)f_i(x) dx-\int_0^\infty Q_i(x)\tilde{\psi}_i(x)f_i(x) dx\leq 0\;.
$$
This concludes the proof. 
\end{proof}
\printbibliography[segment=2, heading=subbibintoc]
\end{refsegment}
\clearpage
%\begin{refsegment}
 %\input{chapters/chapter2_applications}
 %\printbibliography[segment=2, heading=subbibintoc]
%\end{refsegment}
%\begin{refsegment}
%\input{chapters/chapter3_approximation}
%\printbibliography[segment=2, heading=subbibintoc]
%\end{refsegment}
\begin{refsegment}
%!TEX root = ../main.tex
\chapter{Repeated auctions from a seller's standpoint}
\label{Chapter:Seller}
\label{sample_complexity_auction}
\centerline{\colorbox{mygray}{ \begin{minipage}{11cm}
\begin{center} \textbf{First read of this chapter, key concepts and ideas}\end{center}
This chapter focuses on the complexity/cost of learning optimal, or good enough, mechanism using datasets of past values. It contains two sections, that provide the theoretical material required for this chapter and the following one, respectively  Section \ref{SE:StatLearn} and Section \ref{SE:Bandits}. The crucial results of this chapter are Theorem \ref{TH:EmpMonop}, that describes how many samples are required in the symmetric case (or with one bidder) to learn the optimal auctions; Corollary \ref{CO:MyersonComp|} then extends this result to the asymmetric case (combining Theorem \ref{TH:LlevelCompl} and \ref{thm:pseudo_dim_t_level}). Finally, maybe the most important claim of this chapter is that eager second-price auctions with monopoly prices are maybe the better compromise in efficiency vs.\ learning cost, stated in Theorem \ref{thm:eager_2_approx_myerson}.
\end{minipage}
}
}

\bigskip

\section{Motivation} 
%Auctions are used in most Internet platforms to organize interactions between the different stakeholders. Ebay was one of the first big online platforms to use ascending auction to sell objects on the platform. Yahoo and Google are also using auctions to sell ad opportunities on their search engine. For instance, they let advertisers bid on some keywords to get sponsored links above the first results for a certain user query.  Facebook and LinkedIn are also using them to determine which ad to display, Amazon decides which products are going to be sponsored (and/or advertised), etc.  On all these examples, the digitalization of the auction marketplace enabled auctioneers to store all bid history and to  construct estimates $\widehat{F}_1,\ldots, \widehat{F}_n$ of the value distribution $F_1, \ldots, F_n$ of buyers. It allows them to rely on past data, rather than some \emph{a priori} knowledge of these distributions, to optimize the revenue. 

%%%%%%%%%%%% 
% Explain F as information asymmetry , private features

The first large-scale field experiment in production showed how engineers at Yahoo could handle their huge datasets to learn an optimal reserve price per key word 
\citep{OstSch11}. Bidders were assumed to be non-strategic and  to bid truthfully on their platform. In the Ebay case, as buyers are different from one auction to the other, the seller knows that, with running an incentive-compatible auction, bidders will bid truthfully. Hence, the online platform is able to learn an optimal reserve price per object and derive to a revenue-maximizing objective. 
In these two examples, the seller has access to samples from bidders' past values and they aim at exploiting this information to learn a revenue-maximizing auction. The value distributions encompass the variability of values between bidders or between objects sold on the platform. 

The emergence of this setting created numerous bridges between statistical learning and auction theory \citep{bar2002incentive,blum2004online,lavi2004competitive}, the former being used to estimate the quantities (e.g. value distributions) to compute solution for the latter. This chapter casts some light on these links,  and how far the underlying problem of learning revenue-maximizing auctions has been tackled.

\section{Statistical Learning Theory Tools for Revenue Maximization}\label{SE:StatLearn}
We first start with a short reminder on statistical learning theory. After introducing the different notions, we shall indicate to what they correspond in the auction setting. 

A learner is given a set of hypothesis $\mathcal{A}$: it is a set of possible auctions to run -- e.g., second-price auctions with a set of possible reserve prices. She  is also given a set of observations $S_T = \{\textbf{x}_1,\dots,\textbf{x}_T\}$,  sampled independently from the joint product distribution $\bF=F_1\otimes\ldots\otimes F_n$ and belonging to a set of distributions $\cD$ on  a domain $\mathcal{X}$. We emphasize again that $\textbf{x}_t$ is a vector that corresponds to all bidder's value and is sampled according to a distribution $\bF$ whose marginals corresponds respectively to every bidder's value distribution $F_1,\ldots F_n$. For each value vector $\textbf{x}\in\cX$, and each IC auction $a\in\cA$, we denote by $r_a(\textbf{x})$ the revenue of the auction at the truthful equilibrium.

A classical assumption is to consider that for a given distribution $\bF$, there exists an optimal hypothesis (i.e. an optimal auction or an optimal vector of reserve price). This hypothesis is called the \textit{target hypothesis} or \textit{optimal Bayes hypothesis}. For the auction setting, the optimal Bayes hypothesis is defined as
\begin{align}
\label{eq:optimal_auction}
a^*_\cA(\bF) = \argmax_{a \in  \mathcal{A}} R(a) ~~~~~~\text{ where }~~~R(a) = \mathds{E}_{\textbf{x} \sim \mathbf{F}}[ r_a(\textbf{x})]\;,
\end{align}
which is, by definition, the Myerson auction run on $\bF$ if $\mathcal{A}$ represents the whole set of auctions denoted by $\mathfrak{A}$. As it does not depend on a particular class of auctions, we simply denote it $a^*(\bF)$.
The practical objective of the learner is to optimize $R(a)$  accessing only the empirical distribution described by $S_T$ rather than the true distribution $\bF$.

A popular approach is to replace the true distribution $\bF$ in Equation \eqref{eq:optimal_auction}, by its empirical counterpart. This is referred to as Empirical Revenue Maximization (ERM) principle in statistical learning:
$$
\widehat{a}_\cA(S_T) = \argmax_{a \in  \mathcal{A}} \widehat{R}_{S_T}(a) ~~~~~~\text{ where }~~~ \widehat{R}_{S_T}(a) = \frac{1}{T}\sum_{t=1}^{T}r_a(\textbf{x}_t)\;.
$$
The goal is to provide error guarantees on the ERM hypothesis $\widehat{a}_\cA(S_T)$ against the Myerson auction $a^*(\bF)$ depending on the number of samples $T$ and some relevant complexity measure of the hypothesis class $\cA$. Indeed, to make the problem tractable and to avoid overfitting, the learner often restricts the complexity of hypothesis space  $\cA$. This leads to a classical bias / variance trade-off that can be materialized by the following decomposition of the excess-risk between $\widehat{a}_\cA(S_T)$ and $a^*(\bF)$:
 \begin{align*}
 R(a^*(\bF)) - R\left(\widehat{a}_\cA(S_T)\right)  = \underbrace{R(a^*(\bF)) - R(a^*_\cA(\bF))}_{\text{approximation error}} + \underbrace{R(a^*_\cA(\bF)) - R(\widehat{a}_\cA(S_T))}_{\text{estimation error}}
\end{align*}

The challenge for the learner is, given the knowledge of a set of possible distributions $\cD$ and the sample size $T$, to choose a family of auctions $\cA$ that allows to balance these two error terms. In the reminder of this section, we briefly describe classical tools to derive  theoretical guarantees on the estimation error. We also describe why guarantees are not provided for any arbitrarily complex distribution $\bF$ as it would make worst-case guarantees mostly void. Approximation error is usually handled in a more ad-hoc way, as it is very dependent on the hypothesis class.

\medskip

The rates of convergence of approximation  and/or estimation error are formalized through the notion of sample complexity of an algorithm, i.e., a mapping from the class of finite datasets into $\mathcal{A}$.
\begin{definition2}%[Sample complexity] 
Given $\varepsilon \in [0,1]$ and $\delta>0$, the sample complexity of a  batch learning algorithm $\textsc{alg}$, against a class of joint distributions $\mathfrak{F}$ is the smallest number of samples $T$ such that for all distributions $\bF \in \mathfrak{F}$, if $\textsc{alg}$ learns from a dataset $S_{T} \sim \bF^{\otimes T}$ of  $T$  samples, the following holds
\begin{align*}
\mathds{P}\big\{R(\textsc{alg}(S_{T})) \geq (1-\varepsilon)R(a^*(\bF))\big\} \leq 1-\delta\,;
\end{align*}
Stated otherwise, \textsc{alg} is $(1-\varepsilon)$-optimal with probability at least $1-\delta$.
\end{definition2}

\subsection{The Need to Restrict the Space of Distributions}

There are several reasons explaining the importance of restricting the admissible class of value distributions $\bF$ to control these error terms. The first is that worst-case bounds taken over arbitrarily badly-behaved distribution would be almost uninformative. The second is more pragmatical: arbitrarily bad distributions (ex: arbitrary mixtures or Dirac masses) lead to very hard optimization problems. It is not really crucial to provide guarantees on the solutions of problems that can not be solved (at least yet).

To simply explain why the estimation error term cannot be controlled in a satisfying way without further assumptions on the distribution, we focus the remainder of this sub-section on the case where there is only one buyer with value distribution $F$: the posted-price setting. In this setting, we recall that to maximize her revenue $\Pi(r)=r(1-F(r))$, the seller must set the reserve price as  the monopoly price, a root of  bidder's virtual value distribution.

\begin{definition2}%[Empirical monopoly price] 
Let  $\widehat{F}$ denote the empirical value distribution in the dataset (with only one bidder or in the symmetric case). The empirical monopoly price is 
$$\hat{r}^*=\argmax r(1-\widehat{F}(r))\;.$$
\end{definition2}

It is unfortunately impossible to learn a near-optimal revenue-maximizing auctions \citep{roughgarden2016ironing} without additional assumption.

\begin{proposition2}%[Impossibility to learn revenue-maximizing auctions in full-generality]
[\cite{roughgarden2016ironing}]
\label{PR:counter} Consider  any fixed algorithm and fixed dataset size   $T \in \mathds{N}$ . Then for every $\delta >0$ and $\varepsilon>0$, there exists a value distribution such that the auction output by the algorithm is at most $\varepsilon$-optimal with probability at least $1-\delta$.
\end{proposition2}
\begin{proof}
Consider the family of value distributions
 $$ \mathfrak{F} = \{ F_z\, |\, z  \in \mathds{R}^{+*}\} \quad \text{with } \quad   F_z = \begin{cases}
    z^2, & \text{with probability $1/z$}.\\
    0, & \text{with probability $1-1/z$}.
  \end{cases}
 $$
The optimal price of $z^2$ gives an expected revenue of $z$. For any number of samples $T$, and any $\delta >0$, if $z \geq (1-(1-\delta)^{\frac{1}{T}})^{-1}$ then with probability at least $1-\delta$ the dataset will be composed of only 0.  Let $z_T$ be the price posted by the algorithm in that case; the expected revenue of the algorithm is therefore $z_T/z$. As a consequence, if $z \geq \sqrt{\frac{z_T}{\varepsilon}}$ then the algorithm is only $\varepsilon$-optimal.
\end{proof}

Even though the counter-example distribution  $F_z$ involved in the proof does not satisfy the basic assumption of continuity, it is easy to see that smoothing it won't really change the proof (except for additional technicalities). Proposition \ref{PR:counter} implies that some restrictive assumptions on the joint distribution $\bF$ are required, such as regularity of the marginals $F_1,\ldots,F_n$. Another and stronger requirement than increasing virtual value  is a monotonous hazard rate.

\begin{definition2}%[MHR]
 A distribution $F$ has Monotonic Hazard Rate if the hazard rate $\frac{f(x)}{1-F(x)}$ is non-decreasing over its support. 
\end{definition2}
Uniform, exponential and normal distributions satisfy the MHR condition and, obviously, all MHR distributions are regular distributions. The converse is not true since the distribution $F(z) = 1 - 1/z$ is regular but not MHR.  Intuitively MHR distributions have thinner tails than general regular distributions. 
\begin{theorem2}%[Generalization bound of the empirical monopoly price]
[\cite{dhangwatnotai2015revenue,huang2018making}]\label{TH:EmpMonop}
The sample complexity of the empirical monopoly price is of order
\begin{itemize}
\item[--] $\Theta(\varepsilon^{-3/2}\log(1/\varepsilon)\log(1/\delta))$ for MHR distributions and 
\item[--] $\Theta(H\varepsilon^{-2}\log(H/\varepsilon)\log(1/\delta))$ for bounded $[1,H]$ distributions  
\end{itemize}
 \end{theorem2}
 
To get some intuitions, we provide a simple, but sub-optimal, proof for the case of bounded distributions. But first, let us explain why bounded distributions are assumed to lie on $[1,H]$ and not $[0,H]$; this will not transpire in the proof, as we prove a weaker statement (with a quadratic dependency in $H$ but for any bounded distributions). The reason is that usual techniques do not control an error of $\varepsilon$, but an error of $\varepsilon \Pi(r^*)$, which can be arbitrarily smaller if $\Pi(r^*)$ is close to 0. The assumption that the support is included on $[1,H]$, ensures that $\Pi(r^*) \geq 1$. With a simple renormalisation, we can show that the sample complexity of monopoly price scales as by $\rho\varepsilon^{-2}\log(\rho/\varepsilon)\log(1/\delta)$ if the distribution is supported on $[a,b]$, with $\rho = b/a$.
 \begin{proof}
 Let  F be a bounded distribution whose support in included in $[1,H]$ and let us denote by $r^* = \argmax_r r(1-F(r))$  the monopoly price, by $\widehat{F}$ the empirical CDF and by $\hat{r}^* = \argmax_r r(1-\widehat{F}(r))$ the empirical monopoly price.
 
We recall Dvoretzky-Kiefer-Wolfowitz (DKW) inequality \citep{MassartTightConstantDKWAoP90}: if $C_T(\delta)=\sqrt{\log(2/\delta)/2T}$, then
$$
\mathds{P}\{\sup_x |\widehat{F}_T(x)-F(x)|>C_T(\delta)\}\leq \delta\;.
$$
As a consequence, with probability $1-\delta$
\begin{align*}
r^*(1-F(r^*)) - \hat{r}^*(1-F(\hat{r}^*)) &\leq \bigg(r^*(1-F(r^*)) - r^*(1-\widehat{F}(r^*))\bigg) + \bigg(r^*(1-\widehat{F}(r^*)) - \hat{r}^*(1-\widehat{F}(\hat{r}^*))\bigg)\\
& \qquad + \bigg(\hat{r}^*(1-\widehat{F}(\hat{r}^*)) -  \hat{r}^*(1-F(\hat{r}^*))\bigg)\\
&\leq 2H\sqrt{\frac{\log(2/\delta)}{2T}}
\end{align*}
The second term is negative by definition of the empirical monopoly price.

 Choosing $T = 2H^2\varepsilon^{-2}\log(2/\delta)$ gives an $(1-\varepsilon)$-approximation of the revenue-maximizing auction with probability $1-\delta$.
 \end{proof}
 
For the optimal proof, see \cite{huang2018making}. These sample complexities match the lower bounds provided in \citep{huang2018making} up to logarithmic factors. 

Unfortunately, this simple approach does not generalize to regular distributions, especially to heavy-tailed distribution. Intuitively, there exists a constant probability that a few outliers generate an empirical monopoly price arbitrarily large.  This intuition is formalized in the following proposition. 
\begin{proposition2} There exists a regular distribution $F$ and two constants $\eta_0, \delta_0>0$ such that, for any sample size $T$, 
%
%
%If we denote $S_m = \{x_1,...,x_m\} \sim F^m$, $\widehat{F}$ the empirical distribution corresponding to $S_m$, $r^* = \argmax r(1-F(r))$ and $\hat{r}^* = \argmax r(1-\widehat{F}(r))$, there exists a regular distribution F, $\eta > 0$, and $\delta > 0$ such that for all $m\geq 0$, 
$$\mathds{P}_{S_T \sim F^{\otimes T}}\Big\{\Pi(\hat{r}^*) < (1- \eta_0)\Pi(r^*) \Big\}>\delta_0.$$
\end{proposition2}
\begin{proof}
Consider $F(x) = 1-1/x$ for $x < 2$ and $F(x) = 1 - 1/(2(x-1))$ for $x > 2$. Then $F$ is regular since $\psi(x) = 0$ for $x < 2$ and $\psi(x) = 1$ for $x > 2$ and the monopolistic revenue is equal to 1. 

On the other hand, for any sample size $T \in \mathds{N}$, the following holds
$$ \mathcal{P}_{S_T \sim F^{\otimes T}}\{\exists x \in S_T,\ x \geq 4T\} \geq \exp(-1/16).$$
Hence, with a constant probability, $\hat{r}^*$ must verify $\hat{r}^*(1-\widehat{F}(\hat{r}^*)) \geq 4T\frac{1}{T}= 4$ which entails that $\hat{r}^* \geq 4$. This implies in turn that $\Pi(\hat{r}^*)=\hat{r}^*(1-F(\hat{r}^*)) \leq \frac{2}{3}=(1-\frac{1}{3})\Pi(r^*)$.
\end{proof}
This problem is related to the estimation of the mean of heavy tailed distributions. We refer the interested reader to \citep{lugosi2019mean} for a precise survey on algorithms used to estimate the mean of heavy-tailed distributions. 

To handle heavy-tailed regular distributions, a solution introduced in \citep{dhangwatnotai2015revenue} consists in removing the largest samples.

\begin{definition2}%[Guarded Empirical monopoly price] 
Given a dataset $S_T = \{x_t\, |\, t \in [T]\}$,  assuming the $x_t$ are ordered so that $x_1 \leq x_2 ... \leq x_T$, and an accuracy parameter $\kappa >0$, we   denote by  $$S_T^{\kappa} = \{x_t\, |\, t \leq (1-\kappa) T\}$$ the subset of $S_T$ where the highest $\kappa T$ data-points are removed and by $\widehat{F}^\kappa$ the the empirical value distribution on it. Then the  guarded empirical monopoly price is 
$$\argmax_r r(1-\widehat{F}^\kappa(r))\;.$$
\end{definition2}
\begin{theorem2}[\citet{dhangwatnotai2015revenue}]
The sample complexity  of the guarded empirical monopoly price with $\kappa=\varepsilon$ is of order
$\Theta(\varepsilon^{-3}\log(1/\varepsilon)\log(1/\delta))$ for regular distributions.
 \end{theorem2}

This result is another instance of the classical bias-variance tradeoff \citep{dhangwatnotai2015revenue}, as  removing the high values from the dataset reduces the variance of the estimator at the cost of introducing a small bias.
 
 \medskip
 
\begin{figure}
\includegraphics[width=0.7\textwidth]{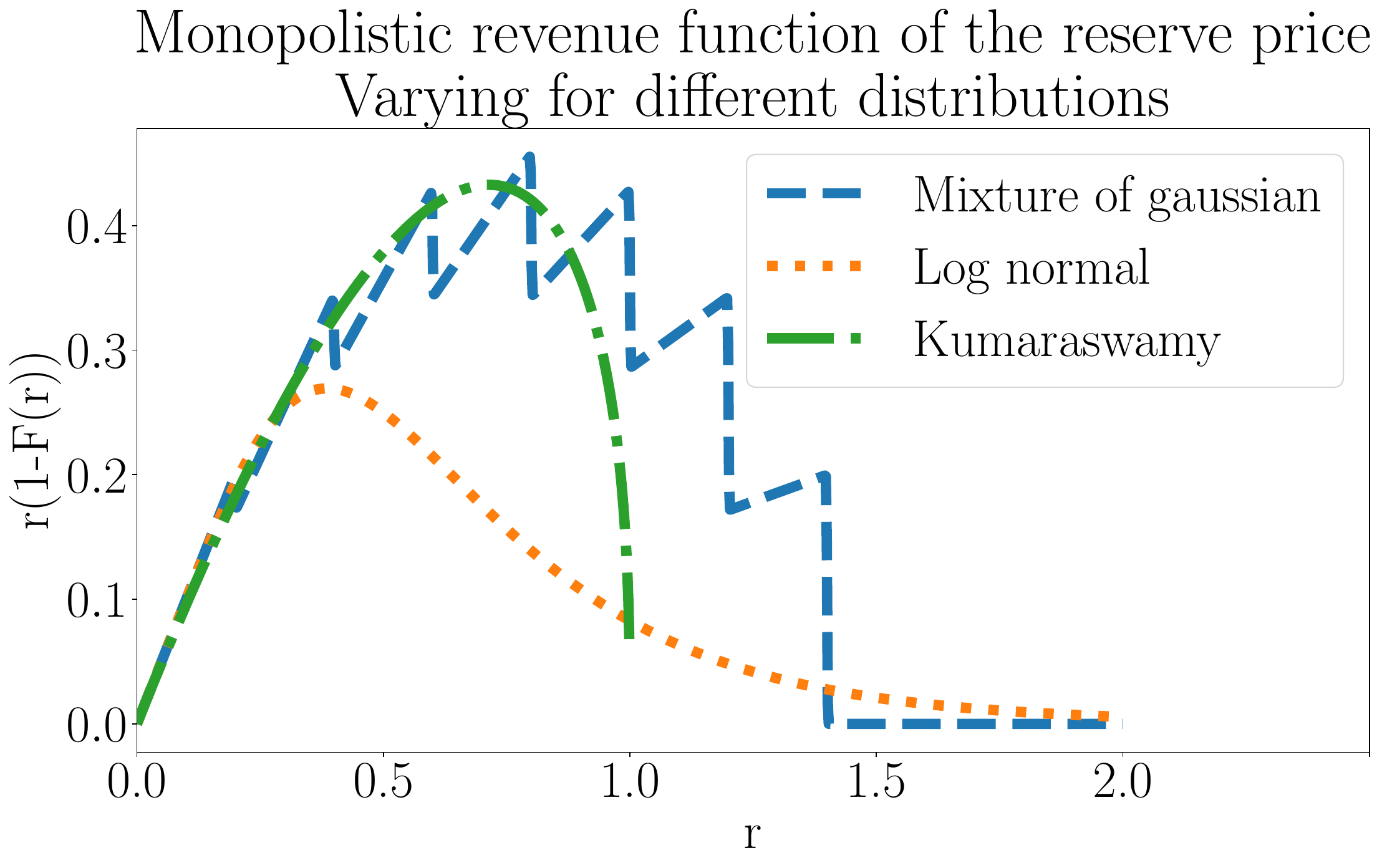}
\caption{Several monopolistic revenues depending on the possible value distribution. The point yellow corresponds to a log normal distribution with mean 0.5 and scale 0.5, the point-dashed green corresponds to a Kumaraswamy distribution with parameters a= 2 and b=10 and the dashed blue one correspond to a mixture of 7 Gaussian with mean equal respectively to (0.2, 0.4, 0.6, 0.8, 1.0, 1.2, 1.4)  and standard deviation 0.001.}
\label{fig:regular_mhr_profit}
\end{figure}

The sample size is smaller for MHR distributions as they induce  strongly concave expected revenue curve, which limits the number of potential candidates around the actual monopoly price. A crucial point in the formal proof is that estimation errors for different possible prices are highly correlated. Indeed, if there are more lower values than expected, the revenue of the true monopoly price will be underestimated but this will be also the case for all prices near this optimal one. From a computational point of view, MHR and regularity  properties also lead to easier optimization problems as illustrated in Figure \ref{fig:regular_mhr_profit}. The monopolistic profit function is represented in three cases: mixture of Dirac, regular and MHR. In general, there are no reasons for the monopolistic profit function to only have one (local) maximum, so that optimizing is challenging. On the other hand,  if $F$ is regular, this profit function is pseudo-concave and has only global (and no local) maxima. As a consequence, the global optimisation is feasible, but at the cost of uniformly good rates as the function can be quite flat in some places. If $F$ is MHR, the profit function is log-concave, its optimization can thus be solved more efficiently.

\subsection{Bounding the Estimation Error}
The estimation term quantifies how far is the performance of the predictor $\widehat{a}_\cA(S_T)$ found by ERM in class $\cA$ from the one of the best predictor in $\cA$, denoted $a^*_\cA(\bF)$. Intuitively, one can expect that under mild assumptions, the estimation error reduces to 0 when $T \to \infty$, the question being the dependency of the speed of convergence in  the size of the hypotheses class $\cA$ and on the sample size $T$. When the class of hypotheses is finite, this estimation error can be controlled with a union bound on some basic concentration inequalities. 
\begin{proposition2} 
Consider a finite class of auctions $\mathcal{A}$ and $S_T \sim \bF^{\otimes T}$ a dataset of $T$ value vectors drawn i.i.d from $\bF$. Assume that the support of $\bF$ is included in $[0,H]^n$.  Then, for all $\varepsilon>0$ and $\delta>0$, if $T \geq  \frac{H^2}{2\varepsilon^2}\Big(\ln\big(\frac{4}{\delta}\big)+\ln\big(|\mathcal{A}|\big)\Big)$, then
$$\mathds{P}_{S_T\sim\bF^{\otimes T}}\Big\{R(\widehat{a}_\cA(S_T)) - R(a^*_\cA(\bF)) \leq \varepsilon\Big\} \geq 1-\delta$$
\end{proposition2}
\begin{proof} 
Given any $a \in \mathcal{A}$,  the following holds
\begin{align*}
R(a)  - R(\widehat{a}_\cA(S_T)) &= R(a)  - \widehat{R}_{S_T}(\widehat{a}_\cA(S_T)) + \widehat{R}_{S_T}(\widehat{a}_\cA(S_T))  - R(\widehat{a}_\cA(S_T))
\\& \leq R(a)  - \widehat{R}_{S_T}(a) + \widehat{R}_{S_T}(\widehat{a}_\cA(S_T))  - R(\widehat{a}_\cA(S_T)) \qquad (\text{by definition of ERM})
\\& \leq 2 \max_{a\in \mathcal{A}} |\widehat{R}_{S_T}(a) - R(a)|
\end{align*}
Then, 
\begin{align*}
\mathds{P}\left\{\exists a \in \mathcal{A},\left|\widehat{R}_{S_T}(a)-R(a)\right|>\varepsilon\right\} &\leq \sum_{a \in \mathcal{A}} \mathds{P}\left\{\left|\widehat{R}_{S_T}(a)-R(a)\right|>\varepsilon\right\}
 \\& \leq 2|\mathcal{A}| \exp \left(\frac{-2 T \varepsilon^{2}}{H^2}\right)
\end{align*}
Setting the right-hand side to $\delta/2$ finishes the proof. 
\end{proof}
This generalization bound is uninformative if the size of the hypotheses space is infinite which happens for large families of value distributions. This simple proof can nonetheless be extended to an infinite set of hypotheses using standard statistical learning tools. The general idea is to reduce the analysis of an infinite class of auctions to a finite set of hypotheses. First, we  introduce below different notions to quantify the complexity on a hypotheses set. 
\begin{definition2}%[Rademacher complexity] 
Let $\mathcal{A}$ be a class of auction and  $S_T=(\textbf{x}_1,\dots,\textbf{x}_T)$ be a fixed dataset of values vector. The \emph{empirical Rademacher complexity} of $\mathcal{A}$ with respect to $S_T$ is 
$$\widehat{\text{Rad}}_{S_T}(\mathcal{A}) = \mathds{E}_\sigma\Big[\sup_{a\in \mathcal{A}}\frac{1}{T}\sum_{t=1}^T \sigma_tr_a(\textbf{x}_t)\Big]$$
where $\sigma = (\sigma_1,\dots,\sigma_T)$ with $\sigma_i$ i.i.d uniform random variables in $\{-1,+1\}$. 

If $\bF$ is a distributions over values vector, its Rademacher complexity is the expectation of the empirical Rademacher complexity over $\bF$
$$\text{Rad}_{T}(\mathcal{A},\bF) =\mathds{E}_{S_T \sim \bF^{\otimes T}}\big[\widehat{\text{Rad}}_{S_T}(\mathcal{A})\big]$$
\end{definition2}

In order to use classical concentration bounds,  possible revenue of an auction should be uniformly bounded over all possible values vectors. Hence, the scope of the next theorem, which is a key result from the learning theory \citep{koltchinskii2002empirical,bartlett2002model}, is restricted to  distributions with support included in $[0,H]^n$ and to the class of ex-post individually-rational auctions. 
\begin{theorem2}
Let $\mathcal{A}$ be a class of ex-post individually-rational auctions and $\bF$ a distribution over values vectors space $\boldsymbol{\mathcal{X}}$ bounded by $H$. Then, for any $\delta > 0$, with probability $1-\delta$ over a dataset $S_T=(\textbf{x}_1,\dots,\textbf{x}_T) \sim \bF^{\otimes T}$, it holds: 
$$\forall a \in \mathcal{A}, \ \ R(a) \leq \frac{1}{T}\sum_{t=1}^T r_a(x_t) + 2\text{Rad}_{T}(\mathcal{A},\bF) + H\sqrt{\frac{\log(\frac{1}{\delta})}{2T}}.$$
\end{theorem2}
As a consequence, computing the Rademacher complexity bounds the estimation error for all the auctions in the hypotheses class. Unfortunately, in most cases, this task is quite challenging. So weaker concepts,  the Vapnik-Chervonenkis (VC) dimension (for the binary-class problem) and the pseudo-dimension (for the real-valued hypotheses classes), were introduced. They can be used to establish generalization bounds and are easier to compute than the Rademacher complexity since they are pure combinatorial notions.  
\begin{definition2}%[Pseudo-dimension]
Let $\mathcal{A}$ be a set of hypotheses on $\boldsymbol{\mathcal{X}}$. A dataset $S_T= (\textbf{x}_1,\dots, \textbf{x}_T) \in \boldsymbol{\mathcal{X}}^T$  is pseudo-shattered by $\mathcal{A}$ if there exists $(\theta_1,\dots,\theta_T) \in \mathds{R}^T$, such that for all $(c_1,\dots,c_T) \in \{-1,1\}^T$, there exists $a \in \mathcal{A}$ such that 
$$\forall t \in \{1,\ldots,T\},\ \mathrm{sign}(r_a(\bx_t) - \theta_t) = c_t.$$
If this $(\theta_1,\ldots,\theta_m)$ exists, it witnesses the shattering.
The pseudo-dimension of $\mathcal{A}$ is the cardinality of the largest set of points in $\boldsymbol{\mathcal{X}}$ that can be pseudo-shattered by $\mathcal{A}$. 
$$P_{\text{dim}}(\mathcal{A}) = \max\Big\{T \in \mathds{N}\ |\ \exists S_T \in \boldsymbol{\mathcal{X}}^T \text{ such that } S_T \text{ is pseudo shattered by } \mathcal{A}\Big\}$$
\end{definition2}
The following result from learning theory  gives a uniform generalization bound in terms of the pseudo-dimension. 
\begin{theorem2}[\cite{haussler1992decision}]\label{Statistical_learning}
Let $\mathcal{A}$ be a class of ex-post individually-rational auctions and $\bF$ a distribution over values vectors space $\boldsymbol{\mathcal{X}}$ bounded by $H$. Then, for any $\delta > 0$, with probability $1-\delta$ over a dataset $S_T=(\textbf{x}_1,\dots,\textbf{x}_T) \sim \bF^{\otimes T}$, for all $a \in \mathcal{A}$, 
$$R(a) = \mathds{E}_{\textbf{x} \sim \bF}(r_a(\textbf{x})) \leq \frac{1}{T}\sum_{t=1}^T r_a(x_t) +  H\sqrt{\frac{2P_{\text{dim}}(\mathcal{A})\log(\frac{T}{P_{\text{dim}}(\mathcal{A})})}{T}} + H\sqrt{\frac{\log(\frac{1}{\delta})}{2T}}.$$
\end{theorem2}
This result is derived from  an upper bound of the Rademacher complexity, as a function of the pseudo-dimension. It can be translated in terms of sample complexity. 
\begin{proposition2}[\cite{haussler1992decision}]
\label{PR:Sample_Comp}
\label{prop:haussler}
Let $\mathcal{A}$ be a class of ex-post individually-rational auctions and $\bF$ a distribution over values vectors space $\boldsymbol{\mathcal{X}}$ bounded by $H$. Then, for any $\delta > 0$, for any $\varepsilon>0$, if $T=\Theta\bigg(H^2\varepsilon^{-2}\big(P_{\text{dim}}(\mathcal{A})\log(\frac{H}{\varepsilon}) +\log(1/\delta)\big)\bigg)$, then
$$\mathds{P}_{S_T\sim\bF^{\otimes T}}\Big\{\exists a \in \cA, |\widehat{R}_{S_T}(a)-R(a)| > \varepsilon\Big\}\leq \delta$$
\end{proposition2}

This result links the sample size $T$ to the "richness" of the auction class $\cA$ on the estimation error. It was first originally applied in learning auction as such \citep{morgenstern2015pseudo}, before being extended \citep{devanur2016sample,gonczarowski2017efficient,guo2019settling}. In Sections \ref{Section:NoApprox} and \ref{Section:Tractability}, we will use the pseudo-dimension to quantify the estimation error through this result.
%We are now equipped with the tools to analyse the error we can expect when learning with different classes of functions, thus we can review some of the methods that have been proposed in the literature. Intuitively, the order in which we present the methods is by increasing tractability and, unfortunately, also increasing approximation error.

\section{Auctions with Asymptotically No Approximation Error}
\label{Section:NoApprox}
We first present families of auctions without asymptotical (as $T\to\infty$)  approximation error. Said otherwise, either there is no approximation error at all or the approximation error can be sent to 0, by controlling a parameter dependent on $T$. 
%\clem{sample complexity should be done next to Haussler.}

\subsection{Approximation of Myerson Auction with Empirical Distributions}
\label{emp_Myerson}
%The first approach consists in trying to estimate the virtual values and to plug-in these estimate in Myerson optimal auction. Unfortunately, it requires a very high number of samples to output a reasonable auction. Indeed, approximating the virtual value is prone to overfitting since it needs to estimate the pdf of the value distribution very accurately (pdf is at denominator). 

With multiple bidders, the most natural approach is to generalize the posted price techniques and to estimate for each bidder the empirical value distribution $\widehat{F}_i$. Based on these empirical distributions, the seller can compute empirical virtual values and run the Myerson auction with these virtual values. Unfortunately, this approach is prone to overfitting and the Myerson auction run on empirical distributions  has  poor revenue guarantees. Intuitively, this comes from the definition of the virtual value; although a quantile $1-F(x)$ can be easily estimated, this is not the case for the associated density $f(x)$ even in not so pathological cases. Then, the virtual value is not precisely estimated which leads to poor performances of the empirical Myerson auction. 
%\clem{I'd like to have more details on empirical Myerson. For instance detail how the pdf is estimated.}

%To go more in details, the approximation error is 0 here, but the estimation error decreases very slowly with the number of samples $m$ There exists a bound on the number of samples needed to insure that the Myerson auction run on the empirical distribution is an $1-\varepsilon$-approximation of the revenue-maximizing auction. 
\begin{theorem2}%[Generalization bound of the empirical Myerson auction]
[\cite{cole2014sample,devanur2016sample}]
With regular value distributions, the sample complexity of the empirical Myerson auction is of order $$\Theta\big(n^{10}\varepsilon^{-7}\log^3(\frac{n}{\varepsilon})\log(\frac{1}{\delta})\big).$$
\end{theorem2}
%{\color{red} VP: Quelqu'un m'a dit que $\log^2(x)$ pouvait etre interprete commme $\log\log(x)$ - et pareil pour $\log^3(x)$. C'est apparemment plus ou moins standard en info. On s'en fout ou bien ?}
This result can be  extended to non-regular distributions \citep{roughgarden2016ironing}. 
The bound can also be improved in the case of distributions with bounded support included in $[1,H]$: the sample complexity its of order $\Theta(n^2H^2\varepsilon^{-2}\log\frac{1}{\delta})$, disregarding the computational burden of ironing.  We mention here that running the Myerson auction requires distributions that are regular, which is not the case of empirical ones. To circumvent that issue, the trick consists in, roughly speaking, ironing the empirical distributions (see  Section \ref{SE:General_OptimalityResults}).

%{\color{red}VP: I think we should include this last result in the Theorem. What about MHR distributions }
\paragraph{Algorithmic Complexity.} The running time of the empirical Myerson auction is also quite high. It takes $\cO(nT\log T)$ operations to compute the empirical cdf. Then, each time the auction is run, computing the attribution and the payment takes $\cO(nT)$.
%\clem{Check this after clarifying how empirical Myerson is actually implemented (with KDE or else).}

\subsection{$L$-level auctions}
\label{$L$-level-auctions}
The class of $L$-level auctions is a generalization of the second-price auctions with reserve price \citep{morgenstern2015pseudo}. Each bidder has $L$ floors, denoted by $r_i^{\ell}$ for $i \in \mathcal{N}$ and $\ell \in \{0,\ldots, L-1\}$. Given a bid $b_i$ of bidder $i$, his index is defined as $$\iota_i = \max\{\, \ell \in \{0,\ldots, L-1\} \text{ such that } b_i \geq r_i^\ell\}, \quad \text{with} \quad \iota_i = - 1\ \text{ if }\ b_i < r_i^0.$$

The winner of the auction is the bidder with the highest non-negative index: if  all bidders have an index equal to $-1$,  the item is not allocated. Ties are broken at the advantage of the highest bidder among those with highest bid, or by standard decision rules (such as uniformly at random).

The payment rule is defined according to Corollary \ref{corollary_charect_DSIC}, and to ensure that the auction is DSIC, it is the lowest winning bid. Formally, with the above breaking-tie rule, the payment  of the  bidder $i$ if he won is equal to 
\begin{itemize}
\item $r_i^{0}$ if all other bidders have index -1,
 \item $\max\{ r_i^{\tau}, b_j \}$ where $\tau>-1$ is the index of the bidder $j$ that would have won without bidder $i$
 \end{itemize}
This class of $L$-level auctions interpolates between the eager second-price auction and the Myerson auction. Indeed, the 1-level auction is equivalent to the eager second-price auction. When $L \rightarrow \infty$, the Myerson auction can be approximated with appropriate reserve prices -- i.e. the approximation error can be made arbitrarily small by taking $T$ arbitrarily large.

\begin{theorem2}%[Approximation error of the class of $T$-level auctions]
[\cite{morgenstern2015pseudo}]\label{TH:LlevelCompl}
Let  $\mathfrak{F}$ be the class of distributions  whose support is included in $[1,H]^n$ and  $L = \Theta(\frac{1}{\varepsilon} + \log_{1+\varepsilon}(H))$, then for any $\bF \in \mathfrak{F}$, there exists a $L$-level auction  with a revenue higher than $1-\varepsilon$ times the revenue of the Myerson auction .
\end{theorem2}

 The idea of this theorem is quite simple, the class of $L$-level auctions, for the above well-chosen value of $L$, is an $\varepsilon$-net for the class of regular value distributions  \citep{morgenstern2015pseudo}. Moreover, the estimation error can be controlled by computing  its pseudo-dimension.
 
\begin{theorem2}%[Pseudo-dimension of the class of $L$-level auctions]
[\cite{morgenstern2015pseudo}]
\label{thm:pseudo_dim_t_level}
Let $\cA$ be the class of $L$-level auctions with $n$ bidders, then its pseudo-dimension satisfies
$$P_{\rm dim}(\cA) = \Theta_{L\to\infty}(nL\log(nL))\,.$$
\end{theorem2}
\medskip

\begin{remark2}
\label{rem:pseudo_dim_t_level}
 Theorem \ref{thm:pseudo_dim_t_level} only provides a scaling of $P_{\rm dim}(\cA)$ with respect to $nL$ for the sake of simplicity. A more precise relation can be derived \citep{morgenstern2015pseudo} since $ 2^{P_{\rm dim}(\cA)} \leq (n P_{\rm dim}(\cA) + nL)^{3nL}$. This will be useful to derive a pseudo-dim for the class of second-price auctions later (i.e. $L=1$).
\end{remark2}

The approximation error will be smaller than $\varepsilon$  by setting  $L =  \Theta(\frac{1}{\varepsilon} + \log_{1+\varepsilon}(H))$ thanks to Prop. \ref{prop:haussler}. Similarly,  the estimation error will be smaller than $\varepsilon$ with $$T=\Theta\Big(H^2\varepsilon^{-2}\big(P_{\rm dim}(\mathcal{A})\log(\frac{H}{\varepsilon}) +\log(1/\delta)\big)\Big)$$ samples, 
 where $P_{\rm dim}(\cA) = \Theta_{L\to\infty}(nL\log(nL))$.
Combining these two claims gives the following.
\begin{corollary2} \label{CO:MyersonComp|}
Let $\mathfrak{F}$ be the class of distributions  with support included in $[1,H]^n$.  For $\varepsilon>0$ and $\delta>0$,  the sample complexity of $L$-level auctions with $L= \Theta(\frac{1}{\varepsilon} + \log_{1+\varepsilon}(H))$ is of order  $$T=\Theta\big(\frac{H^2n}{\varepsilon^3}\log\frac{1}{\delta}\big).$$
\end{corollary2}

%This means that, if bidders have value distributions bounded by H, and the seller has access to a dataset of values of size $m=\Theta\big(\frac{H^2n}{\varepsilon^3}\log\frac{1}{\delta}\big)$, if the seller finds the Empirical Revenue maximizer auction on this dataset in the class of T-auctions with $T=\Theta(\frac{1}{\varepsilon} + \log_{1+\varepsilon}(H))$, then, with probability $1-\delta$, this auction is an $1-\varepsilon$ approximation of the Myerson auction. This gives an upper bound on the sample complexity of learning a revenue maximizing auction in the class of bounded distributions of order $\mathcal{O}\big(\frac{H^2n}{\varepsilon^3}\log\frac{1}{\delta}\big)$.
%\begin{remark2}[Strong improvement compared to the empirical Myerson auction]
%We remark that to get a $1-\varepsilon$ approximation, we only need a number of samples of order $\varepsilon^{-3}$ versus $\varepsilon^{-7}$ for the empirical Myerson auction described in \citep{cole2014sample}.
%\end{remark2}

\paragraph{Algorithmic Complexity.} This improvement over the empirical Myerson auction in terms of sample complexity comes at a cost: tractability. Even though each time the auction is run only requires $nL$ operations (independent of $T$) to compute attribution and payment, the computation of optimal $L$-level auction -- i.e. of the optimal values for $r_i^\ell$ -- is NP-hard  \citep{paes2016field,roughgarden2016minimizing}.

\subsection{Further improvements.}
These results were  improved to reach first a linear dependence in $H$, by preprocessing the data, removing some outliers and running the Myerson auction on this new empirical distribution. \citep{devanur2016sample}. It has been even further improved with a bound in $\mathcal{O}(nH\varepsilon^{-2}\log\frac{1}{\delta})$ for [1,H]-value distributions, by building an ad-hoc empirical distribution  called the dominated empirical distribution \citep{guo2019settling}. State of the art are summarized in Table \ref{sample_complexity_bounds}, taken from \citep{guo2019settling}.

\begin{table}[H]
\begin{tabular}{|l|l|l|}
\hline
Setting   & Lower bound               & Upper bound                    \\ \hline
Regular   & $\Omega(n\varepsilon^{-3})$  & $\mathcal{O}(n\varepsilon^{-3}\log\frac{1}{\delta})$  \\ \hline
MHR       & $\Omega(n\varepsilon^{-2})$  & $\mathcal{O}(n\varepsilon^{-3}\log\frac{1}{\delta})$  \\ \hline
{[}1,H{]} & $\Omega(nH\varepsilon^{-2})$ & $\mathcal{O}(nH\varepsilon^{-2}\log\frac{1}{\delta}$) \\ \hline
\end{tabular}
\caption{Current status of sample complexity bounds in the batch setting depending on the class of value distributions. Table taken from \citep{guo2019settling}.}
\label{sample_complexity_bounds}
\end{table}

The question of finding the optimal sample complexity is more or less settled for different interesting classes of distributions. On the other hand, most of the ``optimal'' (in the sense that some upper-bound matches the associated lower bound, up to logarithmic terms) techniques suffer from their  computational complexity of running the optimal auction (and not learning it), as the empirical Myerson auction method. Indeed, learning the optimal auction has a cost of $\cO(nT\log T)$, but  running it has a fixed cost of $\cO(nT)$ operations to compute each allocation and each payment. There is a clear tension: the smaller the error~$\varepsilon$ (large values of $T$), the larger the running time of the optimal auction. This is simply unpractical for large-scale auctions systems such as the Ebay example. The next subsection describes how to handle revenue maximization on more tractable auctions, at the cost of keeping a bounded, yet incompressible, approximation error.

\section{Tractability at the Cost of  Approximation Error}
\label{Section:Tractability}
%{\color{red}VP: J'ai pas compris la premiere phrase}
%\clem{better ?}
It is quite desirable,  for an auction aiming to be implemented in practice, to have a  complexity independent of $T$ both for  computing the attribution and the payment. This might require searching in a parametric class of auctions $\cA$ rather than in a non-parametric one, like the different versions of the empirical Myerson auction. Further, the number of parameters also has to be independent of $T$, contrarily to $L$-level auctions with $L=\Theta(\frac{1}{\varepsilon} + \log_{1+\varepsilon}(H))$, as $\varepsilon$ is linked to $T$. For instance, a $L$-level auction with $L$ fixed independently of $T$ would be satisfying: the running time of the auction is independent of $T$, at the expense of not having anymore asymptotically zero approximation error. In order to provide simple insights, we are going to focus on simpler families of auctions $\cA$ derived from the second-price auction. 
We are mostly going to focus on controlling the approximation error, so we first introduce new require notations.
\begin{definition2} 
An auction family $\mathcal{A}$ is  a $\kappa$-approximation of the Myerson auction for a class of distribution $\mathfrak{F}$ if
$$\forall \bF \in \mathfrak{F}, \quad R(a^*_\cA(\bF)) \geq \frac{1}{\kappa} R(a^*(\bF))$$
%{\color{red}VP: Not clear. $\bF$ is just one distribution ? or a class of distribution ? and it is at the truthful equilibrium, did we mention this at some point ?}
\end{definition2}

Stated otherwise, the approximation error has the form
\begin{align*}
R(a^*(\bF)) - R(a^*_\cA(\bF)) \leq \left(1 - \frac{1}{\kappa}\right) R(a^*_\cA(\bF)) \,.
\end{align*}

 \subsection{Second-price auctions with anonymous reserve price}
 First, let us consider the simple family of second-price auctions with an anonymous reserve price (which contains the Myerson auction in the symmetric case). From the learning point of view, only one parameter must be learned.
 
\begin{proposition2}%[pseudo-dimension of second price auctions with anonymous reserve prices] 
Let  $\mathcal{A}$ be the class of second-price auctions with anonymous reserve prices. For $n\geq 2$, the pseudo-dimension of $\mathcal{A}$ is
$$P_{\rm dim}(\mathcal{A}) = 2\,.$$
\end{proposition2}
\begin{proof}
First, an auction in $\cA$ is defined by only one parameter,  thus we identify it  with its anonymous reserve price and we  denote it by $a$ for simplicity. Finding a set of cardinality 2 that can be pseudo-shattered by $\cA$ is trivial. We only need to prove that any set of dimension 3 and higher cannot be shattered. 

We remind that a dataset $S_T = (\bx_1, \dots, \bx_T)$ of size $T$ is pseudo-shattered by $\cA$ if there exists $\theta\in\lR^T$, such that for any $c\in\{-1, 1\}^T$, there exists $a\in\cA$ such that $\forall t\in [T]$, ${\rm sign}(r_a(\bx_t)-\theta_t) = c_t$.

Regardless of the number of bidders $n$, the function $a\mapsto r_a(\bx)$ is quasi-concave in the reserve price $a$ and thus can cross (strictly) at most twice any threshold $\theta_t$. Hence, a dataset $S_T$ of size $T$ can only generate a subset of $\{-1, +1\}^T$ of size $2T+1$: when $a$ ranges from $0$ to $\infty$, the vector ${\rm sign}(r_a(\bx_t) - \theta_t)$ changes values at most twice per points in $S_T$. Thus, a set $S_T$ can be pseudo-shattered only if $2^T \leq 2T+1$ which means that necessarily $T \leq 2$.
\end{proof}

Proposition \ref{prop:haussler} yields that the sample complexity is $T=\Theta\left(H^2\varepsilon^{-2}\big(\log(\frac{H}{\varepsilon}) +\log(1/\delta)\big)\right)$ for distribution with bounded support on $[1,H]$. Unfortunately, the approximation power of  such a simple class of auctions is poor and the approximation error remains large. 
 
\begin{theorem2}[\cite{hartline2009simple}]
With regular value distributions, the anonymous second-price auctions are a 4-approximation of the Myerson auction.
 \end{theorem2}
\noindent \textit{Sketch of proof:} The proof relies on the following  Bulow-Klemperer variant lemma. 
\begin{lemma2}[\cite{hartline2009simple}]\label{bulow-klemperer-style}
Consider the following two settings. In the first one, there are $n$ bidders with value distribution $F_i$ and, in the second one, there are $2n$ bidders, the original ones and one independent copy of each one of them. The second price auction without reserve price in the second setting is a 2-approximation of the Myerson auction in the first setting.
\end{lemma2}
\begin{proof}
Let $R_{2n}$ be the revenue generated by the second price auction without reserve price with the original set of $n$  bidders (of values denoted by $x_i$) and their $n$ copies (whose values are denoted by $y_i$). Since a bidder is identical to his copy, they both generate the same revenue to the seller. As a consequence, in this auction, the revenue of the original $n$ bidders is equal to the revenue of their $n$ copies, hence equal to $R_{2n}/2$. Lemma \ref{bulow-klemperer-style} states that $R_{2n}$ is bigger than half the revenue of the Myerson auction. This implies that, overall, the original $n$ bidders generate $1/4$ of the Myerson auction revenue.

In this auction, one of the original $n$ bidder gets the item only if his bid  is the highest and, in that case, he pays the second highest bid amongst the $2n$ ones, which is equal to the maximum between the second highest bid of the original bidders and the highest bid of the copies. As a consequence, the allocation and payment of that bidder is exactly the same as in an auction with only the $n$ original bidder with a random reserve price set as the maximum bids of the copies,  i.e., $\max\{ y_i \ ; \ i \in [n]\}:=y^{(1)}$. So there is a second-price auction with random reserve price that is a $1/4$ approximation of the Myerson revenue. It is crucial to notice that this random reserve price $y^{(1)}$ is independent of the highest and second highest values, denoted respectively by  $x^{(1)}$ and $x^{(2)}$  so that the expected revenue at the truthful equilibrium satisfies
$$
\mathds{E}_{y_i\sim F_i} \mathds{E}_{x_i\sim F_i} \big[\max\{x^{(2)},y^{(1)}\} \mathds{1}\big\{x^{(1)} \geq y^{(1)}\big\}\big]  \leq \max_{y^*} \mathds{E}_{x_i\sim F_i} \big[\max\{x^{(2)},y^*\} \mathds{1}\big\{x^{(1)} \geq y^*\big\}\big] 
$$
by the pigeon-hole principle. As a consequence, setting as reserve price $Y^*$ that attains the maximum on the right generates at least $1/4$  of the Myerson auction revenue.
\end{proof}

 \subsection{Second-price auctions with personalized reserve prices}
 \label{Section:NPhard}
The approximation power of this family of auction can be increased with personalized reserve prices, one for each buyer. Recall that two different families of second price auctions with personalized reserve prices can be considered, either eager or lazy \citep{paes2016field}, see Section \ref{optimal_reserve_for_lazy}. 

% \paragraph{Lazy second-price auction.}
%In a lazy second-price auction, the potential winner is the bidder with the highest bid. If he clears his reserve price (if not, the item is not allocated), he pays the maximum between his reserve price and the second highest bid.
% 
% \paragraph{Eager second-price auction.}
% In an eager second-price auction, the set of potential winners S is the set of bidders who clears their reserve (if empty, the item is not allocated). The winner is the one with the highest bid in the set and he pays the maximum between his reserve price and the second highest bid in the set of potential winners.
%
Recall that in an eager auction, the item is allocated more often than in the lazy one: whenever at least one bidder clears his reserve price versus when the highest bidder clears it. However, it is sold at a lower price in the eager version: the maximum between the reserve price and the highest bid \emph{among the other bidders who cleared their reserve}, rather than \emph{amongst all other bidders}.

In practice, the lazy version of the auction is not very interesting, yet it is a very interesting theoretical tool to  understand eager second-price auctions, generally implemented in practice \citep{drutsa2020reserve,choi2019online}. The reason is that, in a lazy second-price auction, being truthful is only a weakly dominant strategy; however,  bidding 0 when the value is below the reserve price and truthful otherwise is also a weakly dominant strategy. With those bidding strategies, the lazy second-price auction becomes \emph{de facto} an eager second-price auction. 

\subsubsection{Lazy second-price auction}

The main reason explaining the relative simplicity of lazy second-price auctions with respect to the  eager version is the decoupling of the revenue maximization problem across the bidders. In the lazy version, the reserve price of a bidder $j\neq i$ has no influence on whether bidder $i$ gets the item and its cost. Those only depends on the bid of bidder $i$ and his reserve price denoted by $r_i$. Hence, as seen in Section \ref{optimal_reserve_for_lazy}, the lazy second-price auction is optimized by setting the personalized reserve prices to the respective bidders' monopoly prices or rather their empirical estimates when given a dataset $S_T$. Unsurprisingly, this gives $n$ independent estimation problems, so that the pseudo-dimension is then $n$ times larger.

\begin{proposition2}%[pseudo-dimension of lazy second price auctions]
\label{PR:PseudodimLazy}
Let $\mathcal{A}$  be the class of lazy second-price auctions with personalized reserve prices. For $n\geq 2$, its pseudo-dimension 
$$P_{\rm dim}(\mathcal{A}) = 2n\,.$$
\end{proposition2}
\begin{proof}[Sketch of proof]
This is a direct extension of the proof for the second-price auction with anonymous reserve prices as the $n$ estimation problems are independent.
\end{proof}
%\clem{Can someone check my claim on this !! I'm off road from existing results here, but it helps being more didactic.}

Proposition \ref{PR:PseudodimLazy} states that the estimation problem is not much harder than with anonymous reserve prices. Using  again Proposition \ref{PR:Sample_Comp},  the associated sample complexity is $T=\Theta\left(H^2\varepsilon^{-2}\big(n\log(\frac{H}{\varepsilon}) +\log(1/\delta)\big)\right)$ for distribution with bounded support on $[0,H]$. However, this simple modification already greatly improve the guarantee on the approximation error.

\begin{theorem2}[\cite{dhangwatnotai2015revenue}]\label{lazy-approx}
With  regular value distributions, the lazy second-price with monopoly reserve prices is a 2-approximation of the Myerson auction. 
 \end{theorem2}
 \begin{proof} 
We divide the revenue of the Myerson auction in two parts and bound each term by the revenue of the lazy second price with monopoly reserve. 
 
First, based on the Myerson lemma, recalling that $\psi_i$ stands for the virtual value function associated to the distribtuion $F_i$, see Definition \ref{def:VirtualValue}, we get
\begin{align*}
R(\text{lazy}(\bF))& = \mathds{E}_{{\bx} \sim \bF}\Big[\sum_{i=1}^n \psi_i(x_i)\mathds{1}\{i \text{ is winning lazy 2nd-price }\}\Big]
\\& \geq  \mathds{E}_{{\bx} \sim \bF}\Big[\sum_{i=1}^n \psi_i(x_i)\mathds{1}\{i \text{ is winning Myerson auction \& lazy 2nd-price}\}\Big]
\end{align*}
 since if $i$ is winning the lazy second-price auction, his virtual value is non-negative. 
 
The revenue of the lazy auction can also be compared to the revenue of the Vickrey auction (i.e. the second-price auction with no reserve price), as follows
\begin{align*}
 & \mathds{E}_{{\bx} \sim \bF}\Big[\sum_{i=1}^n\psi_i(x) \mathds{1} \{i \text{  wins Myerson aution \& not lazy 2nd-price}\}\Big]
\\& =  \mathds{E}_{{\bx} \sim \bF}\Big[\sum_{i=1}^n\psi_i(x)  \mathds{1} \{i \text{ wins Myerson auction \& not Vickrey auction}\}\Big]
\\& \leq  \mathds{E}_{{\bx} \sim \bF}\Big[\sum_{i=1}^nx_i  \mathds{1} \{i \text{  wins Myerson auction \& not Vickrey auction}\}\Big]
\\& \leq \R(\text{Vickrey}(\bF))
\\&\leq R(\text{Lazy}(\bF))\;.
 \end{align*}
Indeed,  the first equality is a consequence of the fact that if a bidder wins the Myerson auction, he has a non-negative virtual value. Hence, if he does not win the lazy second price with monopoly reserve, he does not win the Vickrey auction. The second inequality is deduced from the definition of the virtual value, the third one from the payment of the Vickrey auction and the last one from the Myerson lemma that shows that the monopoly reserve prices are the optimal ones for a lazy second-price auction. 
 
 As a consequence, 
% \begin{align*}
%R(\text{Lazy}(\bF)) &\geq \mathds{E}_{{\bx} \sim \bF}\Big[\sum_{i=1}^n p_i(x) \mathds{1} \{i \text{ is winning lazy 2nd price \& not Myerson auction}\}\Big]
% \\& \geq \mathds{E}_{\bx \sim \bF}\Big[\sum_{i=1}^n \psi_i(x) \mathds{1} \{i \text{ is winning Myerson auction \& not lazy 2nd price}\}\Big]
% \end{align*}
% and
\begin{align*}  2R(\text{Lazy}(\bF)) &\geq \mathds{E}_{{\bx} \sim \bF}\Big[\sum_{i=1}^n \psi_i(x_i)\mathds{1} \{i \text{ is winning Myerson auction \& lazy 2nd price}\}\Big] 
\\&\qquad +  \mathds{E}_{{\bx} \sim \bF}\Big[\sum_{i=1}^n \psi_i(x_i)\mathds{1} \{i \text{ is winning Myerson auction \& not lazy 2nd price}\}\Big]
\\& = R(a^*(\bF))
\end{align*}
The last equality comes from by the Myerson lemma, and this concludes the proof.
\end{proof}

In a nutshell, if the lazy auction were implementable in practice, it would be a good compromise. Its learning time is $\cO(nT \log T)$, its running time is $\cO(n)$, the sample complexity is in $\varepsilon^{-2}$ and it is a 2-approximation of the Myerson auction. 
%Now, we saw that, in practice, the one that can be implemented is the eager version. So, can we have guarantees at least as good for it? {\color{red}VP LOL, seriously ?}

\subsubsection{Eager second-price auction}

First, in terms of estimation, the problem of optimizing  the eager second price auction is  not much harder than the lazy one, as the pseudo-dimensions are rather similar: they  only differ by a factor $\log(n)$.
\begin{proposition2}%[pseudo-dimension of second price auctions with anonymous reserve prices] 
Let  $\mathcal{A}$ be the class of second-price auctions with personalized reserve prices; its pseudo-dimension satisfies
$$P_{\rm dim}(\mathcal{A}) = \mathcal{O}(n\log(n))\,.$$
\end{proposition2}
 \begin{proof}[Sketch of proof]
 This is a corollary of the pseudo-dimension of $L$-level auctions for $L=1$ (Thm. \ref{thm:pseudo_dim_t_level}). As explained in Remark \ref{rem:pseudo_dim_t_level}, it amounts to plugging in the last line of the proof of $L$-level auctions with $L=1$ and conclude.
 \end{proof}

Similarly, the guarantee on the approximation error is the same, it is a 2-approximation, whether the distributions are MHR or regular. 
%\iffalse
%\begin{theorem2} [\citep{hartline2009simple}]
%With MHR value distributions, the eager second price with monopoly reserve prices is a 2-approximation of the Myerson auction. 
% \end{theorem2}
% \begin{proof} 
% We denote by $q_i$ the allocation rule corresponding to the eager second price with monopoly reserve, by $q_i^{*}$ the allocation rule corresponding to the Myerson auction and by $Rev$ seller's revenue depending on the auction.  
% 
%The eager second price is incentive-compatible and verifies the Myerson lemma. 
%$$Rev_{eager} = \sum_i \mathds{E}_{F_i}(P_i(X_i)) =  \sum_i \mathds{E}_{F_i}(\psi_i(X_i)Q_i(X_i))\;.$$
%A winning bidder pays at least his monopoly price $r_i=\psi_i^{-1}(0)$ when he wins an auction. Hence, 
%$$Rev_{eager} = \sum_i \mathds{E}_{F_i}(P_i(X_i)) \geq  \sum_i \mathds{E}_{F_i}(r_iQ_i(X_i))\;.$$
%For an MHR distribution, 
%$$r + \psi(x) = x + \frac{f(r)}{1-F(r)} - \frac{f(x)}{1-F(x)} \geq x \text{ for } x \geq r\;.$$
%Then, 
%\begin{align*}
%2Rev_{eager} &\geq  \sum_i \mathds{E}_{F_i}((r_i+\psi_i(X_i))Q_i(X_i))  
%\\&\geq \sum_i \mathds{E}_{F_i}(X_iQ_i(X_i)) \geq \sum_i \mathds{E}_{F_i}(X_iQ_i^{*}(X_i)) \geq Rev_{Myerson}
%\end{align*}
%The second inequality is verified since the eager second-price auction allocates the item to the bidder with the highest value among the one who clears their reserve. 
% \end{proof}
% This theorem can be extended to regular distributions with a slightly more involved proof. 
% \fi
  \begin{theorem2}[\cite{hartline2009simple}]
  \label{thm:eager_2_approx_myerson}
With regular  value distributions, the optimal eager second price is a 2-approximation of the Myerson auction. 
 \end{theorem2}
 \begin{proof} 
 We prove that the eager second price with monopoly reserve is a 2 approximation of the Myerson. 
 The proof is very similar to the one of Theorem \ref{lazy-approx}.
 We divide the revenue of the Myerson auction in two parts and bound each term by the revenue of the eager second price auctions. 
 
First, based on the Myerson lemma, 
\begin{align*}
R(\text{eager}(\bF)) &= \mathds{E}_{{\bx} \sim \bF}\Big[\sum_{i=1}^n \psi_i(x_i)\mathds{1}\{i \text{ wins eager 2nd-price}\}\Big]
\\& \geq\mathds{E}_{{\bx} \sim \bF}\Big[\sum_{i=1}^n \psi_i(x_i)\mathds{1}\{i \text{ wins Myerson auction \& eager 2nd-price}\}\Big]
\end{align*}
 since if $i$ is winning the eager second-price auction, his virtual value is non-negative. 
 
The item is allocated in the Myerson auction if and only if the item is allocated in the eager second-price auction. Hence, there exists a one-to-one mapping between a winner in the eager second-price auction and a winner in the Myerson auction. 
 
Consider the case where these two winners are different and denote by $i$ the winner of the eager second-price auction with monopoly reserve and by $j$ the winner of the Myerson auction. Let denote by \textbf{x} the vector of value corresponding to this case and by $p_{eager}(\textbf{x})$ the payment of the eager second price with monopoly reserve for this specific vector of values. 

By definition of the payment rule of the eager second price auction and by definition of the virtual value, 
$$p_\text{eager}(\textbf{x}) \geq x_j \geq \psi_j(x_j).$$
 We conclude the proof with the same reasoning of Theorem \ref{lazy-approx}. Since the eager second price with monopoly reserve is a 2-approximation, the eager second price with optimal reserve is a 2-approximation.
 
 \end{proof}
 
 \paragraph{Algorithmic complexity.} Similarly to the lazy version, running the eager second-price auction has a complexity of $\cO(n)$. The main difference comes in the complexity of learning the set of optimal reserve prices. For the lazy version, the optimal reserve prices are the monopoly prices, that can be computed in $\cO(nT \log(m))$. This is no longer true for the eager version and finding the optimal reserve prices is NP-hard  \citep{paes2016field,roughgarden2016minimizing}, which explain why the more general $T$-level auctions has the same limitation. This seems to be in contradiction with the objective of consider non-zero approximation error to get tractable learning and running complexities. So the question we investigate in the following section is the performance of the eager second-price auctions, but with sub-optimal reserve prices set as the computable monopoly prices. 
\subsubsection{Eager with monopoly reserve prices}
 
We now prove that the eager second-price with monopoly reserve prices generates a higher revenue for the seller than the lazy second price with monopoly reserve prices. This is not obvious as in the eager version, the winning bidder pays the highest second bid in the set of bidders who cleared their reserve price. This second highest bid can be lower than the second highest bid in general which is the one paid in the lazy version. 
  \begin{theorem2}[\cite{fu2013vcg}]%[eager with monopoly reserve dominates lazy]\label{eagerbetterlazy}
With regular value distributions, the revenue of the eager second-price auction with monopoly reserve price is higher than the revenue of the lazy second-price auction. 
 \end{theorem2}
\begin{proof}
 We denote by $r_i$ the monopoly price corresponding to bidder $i$.  We will compare the expected payment of bidder $i$, in the lazy or eager auction, conditioned to the values of all other bidders. First, we  define $x_{-i}^{\text{lazy}} = \max_{j\neq i}\{x_j\}$ and $x_{-i}^{\text{eager}} = \max_{j\neq i, x_j \geq r_j}\{x_j\}$. In particular, this implies that $x_{-i}^{\text{lazy}}\geq x_{-i}^{\text{eager}}$. Moreover,  if $x_{-i}^{\text{lazy}} \leq r_i$, the two auctions are identical for bidder $i$ that pays $r_i$ if $x_i\geq r_i$. 

To compare the two auctions, we can therefore restrict ourselves to the case where $x_{-i}^{\text{lazy}} \geq r_i$. The expected payment of bidder $i$ in a lazy second-price auction is in this case equal to 
$$
x_{-i}^{\text{lazy}}(1-F_{i}(x_{-i}^{\text{lazy}}))
$$ 
For the eager second-price auction, the payment is equal $\tilde{x}_{-i}^{\text{eager}}= \max\{x_{-i}^{\text{eager}},r_i\}$ and the expected payment of bidder $i$ is 
$$
\tilde{x}_{-i}^{\text{eager}}(1-F_{i}(\tilde{x}_{-i}^{\text{eager}}))
$$
Since $F_i$ is regular, then  $x(1-F(x))$ is non-increasing for $x \geq \psi_i^{-1}(0)$. As a consequence, the expected payment in the lazy second price is lower than the expected payment in the eager second-price auction with monopoly reserve prices. 
\end{proof} 

Especially, as a corollary, it means that an eager second-price auction with monopoly reserve is still a 2-approximation of the Myerson auction. Thus, in the end, the eager auction with monopoly reserves provides a good compromise in terms of tractability versus optimality. Indeed, the computational complexity are $\cO(n)$ (running) and $\cO(nT\log(T))$ (learning) and enjoys the pseudo-dimension of learning the monopoly prices of $2n$. The only concern that may remain is the approximation error that is only guaranteed to be less than $\frac{1}{2}R(a^*(\bF))$. In fact, this guarantee is  loose, and the actual error is often much smaller.

\subsubsection{Why the approximation error may not be too large}
We proved in Section \ref{sec-Bulow} that  the Vickrey symmetric auction  with $n$ bidders and regular value distributions is a $\left(\frac{n-1}{n}\right)$-approximation of the Myerson auction. Overall, even though it is not straight-forward to extend it in the non-symmetric cases, this result indicates that when the competition is strong, the approximation error incurred by second-price auctions is not as big as $\frac{1}{2} R(a^*(\bF))$. So, often, the guarantee of the eager second-price auction being a $\frac{1}{2}$-approximation of the Myerson auction is loose and the error is much smaller as shown on Figure \ref{fig:BK}.

\subsection{The boosted second-price auction}
A simple extension to the eager second-price auction has been proposed to empirically improve the seller's revenue by reducing the approximation error: boosted second-price auction \citep{Golrezaei2017}. It relies on the following point of view: the eager second-price auction can be seen as a Myerson auction with approximated virtual value functions $\widetilde{\psi}_i(x) = x-r_i$ where $r_i$ is the personalized reserve price. 

Practically, a drastic improvement in terms of approximation error can be made by adding a slope parameter, different for each bidder. The result is the boosted second-price auction, which is a Myerson auction with approximated virtual values functions $\widetilde{\psi}_i(x) = \beta_i x-r_i$ where $r_i$ is the personalized reserve price and $\beta_i$ the "boost". It turns out that for certain families of distributions (ex: generalized Pareto distributions) the virtual value is affine, making the boosted second-price auction coincide with the Myerson auction. 

It retains the following two good properties of second-price auctions with personalized reserve prices: 1) it is parametric and thus has a reasonable pseudo-dimension and 2) it has a running time independent of the sample complexity $T$. And  it even has a lower approximation error as  the auctions class is strictly larger. Actually, for some families of distributions, it even has 0 approximation error even in an asymmetric setting. The main caveat is the computational complexity of the training. As the eager second-price auction, the global optimization is NP-hard to solve. However, by initializing with an eager second-price auctions ($\beta_i = 1$) with monopoly reserve prices $r_i$ and launching an optimization from there, it proved to empirically perform very well  \citep{Golrezaei2017}.

\subsection{Learning reserve prices in first-price auctions}
\label{learning_reserve_prices_first_price}
As indicated by the second price or Myerson auctions, a crucial step when designing an optimized mechanism consists in estimating the bidders' value distributions. This step is even harder in first price auctions \citep{OptimalNonparametricEstimationOfFirstPriceAuctions2000,AtheyHeile2007ReviewEstimationAuctions} than it is in second price auctions.

The central idea is to assume that  bidder $i \in \mathcal{N}$ optimizes his bids according to the best response strategy described in Equation \ref{eq:bestresponsefirstprice}. Hence - keeping the notations of Subsection \ref{first-price-auction}) -  when his value is $x_i$, he bids $b_i$, a solution of
\begin{equation}\label{eq:estimationValueFromBid}
x_i=b_i+\frac{G_i(b_i)}{g_i(b_i)}\;.
\end{equation}
When there are sufficiently many repeated auctions, the seller can compute/estimate the cdf $G_i$ (and the pdf $g_i$, possibly with  kernel density estimators) from the  distribution of bids of the competition that bidder $i$ is facing. Plugging back this knowledge into Equation \eqref{eq:estimationValueFromBid} gives an estimate of the value $x_i$. Repeating this operation for all bids produces estimate of the value distribution. 

This  approach appeals to stationarity assumptions, a potential drawback. In practice issues may also arise from the fact that bidders and seller may have different estimates of the bid distribution the bidders are facing. Then the seller's estimate of the optimization problem solved by the bidders could be inaccurate.  

\subsubsection{Value distribution estimation and reserve price issues}
Another major difficulty in setting optimal asymmetric reserve prices is that they have a somewhat complex and non-linear impact on the optimal bidding strategies of the buyers and, as a consequence, will affect the allocation probability $q_i(x)$ in a potentially complex fashion at equilibrium. 

However, a natural, yet possibly suboptimal, choice is to set $r_i=\psi_i^{-1}(0)$, i.e. setting the reserve value at the monopoly price $\psi_i^{-1}(0)$, at least for regular distributions. This guarantees that the term under the expectation in Equation \eqref{eq:RevSeller1stPriceWithVValue} is always non-negative -- in a first-price auction,  bidders never bid above their value. This principle is quite similar to one studied in Section \ref{Section:NPhard} for eager second-price auctions, as finding optimal reserve prices is NP-hard \citep{paes2016field} for that type of auctions. However, in first price auctions and other non-DSIC auctions, setting this reserve may induce a change of optimal bidding strategy and hence a different $\beta_i$, making the evaluation of the impact of such choice of reserve price on seller revenue theoretically delicate. To compute the monopoly price, the seller needs to estimate the value distribution of bidder $i$ from his bids, a task we now turn to.

\subsubsection{Setting optimal reserve prices in first-price auctions cannot be done by naive ERM}

In first price auctions, the bidder requires much information about the competition to compute best responses. Even if he knew perfectly the value distribution of the other buyers, it would still be numerically challenging to bid optimally and reach the Nash equilibrium. The situation is even worse when buyers have to estimate the distribution of the competition while bidding. 

On the other hand, setting the reserve prices is also more complicated for the seller. With second price auction, she could gather  data and form a dataset of bids whose  distribution should be close to the value distribution (assuming myopic and non-strategic agent). It is then possible to run an ERM based on this dataset. 

On the contrary, with first price auctions, each bid received has a distribution that depends on the reserve price chosen at the time. And, in the future, choosing another reserve price will induce yet another distribution of bids.  As a consequence, the data from the ``training''  set (past bids) and the ``test'' set (future bids) have different statistical properties and na\"ive empirical risk minimization will not work.

 A seller can however use the theory discussed above to account for the impact of reserve price on bid distributions and  then simulate from these new reserve-price dependent distributions  and measure the effect of different reserve prices in a unbiased way. The difficulty of solving for Nash equilibrium creates nonetheless a hurdle to the practical implementation of such ideas  \citep{feng2020reserve}.

\subsection{New numerical methods for multi-item auctions}
The multi-item framework is more intricate than single-item. Myerson's fundamental result has been extended to specific settings depending on the number of objects and on the properties of the bidders' utility functions  \citep{armstrong1996multiproduct,manelli2007multidimensional,daskalakis2013mechanism,yao2017dominant}.  A general and analytical optimal auction in the multi-item framework has yet to be found. 

Because of the amazingly large variety of different settings, \emph{automatic mechanism design} has been introduced to provide a (numerical) framework for learning revenue-maximizing mechanisms satisfying constraints chosen by the designer \citep{conitzer2002complexity,albert2017automated}. This framework was  complemented by  introducing neural networks  for different instances of the multi-item problem \citep{dutting2017optimal,shen2019automated,golowich2018deep} to take advantage of the large expressivity power of neural networks architectures.

A general algorithmic approach to approximately solve the seller's optimization problem in multi-item, multi-bidder settings has been implemented \citep{dutting2017optimal}. The seller's auction is parametrized by a weight vector $\boldsymbol{\omega} \in \mathds{R}^{n\times m}$ corresponding to two neural networks which take valuations for each item  and each player as inputs and return respectively the allocation probability $q_{\boldsymbol{\omega}}  \in \mathds{R}^{n\times m}$ of each item and each player, and  the payment for each player $p_{\boldsymbol{\omega}} \in \mathds{R}^{n\times m}$ per unit of item. In the case of combinatorial auctions, bidders would submit a bid for each possible bundle (in our setting, a bit $b_s$ then belongs to $\mathds{R}^{n\times 2^m}$). The bidders valuations in a combinatorial setting would not need to be explicitly described as per the large literature on succinct representations of bidding languages for combinatorial auctions. The networks are trained by batches of size $L$, i.e., $L$ vectors $\bx_t  \in \mathds{R}^{(2^m)^n}$ (corresponding to each bundle per bidder) are sampled at each iteration. 

 The first term of the loss function used to train the network is the negated empirical revenue computed on the dataset of valuations $S_T=\{\bx_1,\ldots,\bx_T\}$, 
\begin{equation*}
\mathcal{L}_{\rm Rev} = - \frac{1}{T} \sum_{t=1}^{T} \langle p_{\omega}(\bx_t), q_{\omega}(\bx_t)\rangle
\end{equation*}
To ensure the IC constraint, two different approaches have been considered. The first one is a hard constraint implemented by defining an architecture which is DSIC by design, called MyersonNet. Myerson's lemma  is used to design this architecture that learns the optimal DSIC auction in the single-item setting. However, for each new  setting of the problem, a new architecture must be designed \citep{shen2019automated}.

The second approach, the  RegretNet architecture, uses a soft constraint in a Lagrangian corresponding to the incentive-compatibility objective. For each bidder, the empirical ex-post regret for bidder $i$ is defined as
$$\widehat{\rm Reg}_i(\boldsymbol{\omega}) = \frac{1}{T} \sum_{t=1}^{T}  \max_{{b}^*_{t}} u_i^{\boldsymbol{\omega}}({b}^*_t,{\bb}_{-i,t}) - u_i^{\boldsymbol{\omega}}({b}_{i,t},{\bb}_{-i,t}) $$
This regret is 
%completely different from the online regret we considered in Chapter 3. This is 
the difference between the maximum utility bidder $i$ can get by optimizing his bids  $b^*_t \in \mathds{R}^{n\times 2^m}$ and the utility he gets when bidding truthfully (sort of similarly to the regret introduced in Section \ref{Section:Bandits}, yet the maximum is inside the sum instead of being outside). This quantifies serves as a proxy on how untruthful an auction is: the higher the regret, the less truthful the auction is as  bidders can largely increase their utility by deviating. The augmented Lagrangian method is then used to optimize the Lagrangian function defined as:  
$$\mathcal{L}(\boldsymbol{\omega},\boldsymbol{\lambda}) = \mathcal{L}_{\rm Rev} + \sum_{i=1}^{n} \lambda_i \widehat{\rm Reg}_i(\boldsymbol{\omega}) + \frac{\rho}{2}\bigg(\sum_{i=1}^{n} \widehat{\rm Reg}_i(\boldsymbol{\omega})\bigg)^2
$$
This Lagrangian function is the sum of the negated actual revenue of the mechanism with two penalties which quantify the lack of incentive compatibility, thus insuring that the learned mechanism is approximatively DSIC.  The bids $b_i$, which are maximizing $\widehat{\rm Reg}_i(\boldsymbol{\omega})$, are optimized through gradient descent, making the optimization unfortunately very slow. In some multi-item instances, this approach actually recovers the optimal revenue-maximising mechanisms (when the latter is known theoretically).

This approach can be complemented by introducing a network encompassing the best bids for one specific bidder, avoiding running a gradient descent for each specific value and by trying to take advantage  the continuity of the problem  \citep{rahme2020auction}. The idea is to leverage the fact that if two valuations are close to each other, their optimal bids should be also close to each other. 

These numerical approaches can help theoreticians to identify some good candidates for the revenue-maximizing auctions in more exotic cases when bidders have some budget constraints \cite{feng2018deep}. However, a general theory for designing optimal mechanisms in the general case of multi-item auctions is still out of reach.

\section{Contextual Estimation of Reserve Prices}

In the previous sections, it was more or less implicitly assumed that only one  type of goods were being sold repeatedly. In practice, especially in the motivating examples of repeated auctions like internet advertising, the items sold are  different from one to each other, at least partly. For instance, successive ad slots sold may have same size but different placements, same placement but different size, or may be on different pages of the same website, or on different websites. An easy solution would be to consider all these different items separately, but it would mean only having a small number of samples per items, which would prevent accurate estimation of the monopoly price. To address this issue, some underlying structure and some regularity are required to formalize the idea that samples obtained for one item also informs on the distributions of similar items. 
% Intro
\subsection{Contextual Auctions}
In this section (only), we will assume that an item is described by a public set of $d$ features $\bz\in\lR^d$ such that similar items have similar features (for some distance of $\mathds{R}^d$). By \emph{public}, we mean that $\bz$ is available to both the seller and the bidders. The bidders use this information to estimate their values $\bx=(x_1,\ldots,x_n)$ more accurately;  in particular the distribution of values now depends on $\bz$. For simplicity, we assume in this section that the seller only estimates an anonymous reserve price, the same for all bidders, but that it may depend on the available information $\bz$. The extension to personalized reserve prices is straight-forward when they are independent, like in the case of the lazy second-price auction. Formally, the seller aims at learning a reserve price as a function of $\bz$, hence, a mapping $r^*\in \lR^d \to \lR_+$. For learnability reasons, we will restrict  $r^*$ to belong to some compact sub-class of hypothesis $\mathfrak{R}\subset (\lR^d\to\lR_+)$. The learning of this contextual optimal reserve price relies on the observations of samples from the distribution of values of the bidders. In fact, only the observation of the highest and the second highest value is necessary, so we denote by $x^{(1)}$ the highest value amongst the bidders and by $x^{(2)}$ the second highest value. Further, we denote by $\mathcal{F}$ the joint distribution of $(x^{(1)}, x^{(2)}, \bz)$. Then, $\mathcal{F}(.|\bz)$ is the distribution of the two highest values conditionally to the contextual information $\bz$. In the end, finding the reserve function can be written as follow:
\begin{align}
	&r^* = \argmax_{r\in\mathfrak{R}} \lE_{(x^{(1)}, x^{(2)}, \bz) \sim \mathcal{F}}\left[\phi(r(\bz),x^{(1)}, x^{(2)})\right] \\
	&\text{where}~~ \phi(\rho, x^{(1)}, x^{(2)}) = x^{(2)}\indicator{x^{(2)} > \rho}+\rho \indicator{x^{(2)} \leq \rho \leq x^{(1)}} \nonumber
\end{align}

This problem is quite difficult, both because of the optimization over a set of function $\mathfrak{R}$, and because of the quite complex objective function. Similarly to the monopolistic profit function, for a given fixed $\bz$, if the distribution $\mathcal{F}(.|\bz)$ is regular or MHR, the function $\rho \mapsto \lE_{(x^{(1)}, x^{(2)})\sim \mathcal{F}(.|\bz)\left[\phi(\rho, x^{(1)},x^{(2)})\right]}$ is pseudo-concave or log-concave, but not concave. Unfortunately, a sum of pseudo-concave or log-concave function is in general not pseudo-concave. Thus, whenever considering a parametric class of functions $\mathfrak{R}$ strictly smaller than the whole set of functions, such as linear mappings, the loss marginalized over $\bz$ may not be unimodal in the parameters, leading to hard optimization problems. In the following, we present a high-level overview of methods proposed to solve this learning problem.

\subsection{Linear Reserve Price Function with Surrogate-Based Approaches}
A simple class of functions for $\mathfrak{R}$ are  linear functions: $\cR = \{\bz\mapsto \langle\theta,\bz\rangle: \theta\in\lR^d\}$. However, as  mentioned right before, the objective to optimize, $\theta \mapsto \lE[\phi(\langle \theta,\bz \rangle, x^{(1)}, x^{(2)})]$ is potentially multimodal, as a mixture of \emph{only} pseudo-concave functions.
A first classical direction to tackle such an \emph{a priori} complex learning problem is to try to design a surrogate loss $\widetilde{\phi}$, easier to optimize in expectation (such as concave) than the initial objective $\phi$. Obviously, a key requirement  is that the surrogate problem is \emph{consistent} with the initial one, meaning that the optimization of the former leads to a solution of the latter.  Unfortunately, simplifying the learning problem by finding a consistent concave surrogate formulation is actually impossible, even for the simpler case  where the second highest value $x^{(2)} = 0$ \citep{medina2014learning} . For the sake of notations, we define $\phi_0(\rho, x^{(1)}) = \phi(\rho, x^{(1)}, 0)$.

\begin{theorem}[\cite{medina2014learning}]
Let $\widetilde{\phi} : [0,1]\times[0,1]\to\lR$ be a bounded function, concave in its first argument. If $\widetilde{\phi}$ is consistent with $\phi_0$, then $\widetilde{\phi}(\cdot , x)$ is a constant function for any $x\in[0,1]$.
\end{theorem}

This theorem indicates that learning methods, even based on linear functions for $\mathfrak{R}$, need to rely on non-concave maximization methods. We detail two examples of such methods in the following.

\subsubsection{Solution based on DC-programing}
 A first method is based on the following piece-wise linear surrogate \citep{medina2014learning}:
$$\widetilde{\phi}(\rho,x^{(1)}, x^{(2)}) = \phi(\rho,x^{(1)}, x^{(2)}) - (\rho - (1+\gamma)x^{(1)})\indicator{\rho>x^{(1)}}\indicator{\rho\leq (1+\gamma)x^{(1)}}, ~~~~~\text{for}~ \gamma>0\,.$$
While using a surrogate introduces a bias, as the maximizer in expectation of $\widetilde{\phi}$ will not maximize exactly the expected monopoly revenue, this bias can be made small by taking $\gamma$ close to 0. However, this comes at the cost of making the Lipschitz constant $1/\gamma$ of the surrogate grow significantly. The key idea under this surrogate $\widetilde{\phi}$ is that, because it is piece-wise linear, it is possible to \emph{explicitly} decompose it as a difference of convex functions. This means that the empirical risk $\frac{1}{T}\sum_{t=1}^T \widetilde{\phi}(r(\bz_t), x^{(1)}_t, x^{(2)}_t)$ can, in turns, be explicitly decomposed as a difference of convex functions as soon as $r$ is a linear function. Then, it is possible to optimize this empirical objective on classes $\mathfrak{R}$ of linear function by using DC-programming algorithms, such as  DCA \citep{DCA}. 

\subsubsection{Solution based on Objective Variables}
Another method exploits the idea of the introduction of \emph{objective variables} in a
Bayesian framework \citet{rudolph2016objective}. First, the objective $\phi$ is smoothed using a Gaussian to define the following surrogate:
$$
\widetilde{\phi}(\rho, x^{(1)}_t, x^{(2)}_t) = \log\Big(\lE_{\epsilon\sim\cN(0,\sigma^2)}\Big[\exp(\phi(\rho+\epsilon, x^{(1)}_t, x^{(2)}_t))\Big]\Big), \hspace{0.5cm} \text{for some}~\sigma>0.
$$
As $\sigma$ tends to 0, this surrogate $\widetilde{\phi}$ converges towards $\phi$, meaning it increased the smoothness, but it is still potentially multimodal in expectation, ruling out simple descent methods for the optimization. To tackle this, they introduce a probabilistic model using additional \emph{objective variables} $\eta$ defined as follow:
$$
\eta \sim {\rm Bernoulli}\Big(\exp\big(-(x^{(1)} - \phi(\rho + \epsilon, x^{(1)}, x^{(2)}))\big)\Big), \hspace{0.5cm}  \text{where}~~\epsilon \sim \cN(0, \sigma^2)\,.
$$
This additional variable intuitively represents how satisfying the revenue is for a given auction (a sample) and thus is aimed to be put to 1. Hence, a dataset $\{x^{(1)}_t, x^{(2)}_t, \bz_t, \eta_t=1\}_{t\in [T]}$ where $(x^{(1)}_t, x^{(2)}_t, \bz_t)_{t\in [T]} \sim \mathcal{F}^{\otimes T}$ is used to estimate the parameters of the reserve price function $r$ to fit this probabilistic model. The key point is that the maximum at posteriori (MAP) estimation recovers the parameter that maximizes the initial smoothed objective, the expectation of $\widetilde{\phi}$. The MAP estimation under this model is performed using the Expectation-Maximixation (EM) algorithm. As such, the guarantee is only to improve the solution at every step, but there is no global convergence guarantee. However, it exhibits significant empirical improvements (in reasonable learning time) over the previous method based on DC programming.

\subsection{Using a Bid Prediction}
Another very different approach does not rely on a surrogate of $\phi$ \citep{medina2017revenue}, but on having access to a good prediction of the highest value. For simplicity,  consider the posted-price setting, i.e., when there is only one bidder, or $x^{(2)} = 0$ (and for this subsection, we then drop $x^{(2)}$ from  notations).  Assume the unique bidder has access to a prediction function of the highest value $\hat{x}(\bz)$ with a given squared error -- i.e. such that 
\begin{align}
\lE_{\mathcal{F}}\left[(\hat{x}(\bz) - x)^2\right] = \eta^2\,.
\end{align}
Using this value prediction and a training dataset $\{x^{(1)}_t, \bz_t\}_{t \in [T]} \sim \mathcal{F}^{\otimes T}$, a reserve price function is built as follows. First, the feature space $\lR^d$ is partitioned  by discretizing the image of the value prediction function $\hat{x}(.)$ into $K$ subset $C_1, \dots, C_K$. Formally, this partition is defined by $\tau\in \lR_+^K$ and $C_k = \{\bz\in \lR^d: \tau_k \leq \hat{x}(\bz) < \tau_{k+1}\}$ (where by convention $\tau_{K+1} = +\infty$). This  vector $\tau$ is built to minimize the sum of intra-partition variance of the prediction $\hat{x}(.)$. Then, the reserve price function $r$ is defined in a piece-wise manner on this partition. Denoting $r_k$ the empirical monopoly price computed on the restriction of the dataset to $C_k$, the reserve price is defined as
\begin{align}
r(\bz) = \sum_{k=1}^K r_k \indicator{\hat{x}(\bz) \in C_k}
\end{align}
Further, it is possible to provide a guarantee on the performance of $r(.)$ that depends on the accuracy of the value prediction $\eta$.
\begin{theorem}[\cite{medina2017revenue}]
For $\delta>0$, with probability at least $(1-\delta)$ over the learning samples, it holds
$$
\lE_{\mathcal{F}}\left[r(\bz)\indicator{r(\bz) \leq x}\right] \geq \lE_{\mathcal{H}}\left[x\right] - \cO\left(K^{-\frac{2}{3}}+\eta^\frac{2}{3}+T^{-\frac{1}{6}}\right)\,.
$$
\end{theorem}
The intuition behind this result is clear: the higher the prediction of the value $\hat{x}(\bz)$, the higher the revenue extracted. However, it only provides a guarantee relatively to the expected value $\lE_{\mathcal{F}}\left[x\right]$ rather than the optimal revenue that could be extracted.

\bigskip

Overall, these three algorithms are costly to run on big datasets, highlighting the complexity of the underlying problem. There exists more efficient computation of reserve prices, but usually by sacrificing the objective of revenue maximization for weaker objectives \citep{shen2019learning}.

\section{Cost \& Online Estimation of Auctions} \label{SE:Bandits}
Given the sample complexity of a class $\mathcal{A}$ of (ex-post individuallly-rational) auctions, as given in Proposition \ref{PR:Sample_Comp}, it is possible to compute the global cost of learning the optimal auction in that family. Indeed, recalling the statement of Proposition  \ref{PR:Sample_Comp}, only $\Theta\bigg(H^2\varepsilon^{-2}\big(P_{\text{dim}}(\mathcal{A})\log(\frac{H}{\varepsilon}) +\log(1/\delta)\big)\bigg)$ samples are required to find an $\varepsilon$-optimal auction (within the class $\mathcal{A}$) with probability at least $1-\delta$. 

As a consequence, after $t$ samples, by inverting this equation (and setting arbitrarily $\delta =1/t^2$),  a learner can compute an $\mathcal{O}(H\sqrt{\frac{P_\text{dim}(\mathcal{A})\log(t)}{t}})$-optimal auction  with probability at least $1/t^2$. Summing all the errors from the first ($t=1$) up to the last ($t=T$) auction, we get a total cost of learning the optimal auction of the order of $\mathcal{O}(H\sqrt{P_\text{dim}(\mathcal{A})T\log(T)})$.

Those computations are possible for value distributions with support bounded in $[0,H]$, but similar computations  give, for the case of regular distributions and the class $\mathcal{A}$ of all ex-post IR auctions, a total cost of learning the Myerson auction of the order of $\mathcal{O}(n^{\frac{1}{3}}T^{\frac{2}{3}}\log^{\frac{1}{3}}(T))$.

\medskip

A crucial implicit assumption made for these arguments to hold is that, no matter the auction mechanism chosen at each stage, the seller gets to observe perfectly a sample of the value distribution of each (or at least one in the symmetric case) buyer. In many applications, this is unfortunately not true. Consider for instance the posted price mechanism, then the feedback actually received is only whether the value is above - or below - current price. 
Similarly, if reserve prices in second price auctions are too high, bidders might decide to opt-out the current auction (as in posted price, see also the discussions on lazy vs eager auctions) and/or the seller might only  have   her revenue as  feedback, because she is using some black-box tool to actually run the auction.

This setting is called with \textsl{partial feedback} and are closely related to the multi-armed bandit scenarii, and  therefore similar techniques (quickly recalled in the following section, see \cite{bubeck2012regret,lattimore2020bandit,slivkins2019introduction} for more details) can be used.
\subsection{A quick reminder on multi-armed bandits}
\label{Section:Bandits}
In a bandit problem, an agent faces a sequential decision problem. At each stage $t \in \mathds{N}$, she chooses an action (takes a decision, pulls an arm...) $k_t$ in some finite set $\mathcal{K}$ of cardinality $K$. This generates a reward $X_{k_t,t}$ that belongs to $[0,1]$ -- this assumption can be fairly relaxed -- that is observed by the agent (contrary to the other possible rewards $X_{k,t}$, for $k\neq k_t$, that are \textsl{not} observed). The objective of the agent is to maximize her expected cumulated reward $\sum_{t=1}^T X_{k_t,t}$.

To evaluate the performance of a learning algorithm, the cumulative reward should be normalized; a traditional way is therefore to consider the expected \textsl{regret}, which is the difference between the  cumulative reward obtained by always playing the same action at each stage and the cumulative reward of the agent. 

It remains to describe how  rewards are generated. Basically, there are two extreme distinct possible scenarii that are quite different (and so are the associated algorithms). In the \textsl{stochastic} case, $X_{k,t}$ are i.i.d., of (unknown) expectation $\mu_k$. In the \textsl{adversarial} case, $X_{k,t}$ can be any sequence of values in $[0,1]$, and $X_{k,t+1}$ could even depend on all the past history (up to the previous stage $t$). The regret can be rewritten in the following form
\begin{description}
\item[Stochastic] $\displaystyle R_T = T  \mu_\star - \sum_{t=1}^T \mu_{k_t} = \sum_k \Delta_k N_k(T)$ where $\mu_\star=\max \mu_k$, $\Delta_k = \mu_{\star} - \mu_k$ is called the \textsl{gap} (or the cost) of choosing action $k$ instead of an optimal one, and $N_k(T) = \sum_{t=1}^T \mathds{1}\{k_t = k\}$ is the number of times this decision has been made
\item[Adversarial] $\displaystyle  R_T = \max_{k} \mathds{E}\sum_{t=1}^T X_{k,t} - \sum_{t=1}^T X_{k_t,t}$, where the argument of the maximum might change with time (unlike with stochastic data).
\end{description}
The multi-armed bandit literature focuses on finding algorithms that provably control the regret with  sub-linear growths (both in $T$ and $K$). As mentioned before, the techniques differ quite a lot from stochastic to adversarial data.

\subsubsection{UCB and variants for stochastic data}
Under the stochastic assumption, a natural proxy for $\mu_k$ is the empirical mean $$\overline{X}_{k,t} = \frac{1}{N_k(t)}\sum_{s=1}^t X_{k,s} \mathds{1}\{k_s=k\}.$$
Unfortunately, this quantity is  biased, and usually negatively, because reinforcement algorithms tend to select actions that performed well in the past. Moreover, a negative bias is naturally reinforced with time (i.e., it will not disappear), which is not the case of a positive bias (as algorithms will certainly sample again that action). As a consequence, a celebrated algorithm called UCB, for Upper-Confidence Bound, adds a small error term to the empirical mean so that it will still be biased, but positively (this property is a direct consequence of Hoeffding's inequality). 

UCB algorithm is defined by 
$$
k_{t+1} = \arg\max_k \Big\{\ \overline{X}_{k,t}  + \sqrt{\frac{\log(T)}{N_k(t)}}\ \Big\}
$$
\begin{theorem2}
UCB algorithm has an expected regret bounded as 
$$
\mathds{E} [R_T ] \leq \min \Big\{\ 2\sqrt{2KT\log(T)}  + 2K\, ; \, 4\sum_k\frac{\log(T)}{\Delta_k}+2K \Big\}
$$
\end{theorem2}
Before giving the proof, we should mention that the different universal constants can be improved with slightly more involved proofs. Similarly, it is possible to change the $\log(T)$ term in the definition of UCB by $2\log(t)$. Yet this result is sufficient for us.
\begin{proof}
First, let us recall  Hoeffding inequality. It states that $\frac{1}{t} \sum_{s=1}^t X_{k,s} \leq \mu_k + \varepsilon$ with probability at least $1-\exp(-2t\varepsilon^2)$. This implies that, with probability at least  $\frac{K}{T}$,
$$
\forall t \leq T, \forall k \neq \star, \ \ \overline{X}_{k,t}  \leq \mu_k + \sqrt{\frac{\log(T)}{N_k(t)}}\ \text{ and } \ \mu_\star \leq  \overline{X}_{\star,t}  + \sqrt{\frac{\log(T)}{N_\star(t)}}, \
$$
 It remains to control the regret. To lighten  notations, we introduce $\varepsilon_{k,t} = \sqrt{\frac{\log(T)}{N_k(t)}}$. For the second bound, notice that, by definition of UCB, $k_t=k$, if
$$\overline{X}_{\star,t}  +\varepsilon_{\star,t} \leq \overline{X}_{k_t,t}  +\varepsilon_{k_t,t},
$$
which implies that, on an event of probability at least $1-K/T$,
$$
\mu_\star \leq \mu_k + 2\varepsilon_{k_t,t} = \mu_k + 2\sqrt{\frac{\log(T)}{N_k(t)}}.$$
Inverting the above equation gives that necessarily, on that event, $N_K(T) \leq \frac{4\log(T)}{\Delta^2_k}+1$, and summing over the different actions $k$ gives the second bound.

For the first bound, recall that $R_T = \sum_k N_k(T)\Delta_k$.  Since we proved above that on event of probability at least $1-K/T$, $\Delta_k \leq \sqrt{\frac{4\log(T)}{N_k(T)-1}}$, we get that on this event (using the fact that $\Delta_k$ is also smaller than 1 to avoid dividing by 0)
\begin{align*}
R_T &=\sum_k \Delta_kN_k(T) \leq \sum_{k: N_k(T) \geq 2} \sqrt{\frac{4\log(T)}{N_k(T)-1}}N_k(T)+K \\ &\leq 2\sqrt{2\log(T)}\sum_k \sqrt{N_k(T)} + K \\ &\leq 2\sqrt{2KT\log(T)} +K,
\end{align*}
where the second inequality comes from the fact that $\frac{N}{\sqrt{N-1}}\leq \sqrt{2N}$ as soon as $N \geq 2$ and the last one is a consequence of Cauchy-Schwartz inequality.
% Preuve plus jolie, mais plus complexe
%\begin{align*} 
%R_T &=  \sum_{t=1}^T \mu_\star - \mu_{k_t} = \sum_{t=1} \underbrace{\big(\overline{X}_{\star,t}  +\varepsilon_{\star,t}\big) - \big(\overline{X}_{k_t,t}  +\varepsilon_{k_t,t}\big)}_{\leq0 \text{\ by def of UCB}}\\
%&+ \sum_{t=1}^T\underbrace{\mu_\star - \big(\overline{X}_{\star,t}  +\varepsilon_{\star,t}\big)}_{\leq 0 \ \text{\ with proba\ } \frac{1}{T^3}}  + \sum_{t=1}^T  \underbrace{\big(\overline{X}_{k_t,t}  +\varepsilon_{k_t,t}\big) - \mu_{k_t,t}}_{\leq 2\varepsilon_{k_t,t} \ \text{\ with proba\ } \frac{K-1}{T^3}}
%\end{align*}
%As a consequence, we immediately get that with probability at least $1-\frac{K}{T^2}$
%\begin{align*}
% R_T  &\leq 2\sum_{t=1}^T \varepsilon_{k_t,t}   \leq  2\sum_k \sum_{t=1}^T  \varepsilon_{k,t} \mathds{1}\{k_t=k\}   \\
%& = 2 \sum_k \sum_{s=1}^{N_k(T)-1} \sqrt{\frac{2\log(T)}{s}} +1  \leq 4\sqrt{2\log(T)} \sum_k \sqrt{N_K(T)}   +2K \\
%& \leq  4\sqrt{2KT\log(T)} +2K.
%\end{align*}
%This gives the first bound since $R_T \leq T$ on the remaining event. 

%that the second line above gives that with  probability at least $1-\frac{K}{T^2}$
%\begin{align*}
% \sum_k N_k(T) \Delta_k = R_T  &\leq 4\sqrt{2\log(T)} \sum_k \sqrt{N_K(T)} + 2 K\\
%  &= 4\sqrt{2} \sum_k \sqrt{N_K(T)\Delta_k}\sqrt{\frac{\log(T)}{\Delta_k}} + 2 K \\
%  & \leq 4\sqrt{2} \sqrt{R_T}\sqrt{\sum_k \frac{\log(T)}{\Delta_k}}+2K,
%\end{align*}
%where we used Cauchy-Schwartz inequality at the last line. Solving in $R_T$ gives the result.
\end{proof}
The UCB algorithm has been generalized, extended with many different variants to improve the different dependencies. For our purpose, we might only consider the MOSS algorithm \citep{audibert2009minimax,degenne2016anytime} whose expected regret scales as $\mathcal{O}\Big(\sqrt{KT}\Big)$.

\subsubsection{EXP3 and variants for adversarial data}
Unfortunately, UCB algorithm and its variants only works with i.i.d.\ data (or with strong stationarity assumptions). In the adversarial case, when $X_{k,t}$ can be any sequence, its regret would increase linearly and another class of algorithm (based on optimization concepts instead of reinforcement) has been introduced; it is called EXP.3 (for Exploration and Exploitation with Exponential weights). The basic idea is to always choose actions at random, with a positive probability, so that important sampling techniques can be used to build unbiased estimates of rewards. Let us then denote by $p_{k,t}$ the probability of choosing action $k$ at stage $t$ (that obviously depends on the past history). A classical way to estimate the reward $X_{k,t}$ is then to define
$$
\widehat{X}_{k,t} = 1- \frac{1-X_{k,t}}{p_{k,t}}\mathds{1}\{k_t=k\};
$$
this estimate has the good properties of being both unbiased and always smaller than $1$ (even if possibly arbitrarily small). The EXP.3 algorithm is defined by
$$
p_{k,t+1} = \frac{\exp\Big(\eta\sum_{s=1}^t \widehat{X}_{k,s} \Big)}{\sum_{k'}\exp\Big(\eta\sum_{s=1}^t \widehat{X}_{k',s} \Big)},
$$
where $\eta$ is a parameter to be chosen.

\begin{theorem2}
With the choice of  $\eta = \sqrt{\frac{\log(K)}{KT}}$, the expected regret of EXP.3 scales as
$$
\mathds{E} [R_T] \leq 2\sqrt{K\log(K)T}
$$
\end{theorem2}
\begin{proof}
The proof relies on a careful study of $\Phi(W_t)$ where $W_{k,t}=\sum_{s=1}^t \widehat{X}_{k,s}$ and $\Phi(Z) = \frac{1}{\eta} \log\Big( \sum_k \exp(\eta Z_k)\Big)$.
First, notice that by definition of $\Phi$,
$$
\Phi(W_{t+1}) - \Phi(W_t) = \frac{1}{\eta}\log\Big(\sum_k p_{k,t+1}\exp(\eta\widehat{X}_{k,t+1})\Big).$$
Using the facts that $\exp(\eta x) \leq 1+\eta x +\eta^2 x^2$ if $x \leq 1$, that $\eta \widehat{X}_{k,t} \leq 1$ and finally that $\log(1+x) \leq x$, we get
$$
\Phi(W_{t+1}) - \Phi(W_t) \leq  \sum_k p_{k,t+1}\widehat{X}_{k,t+1}+\eta \sum_k p_{k,t+1}\widehat{X}_{k,t+1}^2
$$
In particular, plugging back the definition of $\widehat{X}_{k,t+1}=1- \frac{1-X_{k,t}}{p_{k,t}}\mathds{1}\{k_t=k\}$, we get
$$
\Phi(W_{t+1}) - \Phi(W_t) \leq  \sum_k X_{k,t+1} \mathds{1}\{k_{t+1}=k\} +\eta \sum_k p_{k,t+1}\big(1- \frac{1-X_{k,t+1}}{p_{k,t+1}}\mathds{1}\{k_{t+1}=k\}\big)^2\,.
$$
Conditionally to the past history, the  middle term satisfies, in expectation, 
$$
\mathds{E}  \sum_k X_{k,t+1} \mathds{1}\{k_{t+1}=k\} =  \sum_k X_{k,t+1}p_{k,t+1} 
$$ 
For the last term, expanding the square yields that
\begin{align*}
&\sum_k p_{k,t+1}\big(1- \frac{1-X_{k,t+1}}{p_{k,t+1}}\mathds{1}\{k_{t+1}=k\}\big)^2\\
&=\sum_k \frac{(1-X_{k,t+1})^2}{p_{k,t+1}}\mathds{1}\{k_{t+1}=k\}-2\sum_k(1-X_{k,t+1}) \mathds{1}\{k_{t+1}=k\} + 1
\end{align*}
Taking expectation, conditionally to the past history, gives that this term is controlled as 
\begin{align*}&\mathds{E} \sum_k p_{k,t+1}\big(1- \frac{1-X_{k,t+1}}{p_{k,t+1}}\mathds{1}\{k_{t+1}=k\}\big)^2\\
\leq &\sum_k (1-X_{k,t+1})^2-2\sum_k(1-X_{k,t+1}) p_{k,t+1}+ 1
\end{align*}
and the latter is always smaller than $K$. To see this, assume, without loss of generality that $X_{1,t+1} \geq X_{k,t+1}$ for all $k \in [K]$ so that
\begin{align*}&\mathds{E} \sum_k p_{k,t+1}\big(1- \frac{1-X_{k,t+1}}{p_{k,t+1}}\mathds{1}\{k_{t+1}=k\}\big)^2\\
\leq &\sum_{k \geq 2} (1-X_{k,t+1})^2 + (1-X_{1,t+1})^2- 2(1-X_{1,t+1})+1\\
\leq &\sum_{k \geq 2} (1-X_{k,t+1})^2 + X_{1,t+1}^2 \\
\leq & K
\end{align*}
Plugging this back in the definition of $\Phi(W_{t+1}) - \Phi(W_t)$, and taking the expectation conditionally to the past history give
$$
\mathds{E} \Phi(W_{t+1}) - \Phi(W_t) \leq  \sum_k X_{k,t+1}p_{k,t+1} + \eta K.
$$
Summing over $t$, and using the fact that $\Phi(0) = \frac{\log(K)}{\eta}$ and $\Phi(Z) \geq \max_kZ_k$, we finally get
$$
\mathds{E}  \max_k \sum_{t=1}^T\widehat{X}_{k,t} - \mathds{E}\sum_m X_{k_t,t} \leq \frac{\log(K)}{\eta}+\eta K T
$$
which gives the result, as $\mathds{E}  \max_k \sum_{t=1}^T\widehat{X}_{k,t} \geq   \max_k\mathds{E}   \sum_{t=1}^T{X}_{k,t} $.
\end{proof}
The EXP.3 algorithm is a standard building block of many online learning algorithms with adversarial data. As UCB, it has been improved in many directions, notably to get rid of the sub-optimal $\sqrt{\log(K)}$ term in the regret bound (yet at the cost of a much more intricate proof). It is also possible to estimate $X_{k,t}$ with $\frac{X_{k,t}}{p_{k,t}}\mathds{1}\{k_t=k\}$. However, this estimate can be arbitrarily large and the variance of EXP.3 cannot be directly controlled. The trick is then to  add a forced exploration term, i.e., to play uniformly at random with probability $\sqrt{K\log(K)/T}$ at each round.

\medskip

Similarly to the stochastic case, it is possible to get rid of the $\sqrt{\log(K)}$ term that arises in the EXP.3 regret analysis with a more involved algorithm (and proof techniques). It is then ``optimal''' in the sense that any learning algorithm must (in some difficult problem instances) have a regret scaling at least as $\Theta\big(\sqrt{KT}\big)$.

\subsection{Auctions learning with partial feedback}

As mentioned before, there are many instances where a seller only has incomplete data on auctions run in the past (in posted price, lazy/eager second price, with a black-box selling mechanism, etc.). However, it is still possible for the seller to learn the optimal mechanism in many different cases, with actually a very small extra-cost compared to the batch-approach.

Consider for instance the online posted price problem. A seller repeatedly posts a price $p_t \in [0,1]$ to sell identical items and buyers sequentially arrives, with private value $x_t \in [0,1]$. The buyer $t$ buys the item if $x_t \geq p_t$, without revealing the true value. As a consequence, the partial feedback available to the seller, before fixing the next price $p_{t+1}$, are all the indicators $\mathds{1}\{x_s \geq p_s\}$ for $s \in [t]$. As before, the objective of the seller is to find, \textsl{as quickly as possible}, the best price $p^*$ or to minimize the regret
$$
\max_{p \in [0,1]} p\sum_{t=1}^T \mathds{1}\{x_t \geq p\} -  \sum_{t=1}^T p_t \mathds{1}\{x_t \geq p_t\} 
$$
We recall that if data are stochastic, i.e., i.i.d.\ with cdf $F$, then $p^*$ is a root of the virtual function (and \textsl{the} root if $F$ is regular). Yet, the analysis carries on with adversarial sequence of price $p_t$.

\medskip

In sequential learning, a first and na\'ive possibility is quite often to discretize the decision space (here $[0,1]$) and to run an UCB or EXP.3 algorithm on the discretization (depending if data are stochastic or adversarial), agnostically to the structure at hand. Given some $\varepsilon >0$, the size of the discretization in the posted price problem is $1/\varepsilon$, leading to a global regret of the order of 
$$
\mathcal{O}(T\varepsilon) + \mathcal{O}(\sqrt{T/\varepsilon})  = \mathcal{O}(T^{\frac{2}{3}}) \quad \text{ with the choice of } \quad \varepsilon = T^{-\frac{1}{3}}.
$$
In the above equation, the first term corresponds to the approximation error due to the discretization and the second to the estimation (or learning) error of the optimal price in the discretized set.

On the other hand,  the regret bound can be largely improved by leveraging the structure of the problem, as least in the stochastic case, when the generating distribution behaves nicely enough (the worst-case learning cost being indeed $T^{2/3}$, see \citep{kleinberg2003value}). For instance, a typical assumption is that the monopoly profit function  $\Pi(p)=p(1-F(p))$ is approximatively quadratic around $p^*$, i.e., that $\Pi(p^*) - \Pi(p) = \Omega(p-p^*)^2$. Using a UCB algorithm with a uniform $\varepsilon$-discretization then yields a total regret of the order of
$$
\mathcal{O}(T\varepsilon^2) + \mathcal{O}\Big(\sum_{k=1}^{\frac{1}{\varepsilon}} \frac{\log(T)}{k^2\varepsilon^2}\Big)
 = \mathcal{O}(\sqrt{T\log(T)}) \quad \text{ with the choice of } \quad \varepsilon = \Big(\frac{T}{\log(T)}\Big)^{-\frac{1}{4}}.$$

This simple technique cannot be improved, even with a stronger assumption: indeed, since if the approximation term is of order $T\Delta$, the estimation error is at least $\log(T)/\Delta$.  However, the problem has a stronger property that can be further leveraged: if a price $p_t$ is accepted, then any lower price would also have been accepted (and reciprocally). In particular, this can be used in the following simplest possible problem, but where the solution is highly counter-intuitive.

Suppose that each buyer has the same exact value $x_t=x$. Then it's clear that the optimal price is $p^*=x$; the remaining question is the learning cost. As mentioned before, the feedback is in that case binary; either ``$x$ is greater than $p_t$'' or ``$x$ is smaller than $p_t$''. In order to find $x$, with the fewest query possible, then a binary search is optimal. However, the binary search is \textsl{exponentially} sub-optimal in terms of learning cost.
\begin{proposition2}\label{PR:Cautiousbinary}
The regret of a binary search can be as large as $\Omega(\log(T))$. On the other hand, there exists a more \textsl{cautious} search whose regret is smaller than $\mathcal{O}(\log\log(T))$.
\end{proposition2}
\begin{proof}
Assume that $x=\frac{1}{2}$. Then a binary search will use $\log(1/\varepsilon)$ posted -- and refused-- prices, to reach the precision $\varepsilon$. Even with the optimal choice of $\varepsilon=1/T$, this gives a $\log(T)$ regret.

\medskip

The  cautious search works in epochs $\ell \in\{0,1,\ldots, \log_2 \log_2(T)\}$ -- let us assume for simplicity here that $\log_2\log_2(T)$ is an integer. At the $\ell$-th epoch, the prices posted increase by $1/2^{2^\ell}$ until such a price is refused and the next epoch begins. Let $p^{(\ell)}_{\star}$ be the last accepted price at epoch $\ell$ and  $p^{(\ell+1)}_{j}$ the $j$-th price posted at epoch $\ell+1$, then $p^{(\ell+1)}_{j} = p^{(\ell)}_{\star}+\frac{j-1}{2^{2^{\ell+1}}}$. At the end of the epoch $\log_2 \log_2(T)$, the cautious binary search posts the last accepted price until the final stage $T$.

To compute the regret of the cautious binary search, notice that at each epoch $\ell$, only one price is rejected, and this rejection has a cost smaller than 1. Moreover, since 
$p^*\in  [p^{(\ell)}_{\star}, p^{(\ell)}_{\star} + \frac{1}{2^{2^\ell}})$, then $p^*- p^{(\ell+1)}_{1} \leq  \frac{1}{2^{2^\ell}}$ and more generally, as long as posted prices are not rejected
$$
p^*- p^{(\ell+1)}_{j} \leq  \frac{1}{2^{2^\ell}} - \frac{j-1}{2^{2^{\ell+1}}}.
$$
Since they are at most $2^{2^\ell}$ posted prices in epoch $\ell+1$, the cumulative cost of errors in that epoch is bounded by 
$$
\sum_{j=1}^{2^{2^\ell}}  \frac{1}{2^{2^\ell}} - \frac{j-1}{2^{2^{\ell+1}}} = \sum_{j=1}^{2^{2^\ell}}  \frac{j}{2^{2^{\ell+1}}} \leq 1.
$$
As a consequence, each epoch has a bounded cost of (at most) 2 which gives the result as only   $\log_2\log_2(T)$ epochs are needed to get an  error on $p^*$ smaller than $1/T$.
\end{proof}

The crucial property  to obtain $\log\log(T)$ regret \citep{kleinberg2003value}  is that the cost function $\Pi(p^*) -\Pi(p)$ is asymetric and decreases much slower on the left that on the right. This property was later used again   in the stochastic case  to generalize Proposition \ref{PR:Cautiousbinary} if the support of the distribution is finite to get a worst case bound of $\sqrt{KT}$ \citep{cesa2019dynamic} and in the adversarial case to lower the parameter dependency in front of the $T^{2/3}$ term y \citep{bubeck2017multi}.

\medskip

The fact that the reward mapping $\Pi(p)$ cannot be \textsl{any} function, but must belong to a specific family, can also be leveraged  in learning the optimal reserve price in symmetric  (repeated) second-price auction \citep{cesa2015regret}. Consider for instance the more complex case where there are not only one but $n$ bidders at each auction, and the feedback to the seller is, as in posted price, the revenue of the auction -- and not the true value. In this specific case, the revenue is either the current reserve price $p_t$ (if the highest bid is above it and the second highest below) or the second highest bid. In both cases, the learner gets information on not only $\Pi(p_t)$ but on the whole function $\Pi(\cdot)$. The learning algorithm somehow combines the idea behind the cautious binary search and UCB. It proceeds by epochs, and at each stage of the epoch $k$ the proposed reserve price $p_k$ is the same and always smaller than $p^*$ (at least with arbitrarily high probability). At the end of an epoch, based on the data collected, a confidence interval of $\Pi(\cdot)$ is constructed - this is possible because it only depends on the distribution of the second highest bid in the symmetric case and only bids above the current reserve price matter -- based on the Dvoretzky-Kiefer-Wolfowitz inequality.  Epoch after epoch, the error on the optimal reserve price decreases and it is possible to control the regret (at the cost of intensive computations).

\begin{proposition2}
Learning the optimal reserve price in symmetric second price auctions has a cost of $\mathcal{O}(\sqrt{T\log(T)})$, where the dependency in the value distribution $F$ is independent of $T$, but hidden in the $\mathcal{O}(\cdot)$ notation.\end{proposition2}

\printbibliography[segment=3, heading=subbibintoc]
\end{refsegment}
\clearpage
\begin{refsegment}
 %!TEX root = ../main.tex
\chapter{Adaptive and strategic learning agents}
\label{Chapter:Learning}
\centerline{\colorbox{mygray}{ \begin{minipage}{11cm}
\begin{center} \textbf{First read of this chapter, key concepts and ideas}\end{center}
This chapter focuses on dynamic settings, where the interactions between bidders and seller are repeated. The first of the three main results are that, if the mechanism is fixed, a bidder can learn his value distribution on the fly at a small cost (or regret), see Proposition \ref{Prop:UCbid}. The two other main results explain how one agent (either a bidder or a seller) can take advantage of the other one. Indeed, the seller can extract the surplus from the bidder if, and only if, she is way more patient than him (Theorems \ref{thm:lower_bound_amin} and \ref{Thm:UpperBounAmin}). On the other hand, if the seller has to commit to some class mechanism, say second price auction with personalized reserve prices, then the bidder can use these repetition to somehow \textsl{manipulate} the seller at his advantage. Theorem \ref{thm:settingReserveValToZeroImprovesPerformance} gives a simple statement of this maybe counter-intuitive fact.
\end{minipage}
}
}

\bigskip

Online auctions are one of the most fundamental tool of the modern economy. A crucial assumption behind the results of Chapter \ref{Chapter:Seller} is that the seller has access to some large sample batch  of bidders' valuations. The objectives were to learn the optimal (or at least the best possible in some class) mechanism based on this dataset. This model particularly fits problems where the bidders are different from one auction to the other (typically such as on Ebay) so that it is legitimate to assume that they bid their values (as long as they were facing an incentive compatible auction) myopically, i.e., non strategically. Unfortunately, this assumption  of facing new bidders for each different item  no longer holds in many important economical situations, such as online advertising. Indeed, in that market the so-called \emph{demand-side platforms} (DSP), that are aggregate of bidders, repeatedly interact with a single seller (called \emph{Supply-side platforms (SSP)}), billions of times a day \citep{choi2019online}. 

The global objective of the seller remains identical: maximize the total revenue, certainly by learning (or trying to) the value distributions of the bidders. Indeed, it is still interesting for SSP   to optimize the reserve prices (personalized per DSP) since the number of participants per auction is relatively low (the median number of bidders is equal to 6 \citep{celis2014buy}). The difficulty is now that bidders are also present in the game for a long period optimizing their cumulative  utility, instead of best-replying myopically to the seller design of mechanism. Thanks to this long-term optimization, it is possible for a bidder to sequentially learn his ``optimal'' bidding strategy, if the mechanism or his opponent's strategy or even his own distribution of valuations are unknown at first. Such a bidder is \emph{adaptive} to his environment, but he could even be \emph{strategic}.
Intuitively, and this will be detailed later on, a strategic buyer might be tempted to modify his bids if he knows that the seller is using them to set reserve prices. It can be much more profitable to face a low reserve price by bidding non-truthfully than facing a high reserve price with truthful bids.

\medskip

It might be worth mentioning here that bidders might have different values for different auctions because the intrinsic value of an ad is strongly related to the probability that the user seeing it clicks on it. As a consequence, the value distribution of bidders in this chapter represents the time-variability of valuations for one specific bidder for the different items that are sold successively. Notice, on the contrary, that in the precedent chapter, the value of a buyer was fixed, but unknown to the seller that only had some prior (the distribution) on its realization.

\medskip

For simplicity, we shall assume in this section that values always belong to $[0,1]$; this assumption can be weakened, but at the cost of technicalities.
\section{Adaptive bidders - Online learning to bid}

As mentioned above, the buyers were assumed to be myopic in the previous chapters (or equivalently, only present for one single auctions), a crucial hypothesis that should be removed, at least in online ads markets. We will on the contrary assume that a single buyer bids at each auction and that he can use these repetitions to \textsl{improve} his bidding strategy by adapting to his unknown environment; there are at least two different aspects that can be learned sequentially, depending on where the lack of information lies.

\begin{enumerate}
\item \textit{Bidding without knowing its own value:}  the bidder does not know the expected value it gives to the item but could learn his own value distribution, by gathering a new value sample each time he wins an auction. This is particularly relevant in online advertising where advertisers have to show ads to potential buyers to understand their propensity to buy a certain product (or at least, the propensity to click on an ad). 
\item \textit{Bidding without knowing the mechanism:} a strategic bidder does not know precisely the mechanism used by the seller (or alternatively, has few or no information on the other bidders' valuation distributions). The bidder can sequentially and incrementally adapt his bidding strategy during  the $T$ successive auctions to maximise his expected  cumulative utility, hopefully achieving (almost) the same performances as if he knew the whole mechanism in advance.
\item \textit{Bidding with budget constraints:} a strategic bidder could have a pre-specified constraint that on the budget he can spend during the $T$ successive auctions. In this setting, at each round, he should bid (and spend some of his budget if he wins the auction) without knowing the exact valuation he will get for the future rounds. The bidder can sequentially and incrementally adapt his bidding strategy during  the $T$ successive auctions to maximise his expected  cumulative utility, without respecting the constraint on his budget.
\end{enumerate}

\subsection{Online learning the value distribution to bid}\label{ssec:learning_to_bid}
In some oversimplified online ad example, a publisher (the website where an ad is displayed) gets paid by the advertiser each time a user clicks on that ad because this is that random event that is relevant (as it might generate a sale afterwards), and not simply the fact that the ad is displayed (as the user might not see it). As a consequence, the actual value of an ad is random -- either 0 or 1 to simplify things again, i.e., whether the user actually clicks or not -- of  expectation $x$, i.e., the probability of click. If $x$ is known before participating in a DSIC auction, then due to the linearity of expectation, the optimal strategy remains to bid truthfully $x$.

Unfortunately for the seller, there are many examples where the probability of click is unknown to the advertiser before starting a new campaign; he would be willing to bid truthfully this probability repeatedly, but it is unknown. This is an example where multi-armed bandit techniques can be used to learn on the fly the probability of clicks based on the feedback received. Notice that an ad that is not displayed will never be clicked, hence the advertiser must acquire some data to estimate this probability.

\medskip

The simplest model to study this sequential learning problem is the following \citep{Weed16}. Bidder $i$ participates in a sequence $t=1,\ldots,T$ of second price auctions (without reserve prices) where the realized values $v_t \in \{0,1\}$ are i.i.d.\ of unknown expectation $x$. Of course, $v_t$ is not observed before participating to the auction $t$, and only if that auction is won. Let us denote by $b^{(1)}_{-i,t}$ the maximal bids of the opponents (that could actually include a reserve price) and by $b_{i,t}$ the bid of bidder $i$ at that stage then the performance of the optimal truthful strategy is 
$
\sum_{t=1}^T \mathds{E} \big[ x \mathds{1}\{x \geq b^{(1)}_{-i,t}\} \big]
$
while  the learning to bid policy has gathered $
\sum_{t=1}^T \mathds{E} \big[ x \mathds{1}\{b_t \geq b^{(1)}_{-i,t}\} \big]
$. As a consequence, the cost of learning to bid is measured in terms of regret 
$$
R_T = \sum_{t=1}^T \mathds{E} \big[ x \mathds{1}\{x \geq b^{(1)}_{-i,t}\} \big] -  \mathds{E} \big[ x \mathds{1}\{b_{i,t} \geq b^{(1)}_{-i,t}\} \big]\, .
$$
As in multi-armed bandit, the empirical average $\overline{x}_t = \frac{\sum_{s \leq t} v_s \mathds{1}\{b_{i,s} \geq b^{(1)}_{-i,s} \} }{ \sharp \{s \leq t\, ;\,  b_{i,s} \geq b^{(1)}_{-i,s} \} }$ is negatively biased. As a consequence, the algorithm $\text{UCB}_\text{id}$ slightly biased it positively by adding a small error term in its bid
$$
b_{i,t+1} = \overline{x}_t + 2\sqrt{\frac{\log(T)}{\omega_{i,t}}} \quad \text{ where } \omega_{i,t}=\sharp \{s \leq t\, ;\,  b_{i,s} \geq b^{(1)}_{-i,s} \},
$$
is the number of auctions won until stage $t$.
\begin{proposition2}\label{Prop:UCbid}The $\text{UCB}_\text{id}$ algorithm has a sublinear regret against any sequence of opponent bids $b^{(1)}_{-i,t}$, as long as they are independent of $v_t$ (conditionally to the  history), as 
$$
R_T \leq \mathcal{O}\big(\sqrt{T\log(T)}\big).
$$
Moreover, if the sequence $b^{(1)}_{-i,t}$ is also i.i.d.\ (of unknown law to bidder $i$), then the regret grows is even much slower as 
$$
R_T \leq \left\{ \begin{array}{ll} c_\alpha T^{\frac{1-\alpha}{2}}\log^{\frac{1+\alpha}{2}}(T) & \text{if}\ \alpha <1\\
c_\alpha\log^2(T) & \text{if}\ \alpha =1\\
c_\alpha\log(T) & \text{if}\ \alpha >1\end{array}\right.
$$
for some constants $c_\alpha$ independent of $T$ and where $\alpha$ is some regularity parameter called ``margin'' defined as follows. There exist  some $C>0$ such that, for any $\varepsilon >0$,  
$$\mathds{P}\big\{b^{(1)}_{-i,t} \in (x,x+\varepsilon)\big\} \leq C \varepsilon^\alpha.$$
\end{proposition2}
\begin{proof}
The proof is a bit technical and mostly sketched; the main ingredients are more or less classical multi-armed bandit techniques. First of all, notice that, as a direct consequence of Hoeffding inequality, with  probability  of the order of $1/T$, all bids are bigger than $x$. As a consequence, we shall  only focus on this event where  regret is only incurred on  auctions such that 
$$
x < b^{(1)}_{-i,t} <  \overline{x}_t + 2\sqrt{\frac{\log(T)}{\omega_{i,t}}}.
$$ 
Indeed, the optimal bid  $b_{i,t}=x$ loses this auction while $\text{UCB}_\text{id}$ overbids and wins it. It unfortunately pays more than its expected value. The net cost  of this specific auction is
$$
b^{(1)}_{-i,t} -x \leq \overline{x}_t + 2\sqrt{\frac{\log(T)}{\omega_{i,t}}} -x \leq 4\sqrt{\frac{\log(T)}{\omega_{i,t}}}, 
$$
where the last inequality holds for all auctions on the event considered (that holds with probability at least $1/T$). Summing over all the auctions $t=1,\ldots,T$  gives the first bound.

The other bounds are derived from careful computations of 
\begin{align*}
&\ \mathds{E} \Big[ (b^{(1)}_{-i,t} -x)  \mathds{1} \big\{ x \leq b^{(1)}_{-i,t} \leq \overline{x}_t  + 2\sqrt{\frac{\log(T)}{\omega_{i,t}}}\big\}\Big]\\
=\ &\  \mathds{E} \Big[ (b^{(1)}_{-i,t} -x)  \mathds{1} \big\{ 0 \leq b^{(1)}_{-i,t}-x \leq \overline{x}_t -x + 2\sqrt{\frac{\log(T)}{\omega_{i,t}}}\big\}\Big]\\
\leq\  &\ C\mathds{E} \Big[ \big( \overline{x}_t -x + 2\sqrt{\frac{\log(T)}{\omega_{i,t}}}\big)_+^{1+\alpha}\Big]\, ,
\end{align*}
where the last inequality is a consequence of the margin definition and $x_+=\max\{x,0\}$. An important fact is that, if some regret is incurred at auction $t \in \mathds{N}$, then this auction is necessarily won hence the counter $\omega_{i,t}$ increases by one. As a consequence, the overall regret can be controlled by 
$$
C\mathds{E} \Big[ \sum_{s=1}^T\big( \overline{X}_s -x + 2\sqrt{\frac{\log(T)}{s}}\big)_+^{1+\alpha}\Big]\, 
$$
where $\overline{X}_s$ is the average value of the first $s$ auctions won.  Hoeffding inequality implies that $\mathds{P}\{\overline{X}_s -x \geq \varepsilon\} \leq e^{-2s\varepsilon^2}$, hence we get that the total regret is smaller than
\begin{align*}
&\ C\mathds{E} \Big[ \sum_{s=1}^T\big( \overline{X}_s -x + 2\sqrt{\frac{\log(T)}{s}}\big)_+^{1+\alpha}\Big]\\
\ \leq &\ C(1+\alpha)\sum_{s=1}^T\int_{- 2\sqrt{\frac{\log(T)}{s}}}^\infty \Big(\varepsilon + 2\sqrt{\frac{\log(T)}{s}}\Big)^\alpha e^{-2s\varepsilon^2}d\varepsilon\\
\ = \ & \ C(1+\alpha)\sum_{s=1}^T \frac{1}{(2s)^{\frac{1+\alpha}{2}}} \int_{- 4\sqrt{\log(T)}}^\infty \Big(u + 4\sqrt{\log(T)}\Big)^\alpha e^{-\frac{u^2}{2}}du\\
\ \leq \ &C(1+\alpha)\sum_{s=1}^T \Big(\frac{16\log(T)}{t}\Big)^{\frac{1+\alpha}{2}}+ C\frac{(1+\alpha)}{2}\sum_{s=1}^T \frac{1}{s^{\frac{1+\alpha}{2}}}\int_{4\sqrt{\log(T)}}^\infty u^\alpha e^{-\frac{u^2}{2}}du\, .
\end{align*}
The result follows from instantiating the above sum over different values of $\alpha$. For $\alpha >1$, the first sum is controlled  using the fact that all terms are necessarily smaller than 1.
\end{proof}
These results only hold if the realized values $v_t$ are i.i.d.\ of unknown expectation $x$, using the basic ideas of stochastic multi-armed bandits and devising a new bidding algorithm based on UCB.  If values $v_t$ can be any sequence, then it is also possible to achieve non-trivial regret bounds by using as a building block EXP.3 algorithm instead of UCB \citep{Weed16}.  Those results and techniques can also be exported to other auctions settings than Vickrey \citep{feng2018learning}.

\subsection{Adaptivity to the mechanism and other bidders}
If a bidder has, at first, not enough information on the auction mechanism and/or the distributions of values of his competitors, he cannot compute an appropriate bidding strategy even if he knows perfectly the values he gives to item. The repeated auction setting can then be helpful for him to learn, on the fly, a strategy. A possible approach is again to use ides from multi-armed bandits, but more precisely on contextual bandits. 

For simplicity, assume that the value distribution of the bidder has a finite support included in $[0,1]$, denoted by $\{x^*_1,\ldots, x^*_L\}$. Then a bidding strategy consists in finding, for each possible values $x^*_\ell$ a corresponding bid. A classical contextual bandits technique consists in discretizing the set of bids $[0,1]$ in $\{b^*_1,\ldots,n^*_K\}$ and  in running independent versions of a base bandit algorithm (such as EXP.3 for instance), one for each possible values, where the set of arms is the discrete set of bids.    Such regret minimizing algorithms have indeed been reported in the online advertising industry \citep{nekipelov2015econometrics}.  

\medskip

In this setting, we can even assume that the opponents change through time, so that their sequence of bids  might be arbitrary (but, conditionally to the past and to the actual value $x_\ell$,  their bids at some given auction $t \in \mathds{N}$ are independent from the bid of player $i$). In this setting, the  oracles to which algorithms are compared to are   stationary strategies, i.e., fixed mappings from values to bid; denote their set by $\mathds{B}_i$. Then the maximal expected cumulative utility  bidder $i$ can get in this class of strategies is 
$$
\max_{\beta_i \in \mathds{B}_i} \sum_{t=1}^T \mathds{E}\big[ u_i((\beta_i(x_t),\bb_{-i,t}),x_t)\big]\,,
$$
where $x_t$ is the value for the item $t$ of bidder $i$, $\bb_{-i,t}$ is the vector of bid of his opponent and $\beta_i$ is a possible strategy. This quantity  serves as a benchmark for a learning to bid policy. Notice that if the mechanism is DSIC, then the optimal mapping in the above equation is $\beta(x)= x$, i.e., bidding truthfully.  On the other hand, the realized cumulative utility of bidder $i$ is equal to 
$$
\sum_{t=1}^T \mathds{E} [ u_i((b_{i,t},\bb_{-i,t}),x_t) ]\,,
$$
where $b_{i,t}$ is his bid at stage $t$, after observing the value $x_t$. As a consequence,  the overall regret of bidder $i$ is the difference between those two terms, the benchmark, and the cumulative utility and reads as
$$
R_T = \max_{\beta_i \in \mathds{B}_i} \sum_{t=1}^T \mathds{E} [ u_i((\beta_i(x_t),\bb_{-i,t}),x_t) ]  - \sum_{t=1}^T \mathds{E} [ u_i(b_{i,t},\bb_{-i,t}),x_t) ]\, .
$$
As mentioned above, if we assume that bidder $i$ is running independent versions of EXP.3 (one per possible value) then using  standard results of multi-armed bandits recalled in Section \ref{Section:Bandits}, we get that 
\begin{equation}\label{EQ:Reg_Contextual}
R_T \leq 2\sum_{\ell=1}^L \sqrt{K\log(K)T_\ell} + \max_{b_\ell \in [0,1]} \sum_{t: x_t=x^*_\ell} \mathds{E} [ u_i((b_\ell,\bb_{-i,t}),x^*_\ell)  ] -   \max_{b^*_\ell \in [0,1]} \sum_{t: x_t=x^*_\ell}^T \mathds{E} [ u_i((b^*_\ell,\bb_{-i,t}),x^*_\ell)  ] 
\end{equation}
where $T_\ell = \sharp \{t\leq T: x_t=x^*_\ell\}$ is the number of times the values was equal to $x^*_\ell$. The first term of Equation \ref{EQ:Reg_Contextual} corresponds to the estimation error, and the second term to the approximation error, because EXP.3 are restricted to bid in the discretization of $[0,1]$ while the optimal bidding strategy does not have this restriction.

We will need some regularity assumption on the mechanism used by the seller; we will assume that it is ``$C$-almost Lipchitz'', for some constant $C>0$, in the following sense. For any value $x_i \in [0,1]$, for any bids $\bb_{-i}$ of the opponents and any bids $b_i$ of bidder $i$, there exists a point $b^\varepsilon_k$ in the $\varepsilon$-regular grid of $[0,1]$ (i.e., $b^\varepsilon_k = k\varepsilon$ for some integer $k$) such that $u_i((b_i,\bb_{-i}),x_i) \leq u_i((b_k^\varepsilon,\bb_{-i}),x_i)+C\varepsilon$. Classical auction mechanisms  (first and second price auctions, with or without reserve prices, Myerson auction...) all satisfy this assumption that ensures the approximation error is of order $T\varepsilon$ if $\{b^*_1,\ldots,b^*_K\}$ is the $\varepsilon$-regular grid (in particular, this implies that $K=1/\varepsilon$). We emphasize here that auction mechanism are usually not Lipschitz (as there are discontinuities around the smallest winning bid).

\begin{proposition2}
If the mechanism is $C$-almost Lipchitz (with $C$ unknown)  and the value distribution has a finite support $L$, then there exists a learning to bid policy whose regret, with respect to the optimal in hindsight bidding strategy smaller than 
$$
R_T \leq(2+C) (LT^2\log(T))^{\frac{1}{3}}
$$
\end{proposition2}
\begin{proof}
One just need to put the definition of $C$-almost Lipschitzness  in Equation \ref{EQ:Reg_Contextual}, as this  gives that regret scales as 
$$
R_T \leq 2\sqrt{\frac{1}{\varepsilon}\log(\frac{1}{\varepsilon})LT} + C T \varepsilon \leq (2+C) (LT^2\log(T))^{\frac{1}{3}}
$$
with the specific choice of $\varepsilon = (L\log(T)/T)^{1/3}$.
\end{proof}

This result only holds for distribution with finite support; for continuous distribution, the trick consists in bucketing the support of value distribution into small bins of size $\varepsilon$ and using, as before, an independent version of EXP.3 per bin. This only works with some regularity assumption on a bin. Specifically, given a small bin $\mathcal{I}$, we shall assume that there exist a constant bid $b$ that is $\varepsilon$-optimal on the set of stages where the values $x_s$ belongs to $\mathcal{I}$, i.e.,
$$
\max_{\beta: \mathcal{I} \to [0,1]} \frac{\sum_{s \leq T: x_s \in \mathcal{I}} u_i((\beta(x_s),\bb_{-i,s}),x_s)}{\sharp \{s \leq T: x_s \in \mathcal{I}\}} \leq \max_{b \in [0,1]} \frac{\sum_{s \leq T: x_s \in \mathcal{I}} u_i((b,\bb_{-i,s}),x_s)}{\sharp \{s \leq T: x_s \in \mathcal{I}\}} +C'\varepsilon.
$$
In particular, this assumption is satisfied if $U_i$ is $C'$-Lipschitz and the bids $\bb_{-i,t}$ does not depend (too much, at most in a Lipschitz fashion) on $x_t$. Once again, balancing the approximation (both in the bid  and the value spaces) and estimations errors gives the optimal choice of $\varepsilon = (\log(T)/T)^{1/4}$ for a regret scalling as 
$$
R_T \leq (2+C+C') (T^3 \log(T))^{\frac{1}{4}}.
$$
If the utility function $u_i$ (and/or the opponents bid $\bb_{-i,t}$) is not Lipschitz but less regular (such as $\beta$-H\"older, which means that $|u_i(x)-u_i(y)|\leq L_\beta \|x-y\|^\beta$ for some constant $L_\beta$), then the rate of regret growth would be impacted as one should find better tradeoffs in approximation vs estimation errors. This would typically lead to a regret scaling as $(T\log(T))^{\mathfrak{b}}$ where ${\mathfrak{b}} \in [\frac{1}{2},1)$ is some parameter depending on the different regularities of the mappings at hand.

\subsection{Adaptivity to budget constraints and pacing options in practice}
In practice, a common problem for  bidders is to find strategies  maximizing some cumulative number of ``events'' (e.g., clicks, views) or some utility, subject to some budget constraint \citep{BalBesWei15,gummadi2012optimal,FernandezTapiaAnalyticalSolution2015,FernandezTapiaGueantLasryOptimalRealTimeBidding2017}. 

As an example, let us describe bidder $i$'s problem in second price auction for utility maximization. Given his budget $\mathfrak{B}_i$, $b^{(1)}_{-i,t}$ the maximum bid of the competition at time $t$ (possibly including reserve prices), $x_t$ his value  of the item  and by $b_t$ at time $t$, the objective  is to solve \begin{gather*}
\max_{\{b_t\}_{t=1}^T} \sum_{t=1}^T \indicator{b_t>b^{(1)}_{-i,t}} (x_t-b^{(1)}_{-i,t})\;,\\
\text{subject to } 	\sum_{t=1}^T b^{(1)}_{-i,t} \indicator{b_t>b^{(1)}_{-i,t}}  \leq \mathfrak{B}\;.
\end{gather*}
Under the assumption that bids are much smaller than total budget, the problem loses much of its stochastic component and is often approximated by its so-called fluid approximation (which replaces both objective and constraint by their expectations, effectively appealing to uniform laws of large numbers \citep{dudley_2014}), turning the problem to 
\begin{gather*}
\max_{\{b_t\}_{t=1}^T} \sum_{t=1}^T  \mathds{E}[\indicator{b_t>b^{(1)}_{-i,t}} (x_t-b^{(1)}_{-i,t})]\;,\\
\text{subject to } 	\sum_{t=1}^T \mathds{E}[\indicator{b_t>b^{(1)}_{-i,t}} b^{(1)}_{-i,t}] \leq \mathfrak{B}\;.
\end{gather*}
Under mild assumptions, and this is a consequence of some strong duality properties,  an optimal bidding strategy is  
$$
b_t=\frac{x_t}{1+\mu^*}\;,
$$
where $\mu^*$ is the optimal solution of the dual problem associated with the constrained optimization mentioned above (Proposition 3.1, \citep{BalBesWei15}). The bidding strategy is similar to the optimal bidding strategy in a second price auction; however the value of each item is discounted by a factor accounting for the constraint. 

This result can be extended to a much broader set of auctions, such as  first price, generalized second price etc... where optimal bidding turns out to be of a similar form to optimal bidding without constraint, the value of each item being linearly discounted by a constant accounting for the budget constraint \citep{gummadi2012optimal}.

\medskip

Three important conceptual ideas emerge from this type of problems. The first one concerns the question of pacing.  For a wide variety of auctions (and payment rules), the optimal strategy to maximize the purchased inventory amounts to spending one's budget smoothly, i.e., at the rate of arrival of auction requests \citep{FernandezTapiaAnalyticalSolution2015,FeedbackSystemsBook}. However, a crucial assumption for this result to hold is that the price paid as a function of win rate does not depend on time and hence the environment is stationary. Not surprisingly, when this assumption does not hold, ``smooth-spending'' at the rate of arrival of auction requests is no longer optimal. These general techniques  can however be used in that more general case to understand the optimal rate of winning auctions and of spending in these more general cases. 

To be implemented in practice, these ideas require of course a forecast for both the arrival rate of auction requests and the price paid at a certain win rate. When the bidder has access to such information, ideas of model-predictive control and re-optimization can be used  \citep{CiocanFariasMPC2012}. 

A third important line of work concerns online estimation of the parameter $\mu$ mentioned above, possibly without forecast \citep{BalseiroGur2019}. The essential idea is to use the fact that $\mu^*$ mentioned above in the solution of an optimization problem and to solve this optimization problem online, using online gradient descent \citep{shalev-shwartz_ben-david_2014}. To be slightly more specific, in the problem mentioned above, the optimal solution for $\mu$ in hindsight is determined through 
$$
\inf_{\mu\geq 0} \sum_{t=1}^T (x_t-(1+\mu)b^{(1)}_{-i,t})_+ + \mu \mathfrak{B}\; \triangleq: \inf_{\mu \geq 0} \sum_{t=1}^T \ell_t(\mu).
$$
The functions $\ell_t$ are observable at time $t$ and  are a sequence of functions arriving in a streaming fashion. As a consequence, the estimate of the parameter $\mu^*$ can be updated in an online fashion, without a forecast, by using for instance the online (sub)-gradient descent rule
$$
\mu_{t+1}=\mu_t - \gamma \partial_\mu \ell_t(\mu_t)\;,
$$
where $\partial$ is the  (sub)gradient operator and $\gamma$ is the stepsize in the online gradient descent algorithm. Various theoretical guarantees about this scheme, under a variety of optimistic and pessimistic assumptions about the amount of information that is known about the environment in which the bidder evolves, can be proved \citep{BalseiroGur2019}. 

The literature on this topic is very large, with many different variations \citep{GhoshEtal20009,ChoiMela2018,LeeAli2013,YuanWang2013,XuKuang2015}. The question of handling the situation where the bid to budget ratio is not close to 0, and hence the fluid approximation is not well justified, is quite open and  appears to be more of a genuine stochastic control type. An interesting approach seems to use ideas coming out of the analysis of the online knapsack and related problems \citep{Arlotto_2019}. 

\section{Mechanism design in front of adaptive bidders \& Full surplus extraction}

As illustrated in the previous section, the standard techniques of bandits and online learning can be used by bidders to choose their bidding strategy and potentially improve it overtime (e.g., by improving the estimation of their own valuation). This adaptivity to an initial lack of knowledge comes at a cost for the bidder, the regret that cumulates over time. In a less straight-forward manner, it also impacts the seller, by making mechanism design more complex. Indeed, even a basic notion such as incentive compatibility is changed. For instance, in the context of learning to bid in a DSIC auction, the bidder does not choose anymore his bid as a best response that optimizes his utility: due to the need to find a compromise between exploration and exploitation, $\text{UCB}_\text{id}$ generally overbids. Even while the auction is DSIC, a bidder using $\text{UCB}_\text{id}$ is not truthful anymore. The following subsections describe how notions such as incentive compatibility and revenue maximization can be modified when facing adaptive bidders.

\subsection{Learning against bidders using zero-regret algorithms}
As illustrated in the previous section, the standard techniques of bandits and online learning can be used by bidders to choose their bidding strategy... assuming that the seller let them do it. Indeed, it was more or less implicitly assumed that the mechanism was fixed during the whole interaction, another assumption that is way too restrictive and should be taken care of, as the seller could actually largely leverage this almost predictive behavior for her own interest. 

Let us consider the simple case where all  bidders are independently running regret minimizing algorithms  \citep{braverman2018selling,deng2019prior}, and more precisely where they are using algorithms such as EXP.3, that base their decisions on the mean reward observed. Their particularity is that they rarely pick an arm whose current mean is significantly worse than the current highest mean; for instance if during the first $t$ stages an arm $k$ has generated an average reward (denoted by  $\overline{X}_{k,t}$) that is smaller than the one of arm $\ell$, then the probability of choosing $k$ over $j$ is exponentially small. More precisely, the difference of these log-probabilities scales linearly with $\overline{X}_{k,t}-\overline{X}_{j,t}$. We will consider in the following a  general class of algorithms that exhibit similar behavior, that are called $\eta$-mean based, but more general than just EXP.3.
 
\begin{definition2}%[Mean-based algorithm] 
In the standard multi-armed problem, an algorithm is $\eta$-mean-based, for some $\eta \in (0,1)$, if at any stage $t \in \mathds{N}$ and for any pair of arms $(k,\ell)$, if $\overline{X}_{k,t} < \overline{X}_{\ell,t}- \eta$ then  the probability $p_{k,t+1}$ that the algorithm pulls arm $k$ at time $t+1$ is smaller than $\eta$. 
An algorithm is (asymptotically) mean-based if $\eta = o_{T\rightarrow\infty}(1)$.
\end{definition2}

In particular, EXP.3 is asymptotically mean-based, and so is $\epsilon$-greedy (for $\eta=\varepsilon$).

To simplify the following statements, we are  going to assume that there is only one bidder, whose  value distribution $F$ is known beforehand (an assumption that can be fairly weakened \citep{deng2019prior}). If the seller was using the same mechanisms at each auction $t \in \{1,\ldots,T\}$, then the optimal one would obviously to post the monopoly price. This generates a total revenue of $T$ times the monopoly revenue (because the bidder will quickly learn the optimal strategy). On the other hand, if the seller knows that the bidder is using a mean-based algorithm, she can generate a much higher revenue \citep{braverman2018selling}, almost as high as the total welfare denoted by $\mathfrak{W}(F) = \mathds{E}_{x\sim F}[x]$; with $n$ bidders, the total welfare would be $\mathfrak{W}(\mathbf{F}) = \mathds{E}_{\bx\sim\bF}[\max_i x_i]$

To achieve this, the mechanism must change through time and possibly be itself adaptive to the sequence of bids of the buyer. We therefore introduce the concept of dynamic mechanism, so that the actual auction rules might change from stage to stage.

Recall that at each stage $t \in\{1,\ldots,T\}$, bidders valuations for the item are sampled through  distributions $F_i$. These distributions are fixed from one auction to the other. We denote by $a_t \in \mathcal{A}$ the auction mechanism chosen at time $t$ by the seller and by $b_{i,t}$ the bid of buyer $i$. We also denote by $\mathcal{H}_t = \{a_1,b_{1,1},b_{n,1}\dots,a_{t-1},b_{1,t-1},b_{n,t-1}\}$ the finite history at time stage $t$, that consists of past auctions and past buyer's bids. 
\begin{definition2}\label{def:dynamic_mechanism}
A dynamic mechanism $\mathcal{D}\mathcal{M}: \bigcup_t \mathcal{H}_t \rightarrow  \mathcal{A}$ is a mapping that associates to any finite history  $\mathcal{H}_t$ an auction $a_t=: \mathcal{D}\mathcal{M}(\mathcal{H}_t)$. A bidder's dynamic strategy, $\mathcal{S}$, is a mapping from $\mathcal{H}_t$ to the set $\mathds{B}$  of strategies (i.e., functions from values to bids). 
\end{definition2}

The following theorem states that a seller can \emph{extract the full surplus} of the system, if  bidders are using na\"ive learning algorithms.

\begin{theorem2}
If the bidder is running a mean-based algorithm, for any $\varepsilon > 0$, there exists a dynamic selling mechanism such that the seller can get $(1-\varepsilon)\mathfrak{W}(\bF)T - o(T)$.
\end{theorem2}

The intuition behind this result is that  the seller can lure the na\"ive algorithms such as EXP.3 by setting low prices during a first (large) period of time and then by increasing drastically the reserve price during a second stage. This is illustrated in the following example with $n=1$ bidder \citep{braverman2018selling}.

Assume the bidder value $x_t$ has the following distribution 
$$
x_t = \left\{ \begin{array}{cl} \frac{1}{4} & \text{with probability }\ \frac{1}{2}\\
\frac{1}{2} & \text{with probability }\ \frac{1}{4}\\
1& \text{with probability }\ \frac{1}{4}
\end{array}\right.
$$ Simple computations shows that corresponding monopoly price is $1/4$ and setting it at each time generate a revenue of $T/4$ after $T$ auctions.

 To fool a mean-based algorithm, the seller can use the following scheme. At any stage, it will only allocate the item if the bid is exactly $1$ (any other bid gives a utility of 0). It remains to define the payment associated to a winning bid of 1. During the first $T/2$ stages, it is equal to 0, while it will be equal to 1 during the last $T/2$ stages.
 
Recall that the bidder runs multiple  independent instances of EXP.3, one for each possible values ($1/4$, $1/2$ and $1$), so we can focus independently on the set of stages where the value is constant, and for the sake of simplicity we are going to assume that this value equals $1/2$ (resp.\ 1) exactly $T/8$ times during the first and second half of the game.
\begin{itemize}
\item On the set of stages where the value is 1, EXP.3 quickly learns that bidding 1 is optimal during the first half of the game. This generates a cumulative utility of $T / 8$ to the buyer. As a consequence, EXP.3 will keep bidding 1 when the value
is 1 at each stage of the second half with exponentially high probability. The revenue generated on those stages by the seller is then  approximatively $T / 8$.
\item When the value is $1/2$, bidding 1 during the first half generates a revenue of $T/16$ to the buyer. During the second half, bidding 1 generates a negative utility of $-1/2$ per stage, so that the cumulative utility of bidding 1 decreases, but remains positive during the whole process (and EXP.3 will keep bidding 1 with arbitrarily high probability for almost all stages).  The revenue generated on those stages by the seller is then also  approximatively $T / 8$.
\item When the value is $1/4$, bidding 1 during the first half generates a revenue of $T/16$ to the buyer.    During the second half, bidding 1 generates a negative utility of $3/4$ per stage, so that the cumulative utility of bidding 1 decreases, but remains positive during $T/12$ additional stages where EXP.3 will bid 1 with high probability (and afterwards stop bidding 1 as the cumulative utility of this bid is negative). The revenue generated on those stages by the seller is then also  approximatively $T / 12$.
\end{itemize}
At the end, the total revenue of the seller is therefore of the order of $T/8+T/8+T/12= T/3$ which is much bigger than $T/4$. The trick for the seller was to make the bidder overpay on many  auctions by exploiting the behavior of mean-based algorithms that keep bidding~1 even when instantaneous negative utilities occur. Unfortunately for the seller, this theorem only holds for mean-based buying algorithms.  Even worse, for any dynamic selling mechanism, there exists a buyer's strategy such that he does not pay more than $T$ times the monopoly price  \citep{braverman2018selling}.

\subsection{Trading off ex-post individual rationality for full surplus extraction}
\label{sub_full_surplus_extraction}
If the buyers are using na\"ive algorithms, and the seller knows this, then we proved that she can extract (almost) the full surplus from the system. This was possible because of the asymmetry of information between agents. There are other settings where this full surplus extraction by the seller is possible. The first example we consider is the case where the individual-rationality assumption of the mechanism is removed. This will induce another strong asymmetry between agents, as bidders are somehow ``forced'' to participate in auctions. The idea is to consider the weaker concept of ex-ante, instead of interim, individual-rationality (see Section  \ref{subsec_prop_auction} for more details on the differences).  In the \textit{ex-ante} setting, the bidder does not know the value he will give to the item before he agrees to take part in the auction - he therefore has the same information as the seller on his private valuation.  An ex-ante individually rational mechanism must give a non-negative expected  utility to the bidder.

\begin{theorem2}%[full surplus extraction]
[\cite{cremer1988full}]
There exists an ex-ante individually rational and incentive-compatible auction where the bidders' utilities are all equal to zero and the seller extracts the full surplus.
\end{theorem2}
\begin{proof}Since the seller knows the bidders' value distributions, she  can compute their expected utilities in a second price auction. 

The mechanism constructed is simple. It consists of an entry fee that must be paid before participating to the auction (stated otherwise, the bidder must pay this amount no matter the outcome of the auction); afterwards, a standard second price auction without reserve price is run. Choosing for the entry fee the expected utility in the second price auction gives an expected  utility of zero to each buyer and the seller extracts the full surplus of the game. This mechanism is of course ex-ante-incentive-compatible. 
\end{proof}
This mechanism is not interim nor ex-post individually rational since for all valuation vectors, the utility of all losing bidders is negative. In order to slightly overcome  this issue, \citep{balseiro2017dynamic} and \citep{mirrokni2016dynamic} refined  this mechanism to ensure that bidders' utilities are not too negative at some point in the game; the trick is to  dispatch the fee on the different time steps instead of being paid at the beginning of the game. They also generalize the original setting to more complex dynamic auctions.

Considering the ex-ante setting makes sense only in auctions where the buyers do not know before participating their own valuation (but only the distribution). It is quite unrealistic in many single item auction, but it could make some sense 
when  $T$ successive auctions are run as in online ad market: indeed, the fee must be paid  before taking part in any of the $T$ auctions. While bidder can compute their distribution of values in the future, they do not know in advance what will be the exact future realizations. This unfortunately requires the bidders to also know perfectly the number $T$ of future auctions. This assumption has been weakened by \citep{agrawal2018robust} that adapted the above mechanisms to bidders that do not believe that there will be $T$ auctions. In this case, they are quite likely to refuse to pay a fee computed on $T$ auctions early in the game. %{\color{red} Ideas of techniques and proof?}

A crucial assumption of this line of work is to assume that bidder's value distributions are known to the seller beforehand; this enables her to compute precisely the extra-fees that can be charged to the bidders without breaking the ex-ante individually rational assumption. Similarly, it implicitly assumes that bidders are able to compute best response to dynamic mechanism (and that they implement them); this assumption is weakened in the following section.

\subsection{Learning against (almost) myopic buyers}

A first attempt to remove the prior knowledge of  bidder's value distributions, and instead to learn them \citep{amin2014repeated,mohri2015revenue,golrezaei2018dynamic}, is to consider mechanisms that are incentive compatible (up to a small number of bids) under the assumption that bidders are \emph{almost} myopic or impatient -- i.e., they have a fixed discount on future utilities. This again introduces  an asymmetry between the bidders with a discounted long-term utility, and the seller with an undiscounted long-term revenue (infinitely patient). 

To simplify the exposure, we will focus on the posted price case. Formally, let us denote by  $p_t$ the price of the item at time $t$ chosen by the mechanism  and by  $d_t$ the decision of the buyer to buy ($d_t=1$) or to refuse the item ($d_t=0$). Since the distribution of values is not known beforehand, $p_{t+1}$ can only depend on the finite history  $\mathcal{H}'_t = \{p_1,d_1,\dots,p_{t},d_{t}\}$. The discounted bidder utility is $\sum_{t=1}^{T}  \gamma_t d_t (x_t - p_t)$, where $\gamma =(\gamma_t)_t$ is a sequence of non-negative weights. In this section, we shall assume for simplicity that values are uniformly bounded by 1.

The objective of the seller is to choose a  dynamic selling mechanism $\mathcal{DM}$, that will maximize her revenue,  against buyers that know  $\mathcal{DM}$ and  respond optimally for them in the long run, i.e., in the objective of maximizing their discounted and expected utilities. Let us denote by $d_t^*(\mathcal{DM})$ the optimal strategy of the bidder and by $p_t(\mathcal{DM})$ the price posted.  The performance of a dynamic mechanism will be measured in terms of ``regret'', whose definition is slightly different than in the previous section. 

\begin{definition2}%[Regret of a dynamic mechanism]
Given a dynamic mechanism $\mathcal{DM}$, the discount sequence $\gamma$ and the value distribution $\mathcal{F}$, the regret of the seller is
$$
R_T(\mathcal{DM},\gamma,F)=%\max_{\mathrm{DM}^*} 
\mathds{E} \sum_{t=1}^T d_t^*(\mathrm{DM}^*)p_t(\mathrm{DM}^*) - \sum_{t=1}^T d_t^*(\mathcal{DM})p_t(\mathcal{DM})
$$ 
\end{definition2}
In this setting, $\mathrm{DM}^*$ consists in posting the monopoly price corresponding to the value distribution $\mathcal{F}$ at each round. We emphasize here that the dependencies in $\gamma$ and $F$ in the regret definition are hidden in the best responses $d_t^*(\cdot)$.

\begin{theorem2}[\cite{amin2013learning}]%[lower bound]
\label{thm:lower_bound_amin}
Let $\gamma_t$  be any positive non-increasing sequence and $\mathcal{D}\mathcal{M}$ be any dynamic selling mechanism. Then, there exists a buyer value distribution $F$ such that the regret $R_T(\mathcal{D}\mathcal{M},\gamma,F)  \geq \frac{1}{12} \sum_{t=1}^T \gamma_t$. In particular, sublinear regret is impossible to achieve if  $\sum_{t=1}^T \gamma_t = \Theta(T)$.
\end{theorem2}
This theorem states that if the buyer is patient enough, the seller cannot learn the monopoly price quickly enough to reach a sublinear regret.  However, when the sequence $\gamma_t$ decreases geometrically, i.e., $\gamma_t=\gamma^{t}$ for some $\gamma \in (0,1]$, sublinear regret is possible as $\sum \gamma_t = o(T)$. In words, this means that if the buyer is much more impatient than the seller, the latter can extract surplus; moreover, this can be achieved with a simple two-phased dynamic mechanism \citep{amin2014repeated}.

\begin{enumerate}
\item Phase  1 (of length: $\alpha T$) : offer a random price, uniformly in $[0,1]$
\item  Phase 2 (of length: $(1-\alpha) T$) : compute the optimal price using some robust estimation procedure and post it until the end.
\end{enumerate}

\begin{theorem2}[\cite{amin2014repeated}]\label{Thm:UpperBounAmin}
With the choice of  $\alpha=T^{-1/3}$, the regret can be bounded as $\mathcal{O}(T^{\frac{2}{3}}\sqrt{\frac{\log(T)}{\log(1/\gamma)}})$ if  the discount sequence satisfies $\gamma_t=\gamma^{t}$, for some $\gamma \in (0,1]$.\end{theorem2}
The formal proof of this statement is a bit long and technical, but the main ingredients are quite easy to understand. First, we are going to assume that the horizon $T \in \mathds{N}$ is known beforehand, otherwise one could just use the doubling trick. 

The key idea is to bound the number of times  the buyer can \emph{lie} by not being truthful, i.e., either by buying the item at a price higher than his value or refusing a lower price. Notice that the net cost of a lie at stage $t \in \mathds{N}$, if the stage valuation is $x_t$ and the price posted $p_t$, is exactly equal to $|x_t-p_t|$. Since the prices $p_t$ are i.i.d., and uniformly drawn on $[0,1]$, then the potential costs $|x_t-p_t|$ are smaller than $\varepsilon$  only $2\varepsilon\alpha T$ times during the first phase (at least in expectation, but we are going to neglect the deviations in this sketch of proof). As a consequence, if the buyer lies $L$ times during this phase, then at least $L/2$ of those lies must have a cost of at least $L/(4\alpha T)$. Recall that the buyer puts weight $\gamma_t=\gamma^t$ to the $t$-th stage, so that the cumulative, discounted cost of those $L$ lies is at least 
$$\sum_{t=\alpha T-L/2}^{\alpha T} \gamma^t \frac{L}{4\alpha T} = \frac{L}{4\alpha T}\frac{\gamma^{\alpha T+1}}{1-\gamma}(\gamma^{-L/2-1}-1)\geq \frac{L}{8\alpha T} \big(\frac{1}{\gamma}\big)^{\frac{L}{2}+1}\frac{\gamma^{\alpha T+1}}{1-\gamma}.$$
It remains to control the total gain of those $L$ lies. As best, they will induce a posted price of $p_t =0$ during the second stage, and a per-stage gain of at most 1 for the buyer. As a consequence, the total cumulative gain of the buyer is at most
$$
\sum_{t=\alpha T +1}^T \gamma^t = \frac{\gamma^{\alpha T +1}}{1-\gamma}(1-\gamma^{T(1-\alpha)}) \leq \frac{\gamma^{\alpha T +1}}{1-\gamma}.
$$
All things put together, the cumulative discounted net gain of lying $L$ times is upper-bounded by
$$
\frac{\gamma^{\alpha T +1}}{1-\gamma}-\frac{L}{8\alpha T} \big(\frac{1}{\gamma}\big)^{\frac{L}{2}+1}\frac{\gamma^{\alpha T+1}}{1-\gamma}=\frac{\gamma^{\alpha T +1}}{1-\gamma}\Big(1- \frac{L}{8\alpha T} \big(\frac{1}{\gamma}\big)^{\frac{L}{2}+1}\Big).
$$ 
A direct consequence of the above inequality is that the number of lies $L$, for them to be profitable, must satisfy
$$
\frac{L}{8\alpha T} \big(\frac{1}{\gamma}\big)^{\frac{L}{2}+1} \leq 1 \Longrightarrow L \leq 2\frac{\log(8\alpha T)}{\log(1/\gamma)}.
$$
As a consequence, this gives a simple upper-bound on the number of lies that can be seen as ``outliers'' from the point of view of the seller, when trying to estimate the optimal price. \

From the point of view of the seller, the regret can be decomposed into the cost of the first phase, bounded by its length $\alpha T$, and the cost of the second phase, bounded by $T\eta$, where $\eta$ is the error on the optimal price computed during the first phase. The remaining question consists in bounding this error; standard robust estimation techniques (such as median\footnote{This technique consists in dividing the full dataset of size $\alpha T$ in $2L$ different datasets and estimating the optimal price on each of them. There necessarily exists a majority of small datasets without outliers that estimate correctly the optimal price. Hence taking the median value is a robust procedure as long as $L = o(\alpha T)$.} of means \citep{Lecue}) or gradient descents with outliers indicate that $\eta$ is of the size of $\sqrt{L/\alpha T}$. Adding both terms and considering the previous bound on $L$ gives a regret scaling as, up to multiplicative constant, 
$$
\alpha T +  \sqrt{\frac{\log(T)}{\log(1/\gamma)}}\sqrt{\frac{T}{\alpha}} \leq T^{2/3}\sqrt{\frac{\log(T)}{\log(1/\gamma)}}
$$
with the choice of $\alpha = T^{-1/3}$. The simple idea behind this algorithm was then refined \citep{mohri2015revenue} and extended to the case of $K$ bidders  \citep{golrezaei2018dynamic}.

\begin{remark2}
Once again, this surplus extraction is possible only because there is an (artificial) asymmetry between the seller and the buyer preventing him to be too strategic. 
 This can also be enforced through another approach, yet it is valid only with several (almost) symmetric bidders -- leading to another type of asymmetry between the seller and buyers. The idea is quite simple: make the computations required (e.g., to determine a reserve price) not as a function of the buyer bids, but as a function of his competitors' bid \citep{ashlagi2016sequential,KanNaz14,epasto2018incentive}.  Unfortunately, this approach cannot handle the existence of any dominant buyer, i.e., a buyer with much higher values than the other bidders \citep{epasto2018incentive}. Therefore, the impact of this technique is quite limited since revenue-optimizing mechanisms are mostly important when the buyers are heterogenous. Moreover, in the main real-world application of online advertising, with asymmetric bidders and no specific asymmetry between seller and buyers on future utilities, none of these mechanisms ends up being able to enforce truthful bidding.
\end{remark2}

\subsection{Incentive Compatibility for Adaptive bidders}
As illustrated by Theorem \ref{thm:lower_bound_amin}, when facing a patient bidder, the seller can't expect to get a revenue equivalent to these of a revenue-maximizing auction, even in expectation. Indeed, in this case, the mechanisms previously described cannot ensure the first crucial need to implement a dynamic selling mechanism: making sure the bidders bid truthfully, at least often enough. In the simpler case, when the bidders fully observe their value before bidding, the seller can implement a simple second-price auction without reserve price (or with a fixed one), which ensures truthfulness and a minimal revenue. However, in the slightly more complex setting (but also more realistic) used in Section \ref{ssec:learning_to_bid}, this is no longer straightforward.

This section focuses on a simple setting to study this problem, the \emph{static bid model}, and extension to more complex settings are explained at the end. This setting describes well the situation encountered when buying online advertising space. At time $t\in [T]$, the value $x_{i,t}$ of bidder $i$ is decomposed into two parts: $x_{i,t} = \alpha_i c_{i,t}$. The first part, $\alpha_i \in[0,\alpha_{\rm max}]$ (e.g., a cost-per-click) is a private knowledge bidder $i$ has from the start, while the second part $c_{i,t}$ (e.g., a click on an ad banner) is a binary random variable with mean $\rho_i$ and which realization is observed, for the winner $i_t$ only, \emph{after} assignment. Because the objective of this section is to restrict our study to DSIC dynamic mechanisms, we can assume the bidders are sending the same bid at any time step and have a formalization where the bidders are sending the bids $\bb$ once, at the beginning (at $t=0$). We don't assume the bids are bounded above. Then, the assignment is sequential, in a similar way as in Definition~\ref{def:dynamic_mechanism}. Indeed, let us denote by $\cH_t = \{\bq_1, c_{i_1, 1}\dots, \bq_{t-1}, c_{i_{t-1},t-1}\}$ the history up to time $t$, where $\bq_t = \bq(b, \cH_t) \in\Delta^n$ is the assignment at time $t$ resulting from the sequential assignment function $\bq:\bcB \times \bigcup_t \cH_t \to \Delta^n$.  We will explicitly denote it $\cH_t(\bb,\bc)$ if/when we want to emphasize that $\cH_t$ is a function of the bids $\bb$ and of the click (potential) realizations $\bc\in\{0,1\}^{T\times n}$.  Finally,  because the bidders  only give one bid at the beginning, we can consider the payment $p:\bcB\times\cH_T\to\lR^n$ is done  at the very end (after $t=T$), sort of  a final billing.

In fact, this setting can be viewed as a multi-armed bandit (MAB) with an "unusual" way to define the reward, the assignment function $q$ being the bandit algorithm. So here, the question will be whether it is possible to recover the performance of optimal MAB algorithm (KL-UCB) or whether restricting to assignments $\bq$ for which it is possible to find a payment that make the mechanism incentive compatibility will lead to a degradation of the performance.
Following different objectives of performance,  the pseudo-regret can be defined in terms of seller's revenue \citep{devanur2009} , or in social welfare \citep{babaioff2014a}. In both case, the comparator of the regret is a weighted second-price auction for which $\rho_i$s are known. Remembering we denote by $i_t$ the winning bidder at time $t$ and by $\smax$ the second-highest element ("second max"), we have
\begin{align*}
R_T^{\mathfrak{W}} = T \max_i \alpha_{i} \rho_i - \sum_{t=1}^T \alpha_{i_t}\rho_{i_t} & & R_T^p = T \smax_i \alpha_{i} \rho_i - p(\bb, \cH_T)
\end{align*}

\paragraph{Deterministic Assignment}
We split the analysis on whether the assignment is randomized or not, as it incentive compatibility imposes very different constraints depending on it, leading to different orders of performance. When assignment is deterministic, it is possible characterize the dynamic mechanisms that are DSIC. We provide this charaterization for the case of two bidders, i.e., $n=2$, for the sake of simplicity, that can be extended to $n>2$ bidders  \citep{babaioff2014a}. As a technical detail, we assume here the sequential assignment is \emph{non-degenerate}, meaning that when a bid $\bb$ generates a given assignment, all bids of the form $(u,\bb_{-i})$ generates the same assignment, for $u$ ranging in some non-degenerate interval containing $b_i$\footnote{This assumption is technical and allows to avoid dealing with exposing results that hold almost surely w.r.t. Lebesgue measure.}.

\begin{theorem2}[\cite{babaioff2014a}]
\label{thm:characterizing_truthfulness}
For $n=2$, given a scale-free\footnote{The scale-free property just means that rescaling the bids doesn't change the outcome of the assignment -- e.g. it does not depend on the currency.} and deterministic dynamic allocation $\bq$, there exists a payment $p$ such that the resulting dynamic mechanism is 0-rational and DSIC iff
\begin{enumerate}
\item (pointwise-monotone) \emph{for any bid profile and for any realization of the history, if bidder~$i$ wins at round $t$, he would still win by bidding higher}, i.e.
\begin{center}
$\forall t \in [T], \forall \bc\in\{0,1\}^{t\times n}, \forall \bb\in\bcB$, if bidder $i$ wins the auction at time $t$,\\ then for $\tilde{b}_{i} > b_{i}$, we have $q_{i}(\tilde{b}_{i}, \bb_{-i}, \cH_t(\tilde{b}_{i}, \bb_{-i},\bc)) \geq q_{i}(\bb, \cH_t(\bb,\bc))$.
\end{center}
\item (exploration-separated) \emph{at any step $t$ whose output impacts a future assignement, the assignment does not depends on the bids $\bb$}.
\end{enumerate}
\end{theorem2}
\begin{proof}
To simplify notation for the proof, we denote $\bq(\bb,\bc) \in \lR^{n\times T}$ the matrix with columns $\bq_t(\bb,\bc)^\top = \bq(\bb, \cH_t(\bb,\bc)))$. Further, we recall that the payment for a truthful mechanism is defined by $p_i(\bb,\bc) = \langle b_i \bc_i, \bq_i(\bb,\bc)\rangle - \int_0^{b_i} \langle\bc_i,\bq_i(u, \bb_{-i},\bc)\rangle{\rm d}u$  \citep{archer2001truthful}. We break the proof in three steps, proving first that 0-rationality and DSIC implies monotonicity, then exploration-separation and finally the converse statement.

\medskip

{\bf DSIC $\Rightarrow$ monotone.} The proof is by contradiction. Assume there exists $t, \bc, \bb, b_i^+$ such that $b_i < b_i^+$ and $q_{i,t}(\bb,\bc) > q_{i,t}(b_i^+, \bb_{-i},\bc)$. W.l.o.g. we can assume there are no clicks at any time $t' \geq t$ (as they do not affect assignment at time $t$) and we denote $\bc' = \bc \oplus \indicator{(i,t)}$, where $\oplus$ denotes the bit change -- i.e. the addition modulo 2. Because buyer $i$ does not win at step $t$ by bidding $b_i^+$, we should have $p_i(b_i^+,\bb_{-i},\bc) = p_i(b_i^+,\bb_{-i},\bc')$. Contradiction will come by proving they are not equal.

We can focus on the integral term of the payment, as the first one does not change between $\bc$ and $\bc'$.
\begin{align*}
\forall u\in[0,\alpha_{\rm max}], & ~\langle \bc_i, \bq_i(u, \bb_{-i},\bc)\rangle \leq \langle\bc_i, \bq_i(u, \bb_{-i},\bc')\rangle & \text{(no clicks after time $t$)}\\ 
& \Rightarrow \int_0^{b_i^+} \langle\bc_i, \bq_i(u, \bb_{-i},\bc)\rangle{\rm d}u \leq \int_0^{b_i^+} \langle\bc_i, \bq_i(u, \bb_{-i},\bc')\rangle{\rm d}u 
\end{align*}
Further, because the assignment is non-degenerate, there exists an interval $\cI$ containing $b_i$ such that for all $u\in\cI$,
\begin{align*}
\bc_i^\top \bq_i(u, \bb_{-i},\bc) = \langle\bc_i, \bq_i(b_i, \bb_{-i},\bc)\rangle < \langle\bc_i, \bq_i(b_i, \bb_{-i},\bc')\rangle = \langle\bc_i, \bq_i(u, \bb_{-i},\bc')\rangle
\end{align*}
Then, it means $\int_0^{b_i^+} \langle\bc_i, \bq_i(u, \bb_{-i},\bc)\rangle{\rm d}u < \int_0^{b_i^+} \langle\bc_i, \bq_i(u, \bb_{-i},\bc')\rangle{\rm d}u$, which is a contradiction with the payments being equal.
\medskip

{\bf DSIC $\Rightarrow$ exploration-separated.} The proof is again by contradiction. Assume there exists $t<t',\bc,\bb$ such that 
\begin{enumerate}
\item $q_{2,t}(\bb,\bc) = 1$, i.e.,. buyer 2 wins round $t$, w.l.o.g., 
\item $\bq_{t'}(\bb,\bc) \neq \bq_{t'}(\bb, \bc')$ with $\bc' = \bc \oplus \indicator{(2, t)}$, i.e., time $t'$ is influenced by output of time $t$,
\item $\exists \bb' \in \bcB, q_{1,t}(\bb',\bc) = 1$, i.e., the assignment at time $t$ depends on bids,
\item $t'$ is minimal (w.l.o.g.) and 
\item there is no click after $t'$ (again, w.l.o.g.).
\end{enumerate}
Since $\bq$ is scale-free, for $b_1^+ = \frac{\tilde{b}_1}{\tilde{b}_2}b_2$, we have $q_{1,t}(b_1^+, b_2, \bc) = 1$, thus, as $\bq$ is pointwise-monotone, $b_1^+ > b_1$. Because the difference between $\bc$ and $\bc'$ is on buyer 2 at time $t$, then $p(b_1^+, b_2, \bc) = p(b_1^+, b_2, \bc')$. Contradiction will come by proving they are not equal.

We focus on the payment of buyer 1 and again, on the integral terms $\int_0^{b_1^+} \langle\bc_1, \bq_1(u, b_2,\bc)\rangle{\rm d}u$ vs $\int_0^{b_1^+} \langle\bc_1, \bq_1(u, b_2,\bc')\rangle{\rm d}u$. Assume w.l.o.g. that $q_{1,t'}(\bb,\bc) < q_{1,t'}(\bb,\bc')$, then by pointwise monotonicity, we have $\forall u < b_1^+, q_{1,t'}(u,b_2,\bc) \leq q_{1,t'}(u, b_2,\bc')$. Then, since $q$ is non-degenerate, the strict inequality holds on a non-degenerate interval, hence $$\int_0^{b_1^+} \langle\bc_1, \bq_1(u, b_2,\bc)\rangle{\rm d}u < \int_0^{b_1^+} \langle\bc_1, \bq_1(u, b_2,\bc')\rangle{\rm d}u,$$which is in contradiction with payments being equal.

\medskip

{\bf exploration-separated + monotone $\Rightarrow$ DSIC.} Since $\bq$ is pointwise-monotone, it is monotone, hence the auction is truthful and 0-rational if it can implement the payment  of a truthful mechanism \citep{archer2001truthful}. The main challenge is to show $p$ is $\cH_t(\bb,\bc)$-adapted, i.e., it can be computed with access to observable information (observed clicks) only, and especially the integral term $\int_0^{b_i} \bc_i^\top \bq_i(u, \bb_{-i},\bc){\rm d}u$. Indeed, for $u < b_i$, buyer $i$ can loose an assignment at some time $t$, which may impact future assignments, potentially requiring to use clicks that were not observed with bids $\bb$ to compute the payment. 
\begin{align*}
p_i(\bb, \bc) &= b_i \langle\bc_i, \bq_i(\bb,\bc)\rangle - \int_0^{b_i}\langle \bc_i, \bq_i(u, \bb_{-i},\bc)\rangle{\rm d}u \\
&= b_i\langle\bc_i, \bq_i(\bb,\bc)\rangle - \int_0^{b_i} \sum_{t=1}^T c_{i,t} q_{i,t}(u, \bb_{-i},\cH_t(i,\bb_{-i},\bc)){\rm d}u ~~~~~~~~~~~ \text{(by definition of $\bq$)}
\end{align*}
As $\bq$ is exploration-separated, then $\bq_t(\bb,\bc)$ only depends on two elements. The first one is obviously $\bb$,  and the second one is  subset $\cE_t(\bc)$ of the set histories $\cH_t(\bb,\bc)$ that is independent of $\bb$. Consequently, we can write $\bq_t(\bb,\cH_t(\bb,\bc)) = \bq_t(\bb, \cE_t(\bc))$ and thus
\begin{align*}
p_i(\bb, \bc) &=  b_i \langle\bc_i,\bq_i(\bb,\bc)\rangle - \int_0^{b_i} \sum_{t=1}^T c_{i,t} q_{i,t}(u, \bb_{-i},\cE_t(\bc)){\rm d}u.
\end{align*}
This implies that the payment is $\cH_t(\bb,\bc)$-adapted.
\end{proof}

Note that Theorem~\ref{thm:characterizing_truthfulness} characterizes the dynamic assignment, as the payment is derived as in Corollary~\ref{corollary_charect_DSIC} \citep{archer2001truthful}.
This theorem confirms earlier results \citep{devanur2009},  based on a dynamic selling mechanism with an explore then commit structure (ETC, \cite{perchet2013multi}). We will describe this specific algorithm for welfare regret  \citep{babaioff2014a} and provide upper-bounds for it as it is possible to derive the guarantees both in terms of welfare and revenue for this algorithm; however, seller's revenue regret can be handled quite similarly \citep{devanur2009}.

For this algorithm, it turns out the final payment is very naturally decomposed as the sum of per-step payments, so we describe it this way. For the first $\tau$ steps, the assignment is a round-robin over the bidders, leading each bidder to win $\left\lfloor \frac{\tau}{n}\right\rfloor$ times and paying 0 each time. This exploration phase allows to build an unbiased estimate $\hat{\rho}_i$ of $\rho_i$. Further, it allows to ensure that
\begin{align}
\lP\Big(\exists i \in \cN, |\hat{\rho}_i - \rho_i | \geq \underbrace{\sqrt{2\left\lfloor\frac{n}{\tau\wedge T}\right\rfloor\log\frac{n}{\delta}}}_{\triangleq r}\Big) \leq \delta \,.
\label{eq:hoeffding_clem}
\end{align}

During the remaining $T-\tau$ steps, the auction is a weighted-second price auction\footnote{\cite{babaioff2014a} proposes a slightly different algorithm, using $\hat{\rho}_i$ instead of $\hat{\rho}_i^+$. It enjoys the same guarantee in terms of welfare, but it is unclear whether it also enjoys the same guarantee in terms of revenue.}
$$q_i(\bb) = \indicator{i = \argmax_j \hat{\rho}_j^+b_j}, \ \text{ and } \  p_i(\bb) =  q_i(\bb) \frac{\smax_{j}{\hat{\rho}_j^+ b_j}\hat{\rho}_i^+}\qquad \text{ where } \hat{\rho}_j^+ = \hat{\rho}_j + r\,.
$$

Both regrets considered, $R_T^{\mathfrak{W}}$ and $R_T^p$, can be upper-bounded by the same rate.
\begin{theorem2}[\cite{devanur2009, babaioff2014a}]
\label{thm:babaioff_devanur_upper}
The  algorithm described above guarantees that $R_T^{\mathfrak{W}} =\cO(n^{1/3}T^{2/3}\sqrt{\log(nT)})$ and $R_T^p=\cO(n^{1/3}T^{2/3}\sqrt{\log(nT)})$.
\end{theorem2}
\begin{proof}
Before beginning, as $\tau$ is a variable to optimize over, we need to handle the case $\tau > T$ properly, in order to avoid vacuous upper-bounds. Indeed, the length of the exploration stage is not $\tau$, but rather $\tau \wedge T$, while there are $(T-\tau)_+$ remaining steps. Further, we will make extensive use of the concentration \eqref{eq:hoeffding_clem}.

We denote $i^* = \argmax_i \alpha_{i}\rho_i$ and $i^+ = \argmax_i \alpha_{i}\hat{\rho}_i^+$. Then with probability at least $1-\delta$, we have,
\begin{align}
(\rho_{i^+} + 2r)\alpha_{i^+} \geq \hat{\rho}_{i^+}^+\alpha_{i^+} \geq \hat{\rho}_{i^*}^+\alpha_{i^*} \geq \rho_{i^*}\alpha_{i^*}
\end{align}
From which we can deduce 
\begin{align}
\rho_{i^*}\alpha_{i^*} - \rho_{i^+}\alpha_{i^+} \leq 2 \alpha_{\rm max} r 
\end{align}
Then the regret $R_T^{\mathfrak{W}}$ can be upper-bounded, denoting $\tau\wedge T=\min\{\tau,T\}$, as follow:
\begin{align}
R_T^{\mathfrak{W}} \leq (\tau\wedge T) \alpha_{\rm max} + (1-\delta) (T-\tau)_+ 2 \alpha_{\rm max} \sqrt{2\left\lfloor\frac{n}{\tau\wedge T}\right\rfloor \log\frac{n}{\delta}} + \delta T \alpha_{\rm max}
\end{align}
Choosing $\delta = \frac{1}{T}$ and $\tau = n^{1/3}T^{2/3}\sqrt{\log nT}$ finishes the proof for $R_W(T)$.\\
We know prove the upper-bound on $R_P(T)$. 
\begin{align*}
\smax_i \rho_i \alpha_{i} - \frac{\smax_i \hat{\rho}_{i}^+\alpha_{i}}{\hat{\rho}_{i^+}^+}\rho_{i^+} 
& = \frac{\smax_i \hat{\rho}_{i}^+\alpha_{i}}{\hat{\rho}_{i^+}^+} \left(\frac{\smax_i \rho_i \alpha_{i}}{\smax_i \hat{\rho}_{i}^+\alpha_{i,}}\hat{\rho}_{i^+}^+-\rho_{i^+}\right) \\
& \leq \alpha_{i^+} \left(\frac{\smax_i \rho_i \alpha_{i}}{\smax_i \hat{\rho}_{i}^+\alpha_{i}}\hat{\rho}_{i^+}^+-\rho_{i^+}\right) & \text{(by definition of $i^+$)}\\
& \leq \alpha_{i^+} \left(\hat{\rho}_{i^+}^+-\rho_{i^+}\right) & \text{(with proba. $1-\delta$)}\\
& \leq \alpha_{\rm max} r
\end{align*}
Then the regret $R_T^p$ can be upper-bounded as follow:
\begin{align}
R_T^p \leq (\tau\wedge T) \alpha_{\rm max} + (1-\delta) (T-\tau)_+ \alpha_{\rm max} \sqrt{2\left\lfloor\frac{n}{\tau\wedge T}\right\rfloor \log\frac{n}{\delta}} + \delta T \alpha_{\rm max}
\end{align}
Choosing again $\delta = \frac{1}{T}$ and $\tau = n^{1/3}T^{2/3}\sqrt{\log nT}$ finishes the proof for $R_P(T)$.
\end{proof}

\paragraph{Lower-bounds.} These rates of $T^{2/3}$ for both regrets are \emph{tight}, as shown by the following result,
\begin{theorem2}[\cite{devanur2009, babaioff2014a}]
\label{thm:babaioff_devanur_lower}
For any deterministic, scale-free sequential assignment $q$ and payment $p$ such that $(q,p)$ is DSIC, there exists a set of bids and distributions over $\bc$ such that  $R_T^{\mathfrak{W}},R_T^p= \Omega(n^{1/3}T^{2/3})$.
\end{theorem2}

\paragraph{Comparison to MAB.} As the social welfare coincide with the reward from a MAB point of view, we can compare this performance to optimal performance on MAB problems. It turns out the DSIC constraint is actually strong, as it implies a degradation of the regret by a factor $T^{1/6}$ -- "the cost of (ex-post) truthfulness" -- from $\cO(T^{1/2})$ for optimal MAB algorithms to $\Omega(T^{2/3})$ when ensuring incentive compatibility. To understand intuitively where this degradation comes from, it is possible to focus on explore-then-commit (ETC) types of algorithms for the case $n=2$, as an exploration-separated assignment rule is a special case of ETC algorithm. In a pure bandit setting, two types of ETC algorithms can have a regret of order $\cO(T^{1/2})$. Either an adaptive ETC that eliminates arms as soon as they are detected to be sub-optimal or a fixed-design ETC, \emph{at the condition of knowing in advance the gap} $\Delta$ of performance between both arms. Unfortunately, none of them is \emph{exploration-separated}. For adaptive ETC, because during the exploration step, the decision (taken at each time step) to eliminate an arm or to keep it, depends on the estimated reward of the arm and thus the bids. For a fixed-design ETC, the problem comes from the need to know in advance the gap: whatever the value of $\Delta$, choosing an exploration period of length $\Delta^{-2}\wedge T$ ensures a regret upper bounded by $\cO(T^{1/2})$. However, because the length of the exploration period depends on $\Delta$ (the gap), which in our case is a function of the bids, such choice of the length of the exploration period makes the assigment \emph{not} exploration-separated. Making an ETC algorithm exploration-separated requires for it to be \emph{fixed-design} with the length of the exploration period set \emph{independently} from the gap, which is known to cause a degradation of the regret, this so-called "cost of truthfulness".

\paragraph{Randomized Assignment.} 
From the previous result, it would seem that the "cost of truthfulness" may come from the strong requirement of ex-post incentive compatibility. However,  this lower bound can be circumvented by considering \emph{non-deterministic} sequential assignments, which allows to ensure ex-post incentive compatibility without restricting the algorithm to be exploration-separated \citep{babaioff2010}. Consequently, a \emph{randomized} dynamic mechanism, ex-post DSIC\footnote{Here, \emph{ex-post} is related to the realization of $c_{i,t}$, but in expectation over the randomness of the algorithm.}, with a $\cO({T}^{1/2})$ regret guarantee in welfare can be constructed. It relies on two ingredients:
\begin{enumerate}
\item a MAB algorithm that leads to an ex-post monotone assignment -- e.g. an adaptive ETC with successive elimination \citep{perchet2013multi},
\item a sampling procedure that modifies the bids that are inputed to the mechanism
\end{enumerate}
Using these properties,  it is possible to obtain a regret in terms of welfare that matches the one of the underlying MAB algorithm and as we mentioned previously, an adaptive ETC reaches the  rate of $\cO(T^{1/2})$, which is optimal \citep{babaioff2010}.

\paragraph{Non-stationary values} As explained at the beginning of the section, we considered a simple setting where the values $\mathbf{\alpha}$, and thus the bids $\bb$, are constant over time. It is possible to consider a more complex setting where the values can change over time. In such case, the bidders submit a different bid at each time-step and the payments need to be per-step instead of a final bill. To keep the parallel with bandit problems, this is not a MAB anymore, but rather a contextual linear bandit with finitely many arms (the contextual part coming from the arm set changing as the values are changing).
As it turns out, the algorithm from \cite{devanur2009} is still valid (remember the payments were designed per-step) and is DSIC even in a strong sense along the sequence. Further, one can notice the proof of Theorem~\ref{thm:babaioff_devanur_upper} for the upper-bounds on the regret never relies on the values being constant over time and perfectly holds in this non-stationary setting. As this setting is strictly more general, the lower bounds from Theorem~\ref{thm:babaioff_devanur_lower} also hold, meaning that deterministic mechanisms still have regrets scaling as $T^{2/3}$.
However, it stays an open question whether moving to randomized mechanisms still allows to recover the rate of $T^{1/2}$ which is also the optimal rate for linear bandits. 

\paragraph{Extensions.} Several other directions can be explored to better model the underlying applications or to slightly relax the mechanism design contraints. A way to relax the constraint of ex-post DSIC, is to consider an asymptotic version, where the benefit of not being truthful vanishes over time \citep{nazerzadeh2008, kandasamy2020mechanism}. It turns out this neither allow to avoid the explore-then-commit structure of algorithms for deterministic assignments nor to avoid the degradation of the regret to $T^{2/3}$, even for more complicated mechanisms than auctions \citep{kandasamy2020mechanism}. %{\color{red}One extension exhibits a strong difference though, the case when buyers can have different potential use of the same good (e.g. they each have different possible ads). \cite{babaioff2013} proved an impossibility result showing that the only ex-post DSIC mechanisms are very na\"ive, i.e. constant w.r.t. the bids. In this setting, the only way to build non-degenerated mechanisms is to use a weaker version of incentive compatibility, such as interim DSIC, which consider truthfulness in expectation over the realizations of the clicks.}

\section{Reversing the asymmetry: Strategic buyer vs.\ myopic seller}
In the previous sections, the asymmetry between the seller and the buyers was always in favor of the former. On the other hand, there are many cases where the converse happens: the seller has her hand tied, while the buyers can and try to exploit this. For instance, the seller must sometimes disclose (and commit to) the learning algorithm she is using to devise her mechanism - for instance, a second price auction with  reserve price, or more generally the Myerson revenue-maximizing auction, based on the distribution of bids received. Let us denote by $\mathcal{M}(F_{B_1},\ldots,F_{B_n})$ the mechanism induced by the past distribution of bids $F_{B_i}$; for instance, it can be a second price auction with personalized reserve price computed from $F_{B_i}$ (and not from the $F_i$ that are not known beforehand).

\subsection{A Stackelberg view}
%The objective of \textit{a long-term strategic bidder} is to choose his bidding strategy $\beta_i \in \mathds{B}_i$ (a mapping from values to bids) to maximize his expected long-term utility. Because of this long-term aspect, we can assume that an arbitrarily large number of bids (it was denoted by $T$ previously) are sent, thus we will consider the following steady state analysis of the general dynamic game.  

 %We recall that the strategy $\beta$ is defined so that when his value is $x_i \sim F_i$, the associated bid is $\beta_i(x_i)$; the (pushforward) distributions of such bids are denoted by $F_{B_i}= \beta_i\#F_i$. 
 %This steady-state  objective is particularly relevant in modern applications as most of the data-driven selling mechanisms are using large batches of bids as examples to update their mechanism. We consider that bidders are maximizing their expected utility since we assume the number of auctions to be sufficiently large in modern applications. 

\medskip

%We will also assume that the seller is myopic (but with respect to a batch of $m$ auctions): her learning algorithm is defined such that she maximizes her myopic revenue in a certain class of mechanism. 

%{\color{red} On a pas déjà dit ça ?? In terms of game theory, these interactions are a Stackelberg game between the seller - whose strategy is to pick a mechanism design that  maps bid distributions to reserve prices - and the bidders - who chose bidding strategies. The bidders are the leader of this game because we assume that bidders choose their strategy knowing the mechanism used by the seller.} Our overarching objective is to derive the best-response, for a given bidder $i$,  to  the strategy of the seller (i.e., a given learning algorithm) and the strategies of the other bidders (i.e., their bid distributions).

%\begin{definition2}[Bidder/Seller Stackelberg game] Stackelberg game in which strategic bidders assume the existence of a seller's learning algorithm $\mathcal{M}$. Each strategic bidder $i$ chooses a strategy $\beta_i$ that induces a (pushforward) bid distribution $F_{B_i} = \beta_i\#F_i$ used as input by the seller's algorithm. The goal of the strategic bidder is to optimize$$\argmax_{\beta_i} {U_i(\mathcal{M}(F_{B_i}),\beta_{i})}\;.$$ \end{definition2}

This problem has been tackled under the assumption of perfect knowledge of the optimization algorithm used by the seller \citep{KanNaz14,tang2016manipulate,nedelec2018thresholding}.  It exploits a conceptual opening in most automatic mechanism design works, i.e., the breakdown  of incentive compatibility for the buyer when the seller optimizes over incentive compatible auctions. In some sense, the computation of  bid distributions $F_{B_i}$ instead of value distributions $F_i$ can be seen as an ``attack'' of the optimization algorithm of the seller (possibly based on deep learning for complex auction systems \citep{dutting2017optimal}). However, we point out now that those attacks  differ from the celebrated adversarial attacks in computer vision. Indeed, the latter generally rely on the lack of local robustness of a classifier. Two other major differences are also quite important: these ``attacks'' do not necessarily yield lower revenues for the seller \citep{nedelec2018thresholding}; and they are also part of a dynamic game between buyers and seller and as such have a dynamic component that is absent from classical and static machine learning frameworks, such as image classification. 

For concreteness, consider the case of second price auctions. In the classical setting of auction theory, the buyer is asked to reveal their bid distribution first; facing ``truthful auctions'', they reveal their value distribution. The seller then optimizes their mechanism based on this information, finding an optimal reserve price for this buyer. This is a Stackelberg game, as the two players do not play at the same time. In this instance, the seller is the leader and the buyer is the follower. Most of the literature on optimal auctions is focused on this version of the Stackelberg game.

Howerver, if the bidder knows that the seller is going to find an \textsl{optimal} mechanism, and hence that she will optimize the auction based on the information given by his bid distribution, he can anticipate this optimization to increase his utility. The order of the Stackelberg game is then reversed. The bidder becomes the leader and the seller the follower: he reveals his bid distribution knowing the optimization problem that she will solve. In second price auctions with reserve prices, the bidder has an incentive to disclose a bid distribution that may be different from his value distribution as he then might be facing a more favorable reserve price. 

More formally, the timing of the game we consider is the following:
\begin{enumerate}
\item the seller chooses a mapping $\mathcal{M}: \mathfrak{F} \to \mathfrak{A}$,  from the set of bid distributions to the set of auction mechanisms,
\item based on this choice of mapping, each buyer picks a bidding strategy $\beta_i$,
\item bidder $i$'s  utility is computed in expectation when $x_i \sim F_i$, he bids $\beta_i(x_i)$ and the outcome of the auction (allocation and payment) is defined by $\mathcal{M}(F_{B_1},\ldots,F_{B_n})$.  With a slight abuse of notations, we will denote by $\mathcal{U}_i(\beta_i)$ his expected utility, assuming the other bidders strategies are fixed. 
\item the seller gets her expected revenue under this mechanism.
\end{enumerate}

This objective is particularly relevant in modern applications as most of the data-driven selling mechanisms are using large batches of bids as examples to update their mechanism.
\subsection{The posted price setting}
Let first consider the posted price setting where $n=1$ bidder plays against one seller. We assume, for simplicity of this introductory example, that bidder's value distributions $F_{i}$ is $\mathfrak{U}[0,1]$, i.e., uniform on the interval [0,1]. Let us initially consider that the bidder is bidding truthfully, i.e, $\beta_i = Id$. In this case, $F_{B_i} = F_{i}$ and the seller will set as reserve price the monopoly price by maximizing the monopoly revenue $r(1-F_{i}(r))$. This monopoly price is equal to 0.5 in the case of $\mathfrak{U}[0,1]$.  Note that this maximization problem is computationally simple as the monopoly revenue is a concave function if the value distribution is regular. The bidder can obviously do better. If he bids all the time zero  (or $\varepsilon$ arbitrarily close to zero), $F_{B_i}$ will be equal to a point mass at zero. Through computing the optimal reserve price corresponding to $F_{B_i}$, the seller chooses zero, obviously maximizing bidder's utility. The problem we consider derives from a simple extension of this example to the case of $n$ bidders. In a lazy second price auction, the optimal reserve price for each bidder is still the monopoly price. Yet, as soon as there is some competition, bidders cannot bid zero as they get zero utility in this case.  They have to tradeoff between beating the competition and decreasing their reserve price. 
\subsection{Improving the truthful strategy for any distributions of the competition}
\label{subsec:strat_bidder_any_distrib}
\citep{nedelec2018thresholding} derives a simple strategy which guarantees to the bidder an increase in utility compared to the truthful strategy for any distributions of the competition. This increase depends on the distribution of the competition. Yet, by playing this strategy, the bidder is sure to do better than by bidding truthfully. This is an important practical result as in many ad platforms, bidders have to bid without knowing the distribution of the competition. This strategy, that they call \textit{thresholding at the monopoly price}, has also the key property of making simple the optimization problem of the seller, i.e., if $F_{i}$ is regular, the bid distribution $F_{B_i}$ induced by
this strategy on  $F_{i}$ is also regular. 

\begin{table}[]
\centering
\small
\begin{tabular}{c|c|c|c|c|c|c|c|c}  
& \multicolumn{4}{c|}{Optimal reserve price} & \multicolumn{4}{c}{Utility} \\ \cline{2-9}
&K=1 &K=2&K=3&K=4  &K=1 &K=2&K=3&K=4\\ \hline
\multirow{2}{*}{Truthful bidding} & \multirow{2}{*}{0.5}& \multirow{2}{*}{0.5}& \multirow{2}{*}{0.5}&  \multirow{2}{*}{0.5}& \multirow{2}{*}{1/8} & \multirow{2}{*}{1/12} & \multirow{2}{*}{11/192}& \multirow{2}{*}{13/320} \\ 
 & &  &  & &&&& \\ \hline
\multirow{2}{*}{Zero bidding} &\multirow{2}{*}{0.0}&\multirow{2}{*}{0.0}&\multirow{2}{*}{0.0}& \multirow{2}{*}{0.0}  & 1/2 & 0.0 &  0.0 & 0.0\\ 
 &&&&  & (+400\%)  & (-100\%)  & (-100\%) & (-100\%) \\ \hline
 \multirow{2}{*}{Divide values by 2} &\multirow{2}{*}{0.25}&\multirow{2}{*}{0.25}&\multirow{2}{*}{0.25}& \multirow{2}{*}{0.25}  & 1/4 & $\approx 0.094$ &  $\approx 0.036$& $\approx 0.015$\\ 
 &&&&  & (+100\%)  & (+13\%)  & (-37\%) & (-63\%) \\ \hline
Thresholded at & \multirow{3}{*}{0.25}& \multirow{3}{*}{0.25}& \multirow{3}{*}{0.25}& \multirow{3}{*}{0.25}  & \multirow{2}{*}{1/4} &  \multirow{2}{*}{$\approx 0.132$}  & \multirow{2}{*}{$ \approx 0.076$}  & \multirow{2}{*}{$\approx 0.048$}\\
the monopoly price & &&&&&&&\\
(Theorem \ref{thm:settingReserveValToZeroImprovesPerformance}) &  &&&& (+100\%)  & (+57\%)  & (+33\%) & (+20\%) \\ \hlineB{4.0}
Optimal  regularity-& \multirow{2}{*}{0.0}& \multirow{2}{*}{0.162}& \multirow{2}{*}{0.204}& \multirow{2}{*}{0.22}  & \multirow{2}{*}{1/2} & \multirow{2}{*}{$\approx 0.147$}  & \multirow{2}{*}{$\approx 0.079$}  & \multirow{2}{*}{$\approx 0.049$}\\
preserving strategies&&&&  &  & &  & \\ 
(Theorem \ref{thm:quantificationOptimalThresholding})&&&&  & (+400\%)  & (+76\%)  & (+38\%) & (+21\%) \\ 
\hline
\end{tabular}
\caption{Comparison of the utility of the strategic bidder between the truthful strategy, the strategy corresponding to bidding zero for any values, the linear strategy dividing values by two, the strategy introduced in Theorem \ref{thm:settingReserveValToZeroImprovesPerformance} and the optimal regularity-preserving strategies for each number of competitors (derived from Theorem \ref{thm:quantificationOptimalThresholding}). The first four strategies are fixed and do not require knowledge of the competition to be computed. The last one is competition-specific and exact knowledge of the distribution followed by the highest bid of the competition is needed to compute it. For this example, bidders' value distributions are $\mathfrak{U}[0,1]$ and  opponents are assumed to bid truthfully.}
\label{table_intro}
\end{table}

\begin{definition} Consider a bidder with a regular value distribution $F_{i}$. A bidding strategy $\beta_i$ is regularity-preserving if the bid distribution $F_{B_i}$ induced by $\beta_i$ on $F_{i}$ is a regular distribution.
\label{def:RP}
\end{definition}

When the reserve price is computed from $F_{B_i}$ -  the bid distribution induced by using $\beta$ on $F_{i}$ - a distinction between the reserve price $r_\beta$ and the reserve value $x_\beta$ must be made.

\begin{definition}
Given  a non-decreasing strategy $\beta$, the reserve value $x_\beta$ is  the smallest value above which the seller accepts bids. In particular, if the bidder bids truthfully, his reserve value is equal to his reserve price; on the other hand, if $\beta$ is continuous and  increasing, and $r_\beta$ is the reserve price associated with the strategy $\beta$, then $x_\beta = \beta^{-1}(r_\beta)$. 
\end{definition}
Consider for instance,  $F = \mathfrak{U}[0,1]$, and the bidding strategy $\beta(x) = x/2$, then $r_\beta = 0.25$ and $x_{\beta} = 0.5$. By dividing bids by two, the strategic bidder decreases their reserve price but does not change the reserve value: it is the same as if they were  bidding truthfully.

\begin{theorem2}\label{thm:settingReserveValToZeroImprovesPerformance}
Suppose the value distribution $F$ has a density $f$, with $f>0$ on the support of $F$ and that the left-end point of its support is 0, and that the other bidders' strategies are fixed. Let $\beta_r$ be an increasing strategy with associated reserve value $r>0$ in a lazy second price auction such that the bid distribution associated with $\beta_r$ has a virtual value.  Then there exists another bidding strategy $\tilde{\beta}_r$ such that: 
\begin{enumerate}
\item A reserve value associated with $\tilde{\beta}_r$ is 0 and $\tilde{\beta}_r$ is increasing.
\item $\mathcal{U}_i(\tilde{\beta}_r)\geq \mathcal{U}_i(\beta_r)$, i.e., the utility of bidder $i$ is higher;
\item $P_i(\tilde{\beta}_r)\geq P_i(\beta_r)$, i.e., the payment of bidder $i$ to the seller is also higher,
\end{enumerate}
The following continuous function fulfills these conditions:
$$
\tilde{\beta}_r(x)=\left(\frac{\beta_r(r)(1-F_{i}(r))}{1-F_{i}(x)}\right)\mathds{1}\{x<r\}+\beta_r(x)\mathds{1}\{x\geq r\}
$$
\end{theorem2}
A reserve value equal to zero means that the seller accepts all bids of the strategic bidder. It also means that the reserve price is equal to the minimum bid of the strategic bidder. This result can be applied to improve any preexisting shading strategy. A very important case is to apply this theorem to the truthful strategy, showing that there exists a strategy improving the truthful strategy regardless of the competition distribution. 
We now explain why we can improve any strategy in this setting without knowing the distribution of the competition.  Myerson's Lemma is a key element in this understanding.

In this setting, it is optimal for the seller to choose as reserve price for bidder $i$ the monopoly price corresponding to her bid distribution, and Myerson lemma implies that the expected payment of bidder $i$ in the optimized lazy second price auction is equal to 
\begin{equation*}
P_i(\beta_i) = \mathds{E}_{b\sim F_{B_i}}\bigg(\psi_{F_{B_i}}(b)G_i(b)\mathds{1}\{b \geq \psi_{B_i}^{-1}(0)\}\bigg)\;.
\end{equation*}
In order to simplify the computation of the expectation and remove the dependence on $B_i$, we rewrite this expected payment in the space of values using the fact that the strategic bidder is using an increasing strategy $\beta_i$. We will only consider increasing strategies in the remaining of the survey and so  we define:
\begin{equation*}
h_{\beta_i}(x) = \psi_{F_{B_i}}(\beta_i(x))  
\end{equation*}
With this new notation, the expected payment of the strategic bidder $i$ rewrites as
\begin{equation*}
P_i(\beta_i) = \mathds{E}_{x_i \sim F_{i}}\bigg(h_{\beta_i}(x_i)G_i(\beta_i(x_i))\mathds{1}\{x_i \geq x_\beta\}\bigg)\;.
\end{equation*}
and  her expected utility can be derived as a function of $\beta_i$, since
\begin{equation}\label{equ:utility}
\mathcal{U}_i(\beta_i) = \mathds{E}_{x_i \sim F_{i}}\bigg((x_i-h_{\beta_i}(x_i))G_i(\beta_i(x_i))\mathds{1}\{x_i \geq x_\beta\}\bigg)\;.
\end{equation}
where $x_\beta$ is \emph{the reserve value}. If $h_{\beta_i}$ crosses 0 exactly once and is positive beyond that crossing point,  $x_\beta =  h_{\beta_i}^{-1}(0)$.  If we call $r_i=\psi_{F_{B_i}}^{-1}(0)$ the \emph{reserve price} of bidder $i$ and $\beta_i$  increasing , the \emph{reserve value} is equal to $\beta_i^{-1}(r_i)$. 

If we consider only increasing differentiable strategies, and we denote by  $\mathcal{I}$ the class of  such functions, the problem of the strategic bidder is therefore to solve  $ \sup_{\beta \in \mathcal{I}} U(\beta) $ with $U$ defined in Equation \eqref{equ:utility}. This equation is crucial, as it indicates that optimizing over bidding strategies can be reduced to finding a distribution with a well-specified  $h_\beta(\cdot)$. Our results extend to the case where the strategies are increasing and differentiable except at finitely many points, as we only need $bF_B(b)$ to be absolutely continuous for the previous result to go through.

A crucial difference between the long-term vision and the classical, myopic (or one-shot)  auction theory is that in this setup bidders maximize expected utility globally over the full support of the value distribution. In the classical myopic setting, bidders determine their bids  to maximize their expected utility at each value. In our setup, the strategic bidder also accounts for the computation of the reserve price, a function of her global bid distribution. He might therefore be willing to sometimes over-bid (incurring a negative utility at some specific auctions/values) or underbid (lose some auctions that he would have won otherwise) if this reduces her reserve price. Indeed, having a lower reserve price increases the utility of other auctions. Lose small to win big. In other words, the strategy trades-off ex-post individual rationality (IR) for higher utility (of course ex-ante IR still holds). This reasoning makes sense only with multiple interactions between bidders and seller.

\subsubsection{Thresholding the virtual value}
\label{subsec:thresholding_virtual_value}
A truthful bidding strategy can easily be improved by a strategic bidder, as illustrated by the following elementary example. Consider that the value distribution of the bidder is  $\mathfrak{U}([0,1])$, uniformly  between 0 and 1. With a truthful bidding, the associated virtual value is negative below $1/2$ and positive above, so that the optimal reserve price is $1/2$, so that no auction is won if the value is smaller than $1/2$.  On the other hand, if the strategic bidder was able to send bids so that the virtual value (of bids) below $1/2$ is exactly 0, then  the seller would not have any incentives to choose a reserve price, because of Myerson lemma. In particular, the latter also implies that since the virtual value is zero below $1/2$, the seller receives exactly the same expected payment as with a truthful bidder. 

This technique is called \textsl{thresholding the virtual value}. We now show formally how to find a bidding strategy such that the virtual value of the induced bid distribution is equal to zero below a certain threshold.
\begin{figure}[t]
\centering
\begin{tabular}{cc}
\includegraphics[width=.40\linewidth]{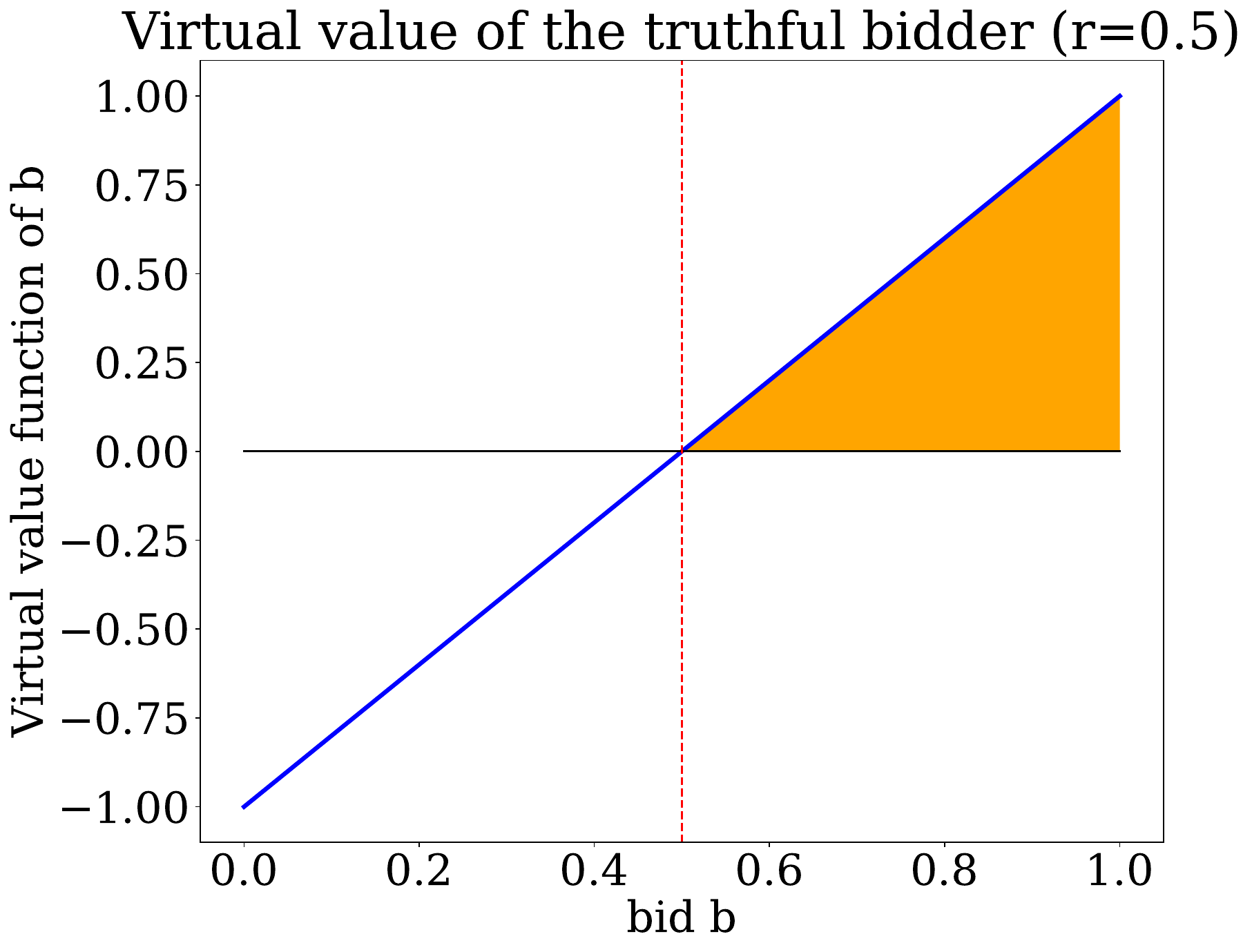} &
\includegraphics[width=.40\linewidth]{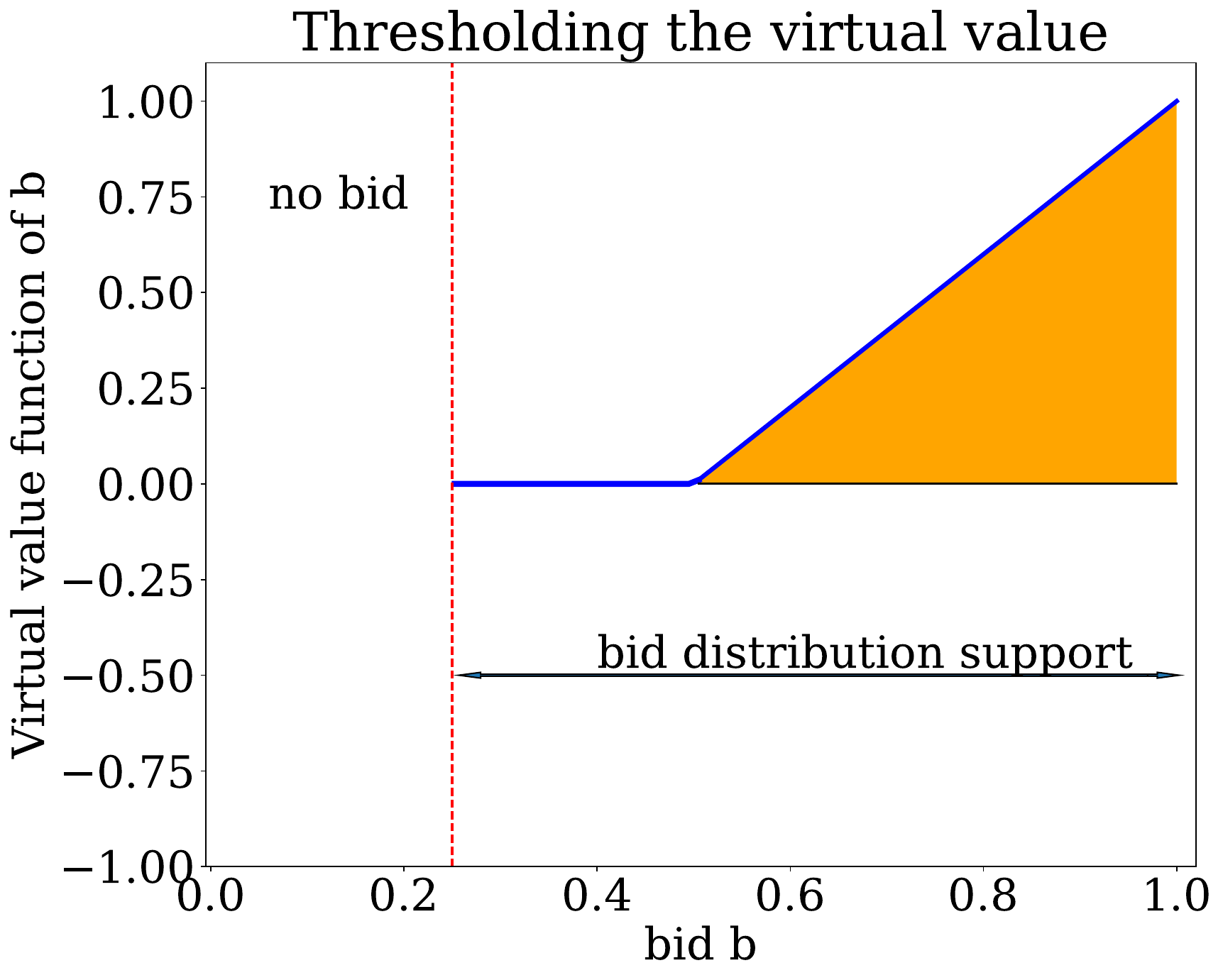}
\end{tabular}
\caption{\textbf{Virtual value of truthful bidder vs. strategic bidder.} The value distribution of the bidder is $\mathfrak{U}[0,1]$, the standard textbook example used for the sake of illustration. Her virtual value is therefore equal to $\psi(x) = 2x -1$, and is represented by the blue line. The dashed red vertical line corresponds to the current reserve price. The green area corresponds to the bidder's payment if we picked $G=1$, i.e., no competition, for the sake of clarity of the plot. The left-hand side corresponds to truthful bidding, the right-hand side to strategic behavior. In both cases, the blue line corresponds to $\psi_{B}$.}
\label{fig:fig1vValueTruthvsStrat} % I can do without the label too
\end{figure}
Before carrying on with reasoning on the virtual value, such as in our motivating example, we need to ensure we can find the corresponding strategy $\beta_i$ that will expose a bid distribution $F_{B_i}$ with the corresponding virtual value to the seller. The two following technical lemmas show how to deduce $\beta_i$ from a given $h_{\beta_i}$.
%To derive the corresponding bidding strategy, we present two important technical lemmas. 
\begin{lemma2}\label{definition_psi}
Suppose $b_i=\beta_i(x_i)$, where $\beta_i$ is increasing and differentiable and $x_i$ is a random variable with cdf $F_i$ and pdf $f_i$, with $f_i > 0$ on the support of $F_i$. Then 
\begin{equation}\label{eq:ODEPhiG}
h_{\beta_i}(x_i)= \beta_i(x)-\beta_i'(x)\frac{1-F_{i}(x)}{f_{i}(x)} = \psi_{F_{B_i}}(\beta_i(x)) \;.
\end{equation}
\end{lemma2}
\begin{proof}
By definition, $\psi_{F_{B_i}}(b) = b - \frac{1 - F_{B_i}(b)}{f_{B_i}(b)}$ with $F_{B_i}(b) = F_{i}(\beta_i^{-1}(b))$ and $f_{B_i}(b) = f_{i}(\frac{\beta_i^{-1}(b)}{\beta_i'(\beta_i^{-1}(b)})$.
Then, 
$h_{\beta_i}(x) = \psi_{B_i}(\beta_i(x)) = \beta_i(x) - \beta_i'(x)\frac{1-F_{i}(x)}{f_{i}(x)}\;.$
\end{proof}
The above results hold when $\beta$ is increasing, continuous, and differentiable except at finitely many points. The second lemma shows that for any function $g$, there exists a function $\beta$ such that $h_\beta = g$.

\begin{lemma2}\label{lemma:keyODEs}
Let $X$ be a random variable with cdf $F$ and pdf $f$, with $f > 0$ on the support of F. Let $x_0$ be in the support of $X$, $C\in\mathds{R}$ and $g:\mathds{R} \rightarrow \mathds{R}$. Define the function $\beta_g$ by
\begin{equation}\label{eq:gAsConditionalExpectation}
\beta_g(x)=\frac{C(1-F(x_0))-\int_{x_0}^{x} g(u) f(u) du}{1-F(x)}\;,
\end{equation}
% \begin{equation}\label{eq:gAsConditionalExpectation}
% \beta_g(x)=\frac{\int_x^{\mathfrak{u}} g(t) f(t) dt}{1-F(x)}=\mathds{E}(g(X)|X\geq x)
% \end{equation}
% is increasing and differentiable on the support of $X$.
then,
$$
h_{\beta_g}(x) = g(x) \text{ and } \beta_g(x_0)=C\;.
$$ 
Moreover, if for some $t \in \mathds{R}$ such that $x_0\leq t$, $g$ is non-decreasing on $[x_0,t]$, then  $\beta_g'(x)\geq (C-g(x))(1-F(x_0))f(x)/(1-F(x))$ for $x\in[x_0,t]$. Hence $\beta_g$ is increasing on $[x_0,t]$ if $g$ is non-decreasing and $g<C$.
\end{lemma2}

\begin{proof}
The result follows by simply differentiating the expression for $\beta_g$, and plugging-in the expression for $h_{\beta_g}$ obtained in Lemma \ref{definition_psi}. The result on the derivative is simple algebra.
\end{proof}
The two technical lemmas \ref{definition_psi} and \ref{lemma:keyODEs} show that for any non-decreasing function $g$, we can find a strategy $\beta_i$ such that the bid distribution induced by using $\beta_i$ on $F_{X_i}$ verifies $\psi_{B_i}(\beta_i(x)) = g(x)$ for all $x$ in the support of $F_{X_i}$.

%
% In Section \ref{subsec:thresholding_virtual_value}, we explained why sending a virtual value equal to zero when the initial one was negative increases the bidder's expected utility. To go from this virtual value to the corresponding bidding strategy $\beta$, we need to solve the simple ODE defined in Lemma \ref{definition_psi}. More formally, the following theorem shows how to improve any strategy assuming the bidder knows the current reserve price or value. This theorem works for non-regular value distributions and in the asymmetric case when the bidders have different value distributions.\nek{Conditions on $\psi$? This looks like an impossible result and will trigger referee backlash}
We explained why sending to the seller a virtual value equal to zero when the initial one was negative increases the bidder's expected utility. To derive the corresponding bidding strategy $\beta$ from the virtual value, the strategic bidder only needs to solve the simple ODE defined in Lemma \ref{definition_psi}.

\begin{figure}[t!]
\centering
\begin{tabular}{ccc}
\includegraphics[width=.40\linewidth]{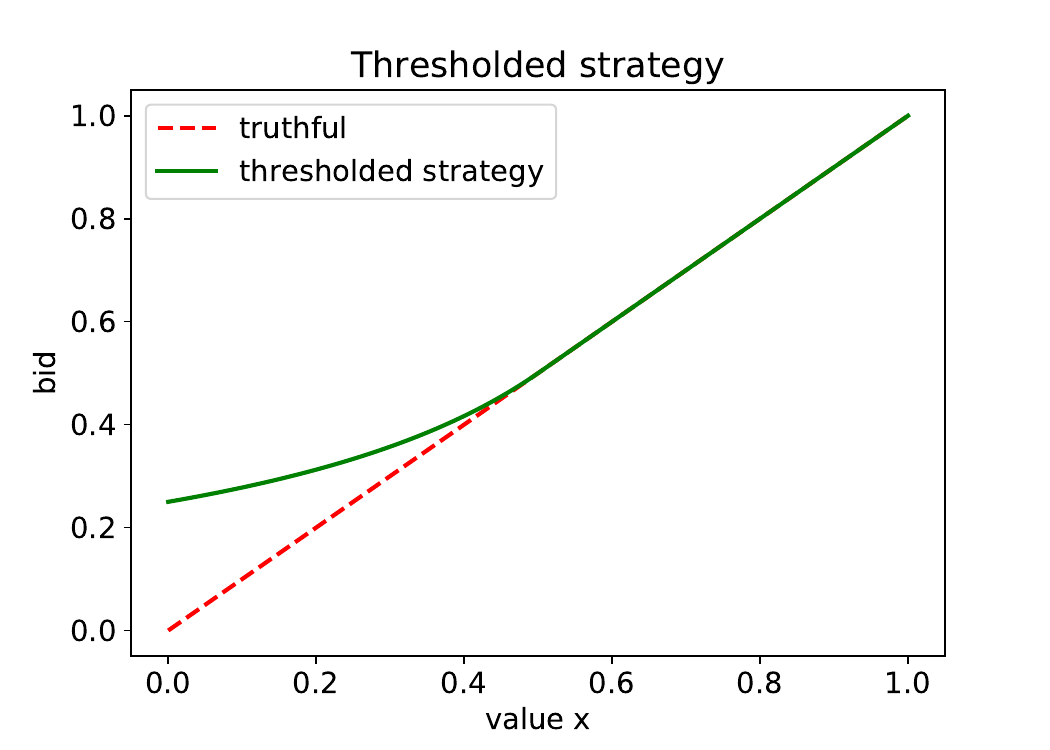} &
\includegraphics[width=.40\linewidth]{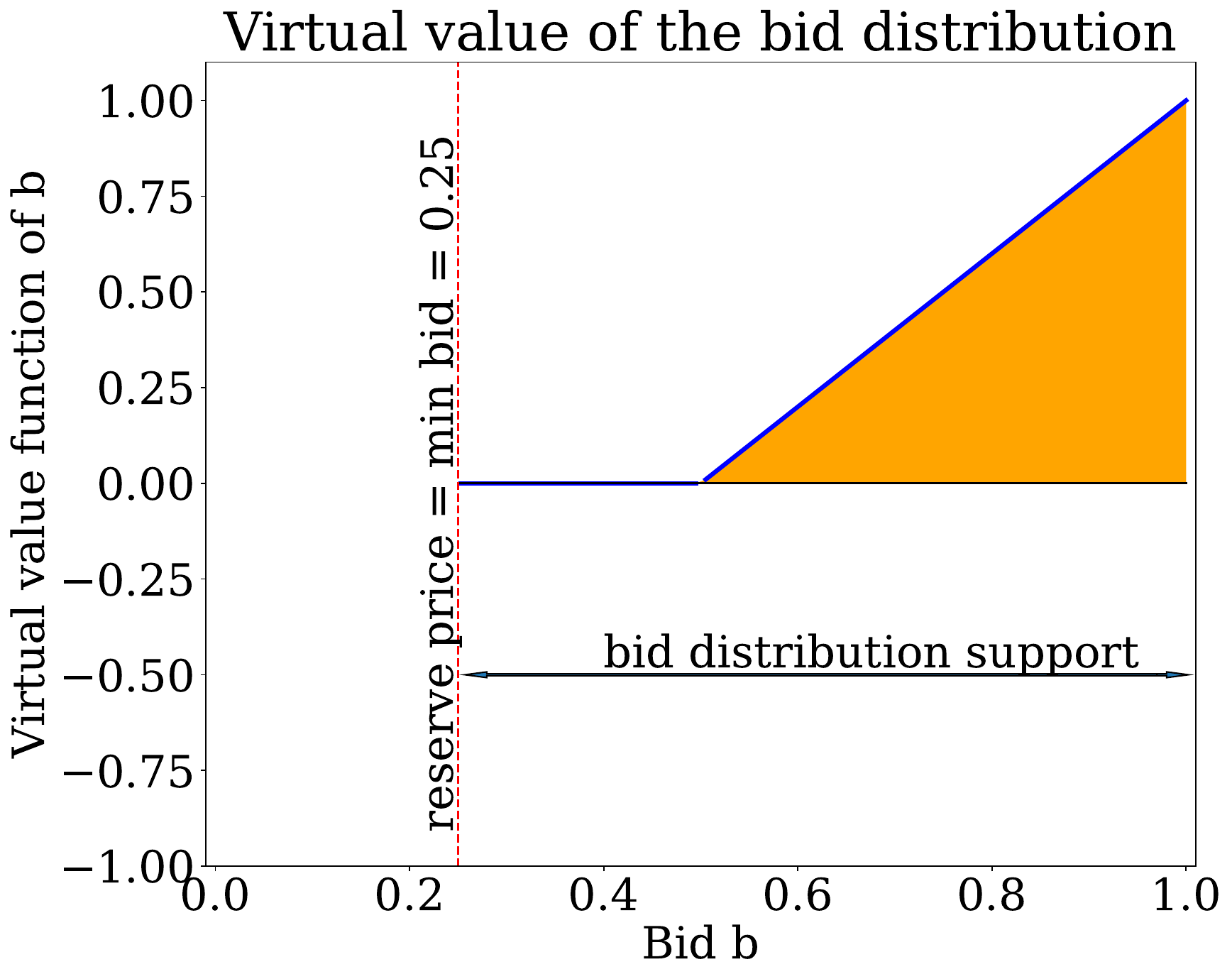}&

\end{tabular}	
\caption{\textbf{The value distribution is $\mathfrak{U}[0,1]$. Left: Thresholded strategy $\tilde{\beta}_{0.5}^{(0)}$ compared to the traditional truthful strategy. Right: virtual value of the bid distribution induced by the thresholded strategy.} The optimal reserve price of the thresholded strategy is equal to 0.25 (corresponding to a reserve value of 0) whereas the reserve price of the truthful strategy is equal to 0.5. (corresponding to a reserve value of 0.5). The green area represents the expected payment corresponding to the thresholded strategy (we assumed $G = 1$ for the sake of clarity).}
\label{fig:fig2} % I can do without the label too
\end{figure}
\begin{figure}[t]
\centering
\begin{tabular}{ccc}
\includegraphics[width=.33\linewidth]{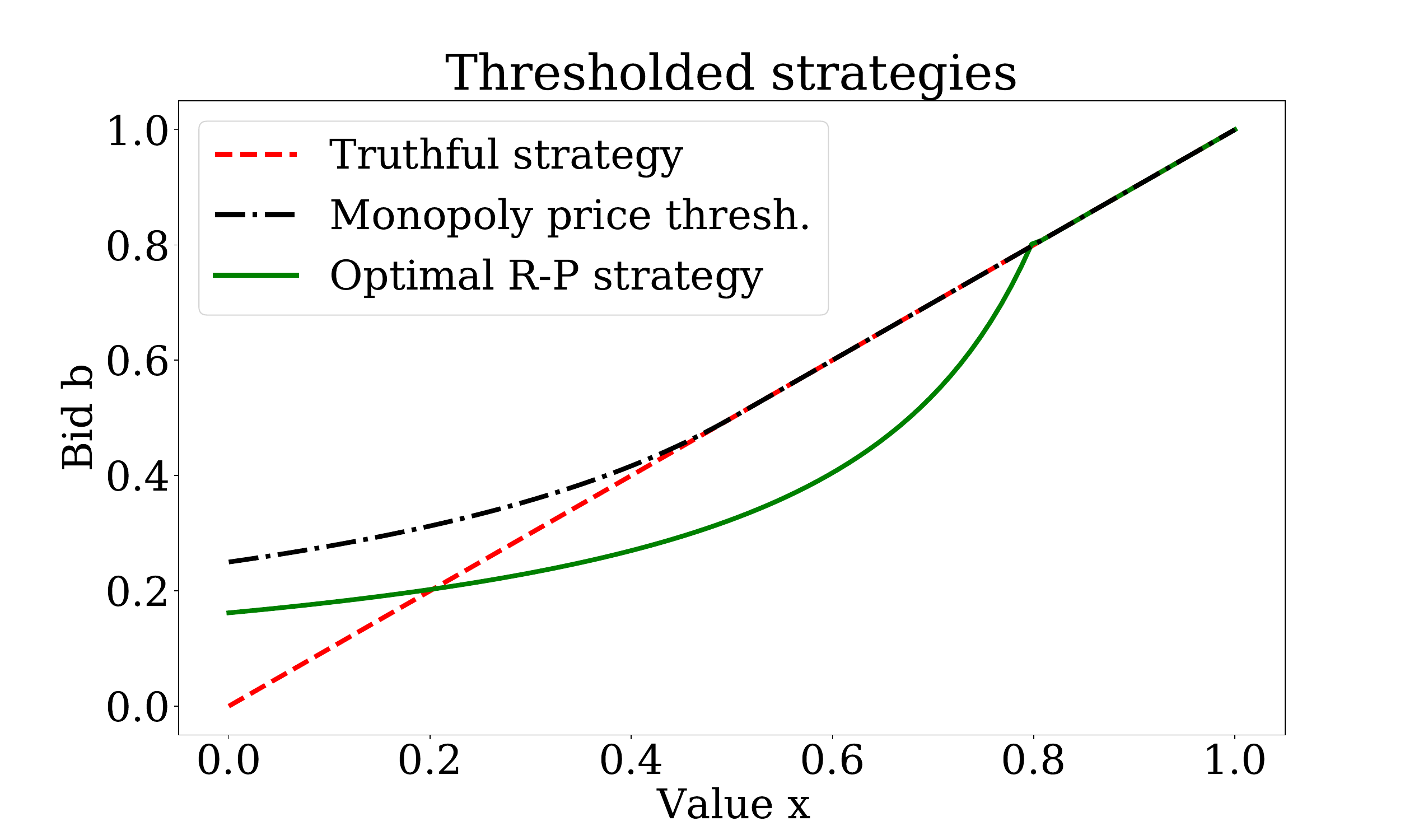} &
\includegraphics[width=.33\linewidth]{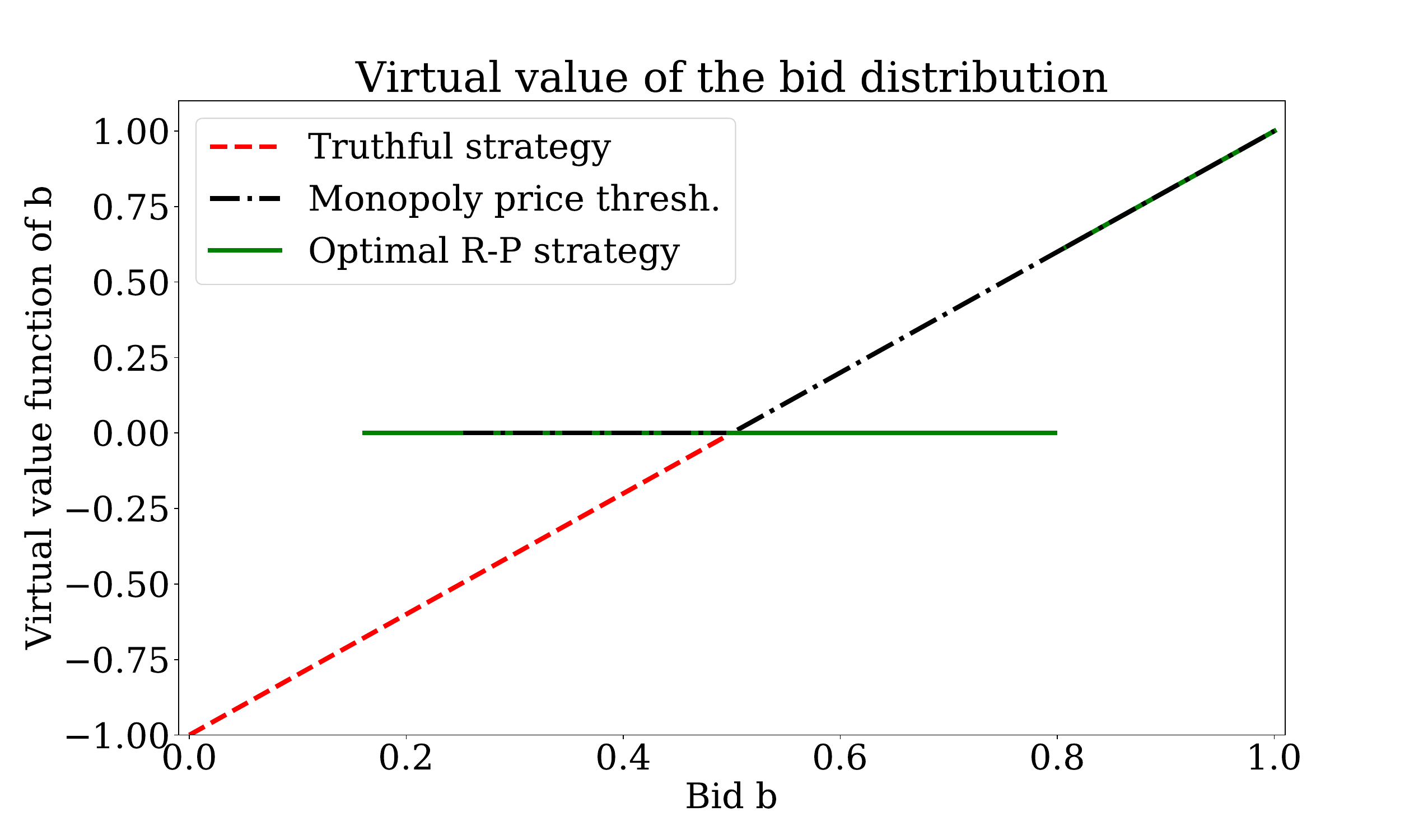}&
\includegraphics[width=.33\linewidth]{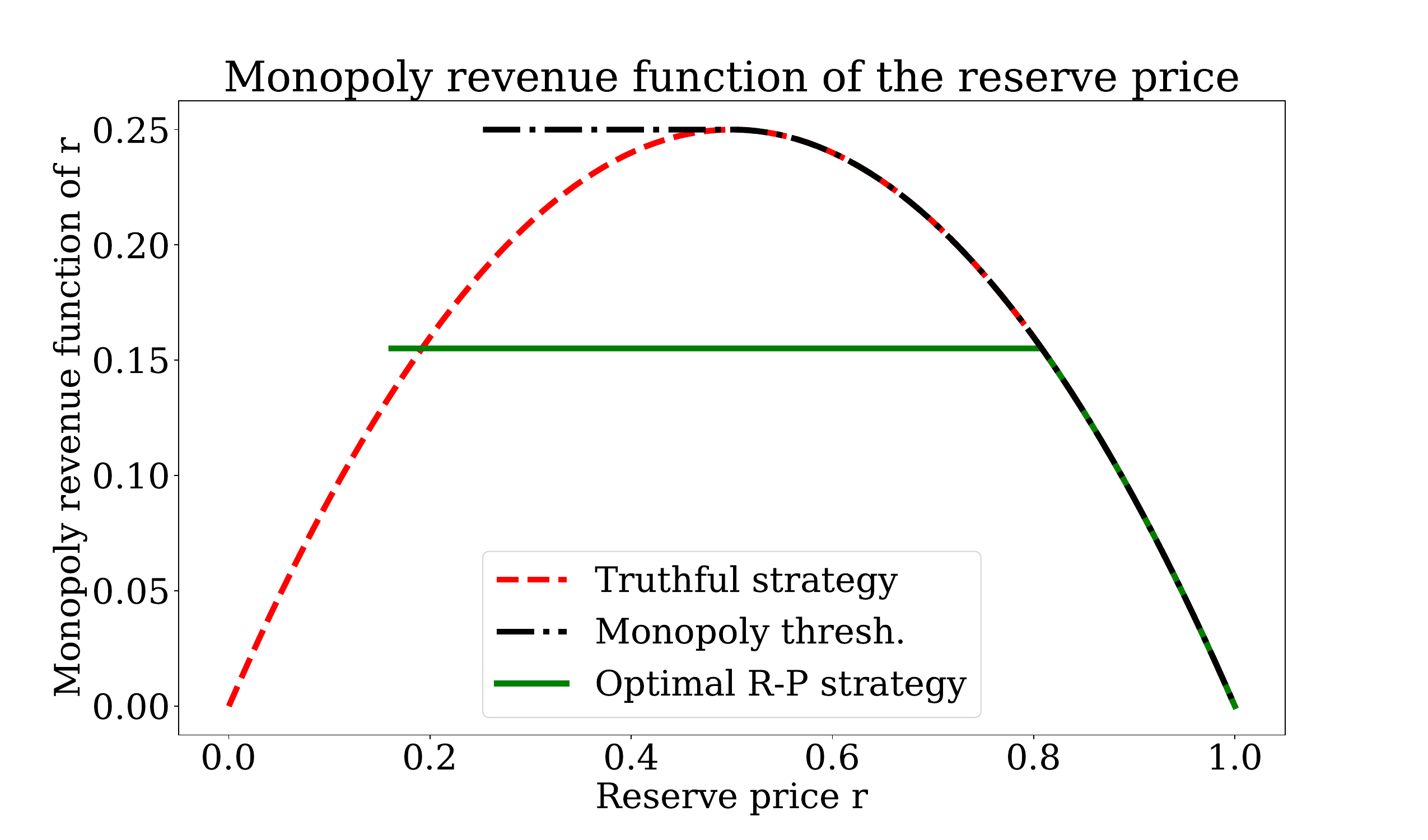} 

\end{tabular}	
\caption{\textbf{The value distribution is $\mathfrak{U}[0,1]$. Left: Thresholded strategy compared to the traditional truthful strategy. Middle: virtual value of the bid distribution induced by the thresholded strategy. Right: monopoly revenue of the induced bid distribution as function of the reserve price.}  }
\label{fig:fig3} % I can do without the label too
\end{figure}

This improvement of bidder's utility does not depend of the estimation of the competition and thus can easily be implemented in practice. 
%This technique is called thresholding the virtual value at the monopoly price.
We plot in Figure \ref{fig:fig3}, the bidding strategy $\tilde{\beta}_{0.5}^{(0)}$ when the initial value distribution is $\mathfrak{U}[0,1]$ and the virtual value of the bid distribution induced by $\tilde{\beta}_{0.5}^{(0)}$ on $\mathfrak{U}[0,1]$. We recall that the monopoly price corresponding to $\mathfrak{U}[0,1]$ is equal to $0.5$. We remark that the strategy consists in overbidding below the monopoly price of the initial value distribution. The strategic bidder is ready to increase pointwise her payment when she wins auctions with low values in order to get a large decrease of the reserve price (going from $0.5$ to $0.25$). Globally, the payment of the bidder remains unchanged compared to when the bidder was bidding truthfully with a reserve price equal to 0.5. Thresholding the virtual value at the monopoly price amounts to overbidding below the monopoly price, effectively providing over the course of the auctions an extra payment to the seller in exchange for lowering the reserve price/value faced by the strategic bidder. This strategy unlocks a very substantial utility gain for the bidder. 

Naturally, a key question is to understand the impact of this new strategy on the utility of the strategic bidder. We compare the situation with two bidders bidding truthfully against an optimal reserve price and the new situation with one bidder using the thresholded strategy and the second one bidding truthfully. We assume, as is standard in many textbooks and research papers numerical examples, that their value distribution is $\mathfrak{U}[0,1]$. 

Then, elementary computations show that in this specific illustrative example, the strategic bidder utility has a 57\% increase, from $1/12$ to $1/12 + (\log(2) - 1/2) / 4 \approx 0.132$, and the welfare has a 8\% increase, from 7/12 to $7/12 + (\log(2) - 1/2) / 4 \approx 0.632$. 
\subsection{Best response for a known distribution of the competition}
\label{sec:best_response_with_competition_knowledge}
%\section{Strategic improvements with increased side information}
We now show, for a specific given distribution of the competition, what is the optimal increasing and regularity-preserving (RP) strategy, as defined in Definition \ref{def:RP}. A direct way to compute the expected utility of a bidding strategy $\beta_i$ when the seller is using a second price auction with personalized reserve price and the other bidders are bidding truthfully has been introduced in Subsection \ref{subsec:strat_bidder_any_distrib}. Indeed, 
\begin{equation}\label{equ:utility_v2}
\mathcal{U}(\beta_i) = \mathds{E}_{x \sim F_{i}}\bigg((x-h_{\beta_i}(x))G_i(\beta(x))\mathds{1}\{x \geq x_{\beta_i}\}\bigg)\;.
\end{equation}
with 
$h_{\beta_i}(x) = \beta_i(x) - \beta_i'(x)\frac{1-F_{i}(x)}{f_{i}(x)}\;$ and $x_{\beta_i} = h_{\beta_i}^{-1}(0)$.
In this section, we assume that the bidder has now access to the distribution of the highest bid of the competition that  denoted by $G_i$, with associated pdf $g_i$
% This form can be computed in many different settings and offers a way to compute optimal bidding strategies in some parametric class of functions by finding zeros of the directional derivatives.
%In the following, unless otherwise stated, the expectation is taken according to the value distribution of the bidder. In order to be able to derive optimal strategies, we make use  of the previous expression and obtain the following lemma (we remove the subscript $i$ as it is now clear we consider bidder $i$).
and that he will optimize his utility among the strategies with thresholded virtual values introduced in Subsection \ref{subsec:strat_bidder_any_distrib}.

\begin{definition2}\label{definition_thresholded_strategies}
A  bidding strategy $\beta$ is thresholded if  there exists $r>0$ such that for all $x < r, h_{\beta}(x) = \psi_B(\beta(x)) = 0$. This family of functions can be parametrized as
\begin{equation*}
\beta_r^\gamma(x)=\frac{\gamma(r)(1-F(r))}{1-F(x)}\indicator{x< r}+\gamma(x)\indicator{x\geq r}\;,
\end{equation*}
with $r \in \mathds{R}$ and $\gamma: \mathds{R} \rightarrow \mathds{R}$ some continuous and increasing mapping.
\end{definition2}

This class of continuous bidding strategies has two degrees of freedom: the threshold $r$ such that for all $x < r, h_{\beta}(x) = 0$ and the strategy $\gamma$ used beyond the threshold. We do not restrict the functions  $\gamma$ that can be used beyond the threshold (beside being continuous and increasing). All the strategies defined in this class have the property that their reserve value is equal to zero, i.e., their reserve price is equal to their minimum bid, when the seller is welfare benevolent and the virtual value of $\gamma$ is positive beyond $r$. We can prove that the optimal regularity-preserving strategy belongs to the class of thresholded strategies.

The following result states that there exists an optimal threshold $r$ for the strategic bidder that depends on the competition and that the optimal strategy to use for $x > r$ is to be truthful. It is derived by computing the directional derivatives of the utility function defined in Equation \eqref{equ:utility_v2}.

\begin{theorem2}\label{thm:quantificationOptimalThresholding}
If $F_i$ is regular, then the optimal increasing and regularity-preserving strategy consists in thresholding at $r^*$ and bidding truthfully beyond $r^*$, defined by: 
\begin{equation*}\label{eq:keyEqThreshStratOneStrategic2}
G(r^*)=\mathds{E}_{x \sim F_i}\bigg(\frac{x}{1-F_i(x)}g\left(\frac{r^*(1-F_i(r^*))}{1-F_i(x)}\right)\indicator{x\leq r^*}\bigg)\;.
\end{equation*}
where $G$ is the distribution of the largest bid of the competition
\label{theorem2}
\end{theorem2}
%This lemma is useful when searching for the optimal shading strategies in various class of shading functions. It can also be used to improve bidding strategies by performing gradient ascent for $\beta$ expanded in a dictionary of functions, such as splines. We ran some of these experiments and got interesting results but do not report on them in this paper.\redtext{is this paragraph needed?}

%When bidder $i$ knows the distribution of the maximum bid of the competition denoted by $G_i$, she can optimize her strategy over various class of functions. 

In Theorem \ref{thm:settingReserveValToZeroImprovesPerformance}, we proved that when the strategic bidder does not know the distribution of the highest bid of the competition, he can use the thresholded strategy at his monopoly price and increases his utility compared to truthful bidding. Theorem \ref{thm:quantificationOptimalThresholding} gives the optimal threshold when the strategic bidder knows $G$.

\paragraph{Some numerical results} We consider the situation where we have 1 strategic bidder, and 1 non-strategic one, both wit in Subsection \ref{subsec:strat_bidder_any_distrib} was to bid truthfully beyond the monopoly price ($r=.5$ here) and using Theorem \ref{thm:settingReserveValToZeroImprovesPerformance} before.  This strategy yields a utility of $0.1316$, a $57\%$ increase over the standard truthful bidding revenue. The optimal strategy coming out of Theorem \ref{theorem2} consists in bidding truthfully beyond $r\simeq .8$ and using the thresholding completion before. The utility is then around 0.1468, a 76\% percent increase in bidder utility compared to bidding truthfully (truthful bidding yields a utility of $1/12\simeq .083$). This second strategy yields a higher utility for the strategic bidder but requires some knowledge of the competition. The optimal strategy in Theorem \ref{thm:quantificationOptimalThresholding} overbids on small values, underbids on intermediate values and is truthful on high values. We also recover that with no competition, the optimal strategy is to bid zero for any possible valuations. In Table \ref{table_intro}, we also notice that the difference in utility is decreasing with the number of players since, with increasing  competition, the strategic bidder cannot lower his bid for values above his monopoly price.

\subsection{Nash equilibrium}
In the previous section, only one bidder was strategic, and the other bidders did not directly react to this strategy. It could be  a reasonable assumption in practice since the number of bidders able to implement sophisticated bidding strategies appears to be limited, but we still investigate the case where all bidders are strategic. For simplicity, we shall assume that they are symmetric, with  the same value distribution $F$. We consider a  large class of admissible bidders strategies: the large set of all thresholded bidding strategies introduced in Definition \ref{definition_thresholded_strategies}. If bidders only use strategies from this set, then there exists a unique Nash equilibrium, and  their  utility is the same as in a  second price auction without reserve price.

\begin{theorem2}[\cite{tang2016manipulate,nedelec2018thresholding}]
\label{thm:nashEquiThresholded} 
Assume that bidders are symmetric, with a  valuations distribution  supported on $[0,1]$, with a continuous positive  density at $0$ and $1$, and such that the virtual value  equals 0 exactly once and is positive beyond. Then there exists a unique symmetric Nash equilibrium in the class of thresholded bidding strategies that can be computed by solving 
\begin{equation}\label{eq:keyEqThreshStratAllStrategic}
\frac{n-1}{r^*(1-F(r^*))}\mathds{E}_{x \sim F}\Big[xF^{n-2}(x)(1-F(x))\indicator{x\leq r^*}\Big] = F^{n-1}(r^*)\;
\end{equation}
 to determine the common reserve price $r^*$. Moreover, at this Nash equilibrium, the revenue of the seller and the utilities of the buyers are the same as in a second price auction without reserve prices. 	
\end{theorem2}
With appropriate shading functions, the bidders can recover the utility they would get when the seller was not optimizing her mechanism to maximize her revenue. The fact that at symmetric equilibrium bidders recover the same utility as in a second price auction with no reserves arguably makes it an even more natural class of bidding strategies to consider from the bidder standpoint. 
\subsection{Perturbation analysis for the Myerson auction}
These precedent results can be extended beyond the lazy second price, and to the Myerson auction (see, e.g.,  \citep{nedelec2019learning}. for more details). The Nash equilibrium has a relatively simple form given in the following Theorem \ref{thm:shadingInSymmMyerson}. 

\begin{theorem2}[\cite{tang2016manipulate,abeille2018explicit}]\label{thm:shadingInSymmMyerson}
%Consider an auction with $K$ independent and symmetric bidders, having value distribution represented by the random variable $X$. 
In the Myerson auction, the symmetric equilibrium strategy   $\shadingFunc_{eq}$  satisfies 
$$
\shadingFunc_{\text{eq}}(x)+\shadingFunc_{\text{eq}}'(x)(\vValue(x)-x)=\beta^{I}(x)\;,
$$
where $\beta^{I}(x)$ is the symmetric equilibrium strategy in a first price auction with no reserve price.  A solution of this equation is $$\shadingFunc_{\text{eq}}(x)=\mathds{E}_{X \sim F}[\beta^{I}(X)|X\geq x]\;.$$
At the equilibrium, the bidders' expected utilities are the same as  in a first price auction without reserve price; in particular, it is strictly greater than their expected payoffs had they bid truthfully. 
\end{theorem2}
\paragraph{\textbf{Discussion}} The intuition behind this result is quite clear. In the Myerson auction, the expected utility of a bidder is the same as in a first price auction where her bids have been transformed through his virtual value function. We call the corresponding pseudo-bids ``virtualized'' bids. Hence, if the bidders can bid in such a way that their virtualized bids are equal to their symmetric equilibrium first price bids, the situation is completely equivalent to a first price auction. And hence their equilibrium strategy in virtualized bid space should be the strategy they use in a standard first price auction with no reserve price. 
\iffalse
Our theorem shows that by adopting such a strategy symmetric bidders can avoid facing a non-zero reserve price. Furthermore, Lemmas \ref{definition_thresholded_strategies} and \ref{lemma:makingSureGIncreasing} show that it is easy for bidders to shade in such a way that their virtualized bids are equal to any increasing function of their value they choose. Also, this shading is specific to each bidder: the corresponding ordinary differential equations do not involve the other bidders. As such it is also quite easy to implement. 
\fi
\subsection{Approximations of the Myerson auction via numerical methods}
This new variational approach unlocks, through numerical optimization, a method to find best-responses to most of the approximated Myerson auction, such as boosted second price auctions \citep{nedelec2019learning}.  
 A straightforward optimization can fail because the objective is  discontinuous as a function of the bidding strategy.  To circumvent this issue,  a new relaxation of the problem which is stable to local perturbations of the objective function and computationally tractable and efficient has been introduced.  This new objective can be  numerically optimized through a simple neural network, with very significant improvements in bidder utility compared to truthful bidding. This simple approach can be plugged in any modern bidding algorithms learning distribution of the highest bid of the competition and we test it on other classes of mechanism without any known closed form optimal bidding strategies. 
 
The major and prohibitive drawback of these approaches is that they require that strategic bidders perfectly know the underlying mechanism design problem (i.e., the revenue maximization problem) solved by the seller, leading to a strong asymmetry between the bidders and the seller, this time in favor of the former.

It is nonetheless possible to remove the prior knowledge on  the exact algorithmic procedure used by the seller to optimize her mechanism by a classical exploration/exploitation trade-off, inspired by reinforcement learning techniques, thus reducing this asymmetry \citep{nedelec2019adversarial}.

\printbibliography[segment=4, heading=subbibintoc]
\end{refsegment}
\clearpage
\clearpage
%\begin{refsegment}
 %\input{chapters/chapter6_strategic_bidders}
%\printbibliography[segment=5, heading=subbibintoc]
%\end{refsegment}
%\begin{refsegment}
%\input{chapters/chapter4_open_problems}
%\printbibliography[segment=5, heading=subbibintoc]
%\end{refsegment}

%\input{chapters/chapter8_first_price_auctions}

%BACKMATTER SEE DOCUMENTATION
\backmatter  % references, restarts sample

\printbibliography[heading=bibintoc]

\end{document}